\documentclass[%trackchanges,%
 preprint]{aastex62}

\shorttitle{IRAS 16562$-$3959 Chemistry}
\shortauthors{Guzm\'an et al.}

\newcommand{\hii}{H{\rmfamily\scshape{ii}}}
\newcommand{\kms}{\mbox{km~s$^{-1}$}}

\newcommand{\Msun}{\mbox{$M_{\sun}$}}
\newcommand{\Lsun}{\mbox{$L_{\sun}$}}
\newcommand{\dagg}{{\dag\!\dag}}
\newcommand{\clump}{IRAS 16562$-$3959}
\newcommand{\hmyso}{G345.49+1.47}
\usepackage{amsmath,amssymb,stmaryrd}
\usepackage[caption=false]{subfig}
\usepackage{chngcntr}
\newcommand{\propyne}{\mbox{CH$_3$CCH}}
\newcommand{\cyc}{\mbox{c-C$_3$H$_2$}}
\newcommand{\hctn}{\mbox{HC$_3$N}}
\newcommand{\hctccn}{\mbox{HC$^{13}$CCN}}
\newcommand{\hccn}{\mbox{HC$_5$N}}
\newcommand{\met}{\mbox{CH$_3$OH}}
\newcommand{\tmet}{\mbox{$^{13}$CH$_3$OH}}
\newcommand{\acet}{\mbox{CH$_3$CHO}}
\newcommand{\htcn}{\mbox{H$^{13}$CN}}
\newcommand{\hntc}{\mbox{HN$^{13}$C}}
\newcommand{\hcqn}{\mbox{HC$^{15}$N}}
\newcommand{\htcop}{\mbox{H$^{13}$CO$^+$}}
\newcommand{\hcdop}{\mbox{HC$^{18}$O$^+$}}
\newcommand{\hcsp}{\mbox{HCS$^+$}}
\newcommand{\ethe}{\mbox{H$_2$CCO}}
\newcommand{\vinyl}{\mbox{CH$_2$CHCN}}
\newcommand{\acry}{\mbox{C$_2$H$_3$CN}}
\newcommand{\acrylonitrile}{\mbox{CH$_2$CHCN}}
\newcommand{\propanenitrile}{\mbox{CH$_3$CH$_2$CN}}
\newcommand{\prop}{\mbox{C$_2$H$_5$CN}}
\newcommand{\cyan}{\mbox{CH$_3$C$_3$N}}
\newcommand{\damm}{\mbox{NH$_2$D}}
\newcommand{\methoxy}{\mbox{CH$_3$OCH$_3$}}
\newcommand{\maser}{${\rm (E_2)}$ \mbox{$J_{K_aK_c}=5_{-1,0}\shortrightarrow4_{0,0}$}}
\newcommand{\metfor}{\mbox{CH$_3$OCHO}}
\newcommand{\form}{\mbox{NH$_2$CHO}}

\defcitealias{Guzman2014ApJ}{Paper~I}

\begin{document}
\title{CHEMISTRY OF THE  HIGH-MASS PROTOSTELLAR MOLECULAR CLUMP IRAS 16562$-$3959}
\author[0000-0003-0990-8990]{Andr\'es E.\ Guzm\'an}
\affiliation{Departamento de Astronom\'ia, Universidad de Chile, Camino El Observatorio 1515, Las Condes, Santiago, Chile}
\affiliation{National Astronomical Observatory of Japan, National Institutes of Natural Sciences, 2-21-1 Osawa, Mitaka, Tokyo 181-8588, Japan}

\author[0000-0003-4784-3040]{Viviana V. Guzm\'an}
%\affiliation{Harvard-Smithsonian Center for Astrophysics, 60 Garden Street, Cambridge, MA, USA}
\affiliation{Joint ALMA Observatory (JAO), Alonso de C\'ordova 3107, Vitacura, Santiago, Chile}

\author{Guido Garay}
\affiliation{Departamento de Astronom\'ia, Universidad de Chile, Camino El Observatorio 1515, Las Condes, Santiago, Chile}

\author{Leonardo Bronfman}\affiliation{Departamento de Astronom\'ia, Universidad de Chile, Camino El Observatorio 1515, Las Condes, Santiago, Chile}

\author{Federico Hechenleitner}\affiliation{Departamento de Astronom\'ia, Universidad de Chile, Camino El Observatorio 1515, Las Condes, Santiago, Chile}

\begin{abstract}
 We present  molecular line observations of  the high-mass  molecular clump 
 \clump\  taken  at 3  mm using the  Atacama Large
 Millimeter/submillimeter Array (ALMA) at $1\farcs7$ angular resolution
 ($0.014$ pc spatial resolution).
 This clump hosts the \added{actively accreting} high-mass young stellar
 object (HMYSO) G345.4938+01.4677, \deleted{which is actively accreting
   while} associated with a hypercompact \hii\ region.
We identify  and analyze emission lines from 22
molecular species (\added{encompassing} 34 \replaced{isotopologues and isotopomers}{isomers})
\added{ and classify them
  into two groups, depending on their spatial distribution within the clump. One of these groups gathers
  shock tracers (e.g., SiO, SO, HNCO) and species formed in dust grains like methanol (\met), ethenone or ketene (\ethe), and acetaldehyde (\acet). The second group collects  species
resembling more the dust continuum emission morphology and  are  formed mainly in the gas-phase, like hydrocarbons (CCH, \cyc, \propyne), cyanopolyynes (\hctn\ and \hccn) and cyanides (HCN and \cyan).}
\deleted{We are able to spatially resolve the line emission of different species within the  distinct environments expected for a molecular clump in this evolutionary stage.}%
Emission from
complex organic molecules (COMs) \added{like \met, propanenitrile (\propanenitrile), and methoxymethane (\methoxy) } arise from gas in the vicinity of
a hot molecular core ($T\gtrsim100$ K) associated with the HMYSO.  \deleted{Other COMs detected toward the
hot core  ($T\gtrsim100$ K) are  methanol (\met), propanenitrile (\propanenitrile), and
methoxymethane (\methoxy).}
 Other COMs such as propyne (\propyne),
 acrylonitrile (\acrylonitrile), and acetaldehyde seem to better
 trace warm ($T\lesssim80$ K) dense gas.
 \deleted{We find that \met,
 \acet,  and ethenone (\ethe) are likely associated with shock
activity, together with more traditional shock tracers like sulfuretted
molecules (SO, OCS), silicon monoxide (SiO), and isocyanic acid (HNCO).}%
In addition, \deleted{we detect} deuterated ammonia (NH$_2$D) \added{is detected}
mostly in the outskirts of \clump\ \added{and} associated with near-infrared dark globules, probably 
gaseous remnants of the clump's prestellar phase. 
\deleted{We detect seven \met\ lines including the 84 GHz class I maser and one vibrationally excited transition.}%
The spatial distribution of molecules in \clump\ \replaced{demonstrates}{supports the view} 
that in \deleted{any given} protostellar clump\added{s}, chemical tracers associated with
different evolutionary stages --- starless to hot cores/\hii\ regions ---
\replaced{coexist}{exist coevally}.
\deleted{Our observations make clear  that an adequate weighting of gas in various stages of chemical  evolution is necessary to interpret, e.g., single dish observations of protostellar high-mass clumps.}
\end{abstract}
\keywords{ISM: individual objects (IRAS 16562$-$3959) --- ISM: molecules --- stars: formation}

{\section{INTRODUCTION}\label{sec-intro}}

High-mass stars and their associated stellar clusters form inside
massive ($\ge500$ \Msun), dense ($n_{\text{H}_2}>10^4$ cm$^{-3}$), and
compact ($R\le0.3$ pc) molecular clumps \citep{Tan2014prpl}. Starting
from a prestellar, IR-dark phase in which no signs of star formation
are apparent, the massive clump eventually fragments and develops at
least one high-mass young stellar object (HMYSO).  High-mass clumps in
this protostellar phase are among the chemically richest regions in
the ISM.  Currently, almost 200 molecules have been detected in the
interstellar medium (ISM) or circumstellar
shells\footnote{\url{https://www.astro.uni-koeln.de/cdms/molecules}}
\citep[see also][]{Tielens2013RvMP}, with the majority of these
molecules being detected toward high-mass star forming regions such as
OMC/Orion-KL or Sgr B2/N-LMH. Most high-mass clumps are characterized
by the presence of complex organic molecules (COMs), which are
comparatively large ($\ge6$ atoms) molecules including carbon and
hydrogen in their composition \citep{Herbst2009ARA&A}. 

High-mass molecular clumps are far from being homogeneous structures,
either in their physical properties or in their chemical abundances.  The
dense inner molecular envelope surrounding the HMYSO (scales few times
$10^3$ AU) reaches temperatures $\ge100$ K, releasing to the gas phase all
the ice mantles from dust grains, and forming what is known as a hot
molecular core (HMC). Copious amounts of UV radiation may arise from the
young high-mass star, ionizing the surrounding gas and forming a small
ultra- or hyper-compact (HC) \hii\ region. The ionizing and dissociating
radiation have the potential of greatly affecting the chemistry of the
illuminated gas. In addition, the shocks induced by the energetic outflows
associated with the HMYSO introduces turbulence, carve outflow cavities
which facilitates radiation to affect farther regions in the clump, and
releases  elements and molecules into the gas phase through dust heating
and sputtering.  All these processes indicate us that the chemical
composition of high-mass clumps depends critically on their evolutionary
state, allowing us to estimate the latter by measuring molecular abundances
and ratios. A good understanding of the chemistry is key for this purpose.

Performing a systematic study of such a complex system entails discerning
the different physical and chemical environments which dominates the
emission of each molecule. For example, it is important to distinguish
whether the emission comes from regions close to the HMC, from outflows,
or from more quiescent and colder gas located farther
from the HMYSO. Single dish observations, associated with angular
resolutions $\gtrsim15\arcsec$, allow us to peer into the sub-clump and
``core'' scales ($0.01$--$0.1$ pc) only for nearby high-mass clumps, most
notably the Orion-KL/IRc2 region ($d\approx414$ pc) and Cepheus A ($d\approx725$ pc).
Without spatially resolved observations, in order to establish
if and what species trace the same parcel of gas\added{,} astronomers resort
to differences in the line kinematics \citep[e.g., ][]{Blake1987ApJ},
excitation temperature, and sometimes source sizes derived from
filling factor estimations \citep[][]{Gibb2000ApJ}.
While this approach has been very useful, the advantage of directly
observing the spatial differences between the emission of \replaced{different}{distinct}
species is patent. It allows us to distinguish
unambiguously the regions traced by the different molecular species and family of molecules, 
facilitating further insight into their formation and chemistry.

Sub-millimeter interferometers have typically provided molecular line
images with beam sizes between $1$ and $10$\arcsec\ and  noise levels \added{of} $10$--$100$ mJy beam$^{-1}$ per $1$
\kms\ channel width. Because of these sensitivity limitations, most of the
studies have focused on clumps hosting one or more HMC and they are usually
directed toward the brightest peaks of emission.  This implies that the
chemistry of the bulk of the clump gas is not generally probed.

High spatial resolution studies of high-mass star forming regions show
the expected chemical variations on molecular core scales ($<0.05$ pc).
Among the high resolution studies focused on HMC\added{-}containing
clumps --- other than Orion KL and Sgr B2/N-LMH --- are
\citet[on IRAS 18182$-$1433]{Beuther2006AA}, \citet[on
  G34.26+0.15]{Mookerjea2007ApJ}, \citet[on
  G19.61$-$0.23]{Qin2010ApJ}, \citet[on
  AFGL2591]{Jimenez-Serra2012ApJ}, and \citet[on G35.03+0.35 and
  G35.20$-$0.74N]{Allen2017AA}.  Other studies aim to the chemistry
induced by shocks produced by the very common outflows, for example,
\citep[on G34.43+0.24MM3]{Sakai2013ApJ} and \citet[on IRAS
  20126+4104]{Palau2017MNRAS}. There are also some high-resolution
studies which include several sources, generally concentrating on
determining evolutionary chemical differences between them
\citep[e.g.,][]{Beuther2009AJ,Wang2011AA,Immer2014AA,Oberg2014FaDi}.
Finally, a few high resolution chemical studies avoid HMCs by
targeting colder and apparently younger high-mass clumps
\citep{Sanhueza2013ApJ,Fayolle2015AA}.
However, by large, the best studied high-mass star formation region is
Orion-KL. Interferometer mm/sub-mm studies
\citep[e.g.,][]{Blake1996ApJ,Wright1996ApJ,Beuther2005ApJ,Widicus-Weaver2012ApJ,
Friedel2012ApJ,Feng2015AA,Gong2015AA}
have determined the presence of a large number of COMs with a rough
separation between N-bearing and O-bearing molecules, tracing the HMC
and the so called compact ridge, respectively.
%% Kinematically, the
%% compact ridge is believed to trace the place where outflowing gas
%% \citep[characterized by large linewidths,][]{Blake1987ApJ}
%% impinges the dense ambient material. The geometry and nature of the
%% Orion-KL region is also greatly affected by an ``explosive'' outflow
%% $\approx540$ yr ago \citep{Bally2017ApJ,Luhman2017ApJ}, which is
%% thought to have considerably affected the excitation, chemistry, and
%% kinematics of the whole region \citep{Zapata2011AA}.

In this work, we study the chemistry and kinematics at small scales
($0.014$ pc) associated with the molecular emission from \clump,
a massive dusty molecular clump located in the giant molecular cloud
G345.5+1.0 \citep{Lopez-Calderon2016AA}. This molecular clump is
located at $\sim1.7$ kpc, it has a mass of $\approx900\Msun$, 
and it harbors the HMYSO G345.4938+01.4677
(\hmyso\ hereafter) associated with a hypercompact (HC) \hii\ region
\citep[hereafter Paper I]{Guzman2014ApJ}.  This HMYSO is actively
accreting as evidenced by the presence of an ionized protostellar jet
\citep{Guzman2016ApJ}, a rotating molecular core/disk
\citep{Beltran2016AARv}, and molecular outflows \citep{Guzman2011ApJ}.
\citetalias{Guzman2014ApJ} presents results on the continuum,
sulfuretted species, and of the hydrogen recombination lines. Here, we
analyze the rest of the rich spectrum and emission maps associated
with other chemical species detected toward \clump.  Employing
high angular resolution and sensitivity observations we are able to
study molecular emission from the dense and warm protostellar cores 
and  of the   quiescent gas forming more extended structures.

The paper is structured as follows. Section \ref{sec-obs} briefly presents
the data and observations (see \citetalias{Guzman2014ApJ} for a detailed
description of the observations and data reduction). Section
\ref{sec-results} gives the observational results and describes some
relatively novel efforts on reducing and systematizing the presentation of
these large datasets. Section \ref{sec-ntex} use simple models --- mostly
based on LTE assumptions --- to extract physical parameters from the
molecular lines. One of the distinct aspects of the molecular emission
toward \clump\ is the large number of detected \replaced{isotopomer}{isomer} groups. We focus
on emission from these to determine isotopic fractionation and
isomerization in Section \ref{sec-iso}.  In Section \ref{sec-discussion},
we compare the ALMA data with near-infrared images (NIR) data and discusse
chemical implications of the observational results. Finally, Section
\ref{sec-summ} summarizes our main results and conclusions.

{\section{OBSERVATIONS}\label{sec-obs}}

The interferometer data used in this work were taken using the Atacama
Large Millimeter/submillimeter Array
\citep[ALMA,][]{Wooten2009IEEEP}. \citetalias{Guzman2014ApJ} describes in
more detail the observations and its calibration.  Briefly, the data were
taken during 2012 using the 12m array\footnote{\raggedright A complete
  description of the ALMA Cycle 0 capabilities is given in the Technical
  Handbook: \url{http://almascience.nrao.edu/documents-and-tools/cycle-0}.}
toward the center of \clump\ with a configuration including
baselines between 21 and 453 m. The lack of shorter baselines or single
dish complementary data prevents the adequate recovery of structures larger
than $\sim19\arcsec$. Therefore, in the rest of this work we avoid
analyzing structures larger than this limit.  In Appendix \ref{sec-ssf} we
investigate in more detail the effects of this lack of short baselines by
comparing the CS line data cubes with independent, single dish data. We
find that the simple approach of adding a constant offset per channel
in order to ensure positive intensities recovers more than $80\%$ of the
line single dish flux and $\sim60\%$ of the peak.  We conclude that the
negatives obtained in \replaced{these}{the interferometric} images are mainly caused by spatial filtering.

The data covers four spectral windows (SpW)
of 1.875 GHz wide each, centered at 85.4, 87.2, 97.6, and 99.3 GHz with a channel
width of 488 kHz ($\approx1.6~\kms$), which due to Hanning smoothing is
equivalent to an effective spectral resolution of about two times the
channel size.
The spectral setup choice was motivated by the main science goal of the observations, which was studying hydrogen recombination lines. The additional molecular line information was obtained gratuitously. For this reason, the dataset does not cover lines from some relevant chemical species in the band like $\text{CH}_3\text{CN}$ and $\text{H}_2\text{CO}$.
Typical synthesized beam is approximately
$2\farcs3\times1\farcs3,\text{ P.A.}=97\arcdeg$, with a  noise per channel of $0.8$--$1.0$
mJy beam$^{-1}$. Channels associated with strong emission lines (like masers) are usually noisier, with a dynamic range limit of $\sim1000$. Even
in these channels, noise levels do not exceed $7$ mJy beam$^{-1}$.

We re-imaged the continuum subtracted data using the task \texttt{tclean}
of the Common Astronomy Software Applications (CASA) using similar parameters as in
\citetalias{Guzman2014ApJ}. That is, we performed clean iterations with no masks until a
threshold of $2\sigma\approx2.5$ mJy was reached in the
residuals. In channels with line intensities higher than $0.5$ Jy
beam$^{-1}$ (e.g., SO, $J,N=3,2\shortrightarrow2,1$, CS, $J=2\shortrightarrow1$, and the
masering CH$_3$OH, \replaced{$5_{-1,0}\shortrightarrow4_{0,0}$}{\maser} transitions) side lobes and
possible aliasing from off beam emission decrease the quality of the
images. In these channels the cleaning iterations reached a fixed limit
number. We determine that a  limit of $5000$
iterations is  sufficient to reach stability of the cleaned flux.
%within a 15\% variations.
No further  improvement  in the quality of the images was detected by 
 performing a larger number of iterations. Fully reduced
spectral cubes and images are publicly available  \citep{data_Guzman2017}.

%%%%%%%%%%%%%%%%%%%%%%%%%%%%%%%%%%%%%%%%%%%%%%%%%%%%%%%%%%%%
%%%%%%%%%%%%%%%%%%%%%%%%%%%%%%%%%%%%%%%%%%%%%%%%%%%%%%%%%%%%
{\section{OBSERVATIONAL RESULTS}\label{sec-results}}
%%%%%%%%%%%%%%%%%%%%%%%%%%%%%%%%%%%%%%%%%%%%%%%%%%%%%%%%%%%%
%%%%%%%%%%%%%%%%%%%%%%%%%%%%%%%%%%%%%%%%%%%%%%%%%%%%%%%%%%%%

Analyzing the observational results obtained toward \clump\ entails 
identifying the observed spectral features. Section \ref{sec-lin} describes
\added{how we identified} to what species the line emission corresponds to.  In order to take
advantage of the spatial information, we calculate zero moment maps of a
representative line (or lines) of each molecule.  Section \ref{sec-morph}
describes the main morphological features of \clump\ which are displayed by several molecules. We group together
species whose emission have similar morphological characteristics
according to a quantitative criterion defined in Section \ref{sec-corr}. In
Section \ref{sec-m1} we briefly study the velocity spatial information
through first moment images. Finally, Section \ref{sec-desc} describes in more detail the main morphological features per molecular species.

{\subsection{Line Identification}\label{sec-lin}}

Within the observed bands we detect $\sim100$ spectral features, which we identify as emission lines associated with 22 molecular species corresponding to 34 \replaced{isotopomers}{different isomers} (besides the hydrogen recombination lines).  Table \ref{tab-lines} lists all these spectral features together with the observed frequency, molecular line, and the equivalent temperature of the upper state energy. Table \ref{tab-lines} gives the name of each species as commonly used in the astronomical literature and alternative names recommended by the \emph{International Union of Pure and Applied   Chemistry}. In general, we prefer to use the shortest chemical names
avoiding denominations with systematic numbering of carbon atoms.

\begin{figure}
\includegraphics[angle=90, ext=.pdf, width=\textwidth]{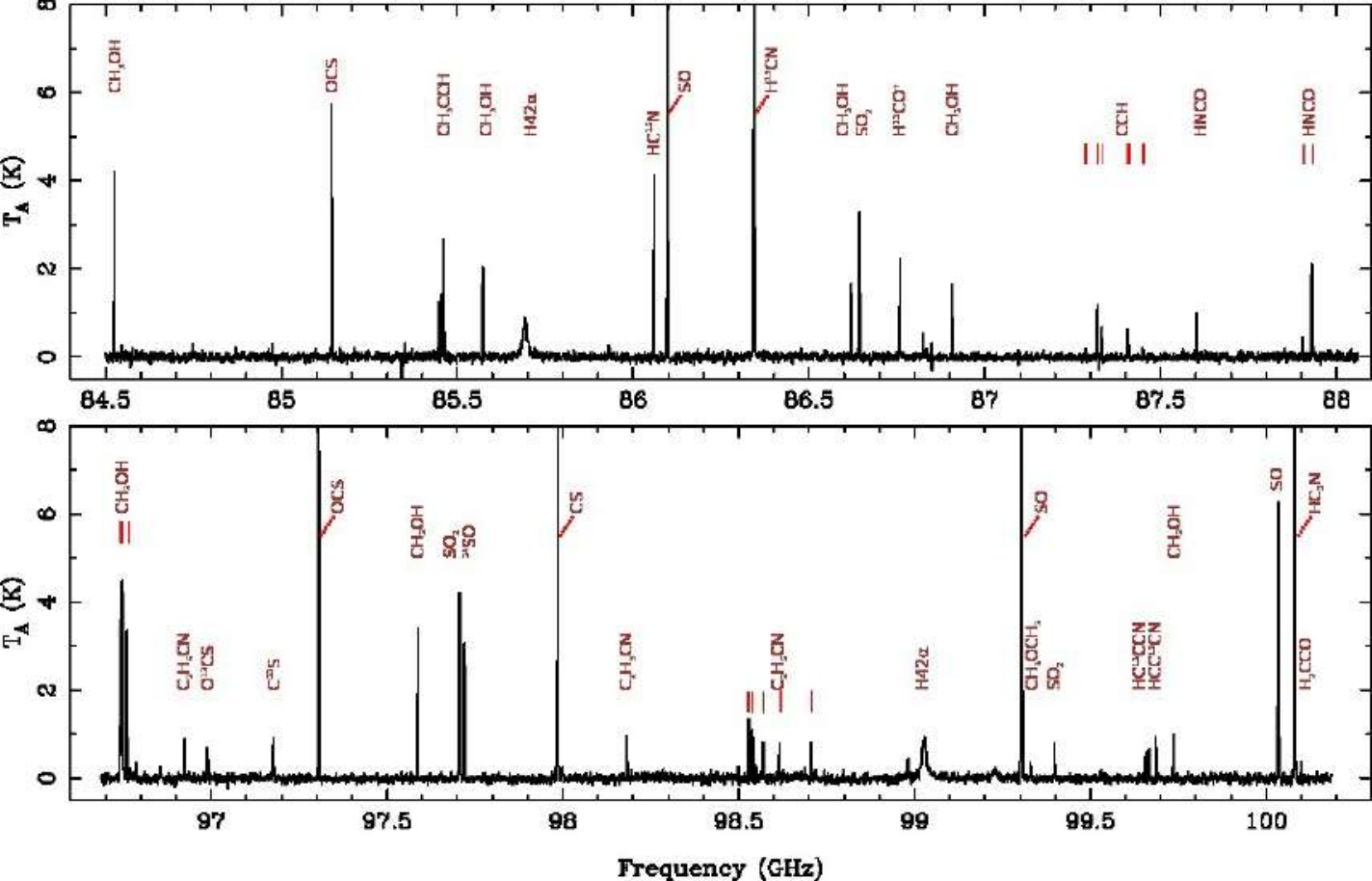}
\caption{Continuum subtracted spectrum taken toward G345.49+1.47. Species associated with the most conspicuous lines are indicated. \label{fig-spec}}
\end{figure}

Figure \ref{fig-spec} shows a one-pixel spectrum observed toward \hmyso\ (the richest position in molecular lines), where we have merged in two panels the lower- and upper-sideband spectral windows.  While this spectrum shows many lines, the moderate line density  allows  an adequate determination and subtraction of the continuum level. 
Despite most of the lines appearing toward the HC \hii\ region G345.49+1.47, there is no a single position in the field which displays emission in all of the species listed in Table \ref{tab-lines}.

To identify different molecular species we used CASSIS together with the Jet Propulsion Laboratory \citep[JPL,][]{Pickett1998JQSRT} and Cologne Database for Molecular Spectroscopy \citep[CDMS,][]{Muller2005JMoSt} spectroscopic databases. For each conspicuous spectral feature, we determine which species in the subset of detected ISM species given by CASSIS have  $V_{\rm LSR}$ between $-6$ and $-17$ \kms. This $V_{\rm LSR}$ interval is within $5.5$ \kms\ from  the ambient $V_{\rm LSR}\approx-11.5$\added{ \kms}. We revise other lines with comparable or \replaced{smaller}{lower} energy upper energy levels and comparable or higher Einstein spontaneous \replaced{radiation}{emission} coefficients in order to discern which is the most likely molecule responsible of the examined lines.  
Once a candidate is assigned to a spectral feature, we compared the data with local thermodynamic equilibrium (LTE) models in order to confirm or reject the identification.
Because the spectrum is not too populated,
there is often only a single  candidate molecule, and mostly two. 
Probable candidate species are also determined by previous detection
toward other star formation regions \citep[e.g.,][]{Blake1987ApJ,Gibb2000ApJ}.
To identify the line we follow the criteria defined in \citet[][\S3.3]{Herbst2009ARA&A}.  For the COMs, we were able to identify at least three lines of the main isotopologue per species.  An anti-coincidence --- that is, a line that should be detected according to the predictions of the LTE model but is not observed --- is taken as a strict rejection criterion.

In addition to the lines listed in Table \ref{tab-lines}, tentative detections of
CH$_3$OCHO (methyl formate, sometimes also written as HCOOCH$_3$) 
and \form\ (formamide) are presented.
% formamide & $    100\pm     0$ & $  13.99\pm  0.04$ & $ -15.29\pm   0.3$ & $  5.299\pm   0.6$ &
% 
We do not include these in Table \ref{tab-lines} because the low
signal-to-noise ratio of these lines ($\lesssim3$ at peak) impedes us from
claiming detection of either species.  In later sections we derive
quantitative upper limits of these two and other undetected molecules.

%%%%%%%%%%%%%%%%%%%%%%%%%%%%%%%%%%%%%%%%
{\subsection{Main Morphological Features of \clump}\label{sec-morph}}

In this section we analyze the morphology of the emission from  different species detected toward \clump\ using  the zero moment maps of the most prominent transitions of each species. To generate the zero moment maps  we use the moment masking algorithm described in \citet{Dame2011arXiv} and the CASA task \texttt{immoments}.
%Figures \ref{fig-cen1} to \ref{fig-oth} show maps of representative transitions of each species in Table \ref{tab-lines}.
\replaced{Fits}{FITS} files of  the zero moment maps are publicly available in \citet{data_Guzman2017}.

\begin{figure}
  \includegraphics[angle=-0, ext=.pdf, width=\textwidth]{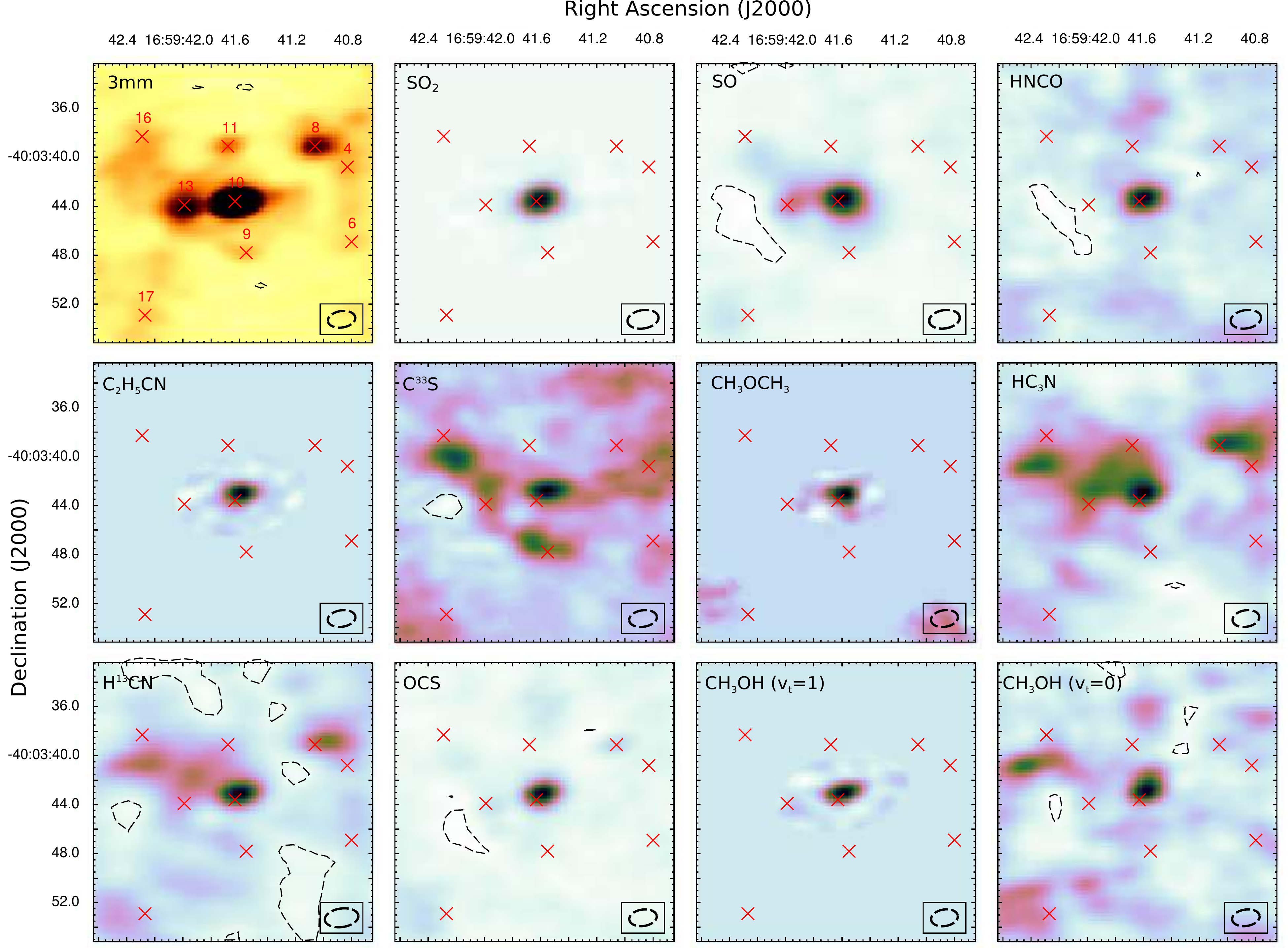}
\caption{Zero moment maps of the emission from  representative transitions from  the central
  23\arcsec$\times$23\arcsec region of \clump.  All these transitions
  show a strong, dominating source consistent with the position of
  \hmyso\ (Source 10).  The upper left panel shows the continuum measured at 3 mm
  in a different color scale, together with the sources identified by
  \citetalias{Guzman2014ApJ}. The name of several continuum
  sources are indicated in the first panel, and their positions are marked in the  rest. The
  dashed contour level in the panels is $-0.03$ Jy beam$^{-1}$ \kms,
  corresponding to approximately $-5\sigma$. The dashed contours in the
  upper left panel correspond to those presented in
  \citetalias{Guzman2014ApJ}.\label{fig-cen1}}
\end{figure}

\begin{figure}
\includegraphics[angle=-0, ext=.pdf, width=\textwidth]{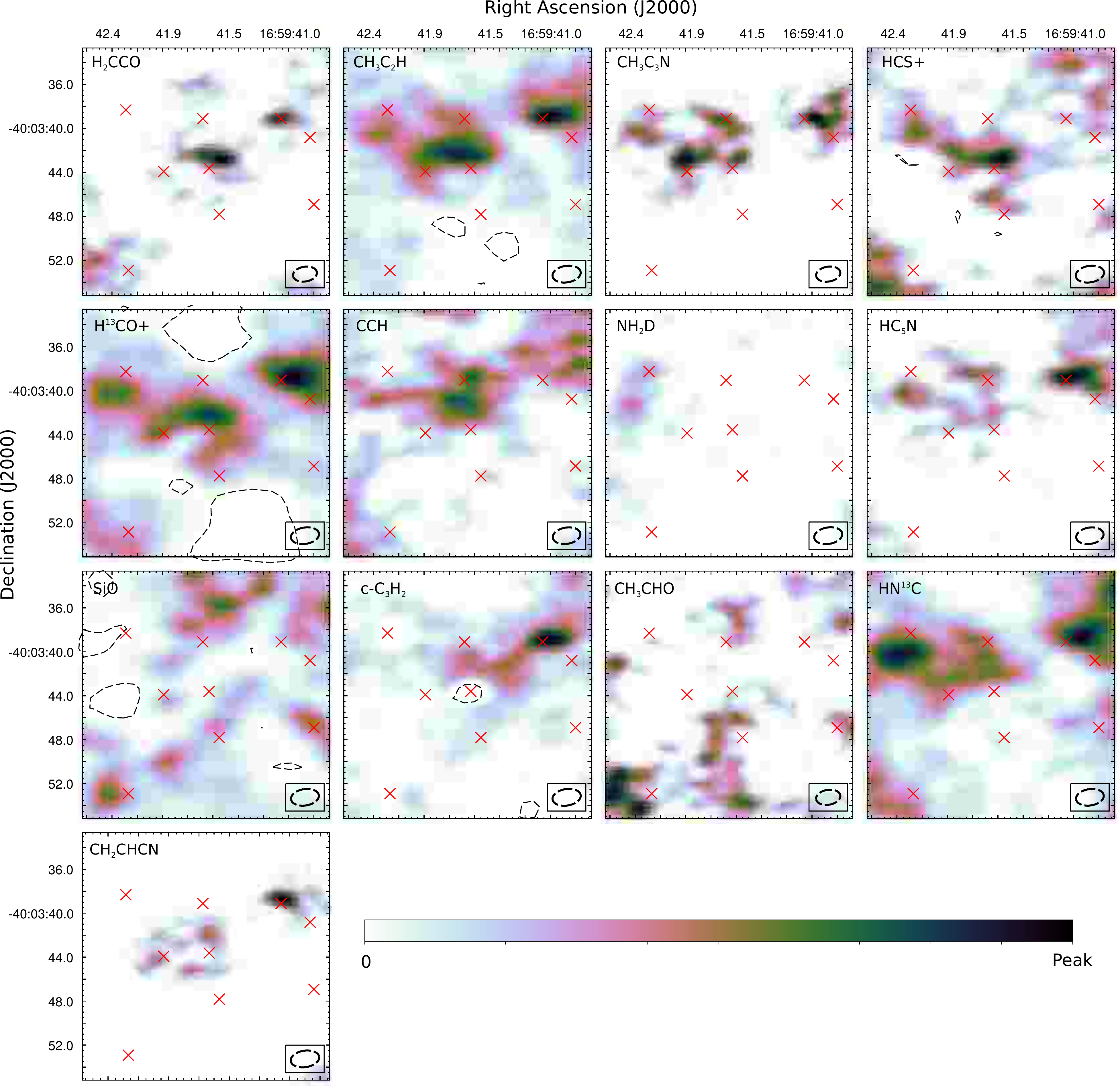}
\caption{Same as Figure \ref{fig-cen1}. Starting from the top left panel to
  the right and from top to bottom: the central source component becomes
  less and less prominent. Some transitions do not display a central
  dominating component at all, like the ones in the \added{two} bottom
  \replaced{row}{rows}. \label{fig-cen2}}
\end{figure}

We note here  that the most prominent feature of \clump\ --- in continuum and in molecular line emission --- is   the central emission associated with the HC \hii\ region and HMYSO \hmyso.
Figures \ref{fig-cen1} and \ref{fig-cen2} show zero moment images toward the inner $23\arcsec\times23\arcsec$ of \clump, centered in \hmyso, of representative transitions of all 22 molecular species detected (see Table \ref{tab-lines}).
Figure \ref{fig-cen1} and \ref{fig-cen2} present the maps organized by visual inspection from those which show a clear strong source associated with G345.49+1.47 (like the sulfuretted molecule transitions) to those  in which there is less obvious emission from the central HMYSO (e.g., SiO).

\begin{figure}
\includegraphics[angle=-0, ext=.pdf, width=\textwidth]{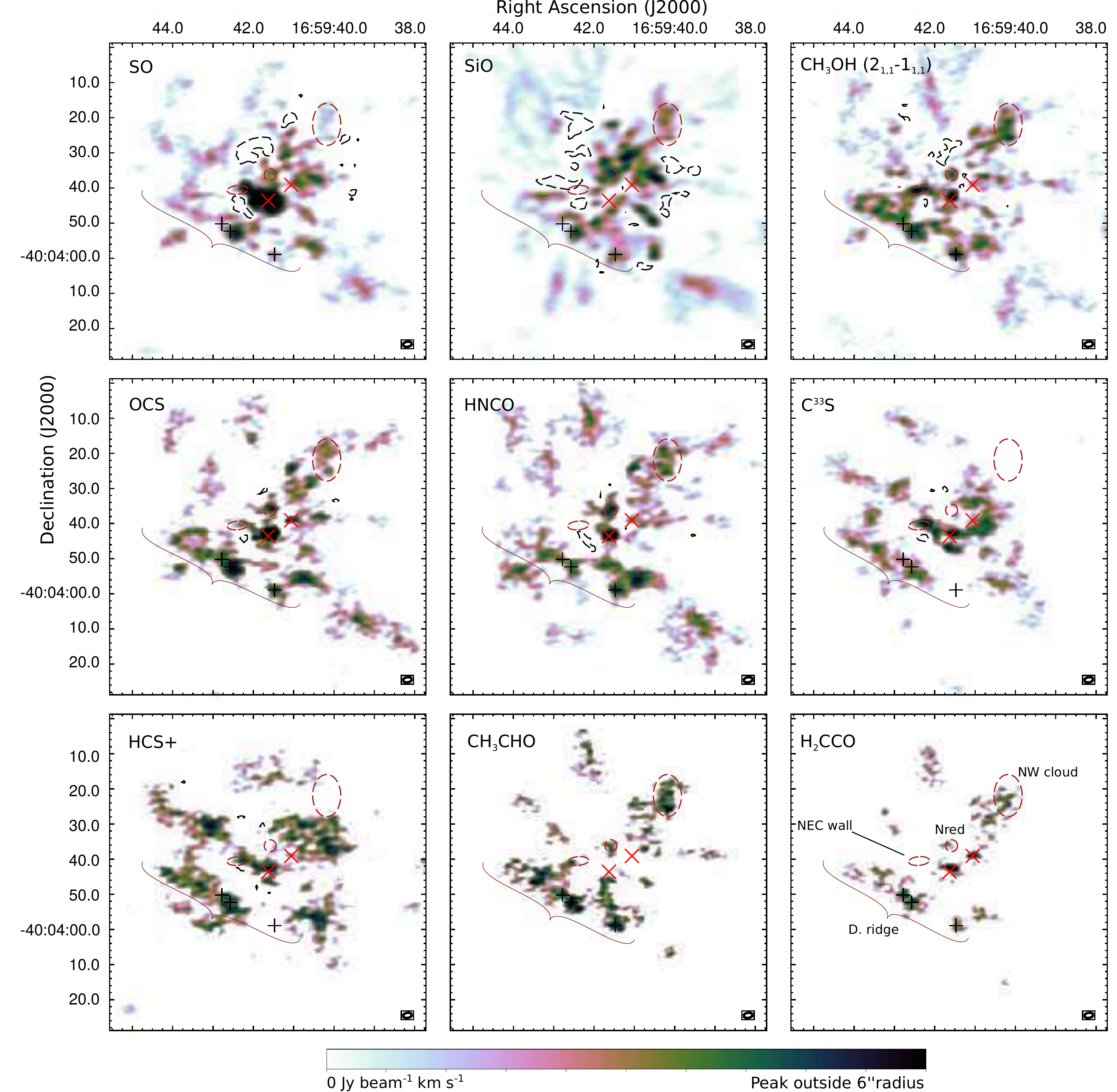}
\caption{Zero moment maps of representative transitions of molecular
  species related with the Shock group (see Section \ref{sec-corr}). The
  dashed contour level in the panels is $-0.03$ Jy beam$^{-1}$ \kms,
  corresponding to approximately $-5\sigma$. The two red crosses mark the position of
  Sources 8 and 10. To ease comparison, we show the NW-cloud, the N-red cloud, the NEC-wall, and the position of the Diffuse Ridge  together with the DR(a), (b), and (c) positions (black crosses) defined in Section \ref{sec-morph} and Figure \ref{fig-consp}.
  \label{fig-shock}}
\end{figure}

\begin{figure}
\centering\includegraphics[angle=-0, ext=.pdf, height=0.94\textheight]{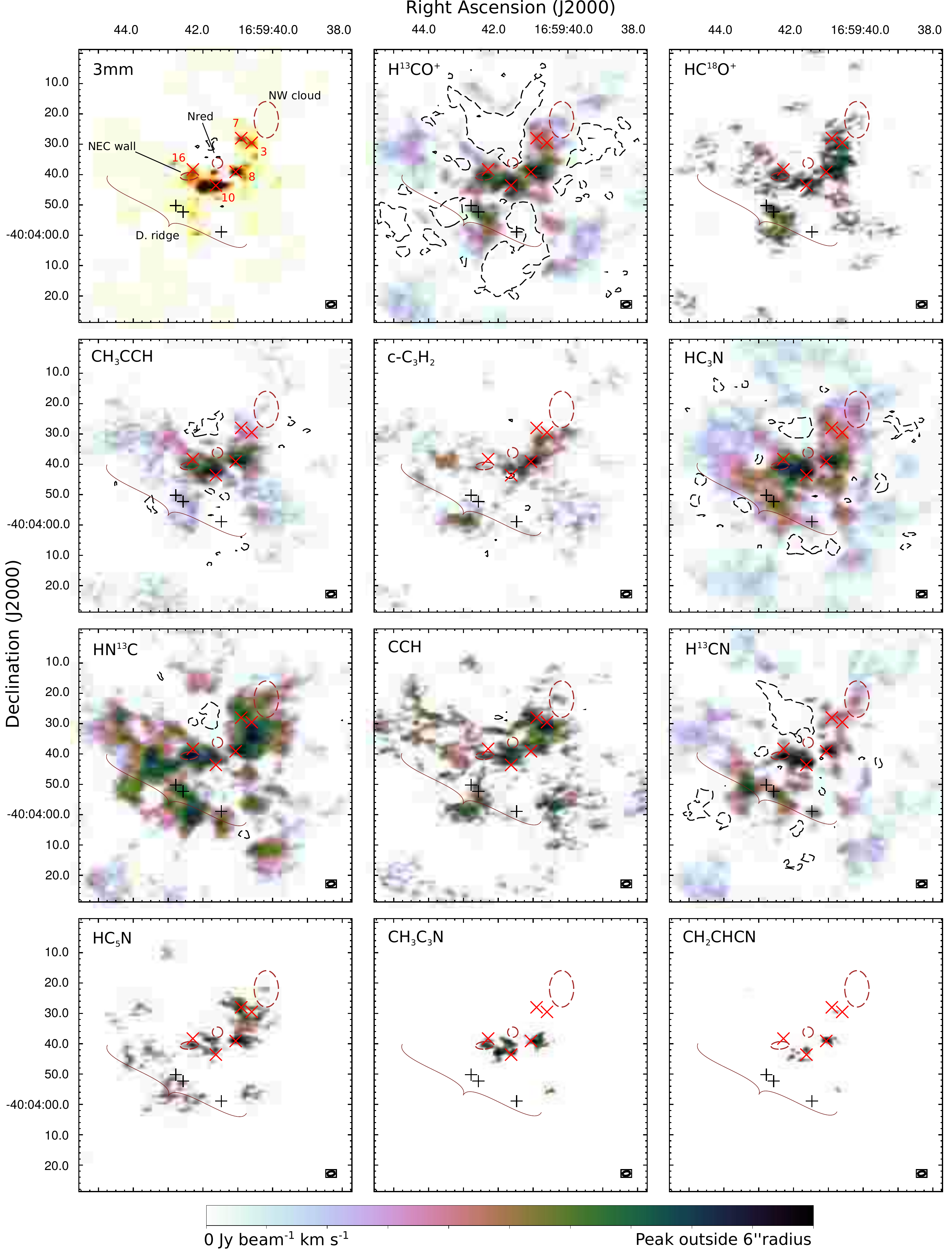}
\caption{Same as Figure \ref{fig-shock}, but with species in the Continuum
  group (see Section \ref{sec-corr}). The top left panel shows the 3 mm
  continuum. Red crosses indicate the continuum sources 3, 7, 8, 10, and 16.\label{fig-cont}}
\end{figure}

\begin{figure}
\centering\includegraphics[angle=-0, ext=.pdf, width=.7\textwidth]{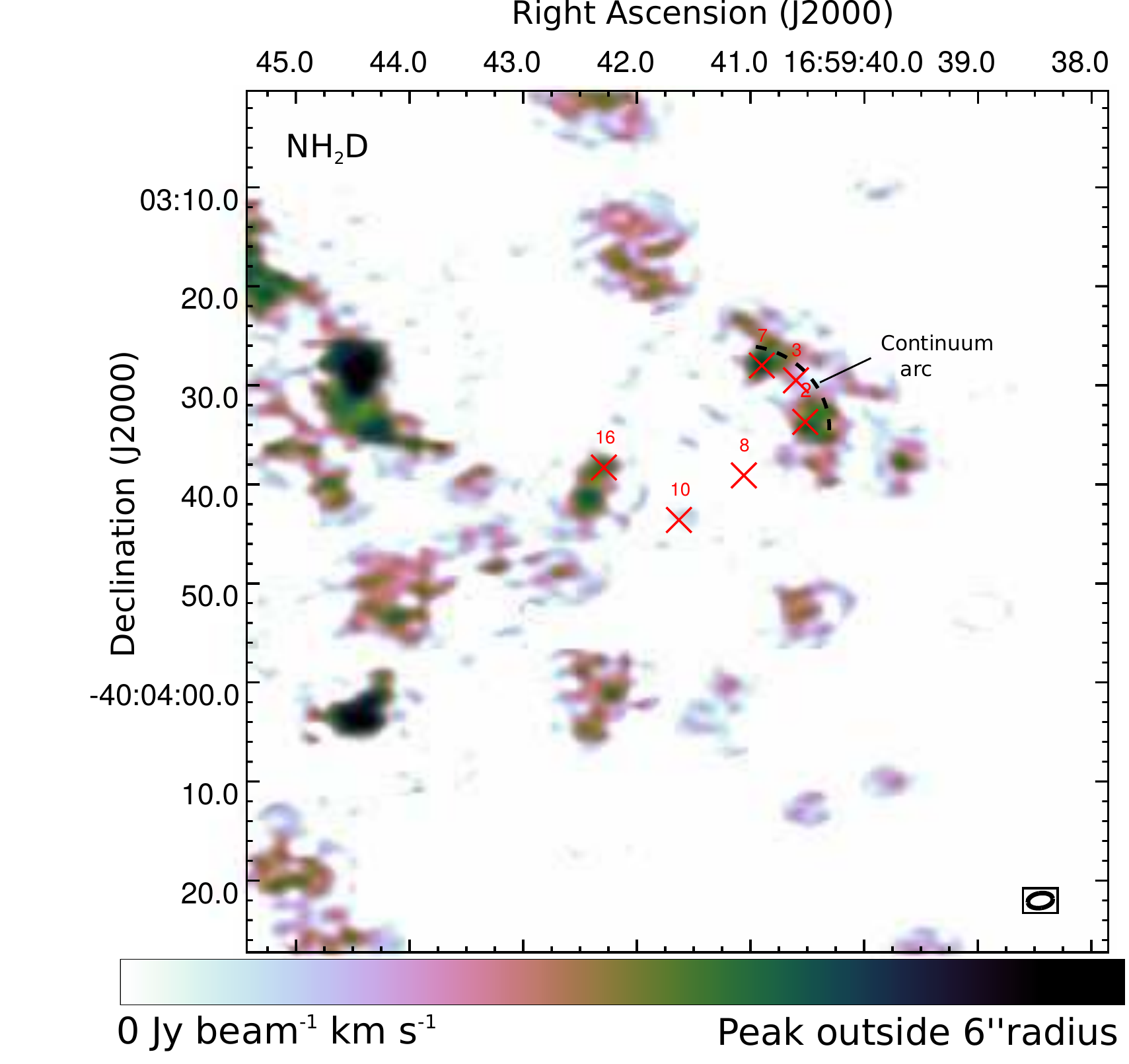}
\caption{Zero moment map of the NH$_2$D $(1,1)$ transition. This species shows no evident
  similarities with either the Shock or Continuum group.
  We indicate the structure referred to as the ``Continuum Arc'' in Section \ref{sec-desc}.\label{fig-oth}}
\end{figure}

Figures \ref{fig-shock} and \ref{fig-cont} show zero moment maps  centered in \hmyso\ but on a larger scale ($\sim1\farcm5\times1\farcm5$). These two figures show, respectively, the maps of two groups of species which display similar morphologies, as estimated from  their 2D cross-correlations. The grouping and correlation calculations are described in more detail in Section \ref{sec-corr}.  Figure \ref{fig-oth} shows the zero moment map of the \replaced{\damm\ emission}{NH$_2$D $(1,1)$ transition}, which is not part of neither of the two previous groups and displays a unique morphology. Finally, the CS, $J=2\shortrightarrow1$ zero moment map (shown in Figure \ref{fig-cs})  is not used in the analysis of this section because the emission is  doubtless optically thick  and  heavily affected by short baseline filtering.

Besides CS, maps displaying strong negative features are those of \htcop, \htcn\ (Figure \ref{fig-cont}), and SiO (Figure \ref{fig-shock}) and they should be interpreted with caution. In this work, we avoid extracting integrated fluxes from regions larger than $10\arcsec$. In any case, note that when measuring the flux arising from a compact source --- like a molecular core --- using the filtered map may be more adequate than using the map without short baseline filtering. This is because in the second case the intensities so measured include emission from material more homogeneously distributed within the clump, which we would not consider it to be part of the core.

\begin{figure}
\includegraphics[angle=-0, ext=.pdf, width=\textwidth]{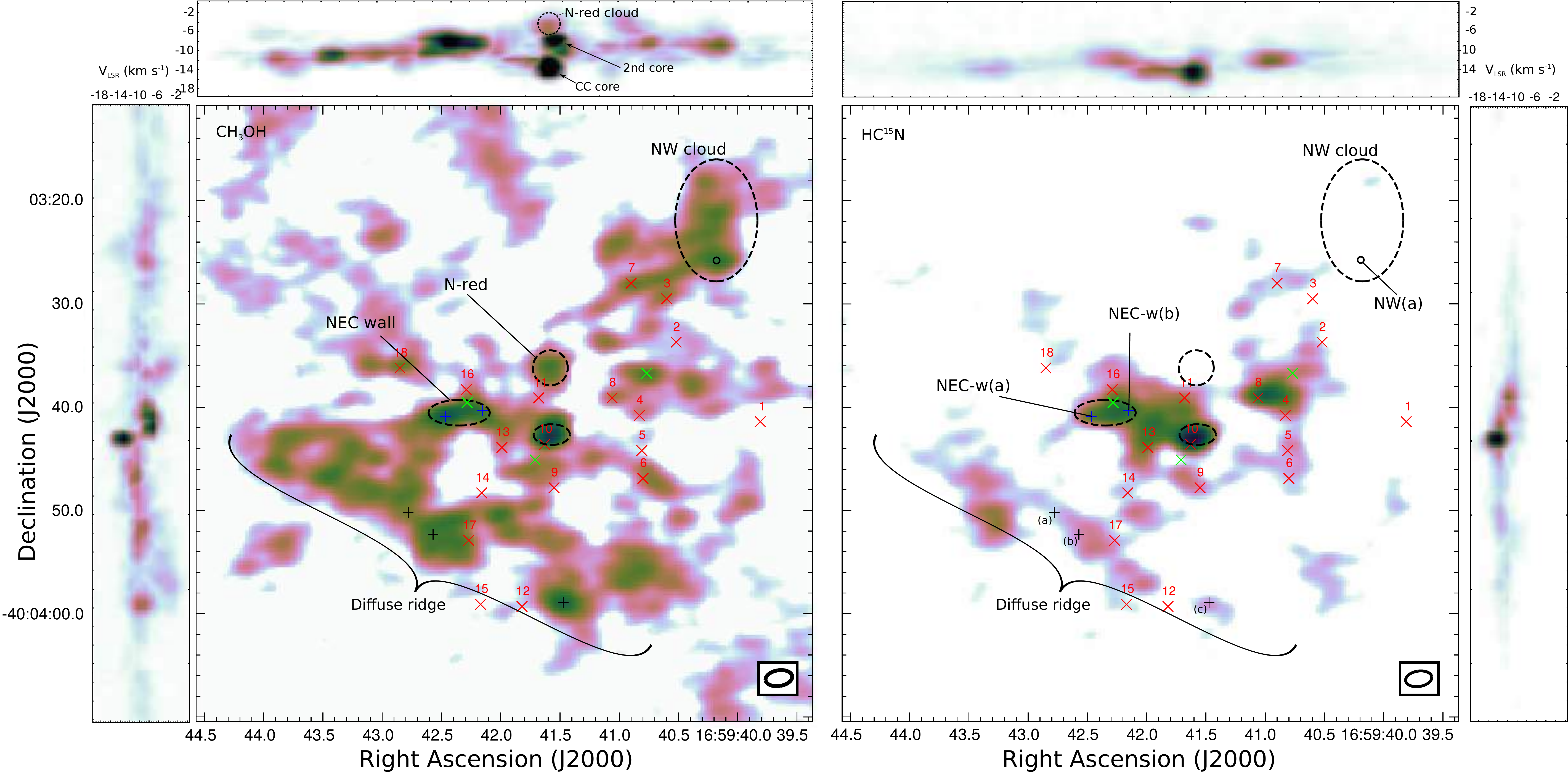}
\caption{Panels a) and b) show zero moment maps of the \met,
  $({\rm E_1})$ $2_{1,1}\shortrightarrow1_{1,0}$ ($E_\textrm{up}=21.6$ K) and the
  HC$^{15}$N, $1\shortrightarrow0$ ($E_\textrm{up}=4.1$ K) transitions,
  respectively. Red circles mark the continuum sources identified at 3
  mm in \citetalias{Guzman2014ApJ}. Green crosses show the position of
  the \met, $({\rm E_2})$ $5_{-1,0}\shortrightarrow4_{0,0}$ maser spots. Black crosses
  mark the DR(a), (b), and (c) positions in the Diffuse Ridge,
  the NEC-wall (a) and (b) positions, and the NW-cloud (a) position
  (Section \ref{sec-cdt}).  Black-dashed ellipses mark the CC core,
  the  N-red cloud, and the NW cloud.  To the left and on top of panel
  a) we show R.A.-velocity and Declination-velocity pv-diagrams,
  respectively. In the top R.A.-velocity pv-diagram we mark the N-red
  cloud with a dashed circle, and two arrows indicate the sources
  which compose the blended CC core emission in the \met\ zero moment
  map. PV-diagrams of the HC$^{15}$N emission are shown to the right
  and on top of panel b). \label{fig-consp}}
\end{figure}

Figure \ref{fig-consp} shows maps of \hcqn\ and \met\ and displays morphological features which are common to several molecular species.
As explained in more detail in Section \ref{sec-corr}, these two molecules are representative of the two distinct groups we define to separate the species according to their morphology. The maps show  emission from  relatively low excitation lines, $E_\textrm{up}=21.6$ K for \met\ and $4.1$ K for HC$^{15}$N.  We also show position velocity diagrams (pv-diagrams) of the data cubes in the lateral and upper panels. The pv-diagrams display, for each position, the maximum intensity measured cutting through the other dimension of the cube. We find that this way of displaying was more effective in separating and identifying structures than the integrated intensity of the cubes collapsed in R.A.\ or declination.
The main morphological features we identify are:
\begin{itemize}
\item{\textit{A central compact core (CC core).} The CC core is the most
  conspicuous feature in many of the transitions detected toward
  \clump. The zero moment emission peaks near the central HC
  \hii\ region/HMYSO \hmyso, marked as Source 10
  ($\text{R.A.}=16^\text{h}59^\text{m}41\fs63$,
  $\text{decl.}=-40\arcdeg03\arcmin43\farcs6$) in Figure \ref{fig-cen1},
  but displaced from it --- as noted in \citetalias{Guzman2014ApJ} ---
  between $0\farcs4$ to $1\farcs4$ in the northwest direction, depending on
  the specific tracer.  \added{Consistently,} \replaced{Continuum}{continuum} at 218 GHz and CH$_3$CN,
  $J=12\shortrightarrow11$ (methyl cyanide) observations of \clump\ presented
  by \citet{Cesaroni2017AA} confirm \deleted{the sources identified in Paper I, but find} that Source 10 actually consists of
  two peaks: one associated with the HMYSO/HC \hii\ region (\hmyso) and
  another associated with a small ($\sim0\farcs5$), continuum source
  located $0\farcs8$ northwest of \hmyso.  Figure \ref{fig-cen1} shows zero
  moment maps of molecules for which emission from the CC core is clearly
  distinguished.  Some care is needed for the interpretation of the
  position based solely on these maps because, as shown in the pv-diagrams
  of Figure \ref{fig-consp}, the zero moment emission in the direction of
  the CC core is composed of two distinct cores with different radial
  velocities.  We identify one of these cores as the CC core because it is
  centered at $V_{\rm LSR}=-15.4$ \kms, which is closer to the central
  velocity of the hydrogen recombination lines arising from the HC
  \hii\ region \citepalias{Guzman2014ApJ} associated with \hmyso.  The
  \met\ pv-diagrams show that the second core (marked as `2nd core' in
  Figure \ref{fig-consp}) is located $\approx1.3\arcsec$ to the northwest
  (2\arcsec\ from Source 10 in the $\textrm{P.A.}=-60\arcdeg$) and centered
  at $V_{\rm LSR}=-9.4$ \kms, that is, redshifted respect to \hmyso.
  Figure \ref{fig-consp} shows these two cores in the R.A.-velocity diagram
  of \met. Some emission from the secondary redshifted core is detected in
  HC$^{15}$N, but much fainter than the proper CC core centered at
  $\approx-15.4$ \kms. This secondary redshifted core is likewise evident
  in emission from other species such as \htcop, HC$_3$N, and HN$^{13}$C.}
\item{\textit{Source 8 core (C8).} Many molecules show strong and extended emission near
  the source marked with number 8 in Figure \ref{fig-cen1}. We note,
  however, that the molecular peak is not coincident with the
  continuum peak.  Usually, the molecular emission is strongest between
  $0\farcs5$ and $1\farcs5$ in the $\textrm{P.A.}=-15\arcdeg$ direction
  from Source 8. This is the case of COMs like \hccn, \cyan,
  \propyne, and \acrylonitrile, where the emission usually spreads toward the
  NE direction forming an elongated cloud.  A methanol maser seems to be associated with the outer rim of this molecular envelope.}
\item{\textit{A Diffuse Ridge.}  It is shown in Figure \ref{fig-consp} as
  the elongated methanol emission feature crossing the south east part of
  \clump.  This structure appears most prominently in \met, CH$_3$CHO, and
  in most other tracers shown in Figure \ref{fig-shock}.  The Diffuse Ridge
  extends for $\approx40\arcsec$ in the $\textrm{P.A.}=65\arcdeg$ direction
  from the position
  $\Delta\textrm{R.A.}=-10\arcsec,\quad\Delta\textrm{Dec.}=-15\arcsec$
  respect to \hmyso. Within the Diffuse Ridge there are two distinct
  emission cores. One is marked with a '(b)' in Figure \ref{fig-consp} and
  is centered at $\Delta \text{R.A.}=+10\arcsec,
  \Delta\text{dec.}=-9\arcsec$ from Source 10, with a diameter of
  $\approx5\arcsec$.  The other, marked with a '(c)', is located at
  $\Delta  \text{R.A.}=-2\arcsec, \Delta\text{dec.}=-15\farcs5$ from Source 10,
  slightly elongated in the direction $\text{P.A.}=75\arcdeg$ with
  diameters of $\approx5\arcsec\times3\arcsec$. 
  Position '(a)', on the other hand, targets
  more diffuse  gas forming the body of this ridge or filament. Hereafter,
  we refer to these three positions as DR(a), DR(b), and DR(c).
  DR(b) is conspicuous in \met\ and  likewise in CH$_3$CHO, SiO, SO, \hcsp, OCS,
  CS, HC$_3$N, and HNCO. It is less noticeable in HN$^{13}$C, H$^{13}$CN,
  and C$^{33}$S, but still present. DR(c) is as well
  distinguished in CH$_3$CHO, SiO, SO, OCS, CS, HNCO, and
  H$^{13}$CN. Emission from DR(c) is also
  detected in \htcop\ and HC$_3$N. In contrast with DR(b), we do
  not detect HN$^{13}$C nor \hcsp\ emission associated with DR(c).}
\item{\textit{A northeast outflow cavity wall (NEC-wall).} This
  molecular feature located $\approx2\arcsec$ south of Source 16
  matches an illuminated section of the outflow cavity wall seen in
  NIR. The emission is conspicuous in \met, CH$_3$C$_3$N, CCH,
  HC$_3$N, HC$_5$N, H$^{13}$CN, HN$^{13}$C and \htcop\
  transitions.  Other molecules display emission farther from the
  cavity wall position, but still associated with it, like
  c-C$_3$H$_2$, C$^{33}$S, CH$_3$CHO, OCS, and maybe SO.  The large
  scale emission from SiO and HNCO which surrounds the NEC-wall area
  is  not clearly associated with it.}
\item{\textit{A northwest cloud (NW cloud).} This emission cloud has approximately $5\arcsec\times3\arcsec$ of size and it is centered around the position
  located approximately 25\arcsec\ in the $\textrm{P.A.}=-50\arcdeg$
  direction respect to the CC core. This feature is displayed by most of the
  molecules shown in Figure \ref{fig-shock}, including \met, SiO, CH$_3$CHO, HNCO, and
  CS.  There is also strong emission associated with HC$_3$N. Less
  prominent emission  arises  from \htcn, \hntc,
  \htcop, CCH, H$_2$CCO, \hccn, OCS, and SO as well.}
\item{\textit{A northern redshifted  cloud (N-red cloud).} This emission
  feature of $\approx1.5\arcsec$ radius is  specially conspicuous
  in transitions shown in Figure \ref{fig-shock}. It is one of the
  emission cores with the largest differences in radial velocity
  respect to the clump. In methanol, its radial velocity is centered
  at $-7.5$ \kms, that is, redshifted by $4.1$ \kms\ respect to the $V_\textrm{LSR}$
  of \clump. Molecules which display strong emission toward the N-red cloud
  are (besides \met) SiO, HNCO, OCS, CS, and SO. Less prominent
  emission is detected from \hctn, CH$_3$CHO, and H$_2$CCO. In each
  case, the emission is consistently redshifted respect to the
  systemic radial velocity of the clump. The N-red cloud is located $\sim7\arcsec$ north of the CC core and is not to be confused with the `2nd redshifted core' we mentioned previously.}
\end{itemize}

%\clearpage
{\subsection{Zero Moment Cross Correlations and Grouping}\label{sec-corr}}

In order to evaluate quantitatively how similar are the zero moment maps of
different species, we calculate the cross correlation
between each pair of maps according to
\begin{equation}
\rho_{12}=\frac{\sum_{i,j}I_{1,ij}I_{2,ij}w_{ij}}{\left(\sum_{i,j}I^2_{1,ij}w_{ij}\sum_{i,j}I^2_{2,ij}w_{ij}\right)^{1/2}}~~,\label{eq-corr}
\end{equation}
where the sums are taken over each pixel position, $I_{1,ij}$ and
$I_{2,ij}$ are the values measured for each image at pixel $i,j$, and
$w_{ij}=0,1$ defines the  masked region used for the calculation. Note that by definition
$\rho_{12}=\rho_{21}$ and $|\rho_{12}|\le1$. A value of $\rho_{12}=1$
implies that $I_1=\alpha I_{2}$, with $\alpha$ a positive constant.
Therefore, the more similar the zero moment spatial distributions are,
the closer their cross correlation is  to 1.

According to the definition of Equation \eqref{eq-corr}, brighter sections
of the image weight more into the calculation of $\rho_{12}$. In order to
avoid the correlation being dominated entirely by the central source, we mask the
pixels (that is, we set $w_{ij}=0$) inside an inner circle of
6\arcsec\ radius centered at G345.49+1.47. We leave outside this specific
analysis some molecules which are only detected toward the center of the
field, e.g.,  CH$_3$CH$_2$CN, $^{13}$CH$_3$OH, and SO$_2$.

\begin{figure}
\centering\includegraphics[angle=90, ext=.pdf, height=.8\textheight]{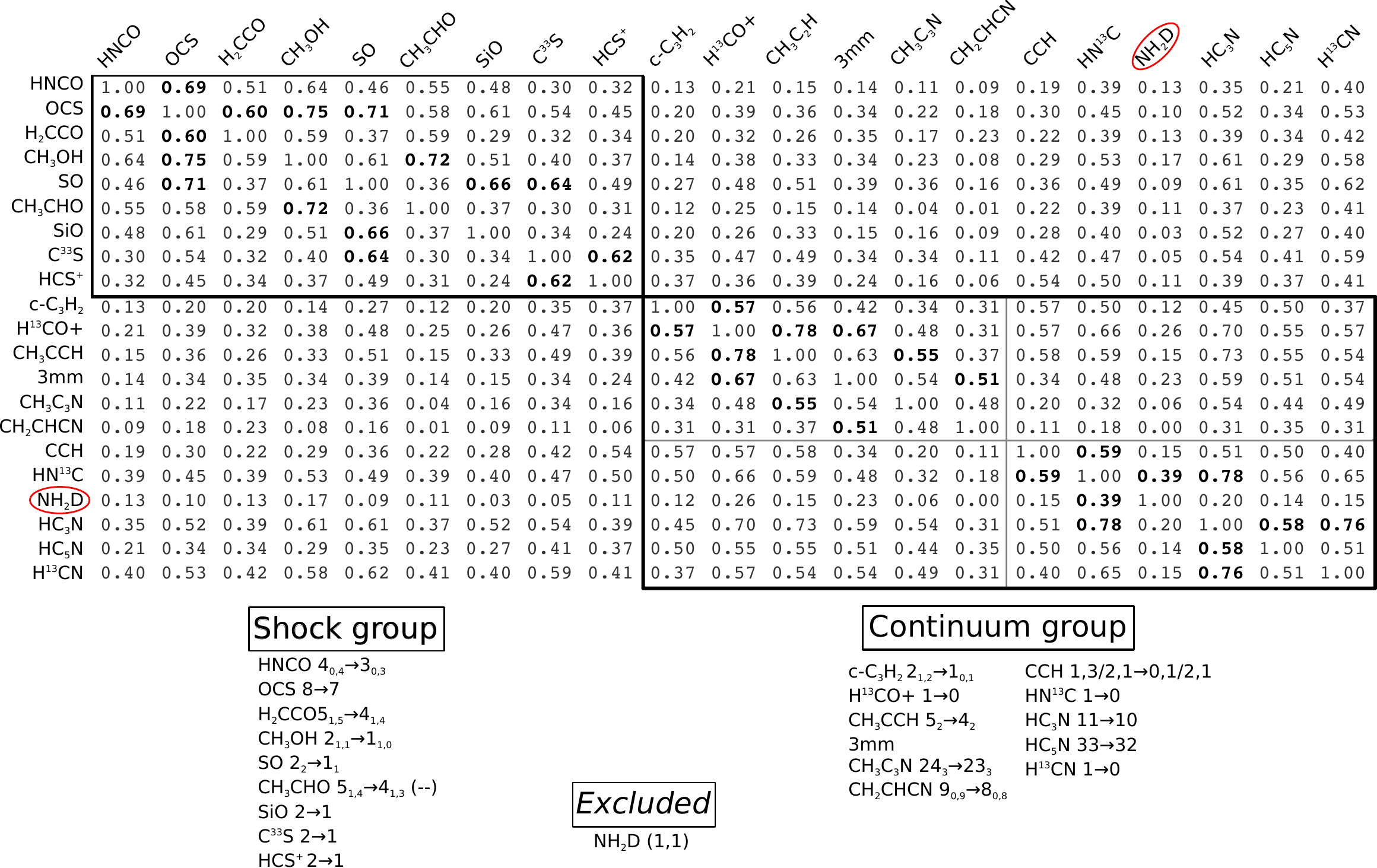}
\caption{Cross correlations between zero moment maps of molecules with
  significant extended emission, and with the continuum (marked with 3 mm).
  For each molecule, the maximum correlation is marked in boldface. Each
  molecule in any of the two enclosed frames has their maximum
  correlation partner in the same group they pertain. The top left and
  bottom right framed groups correspond to the so called Shock and
  Continuum groups, respectively. Below the cross correlation matrices
  we list explicitly the  members  and the
  specific transition used. For completeness we include \damm\ despite this molecule not being part of either group. 
\label{fig-corrs}}
\end{figure}

%method
Figure \ref{fig-corrs} displays the value of the cross-correlations between the zero moment maps of different molecules.   We include in this analysis the continuum image at 3 mm presented in \citetalias{Guzman2014ApJ}. In order to gather together molecular emission based on their spatial distributions, we will use the $\rho_{12}$ value as a measure of morphological similarity.  Cross correlations have been used to this end (although comparing only with the continuum) by \citet{Lu2017ApJ}. For each molecule, we determine which one is the other species (among the set of molecules in this study) with  the highest cross correlation. We call this species the maximum correlation partner (MCP) of the considered molecule. We stress that this relation is not necessarily symmetric: for example, while OCS and \met\ are MCPs of each other, OCS is the MCP of SO but not vice-versa.  The cross correlation values of MCPs are marked in boldface in Figure \ref{fig-corrs}.

%results
Figure \ref{fig-corrs} shows that molecules gather in two groups by linking them together with their MCPs. That is, the MCP of each molecule in any of these groups is within that same group.  In this way, we gather together molecules with similar zero moment morphologies. Visual inspection of the maps confirm this interpretation, except in the case of NH$_2$D, which has the lowest cross correlations with its MCP (0.39). Because the zero moment map of this molecule (Figure \ref{fig-oth}) is so different compared to the rest, we do not include it in the grouping and we analyze it independently.
We estimate the uncertainty of $\rho_{12}$ by  adding  simulated noise to each zero moment map and measure the dispersion of the values of the cross correlations thus obtained. In all cases, the uncertainty due to random noise is $<0.01$, making no discernible effect in the classification.

We denominate  ``Shock group'' the one in which traditionally shock activity tracers  such as SiO, HNCO and SO gather.
We refer to the second group as the ``Continuum group'' because it gathers species better correlated with the 3 mm continuum.
As Figure \ref{fig-corrs} shows, the MCP of the continuum image is the zero moment of the \htcop, $J=1\shortrightarrow0$ transition.
%However, we caution that the cross correlation between these two images reaches a value of 0.67, which is not particularly high.

%possible caveats
We warn that the specific numerical value of the correlation is probably affected by instrumental and observational effects such as the uv-coverage, the pointing within \clump, and the primary beam shape (though not  by constant calibration factors).  This means that how these cross correlations compare between each other is more important than the specific values presented in the matrix shown in Figure \ref{fig-corrs}.  
One possible caveat of this way of calculating similarities between the molecules is the non inclusion of kinematic data, that is,
it is possible that the zero moment of two molecular transitions are very similar, but their lines having very different central velocities. Thus, we calculate the cross correlations using the data cubes of the lines in order to test whether the grouping described depends on ignoring the kinematic information. We find that the same grouping as described in Figure \ref{fig-corrs} occurs, so we keep the simpler approach of calculating cross correlations using the zero moment images, which also allows us to calculate correlations between images of the molecular lines and the continuum.

%%%%%%%%%%%%%%%%%%%%%%%%%%%%%%
{\subsection{Kinematics}\label{sec-m1}}

\begin{figure}
\includegraphics[angle=-0, ext=.pdf, width=\textwidth]{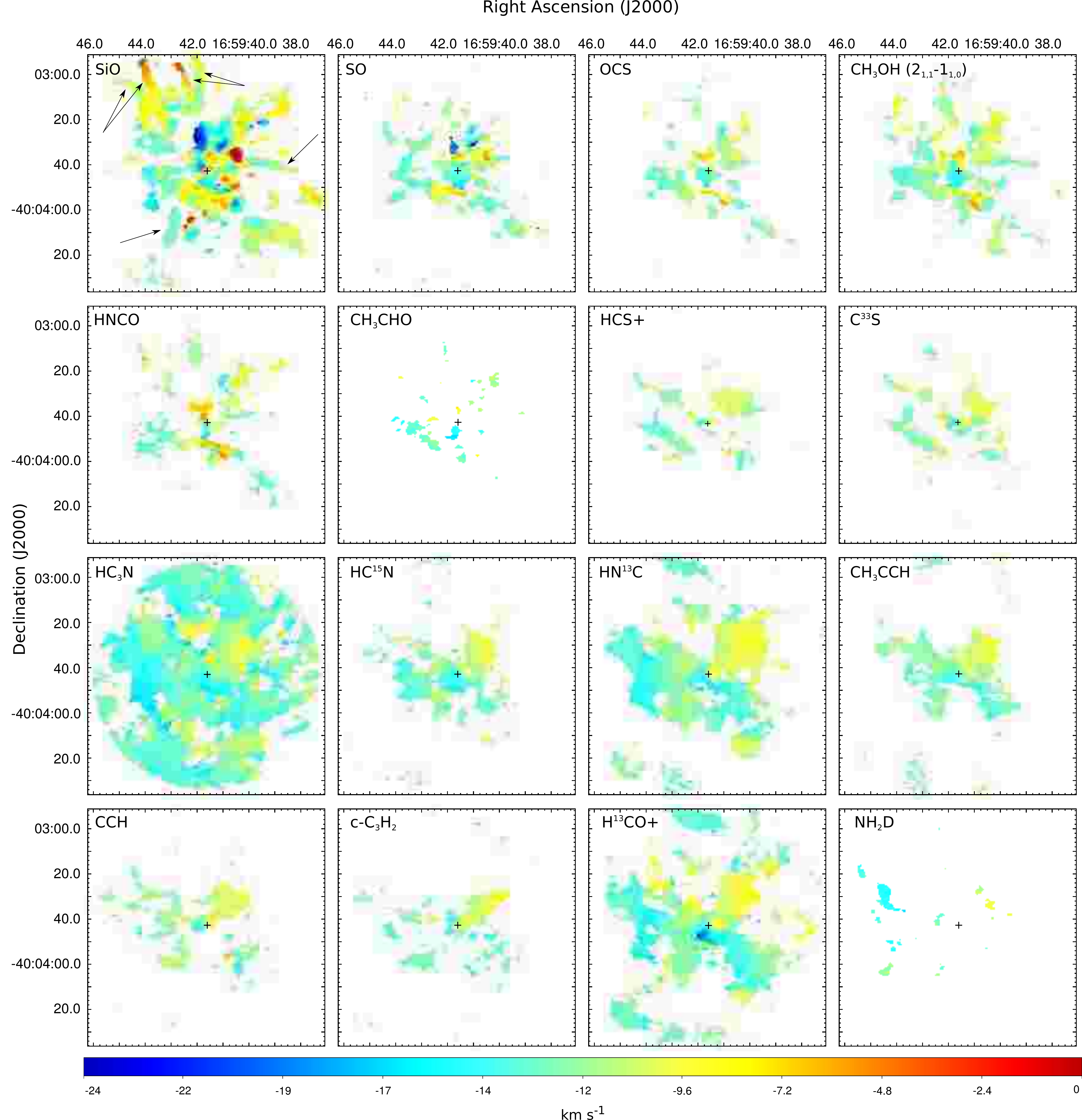}
\caption{Large scale images of the first moment from  molecular transitions with extended emission. To calculate the first moment, we masked emission below 5 mJy beam$^{-1}$ ($\sim 5\sigma$). Black crosses indicate the position of \hmyso. Black arrows in the SiO panel indicate some of the so-called ``fingers.''\label{fig-m1}}
\end{figure}

We analyze the kinematics of \clump\ using first moment maps. Figure
\ref{fig-m1} shows first moment maps of those molecules with resolved
kinematic features and strong extended emission.  We leave out molecules
with only emission toward the CC core like \methoxy, \propanenitrile, and
SO$_2$:  first moment  maps of the first two are rather uninformative because they
display no velocity gradients and seem to be well characterized by a single
$V_{\rm LSR}$. The SO$_2$ map, on the other hand, does show a velocity
gradient characteristic of rotation which was analyzed in detail in
\citetalias{Guzman2014ApJ}.  We also leave out species with only faint
extended emission like \hccn, \cyan, \acrylonitrile\ and \ethe\added{\ also written as CH$_2$CO}.  The first
moment maps of these molecules display similar characteristics as those
shown in Figure \ref{fig-m1}, but with  lower signal-to-noise ratios.

We stress that the kinematic analysis is somewhat hindered by the modest
spectral resolution of the data.  Figure \ref{fig-m1} shows that the
velocities range between $-8$ and $-15$ \kms\ for most molecules, with the
ambient cloud velocity around $-12$ \kms. This is consistent with previous
single dish studies on
\clump\ \citep{Bronfman1996AAS,Urquhart2007AA13CO,Miettinen2006AA}.

On a large scale, most molecules display redshifted velocities ($\sim-10$ \kms, that is, redshifted respect to the ambient cloud) toward the north west section of \clump, and blueshifted velocities ($\sim-14$ \kms) toward the south east. This trend is reminiscent of the general orientation of the jet and CO outflow detected toward \hmyso\ \citep{Guzman2011ApJ}.  This general kinematic trend is well illustrated by the \htcop\ first moment map, but it is also evident in \hctn, \hcqn, \hntc, \propyne, \cyc, and C$^{33}$S. The trend is less evident --- but still tantalizing --- in SO, \met, \hccn, CCH, and even NH$_2$D . The \htcop\ map likewise displays conspicuous blueshifted emission located about $4\arcsec$ south of \hmyso, which is due to gas with $V_{\rm LSR}\approx-19.1$ \kms. This emission is apparent in \hcdop\ and SO, but it is not clearly seen in any other molecule.

By far, the transition showing the largest velocity variations
across the field is SiO, $J=2\shortrightarrow1$.
These variations, illustrated in the first panels of Figure \ref{fig-m1},  
span $20.9$ \kms, whereas for the rest
of the molecules the velocity span is $\lesssim10$ \kms. Interestingly
enough, the only exception to this trend is SO, whose velocity span is
$\sim14.8$ \kms. This kinematic
feature of SO is consistent with the similar morphology
observed between its zero moment map and that of SiO.
%Velocity span
%ls -1 *fits | perl -lane '$f=$_;use PDL;$i=rfits($f);$i=$i->where($i**2>0)->qsort;$n=int(nelem($i)/200);$i=$i->slice("$n:-$n");($mean,$prms,$median,$min,$max,$adev,$rms)=stats($i);print "$f ".$max." ".$min." ".($max-$min)'

Another kinematic feature unique to SiO are the conspicuous filaments or ``fingers'' populating the north east region of \clump, which extend roughly in the radial direction from \hmyso. As shown in Figure \ref{fig-m1}, these features are associated with velocity gradients of the order of $\sim10$ \kms\ across a distance of 30\arcsec\ (0.25 pc at 1.7 kpc). Four discernible ``fingers'' are located toward the north east region of the clump. Additional fingers are apparent toward the south east and  western parts of \clump. These fingers of SiO emission are presumably outflows or outflow cavities generated by other YSOs in \clump. However, the structures are as well reminiscent of the ``explosive'' streamers found toward OMC1 \citep{Bally2017ApJ} and DR21 \citep{Zapata2013ApJ}, in the sense that many of them seem to point toward the center of the clump.  Further investigation will determine what is the true nature of these structures only apparent in SiO emission.

Finally, we note that besides SO, SO$_2$ --- and isotopologues --- there is
no other molecule associated with the velocity gradient characteristic
of the rotating core around \hmyso\ detected in \citetalias{Guzman2014ApJ}.

%%%%%%%%%%%%%%%%%%%%%%%%%%%%%%%%%%%%%%%%
{\subsection{Morphology of the Emission by Molecule}\label{sec-desc}}

In the following paragraphs we describe the main characteristics of
the emission of representative transitions of each
species.
This section expands the description of Section \ref{sec-morph}
focusing on the specific morphology  per molecule.
\citetalias{Guzman2014ApJ} already analyzed in detail the
morphology of most of the sulfuretted molecules emission near the
central HMYSO, so we refer to their analysis for these species (SO,
SO$_2$, OCS, CS and their isotopologues).  To organize the upcoming
discussion we gather the molecules according to their composition and
number of atoms.

{\subsubsection{Simple molecules}\label{sec-simpleMolecules}}

With simple molecules we refer to molecules with 5 atoms or less.

\smallskip\noindent\emph{\underline{Nitrogenated.}} We include here \replaced{HCN}{the HCN isomers}, \hctn\ (cyanoacetylene) and \added{its isotopologues}\deleted{isotopomers}, and \damm. Just for the sake of exposition, we  choose to include HNCO.  The regions where \hctn\ (and the $^{13}$C isotopologues) emission is most intense are the CC core, the NEC-wall, and \deleted{the} C8. The C8 emission peaks $\sim0\farcs7$ to the west from Source 8.
An extended, distinct \hctn\ emitting region is a triangular-like  structure
between sources 10, 11, and an emission peak located $1\farcs3$ north of source 13 (see Figure \ref{fig-cen1}). Emission from the $^{13}$C isotopologues displays the same morphological features.  On a clump scale,  emission from \hctn\ is most similar, as evaluated from the cross correlations, to \hntc\ and \htcn. Figure \ref{fig-cont}  suggests in addition \added{a} good \replaced{correlation}{match} between the \hctn\ zero moment map and those of \htcop\ and \propyne, which is confirmed by the cross-correlations.

Other nitrogenated simple molecules are the isotopologues H$^{13}$CN and HC$^{15}$N.  Their zero moment maps are very similar, with a cross correlation of 0.9, which is higher than any correlation between different species ($\le0.8$). The correlation calculated masking the  central dominant source (Section \ref{sec-corr}) is $0.86$. Kinematically, the molecules are similar as well. The linewidths range between the 2 and 6 \kms, with most of the emission having  linewidths between 3 and 4 \kms.

An evident difference between the line profiles of \htcn\ and \hcqn\ is the hyperfine splitting of the $J=1\shortrightarrow0$ \htcn\  transition produced by the nuclear quadrupole of the $^{14}$N nucleus. We fit Gaussians to the three hyperfine components in order to explore further the observed splitting. The blue and redshifted components  are located at velocities of $-7.3$ and $+4.9$ \kms\ respect to the central component, which is the brightest. Theoretically, the line intensities should be in the
$1:5:3$ ratio in optically thin conditions.
We performed the fitting in every pixel where the peak intensity exceeds 5 $\sigma$. The observed intensity ratios of the blue and redshifted components respect
to the central one are $1\pm0.5:5:2\pm1$ respectively, where the
uncertainties represent the standard deviation of these quotients measured
throughout the field.   The blueshifted
versus the central component ratio appears to concur with theory, while the
redshifted component ratio is slightly lower than expected, but consistent with the observed 1-$\sigma$ variations.  We conclude that our results are consistent --- within
the uncertainties --- with the local thermodynamic equilibrium and
optically thin predictions.
%Note that optically thick conditions would increase both ratios toward unity, so a decrement like the one observed in the redshifted component is not explicable in this way.  Anomalies of the relative intensities of the hyperfine components of HCN are commonly observed and have previously been reported \citep{Walmsley1982ApJ}. \citet{Loughnane2012MNRAS} presents some explanation for the cause of these anomalies, but there is still no consensus in the literature.

The morphology of the HN$^{13}$C emission resembles that of H$^{13}$CN: they both display strong emission associated with the C8 and the NEC-wall. However, there are two important differences, the most relevant being the complete absence of HN$^{13}$C emission from the CC core while \htcn\ is bright there. The other noticeable difference is the HN$^{13}$C emission associated with the arcuate continuum structure joining sources 7, 3, 2, and 8.  Because this arc of emission appears conspicuously in a few other  transitions,  we refer to it hereafter as the Continuum Arc.

HNCO (isocyanic acid), on the other hand, is a molecule with a strong CC core component. Its zero moment maps are well correlated with those of OCS and \met, showing strong emission associated with the Diffuse Ridge, the NW cloud, and the N-red cloud. However, there is not much HNCO emission associated with the C8 or the Source 8 itself, and no emission associated with the NEC-wall. Isocyanic acid is the only nitrogenated species in the Shock group. 

Finally, \damm\ is a special case.  By and large, the zero moment of \damm\ (Figure \ref{fig-oth}) displays little correlation with  features conspicuous in other molecules, and we analyze it separately from the Shock or Continuum groups. The peak of the \damm\ emission is located $36\arcsec$ from \hmyso, in the $\text{P.A.}=65\arcdeg$ direction.  This \damm\ core is apparently part of a filament which extends for $\approx30\arcsec$. Molecules which have emission related with the location of this core are \hntc, CCH, and possibly \hcsp.  A secondary peak of \damm\ emission is associated with a core located $38\arcsec$ south east of \hmyso, in the $\text{P.A.}=122\arcdeg$ direction. This core is not related with any discernible structure in any other molecule, but, as seen below, it seems related with a NIR-dark globule. Less intense emission is located associated with the NEC-wall and with Source 16, and there is also emission connecting both positions. This morphology resembles the continuum \citepalias{Guzman2014ApJ}, which  shows that Source 16 is embedded in an envelope extending to the south until approximately the region we identify in this work with the NEC-wall. Other features clearly associated with  \damm\ emission are  Sources 7 and 2 and some diffuse emission apparently tracing the  Continuum Arc. As we will see below, it does show some faint emission associated with the CC core, barely noticeable in Figure \ref{fig-cen2}.

\smallskip\noindent\emph{\underline{Sulfuretted.}} A rather complete analysis of the morphology of the  emission from the sulfuretted molecules, specially in the central region  of \clump, is given in \citetalias{Guzman2014ApJ}. They concluded that sulfur  oxides (SO, SO$_2$, and isotopologues) are well associated with  \hmyso, in contrast to other carbon-sulfur species such as OCS and  CS. 
All sulfuretted species whose emission extends on scales comparable with the clump size --- that is, all of them except SO$_2$ ---  are in the Shock group. From Figures \ref{fig-shock} and \ref{fig-corrs} it appears that SO and OCS  correlate more with \met\ and (ignoring the CC core) SiO than with other sulfuretted molecules like C$^{33}$S and \hcsp. The zero moment maps of these last two molecules are similar between each other, and they  display some features observable in some molecules of the Continuum group (like CCH, see below).

The only sulfur bearing molecule which was not included in \citetalias{Guzman2014ApJ} is \hcsp. Figures \ref{fig-cen2} and \ref{fig-shock} show the  zero moment of the \hcsp, $J=2\shortrightarrow1$ line. Figure \ref{fig-cen2} shows  that there is emission apparently related with the CC core, but more extended  and located  $\sim0\farcs4$ farther to the north from \hmyso\ compared to than that  of, e.g., methanol. On a larger scale, Figure \ref{fig-shock} shows that  the \hcsp\ emission is most similar to that of C$^{33}$S. This similarity  is reflected in the velocity distribution of both lines.  

Two of the most remarkably similar zero moment maps are those of SO and SiO (see Section \ref{sec-corr}), whose similarity is also observed in the first moment  maps (see Section \ref{sec-m1}). Several common features are recognizable in these two maps (Figure \ref{fig-shock}), but one equally remarkable \emph{difference} is the strong emission from SO associated with the CC core, which is absent in SiO. That is, the SO emission in \clump\ can be characterized by a strong, rotating CC core \citepalias[described in][]{Guzman2014ApJ} surrounding \hmyso, plus spatially extended emission well correlated with SiO. This behavior of SO is somewhat reflected in the rest of the sulfur bearing molecules, all of them  part of the Shock group. 

Finally, we note that the CS zero moment map (Figure \ref{fig-cs}) is less similar than expected to that of C$^{33}$S. We attribute this to the high optical depth of CS compared to that of the C$^{33}$S.

\smallskip\noindent\emph{\underline{Small carbon chains.}} In this category we include CCH and \cyc, the latter being a \replaced{cyclical}{cyclic} molecule. \deleted{Both these molecules are free radicals (with unpaired valence electrons) and they are part of the Continuum group.} Neither of these two molecules have emission associated with the CC core, with \cyc\ displaying an strong absorption feature toward the location of \hmyso, centered at $-13.6\pm0.1$ \kms\ with a FWHM of $4.6\pm0.3$ \kms. One important feature of the  CCH,  $N=1\shortrightarrow0$ transition line\footnote{$N$ represents the \replaced{total}{pure rotational} angular momentum \deleted{excluding spin} \citep{Gottlieb1983ApJ}.} is that it splits in six hyperfine components (discernible in Figure \ref{fig-spec}). The relative observed strength of these components (whose temperature dependence is negligible) is in good agreement with the ratio expected for optically thin emission. CCH emission is strongest in the Continuum Arc, specifically, just below Source 3. Its is likewise strong between the CC core and N-red cloud, in the NEC-wall, and in a cloud located $\sim20\arcsec$ to the south east of \hmyso\ ($\textrm{P.A.}=135\arcdeg$). This last emission feature is also conspicuous in the zero moment map of \cyc\ and it has no obvious counterpart in any other molecule. 

Emission from \cyc\ seems to be less extended compared to that of CCH. It is strongest in the C8 region, extending somewhat toward the Continuum Arc. As pointed out before, there is  clear emission associated with the south east cloud mentioned above. Contrary to CCH, there is no strong \cyc\ emission arising from the NEC-wall. Less prominent \cyc\ emission  is also detected from a cloud located  at $\sim22\arcsec$ to the east of \hmyso\ ($\textrm{P.A.}=80\arcdeg$). We note that there is evident CCH and C$^{33}$S  emission from this region as well. 

Additional similarities between CCH and C$^{33}$S include a filament of emission extending for $\sim15\arcsec$ roughly in  the E-W direction,  located $30\arcsec$ north east of \hmyso\ ($\textrm{P.A.}=45\arcdeg$). We note that this filament is also visible in  \hcsp. 

\smallskip\noindent\emph{\underline{Oxygenated.}} In this category we analyze simple molecules composed of carbon and oxygen, that is, the isotopologues \htcop\ and \hcdop\ and \ethe\ (ethenone\added{, also called ketene}). 
The \hcdop\ map is very similar to that of \htcop\ (correlation coefficient of 0.79), with the \hcdop\ line being more optically thin \citep[$\approx8$ times less abundant according to][]{Wilson1994ARAA} compared with \htcop. 
Figure \ref{fig-cen2} shows that the emission from \ethe\ and \htcop\   is   associated with the CC core and C8. The \ethe\ zero moment map displays a more compact distribution around these two locations compared to that of  \htcop\ or \hcdop. 

On a large scale, both species are different: \ethe\ and \htcop\ were classified in the Shock and Continuum groups, respectively. Ethenone displays a good resemblance to the OCS (its MCP) and \acet\ maps, and the cross correlations given in Section \ref{sec-corr} are practically equal (0.60 and 0.59, respectively). We can identify clearly in the \ethe\ map, emission  coincident with the  Diffuse Ridge, the  NW cloud, and the N-red cloud. In addition, the \ethe\ map shows diffuse emission located $30\arcsec$ north of \hmyso, which is  seen clearly in OCS, \acet, HNCO, and \met.

The \htcop, on the other hand, is the MCP of the 3 mm continuum map. That is, the \htcop\  zero moment map (Figure \ref{fig-cont}) is the  one which better correlates with the continuum away of the central source,\deleted{, which,  according to Paper I, is} dominated by thermal dust \added{\citepalias{Guzman2014ApJ}}.
\htcop\ is one of the few molecules with counterpart emission associated with Source 13 (Figure \ref{fig-cen2}). The zero moment map also shows a source located  $\sim4\arcsec$ south of \hmyso\ whose emission is significantly blueshifted respect to the ambient material (see Section \ref{sec-m1}). Other regions associated with strong \htcop\ emission are the NEC-wall and the C8. The C8 emission extends to the Continuum Arc.  In addition, there is a south east diffuse feature  which correlates roughly with the position of continuum Sources 12, 15, and 17.

Finally, we note that the \htcop\ emission, as it is  the case for \hctn\ and \propyne, is more compactly distributed. There is little \htcop\ emission more than $20\arcsec$ away from \hmyso, which is in stark contrast compared to the molecules of the Shock group.

{\subsubsection{Complex Organic Molecules (COMs)}\label{sec-dcom}}

COMs detected in this work are molecules consisting of carbon and hydrogen atoms plus, except \propyne, either one oxygen or one nitrogen atom. The only molecule detected with an O and a N atom is HNCO, and we do not detect any molecule with more than one N or O atom.  In the following, we call a molecule saturated if  all its carbon-carbon bonds are single (C{\sbond}C). Conversely, unsaturated COMs have double or triple carbon-carbon bonds (C{\dbond}C or C{\tbond}C). All simple molecules with carbon-carbon bonds detected in this work (CCH, \cyc, \hctn, and \ethe) are unsaturated, but this is likely due in part  to  a selection effect: saturated hydrocarbons usually have more atoms.

\smallskip\noindent\underline{\emph{Propyne} (\propyne).} This COM was previously observed toward \clump\ using single dish by \citet{Miettinen2006AA}. The central region (Figure \ref{fig-cen2}) shows some correlation with the CC core, but with the bulk of the emission arising north of \hmyso.  The propyne zero moment peak is located $2\arcsec$ north-east of \hmyso. Propyne is also brightly associated with the C8: a secondary emission peak is located less than $1\arcsec$ west of Source 8. On a larger scale (Figure \ref{fig-cont}), \propyne\ emission displays a rather smooth distribution, with a very good correlation with \htcop, although less bright. These two molecules are MCPs of each other. Practically all the bright features distinguished in the \htcop\ zero moment map have a \propyne\ counterpart.  Propyne  shows a good correlation with \hctn\ as well, excluding the central region.
Due to the difficulty in separating the different  \propyne\ lines due to blending, the zero moment maps shown include the sum of all the \replaced{$5_K\shortrightarrow4_K$}{$5,K\shortrightarrow4,K$} transitions.
%Propyne (or methyl acetylene), as a
%%symmetric rotor, can be used as a reliable temperature estimator
%\citep{Bergin1994ApJ} and in this context its emission will be analyzed in
%more detail later.

\smallskip\noindent\underline{\emph{Nitrogenated} (\propanenitrile,
  \cyan, \acrylonitrile, and \hccn).}  
Only one of these COMs, propanenitrile (\propanenitrile, hereafter \prop), is unequivocally  linked with the CC core. Figure \ref{fig-cen1} shows that the emission from this COM peaks close to \hmyso, but displaced from it by $\sim0\farcs5$ compared to  SO$_2$ or  HNCO. Faint emission of \prop\ (not evident in the zero moment map image) is  detected also toward the C8 (see Section \ref{sec-discussion}). We do not detect this COM toward any other location in the clump. 

As shown in Figure \ref{fig-cen2}, the other three COMs have less correlation with the CC core, and none of their peaks actually correspond with the CC core position. Among these COMs, \cyan\ (cyanopropyne) is the only one which shows some CC core  counterpart. The emission from \cyan\ has two peaks, associated with the NEC-wall and C8, with practically the same intensity. The C8 \cyan\ emission peaks approximately $1\arcsec$ east of Source 8, and extends somewhat to the Continuum Arc.  The NEC-wall emission from this COM apparently joins with the CC core emission  and with emission detected south of Source 16. The overall appearance of this structure is similar to what is observed in other molecules, for example, \hcsp\ (same Figure \ref{fig-cen2}). There is  emission correlated with the position of Source 11, which is a feature seen in a few other molecules (e.g., CCH, \propyne, and \hntc).
Figure \ref{fig-cont} shows the \cyan\ zero moment map on a larger scale. There is not much more emission compared with what is shown in Figure \ref{fig-cen2}. However, we note a small cloud located about 
$18\arcsec$ south west from \hmyso\ in the $\text{P.A.}=-135\arcdeg$ direction. Emission from this location is clearly detectable in CCH, \propyne, \htcn, \hntc, \htcop, \hcqn, \hctn, \hcsp, and 
\acrylonitrile. It is  located near the south west end of the Diffuse Ridge, thus,  we can  better discern it  in molecules without Diffuse Ridge emission like CCH and \propyne. 

Emission from acrylonitrile (\acrylonitrile, hereafter \acry) is better displayed in Figure \ref{fig-cen2}, because most of the emission arises from regions not farther than $15\arcsec$ from \hmyso. The peak of \acry\ emission is clearly associated with C8. Acrylonitrile is one of the faintest molecule we claim detection in this work. With the exception of the peak emission, the rest of the features shown in Figure \ref{fig-cont}  are apparently  real mostly because their location is consistent with emission  seen in other molecules. There is also emission somewhat consistent with the CC core and with the south west cloud described at the end of the previous paragraph.

Finally, Figures \ref{fig-cen2} and \ref{fig-cont} show the zero moment images of the cyanopolyyne \hccn\ (cyanodiacetylene). This molecule does not show strong emission associated with the CC core. Weak emission located $\approx1\arcsec$ north of \hmyso\ joins to the east with emission arising from the NEC-wall, as seen in many other molecules. As it is likewise the case for \propyne, \cyan, CCH, and \hntc, there is \hccn\ emission associated with the location of Source 11. The \hccn\ peak is clearly located in C8. Diffuse emission extends from the C8 to the north following somewhat the Continuum Arc. Relatively intense, diffuse emission, is also associated with Source 7. 

\smallskip\noindent\underline{\emph{Oxygenated} (\met, \acet, and \methoxy).} 
By far, the \met\ (methanol) transitions display the brightest and richest spatial structures of all COMs. Among the eleven detected methanol lines there is one rotational line from a vibrationally excited state and a class I maser transition (\maser). Methanol emission (Figures \ref{fig-cen1} and \ref{fig-shock}) is in every line dominated by strong emission detected toward the CC core, except for the maser transition which is dominated by three bright maser ``spots.'' 

\begin{figure}
\includegraphics[angle=-0, ext=.pdf, width=\textwidth]{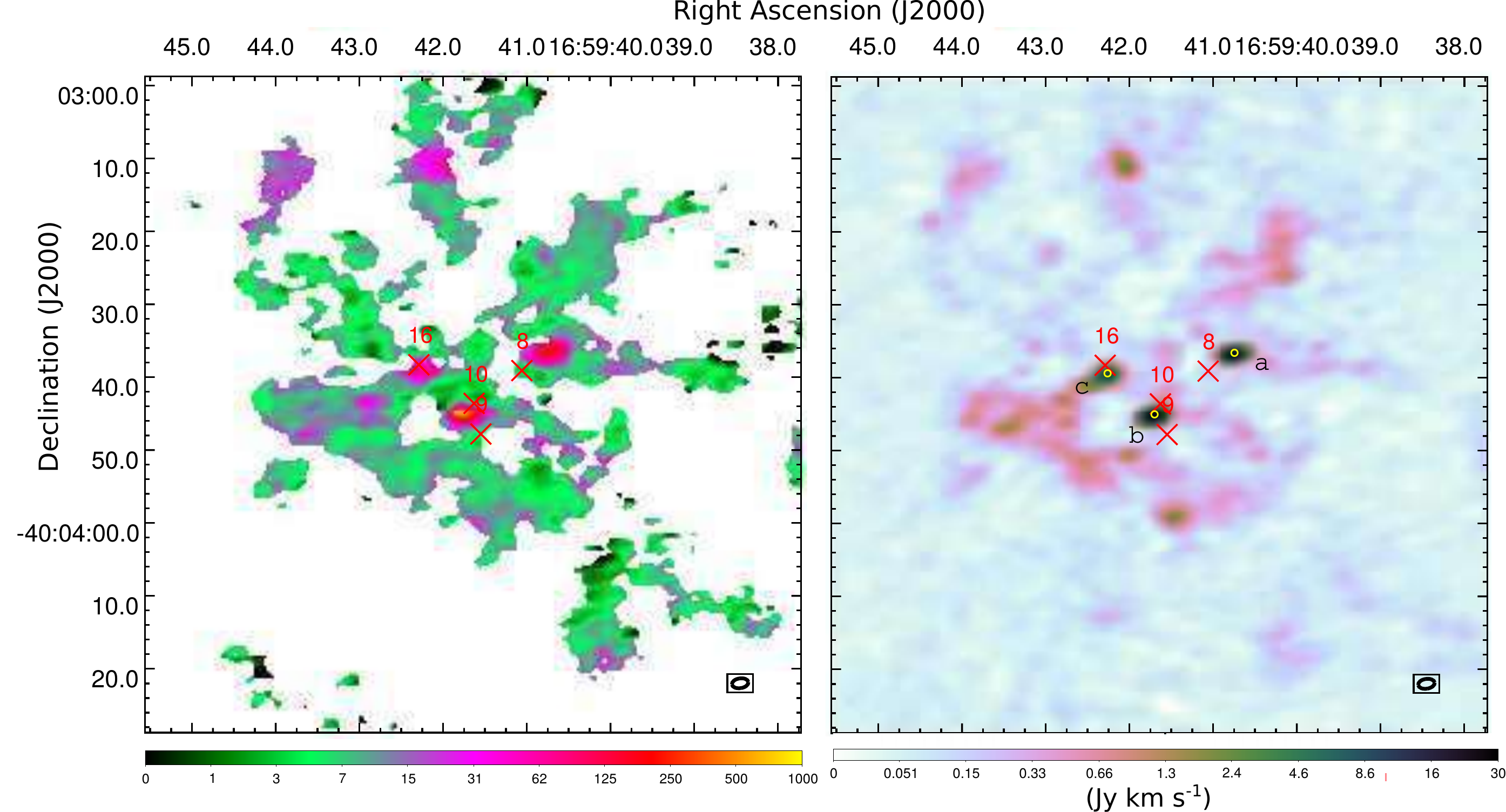}
\caption{\emph{Left panel:} Ratio between the  zero moment of the \met, \maser\  and \added{$({\rm E_1})$} $2_{1,1}\shortrightarrow1_{1,0}$  lines. \emph{Right panel:} Zero zero moment of the   \met, \maser\ emission.  The three masers (yellow circles) correspond   to the three strongest sources (whose coordinates are given in Table \ref{tab-mas}), conspicuous even in the logarithmic color  stretch.  Red crosses and numbers show continuum 3 mm sources nearby the masers.\label{fig-mas}}
\end{figure}

Figure \ref{fig-mas} shows in the left panel the quotient between the zero moment maps of the \maser\ and \added{$({\rm E_1})$} $2_{1,1}\shortrightarrow1_{1,0}$ \met\ transitions. The three maser spots marked \texttt{a}, \texttt{b}, and \texttt{c} in the right panel of this figure (which shows the zero moment of the maser transition in logarithmic color stretch) are associated with line fluxes more than 100 times larger than those of the typically thermal \added{$({\rm E_1})$} $2_{1,1}\shortrightarrow1_{1,0}$ line. Table \ref{tab-mas} indicates the parameters of the maser emission. The strongest maser is \texttt{b} followed closely by \texttt{a}.
Maser \texttt{a} is located $4\arcsec$ in the $\text{P.A.}=-53\arcdeg$ direction from Source 8 and it is associated with bright methanol emission in the rest of the lines.
Other molecules which show clearly emission consistent with the location of Maser \texttt{a} are \acet\ and \ethe, and perhaps less clearly \hcsp\ and OCS. Methanol lines associated with low upper energy levels ($<50$ K) are characterized by brighter emission toward the Maser \texttt{a} location than toward Source 8, whereas the opposite is true for  high energy transitions ($>50$ K). Source 8 is also bright in \ethe\ and OCS but not in \acet.
Maser \texttt{b} is located $1\farcs6$ to the south east($\text{P.A.}=150\arcdeg$) from \hmyso. Due to the proximity of the CC core, this region is associated with diffuse emission in several molecules. However, in contrast with Maser \texttt{a}, there is no distinguishable feature in either methanol or any other molecule coincident with the position of this maser.  Maser \texttt{c}, located $1\farcs4$ south of Source 16, seems to be associated with the NEC-wall. 

\begin{figure}
\includegraphics[width=\textwidth,ext=.pdf]{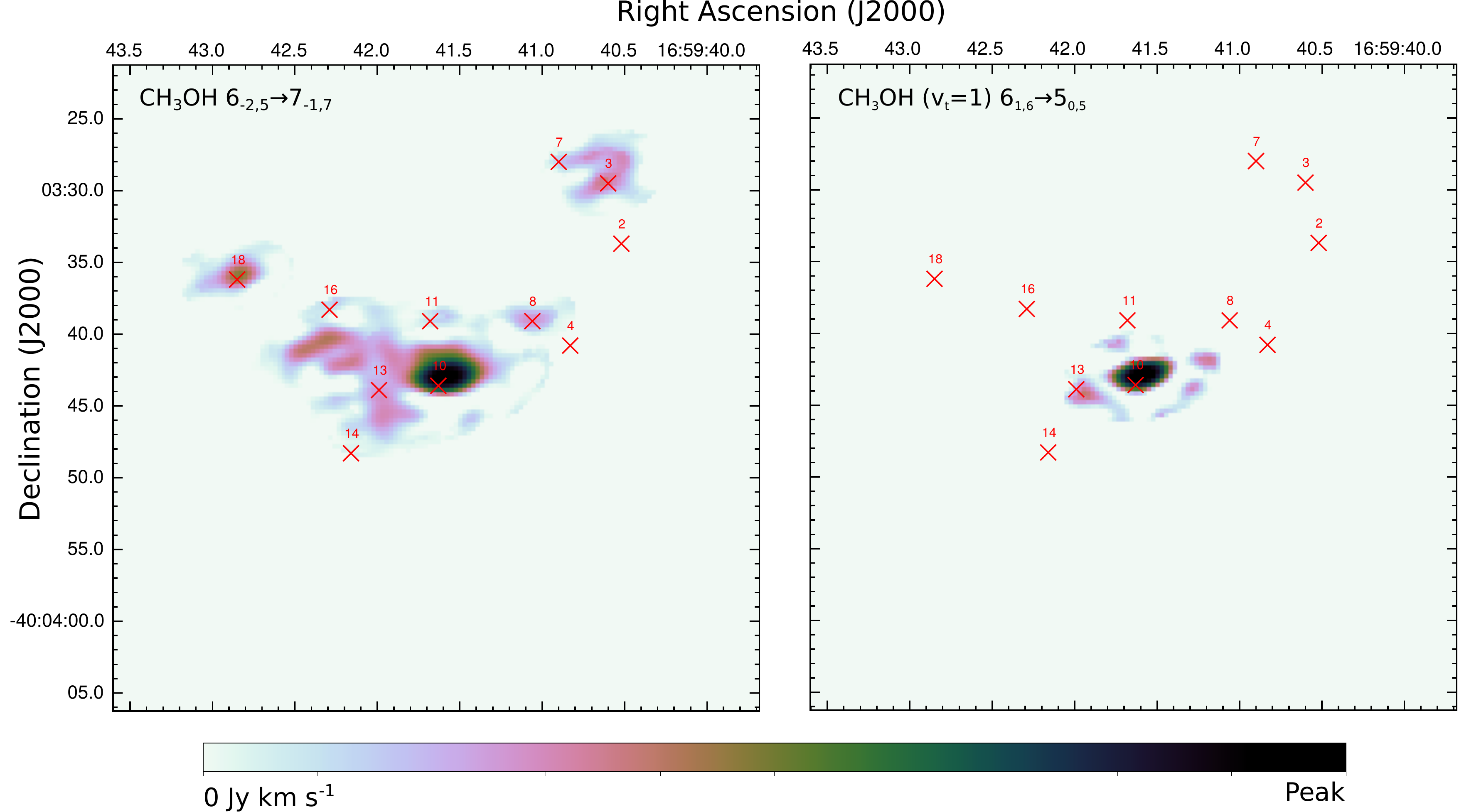}
\caption{Left and right panels show, respectively, the zero moment maps of the \met, \added{$({\rm E_2})$} $6_{-2,5}\shortrightarrow7_{-1,7}$ and $6_{1,6}\shortrightarrow5_{0,5}$ ($v_t=1$) transitions. Red crosses and numbers show the position of continuum 3 mm sources as defined in\citetalias{Guzman2014ApJ}. The upper energy levels of these two transitions are $74.7$ and $340.1$ K, respectively. \label{fig-himet}}
\end{figure}

Partly because of the intricate details observed in methanol we used  the zero moment map of \met, \added{$({\rm E_1})$} $2_{1,1}\shortrightarrow1_{1,0}$ to define some of the most noticeable  of \clump\ (Section \ref{sec-morph}). In that analysis we used a comparatively low energy transition ($E_\textrm{up}=21.6$ K). Higher energy transition zero moment maps are expected to be less rich, but they emphasize  different features.  Figure \ref{fig-himet} shows the zero moment maps of two higher energy transitions, one with $E_\textrm{up}=74.7$ K and the other $340.1$ K. The latter shows only the CC core, and perhaps some emission associated with Source 13. The CC core emission in this highly excited state of \met\ supports the view that this is a HMC.  The $6_{-2,5}\shortrightarrow7_{-1,7}$
($E_\textrm{up}=74.7$ K ) transition does show more structure: besides the CC core, which is the dominating source, there is diffuse emission connecting the northern part of the CC core with the NEC-wall. There is in addition diffuse emission extending south $\approx5\arcsec$ at the R.A. of Source 13. We note that while the low energy methanol (Figure \ref{fig-consp}) shows extended diffuse envelopes blending in with the continuum sources, the left panel in Figure \ref{fig-himet} shows a much less ambiguous correlation with Sources 7, 3, 8 and 18. Structures which are not apparent in the high energy \met\ transitions are the Diffuse Ridge, the NW cloud, and the N-red cloud.

Acetaldehyde (\acet) is another oxygenated COM we detect toward \clump.  Zero moment maps of this molecule are shown in Figures \ref{fig-cen2} and \ref{fig-shock}. The central region (Figure \ref{fig-cen2}) immediately shows that \acet\ is not associated with the CC core. In fact, there is no clear emission associated unambiguously with any continuum source.  On a larger scale (Figure \ref{fig-shock}) \acet\ emission shows features characteristic of the Shock group: its MCP is \met, with a correlation coefficient between them of $0.72$ (we emphasize that Section \ref{sec-corr} uses a low energy \met\ transition for the analysis).  Acetaldehyde is bright toward the Diffuse Ridge, the N-red cloud, the NW cloud, and Maser \texttt{a} position. In general, there is \acet\ emission toward bright methanol regions, e.g.: the diffuse \acet\ emission south of the CC core has a similar morphology as \met; a small cloud located $1\farcs5$ west of Source 6 is bright in many Shock molecules, including \acet; and the cloud located $30\arcsec$ north of \hmyso\ already mentioned in the \ethe\ description.

Finally, \methoxy\ (methoxymethane) is detected only toward two locations: the CC core and Source 8. The CC core \methoxy\ emission is evident in Figure \ref{fig-cen1}, but that of Source 8 is very faint and more evident in the data cube of the $4_{1,4}\shortrightarrow3_{0,2}$ transition\added{ instead of the zero moment map}.

% central regions 
%%%%%%%%%%%%%%%%%%%%%%%%%%%%%%%%%%%%%%%%%%%%%%%%%%%%%%%%%%%%
%%%%%%%%%%%%%%%%%%%%%%%%%%%%%%%%%%%%%%%%%%%%%%%%%%%%%%%%%%%%

{\section{COLUMN DENSITIES AND EXCITATION TEMPERATURES}\label{sec-ntex}}

In this section we model the data and results presented in the previous
sections and determine physical parameters like column densities and
temperatures, mainly from LTE models. Section \ref{sec-prop} focuses 
on the \propyne\ emission and LTE modeling. Section \ref{sec-cdt} makes more detailed models of the emission of the species detected toward several sources in \clump.

While the focus of the present work is on the detected species, as a general remark for the upcoming discussion we mention some noticeable non-detection and molecules whose lines were not covered by our observations. In addition to \metfor\ and \form\ (mentioned in Section \ref{sec-lin}), other molecules commonly associated with hot-cores \citep[e.g.,][]{Gibb2000ApJ} which were observed but not detected are HCOOH (formic acid) and CH$_3$CH$_2$OH (ethanol). These non-detection allow us to estimate upper limits on the column densities of the respective species (Section \ref{sec-cdt}). On the other hand, neither H$_2$S (hydrogen sulfide), CH$_3$CN (methyl cyanide), nor their \replaced{isotopomers}{isotopologues} were covered by our observations. The spectral setup does not efficiently cover the H$_2$CO (formaldehyde) or the HDCO lines either because it only samples transitions predicted to be faint (high $E_{\rm up}$ or very low Einstein $A$-coefficients).

{\subsection{Propyne Temperature and Column Density}\label{sec-prop}}

Propyne (\propyne) is a symmetric top molecule, with its dipole moment aligned with the symmetry axis of the molecule. This implies that radiative rotational transitions do not change the projection $K$ of the angular momentum $J$ onto the symmetry axis \citep[note that $K\le J$]{Townes1975MS}. Levels with different $J$ and the same $K$ are sometimes refer to as $K$-ladders. Thus, radiative transitions only connect \replaced{$J_K\shortrightarrow(J-1)_K$}{$J,K\shortrightarrow(J-1),K$} states and the relative population between different $K$-ladders is determined by collisional excitation equilibrium.  
Therefore, the rotational temperature of \propyne\ between different $K$-ladders is a good indicator of the kinetic temperature $T_K$ of the gas \citep{Bergin1994ApJ}. Indeed, propyne has been used to estimate the kinetic temperature of high-mass star forming clumps  \citep[e.g.,][]{Molinari2016ApJ,Giannetti2017AA}.
Cyanopropyne is another symmetric top detected in our observation, but it is  much rarer than propyne.
Our spectral setup covers the \propyne, $J=5\shortrightarrow4$ transitions connecting the 0- to 3-ladder transitions  (4-ladder detection is marginal).  We calculate the temperature and column density of \propyne\ by fitting Gaussians to the four $K$ components and modeling the rotation diagrams assuming LTE and optically thin conditions. The latter is justified because the line's optical depth never exceeds $0.1$. We perform this fitting on each pixel where we detect at least one \propyne\ line over $5\sigma$, where $\sigma=1.2$ mJy beam$^{-1}$ is the rms noise. It is necessary to do this Gaussian fitting in order to calculate each line's integrated intensity because the linewidths usually imply that the  \replaced{$5_K\shortrightarrow4_K$}{$5,K\shortrightarrow4,K$} $K=0,1,2$ transitions are blended.  We obtain best fitting parameters (column density, temperature, central velocity and FWHM of the line) by minimizing the squared differences weighted by the inverse variances. We minimize and calculate formal uncertainties following the procedure described in \citet{Lampton1976ApJ} implemented by the package \emph{Minuit} within the \emph{Perl Data Language}.

While the temperature characterizing the excitation of different $K$-levels is close to $T_K$, this is not necessarily  the case for the relative $J$ populations and these may be characterized by non-LTE equilibrium. However, following \citep{Bergin1994ApJ} and estimating the collisional cross section of \propyne\ using CH$_3$CN, we conclude that the critical density for the \replaced{$5_K\shortrightarrow4_K$}{$5,K\shortrightarrow4,K$} transitions is $\approx1.5\times10^4$ cm$^{-3}$. Using the density profile $n(r)=1.4\times10^5(0.1\text{pc}/r)^{1.9}$ cm$^{-3}$ proposed for \clump\ by \citet[][$r$ being the radius from \hmyso]{Guzman2010ApJ}, we determine that the density of the clump is above the critical density for $r<0.34$ pc or for projected angular radii $\le40\arcsec$, that is, encompasing all detected propyne emission. Thus, to calculate the total column density of \propyne\ we use $T_K$ and assume LTE conditions.

\begin{figure}
\includegraphics[angle=-0, ext=.pdf, width=\textwidth]{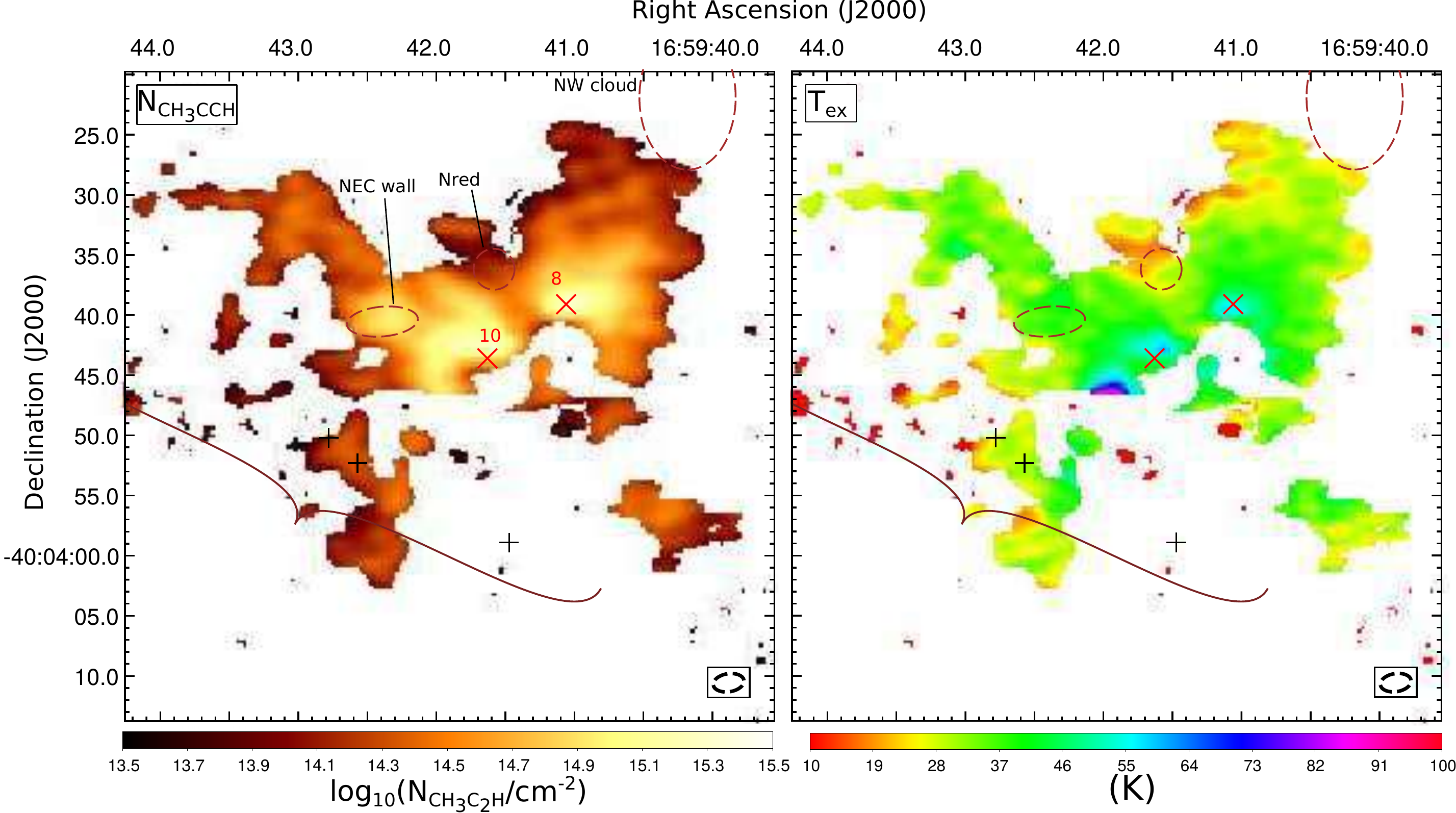}
\caption{Propyne (CH$_3$CCH) physical parameters toward \clump. Left and
  right panels show the column density and excitation temperature,
  respectively.  We show in each panel the location of the NW cloud, the
  N-red cloud, the NEC-wall, the Diffuse Ridge, the position of DR (a),
  (b), and (c) (black crosses); and continuum sources 8 and 10 (red
  crosses). \label{fig-prop}}
\end{figure}

Figure \ref{fig-prop} shows the results of the fitting to the \propyne\ lines. Uncertainties are shown in Figure \ref{fig-errorPropyne}.
Both the excitation temperature and column densities are consistent with previous single dish observations of the \replaced{$5_K\shortrightarrow4_K$}{$5,K\shortrightarrow4,K$} lines by \citet{Miettinen2006AA}.  They found a temperature of $35.9$ K and a column density of $1.5\times10^{15}$ cm$^{-2}$ toward \clump, which was the largest among their sample of 15 high-mass clumps. In general, we find the propyne column densities consistent with that given by   \citet{Miettinen2006AA}.
%Despite the column density reported by \citet{Miettinen2006AA} is beam diluted, we do not find much difference between our values given in Figure \ref{fig-prop} and theirs. It is possible this is  a consequence of the loss of spatially extended flux in our data.

As shown in Figure \ref{fig-cen2} and described in Section \ref{sec-dcom}, \propyne\ is not particularly intense toward the CC core.  Regions north of CC core and C8 are associated with the highest column densities of $\approx2\times10^{15}$ cm$^{-2}$. These regions   are associated with temperatures of typically $60\text{--}65\pm5$ K. The highest temperatures are $\gtrsim85\pm30$ K  and they are detected toward low column densities regions. Generally, we detect warmer temperatures to the center of the clump than toward the outskirts. Averaging the \propyne\ temperature in annuli of $2\arcsec$ starting from \hmyso\ we find that the radial temperature profile is well characterized by a decaying power law given by $80~\text{K}(r/0.01\text{ pc})^{-0.3}$. This temperature dependance is shallower than the one suggested for the dust temperature by \citet[][$\propto r^{-0.4}$]{Guzman2010ApJ}, which may reflect dust and gas requiring densities above $10^{5}$ cm$^{-3}$ to be thermally coupled. 

Our results are in general agreement with what other studies have
found toward high-mass star forming regions. \citet{Gibb2000ApJ},
using single dish data taken toward the HMC G327.3$-$0.6, determined
\propyne\ temperatures of $72$ K and column densities of
$2.9\times10^{15}$ cm$^{-2}$, respectively. Therefore, they 
suggest that this hydrocarbon better traces the warm, extended
component rather than the hot gas. Based on the rotational temperatures derived toward seven HMYSOs,
\citet{Bisschop2007AA} classified propyne as a ``cold'' molecule.
%whose formation path could be entirely attributable to gas chemistry.
Interferometer studies toward three ``organic poor'' HMYSOs --- which are expected to be younger and less chemically evolved than  HMCs --- indicate that \propyne\ is as abundant far from the HMYSO as close to it, thus classifying it as an envelope molecule \citep{Oberg2013ApJ,Fayolle2015AA}. Propyne is also characterized by temperatures between 40--60 K\replaced{which follow a profile (at least in NGC 7538 IRS9) roughly $\propto r^{-0.33}$ \citep{Oberg2013ApJ}}{and, in NGC 7538 IRS9, by a temperature profile which roughly follows $\propto r^{-0.33}$.}. 
 
Being an unsaturated hydrocarbon, there are in principle efficient ion-neutral gas formation routes for \propyne\ \citep{Schiff1979ApJ}.
%% Because \clump\ is already somewhat evolved and characterized by temperatures $30$-$35$ K, it is likely that several mechanisms are currently at work.
%% Particularly, these temperatures increase greatly the mobility of free radicals on the dust ice layers, boosting their chemical complexity \citep{Garrod2008ApJ}.
Concordantly, there is no evidence that in \clump\ the formation of a significant fraction of \propyne\ has occurred in dust grains.  First, the zero moment of all unsaturated molecules (except \ethe) including \propyne\ are classified together within the Continuum group, consistent with ion-neutral gas reactions which tend to form unsaturated species.  Second, let us assume that a large fraction of \propyne\ is formed on dust grains and liberated afterwards to the gaseous phase.  This would imply that \propyne\  should be well correlated with other molecules formed in grains, of which one of the best established examples is methanol. However, methanol emission does not correlate well with propyne  as shown in Section \ref{sec-corr}.   Finally, we note that the continuum correlates well with propyne in \clump\ \deleted{(\propyne\ is the continuum's MCP)}\explain{This is a mistake. \htcop\ is the MCP of the continuu,, as stated before.} and in other high-mass clumps \citep{Giannetti2017AA}.  This is expected if its formation is dominated by gaseous ion-neutral reactions because these depend crucially on the high-energy ($\ge100$ MeV) cosmic ray ionization, which is homogeneous throughout the clump \citep{Herbst1973ApJ}.  That is, propyne's abundance seems to depends more on the total column density of material rather than other circumstances like the presence of shocks, a higher temperature, or special illumination.  We conclude that the good correlation of \propyne\ with the rest of the unsaturated species and the continuum, as well as the lack of correlation with \met\ and with shock tracers like SiO, are consistent with the ion-neutral gas reactions forming a significant fraction of propyne in \clump.

{\subsection{Column Densities and Excitation Temperatures}\label{sec-cdt}}

To determine  excitation conditions and column densities we fit the molecular  emission lines using simple models. It is possible to constrain the excitation state of species with several observed lines such as \met\ and \propyne.  Other molecules such as SO, SO$_2$, OCS, HNCO, \prop, \cyc, \hccn, \cyan, \acry, and \damm\ also have several transitions which help determining their excitation conditions, but these higher excitation lines are only detected toward specific sources.  For the rest of the molecules  we detect either only one  transition or the observed lines are unsuitable for discerning the excitation conditions of the gas (e.g., CCH and \htcn).

The sources for which we model the emission spectra correspond to those features identified in Section \ref{sec-morph}. Spectra for each of the sources are obtained by taking the primary beam corrected intensity (in K) versus frequency either toward specific directions  or spatially averaging the intensity in the solid angle of the source. Spatial integration can improve the signal-to-noise ratio of the spectra if the  physical conditions of the gas in the integrated area do not vary too much and the identified feature form a coherent physical structure. Otherwise, spatial integration may complicate the interpretation of the spectra and even smear out faint lines.

Spectra of the following  sources are analyzed:
\begin{itemize}
\item{\emph{CC core.} Its spectrum is taken toward the peak methanol position, that is, $0\farcs7$ in the $\text{P.A.}=-64\arcdeg3$ direction from \hmyso. Emission from this position avoids most of the red-shifted core nearby (see Section \ref{sec-morph}).}
\item{\emph{C8.} We judge  C8 not being a completely coherent structure, with significant differences between positions closer to Source 8 and those closer to the maser \texttt{a}. Hence, we split the emission in two sources: a 1\farcs5 radius circular region around Source 8 and  the maser \texttt{a} position.}
\item{\emph{N-red cloud.} We spatially average the  emission in the region marked in Figure \ref{fig-consp}.}
\item{\emph{NW-cloud.} We spatially average the  ellipse  marked in Figure \ref{fig-consp}. Because this source is rather large, in order not to fade out some weak lines we also take the \deleted{the} spectrum toward the position marked with NW(a) in Figure \ref{fig-consp}. NW(a) correspond to the peak position of methanol \added{$({\rm E_1})$} $2_{1,1}\shortrightarrow1_{1,0}$ in the NW-cloud. }
\item{\emph{NEC-wall.} Toward this source we consider the spectra in two locations corresponding to the methanol and sulfur monoxide peaks, marked in Figure \ref{fig-consp} with NEC-w(a) and NEC-w(b), respectively.}
\item{\emph{Diffuse Ridge.} The Diffuse Ridge is a much more elongated feature with varying characteristics along its extension. The size of the Diffuse Ridge also means it is likely affected by short baseline filtering.  We select three positions to analyze the Diffuse Ridge, marked from (a) to (c) in Figure \ref{fig-consp}.  Positions DR(b) and DR(c) correspond to the location of two cores in the Diffuse Ridge which have counterparts in several molecules. DR(a) is located on more diffuse gas forming the body of this ridge or filament.}
\end{itemize}

%For the N-red cloud spectrum we integrate the $1\farcs7$ radius  circle indicated in Figure \ref{fig-consp}.

We model the emission of all molecules except \met\ using a single excitation temperature (SET) model \citep{vanderTak2011IAUS}, that is, we assume one excitation temperature per line of sight.  Due to the many  detected lines of \met\ and because they are usually affected by non-LTE excitation, we model its emission using Radex  \citep{vanderTak2007AA}.
%We do not assume that the lines are optically thin, but in practice, most of the emission lines are compatible with this assumption.
We assume, unless explicitly stated, that the beam filling factor of the emission is 1. Therefore, derived column densities  are beam-averaged. Of course, in the cases of extended sources with spatially averaged spectra (e.g., the N-red cloud) these are source-averaged.
For the line profiles, we model them as Gaussians with a single  central velocity ($V_{\rm LSR}$) and FWHM ($\Delta V$) for all transitions from a specific molecule. We stress that due to the Hanning
smoothing  of the ALMA data, the effective spectral resolution is  $976$ kHz. That is,
the instrumental broadening amounts to $\approx3.0$ \kms.
In the SET model, the excitation temperature ($T_\text{ex}$), column density ($N$), $V_{\rm LSR}$,  and $\Delta V$ are free parameters.  For methanol, free parameters are $T_K$, the column density, $V_{\rm LSR}$,   $\Delta V$, and density of the main collision partner (assumed H$_2$). For simplicity, we assume  equal abundances  of the   E- and A-\met\ symmetry states  because the kinetic temperature (Section \ref{sec-prop}) is always larger than the 7.9 K energy difference (in $k_B$ units) between the ground states of E- and A-\met\ \citep{Friberg1988AA}. The use of Radex for other molecules is hindered by the detection of only one or two transitions, which makes the modeling unreliable. We find optimal parameters by minimizing the squared difference  between the data and the model, weighted by $\sigma^{-2}$, where $\sigma$ is the uncertainty  of the primary beam corrected data (in K). Typical $\sigma$ for single pixel spectra (that is, not for spatially integrated) ranges in $0.04$--$0.06$ K. To minimize and  calculate formal  uncertainties we use \emph{Minuit} and follow the prescription in \citet{Lampton1976ApJ}. Tables \ref{tab-met} and \ref{tab-cdt} and Figures \ref{fig-SpecCCcore1} to \ref{fig-SpecS18} show the results of the SET and Radex modeling.

Table \ref{tab-met} shows the results of the Radex modeling of the \met\ lines. Columns (1) to (5) indicate the source, $T_K$,  the logarithm of the column density in cm$^{-2}$ ($\log\left(N\right)$), $V_{\rm LSR}$, $\Delta V$, and H$_2$ density, respectively. Column (6) remarks some noticeable characteristics of the fittings or of the data. For some sources (NEC-wall, C8, and NW-cloud) we exclude from the fitting the \met, \maser\ maser transition because this strong, non-thermal line would require opacities $<-1$.  These are characteristic of strong masers and cannot be adequately modeled by Radex \citep{vanderTak2007AA}. In other sources, for example toward the CC core, lines are well modeled assuming LTE conditions, which implies a lower bound on the density. According to Radex, densities $\gtrsim10^{10}$ cm$^{-3}$ are needed to thermalize the masering \maser\ line. Densities $\ge10^7$ cm$^{-3}$ are usually enough to thermalize the rest of the observed methanol transitions.  For those sources in which all methanol lines are thermalized except the \maser\ we give a range of compatible densities.

Comparing the \maser\ maser transition with the expected LTE intensities is useful to confirm the strong non-LTE effects on these lines.
The maser spots \texttt{a}, \texttt{b}, and \texttt{c} identified in Figure \ref{fig-mas} and whose parameters are given in Table \ref{tab-mas} are associated with  antenna temperatures between 200 and 400 K. These values are larger than the expected LTE emission by factors of 200, 1300, and 200, respectively.  Assuming the angular size of the maser emitting regions covers less than a third  of the beam size we obtain brightness temperatures ranging between 1500 and 3000 K. In addition, the linewidths of masers \texttt{a} and \texttt{c} given in Table \ref{tab-mas} are also slightly narrower by $\approx0.6$ than the linewidths of the rest of the \met\ lines, which is also a characteristic of masers. The \met, \maser\ line corresponds to a class I maser, that is, it is collisionally excited followed by  spontaneous radiative decay \citep{Cragg1992MNRAS}. These are the first class I methanol masers detected toward \clump; the only \met\ maser detected previously is the class II (radiatively excited) 6.7 GHz maser MMB345.498+1.467 \citep{Caswell2009PASA} detected toward Source 18. We note that the \met\ class II maser is not associated with the most luminous source \hmyso, but with the apparently more evolved and less embedded Source 18. In fact, neither class I nor II \met\ masers are  associated directly with \hmyso, but they appear scattered throughout the clump. This is also the case for the masers observed toward IRAS 16547$-$4247 (including the  \maser\ transition), another clump believed to be in a similar evolutionary stage as \clump\ \citep{Voronkov2006MNRAS}. Interestingly, for both clumps, not the methanol but the  OH masers are associated with the central dominating HMYSO.

Table \ref{tab-cdt} shows the best-fit SET parameters for the rest of the molecules toward each source. Columns (1) to (5) indicate the molecule, $T_\text{ex}$ ($T_K$ for \met), $\log\left(N\right)$, $V_{\rm LSR}$, and  $\Delta V$, respectively. In general, the $V_{\rm LSR}$ of different molecules are not the same and  it is not rare to find differences of $\sim3$ \kms\ or more between different molecules for the same line of sight.  For several sources,  it is possible to gather together  sets of molecules which share similar velocities. 
Column (6) in Table \ref{tab-cdt} identifies the number of the group to  which each particular molecule belongs. Considering the limited velocity resolution of the data,  two groups are sufficient to account for the $V_{\rm LSR}$ variations in each source. As a criterion for separating the two groups of  $V_{\rm LSR}$\deleted{s}, we require  the  internal standard deviation of each group being less than half of the standard deviation of the set of all the $V_{\rm LSR}$\deleted{s} of the source. \added{Molecules with a very different $V_{\rm LSR}$ compared with either group (e.g., SiO or optically thick CS)  are not classified.} We find that two groups describe adequately the $V_{\rm LSR}$ distribution of the  CC core, the N-red cloud, the NEC-wall(b),  DR(b), the  NW cloud, and of  the  NW(a). For the rest of the sources, they are either well characterized by a single $V_{\rm LSR}$ or the velocities have a large dispersion which cannot be grouped in two well differentiated sets.  \deleted{Molecules with a very different $V_{\rm LSR}$ compared with either group (e.g., SiO or optically thick CS)  are not classified.}

Table \ref{tab-cdt} gives formal uncertainties for the best-fit parameters, except for those which have been assumed or kept fixed during the minimization. Note that most of the $T_\text{ex}$ are assumed for the SET fitting in Table \ref{tab-cdt}. The procedure to assign the assumed $T_\text{ex}$ for each molecule start with molecules for which it is possible to derive a temperature (generally \met, \propyne, and SO). The \met\ and \propyne\ temperatures  are  estimators of the kinetic temperature of the gas. SO, on the other hand, is  associated with rather low $T_\text{ex}\le20$ and thus it is likely sub-thermally excited. In most cases, the assumed temperature for molecules without an independent $T_\text{ex}$ determination is one of these three. 

In order to assign a sensible $T_\text{ex}$ to molecules for which an independent temperature estimation is not possible, we first check its $V_{\rm LSR}$ group. Two molecules having very different $V_{\rm LSR}$\deleted{s} is evidence that they trace different gas and, therefore, assuming the same $T_\text{ex}$ is not justified. In addition, we asses the critical densities\footnote{Collision and Einstein   coefficients have been obtained from the \emph{Leiden Atomic and Molecular Database}  \citep[][\url{http://home.strw.leidenuniv.nl/~moldata/}]{Shoier2005AA}} 
associated with the lines.  Considering that the critical density of the detected SO transitions ranges between $1.6$--$2.9\times10^5$ cm$^{-3}$, we assign the same temperature as SO to all molecules whose transitions have critical densities above or equal to that of SO, that is, to \hctn, HNCO, SiO, CCH, HCN, CS, and to their \replaced{isotopomers}{isotopologues}.  An exception to this rule occurs when the derived CS peak optical depth --- assuming $T_\text{ex}$ from SO  --- is above 2. Because the CS, $J=2\shortrightarrow1$ line has a similar critical density as SO, high optical depths could reduce the effective critical density and bring the CS line closer to LTE due to radiative trapping \citep{Shirley2015PASP}.  We assume that molecules with transitions associated with critical densities lower than that of SO (like \htcop, \damm, \hcsp, and OCS) are closer to LTE, and therefore, we use either the \met\ or \propyne\ derived $T_\text{ex}$, depending on which one  is closer in $V_{\rm LSR}$.
We use \met\ or \propyne\ temperatures for molecules without collisional excitation parameters (like several COMs) and in the case SO is not detected.
%%%%%For \propyne\ we assume a critical density of  $1.5\times10^4$ cm$^{-3}$ as in section \ref{sec-prop}. 

Column (7) gives additional information about the fitting and noticeable characteristics of the spectra, for example, whether the line has absorption features or if it is associated with  high optical depths. In Table \ref{tab-cdt},  the $T_\text{ex}$ entries which have not been derived from the fit have a superscript ($^\dag$, $^\ddag$, or $^{\dagg}$). This marker  indicates  the origin of the assumed temperature for the SET fitting: the same marker appears in column (7) of the molecule from which $T_\text{ex}$ was adopted. \replaced{In this column}{In a few cases and lacking a better alternative,} we also indicate \added{in this column} whether we use   $T_\text{ex}$ from a molecule within  a different velocity group or from another source. 

Note that Table \ref{tab-cdt} includes the \met\ parameters of Sources 3 and 18.  Source 3 is 
a continuum source with a spectral index characteristic of dust emission \citepalias{Guzman2014ApJ}. The \met\ fitting is of a rather discrete quality, but we emphasize that this is one of the few lines of sight with emission in the \deleted{$E_\text{up}=340.1$ K} \added{$({\rm E_1}) J_{K_aK_c}=6_{1,6}\rightarrow5_{0,5},\,v_t=1$} transition \added{($E_\text{up}=340.1$ K)}. 
Source 18 was identified in \citetalias{Guzman2014ApJ} as an HC \hii\ region associated with a HMYSO less massive than \hmyso. Toward Source 18  we detect \met\ lines including  high energy ($E_{\rm up}>100$ K) transitions. Source 18 is one of the uncommon cases of a relatively isolated continuum source with an unambiguous line counterpart and thus a YSO with a reliable $V_{\rm LSR}$.

Figures \ref{fig-SpecCCcore1} to \ref{fig-SpecS18} (appendix \ref{sec-figSpec}) show the SET and Radex best-fit models  and the spectra in primary beam corrected K versus frequency. 
Individual panels show usually one transition each, but some panels show a few 
closely spaced  lines (e.g., for \propyne) of the same molecule. The name of the molecule and the upper energy level of the  transition are displayed on top and in the top left corner of each panel, respectively.
Models and data are shown in red and black, respectively. The green bar  indicates the frequency range used to calculate the squared difference between data and model. Some panels with faint detections show  the $\pm2.5\sigma$ level, where $\sigma$ depends on the specific spectrum and it is given in the caption.

\begin{figure}
\centering\includegraphics[height=0.23\textheight]{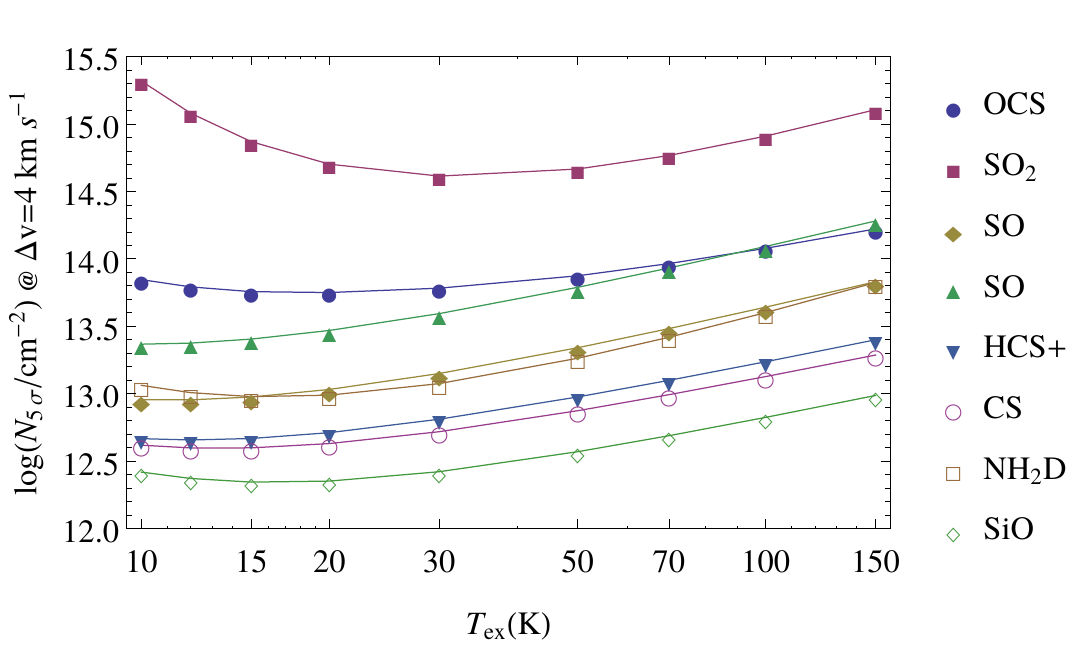}%
\includegraphics[height=0.23\textheight]{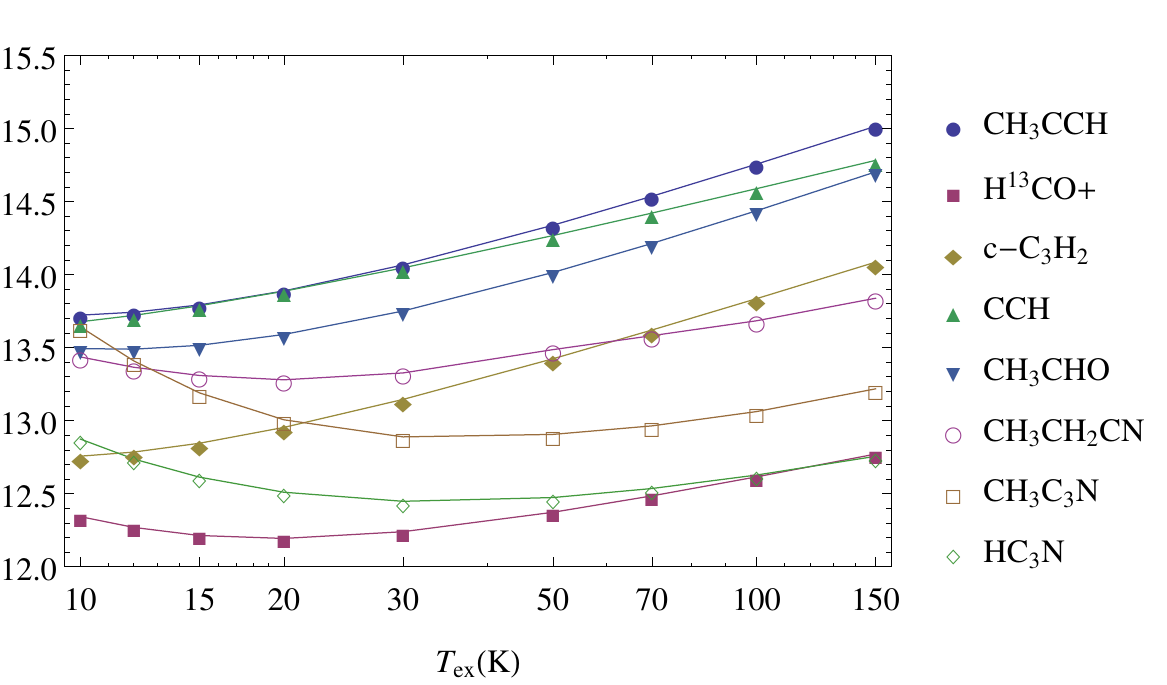}
\includegraphics[height=0.23\textheight]{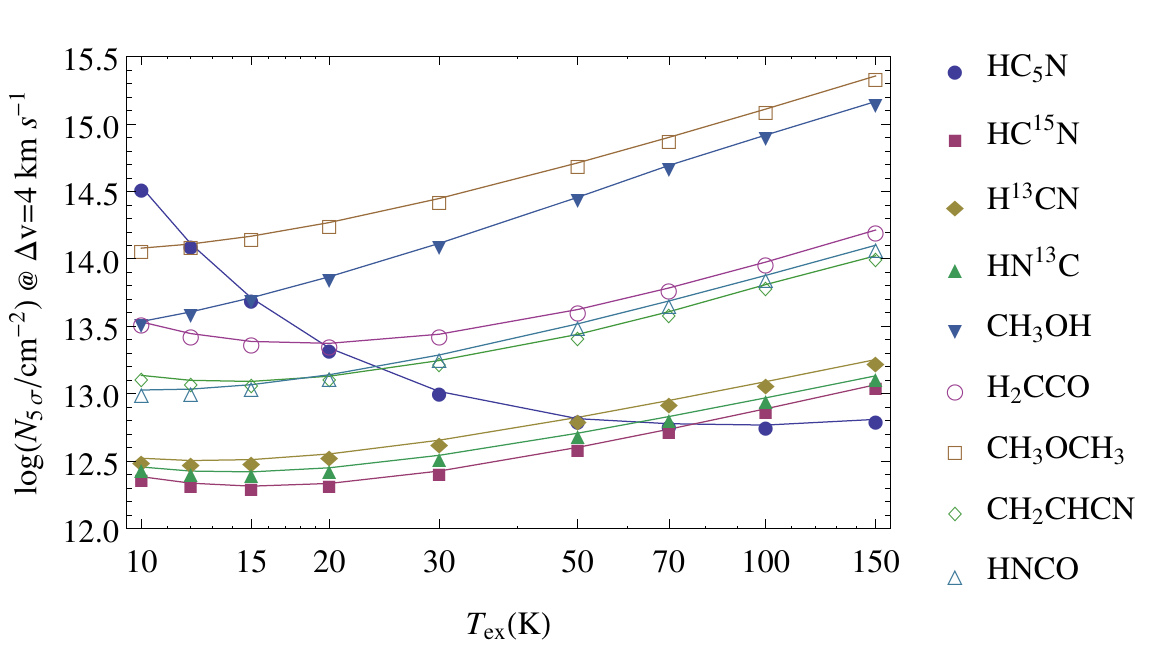}%
\includegraphics[height=0.23\textheight]{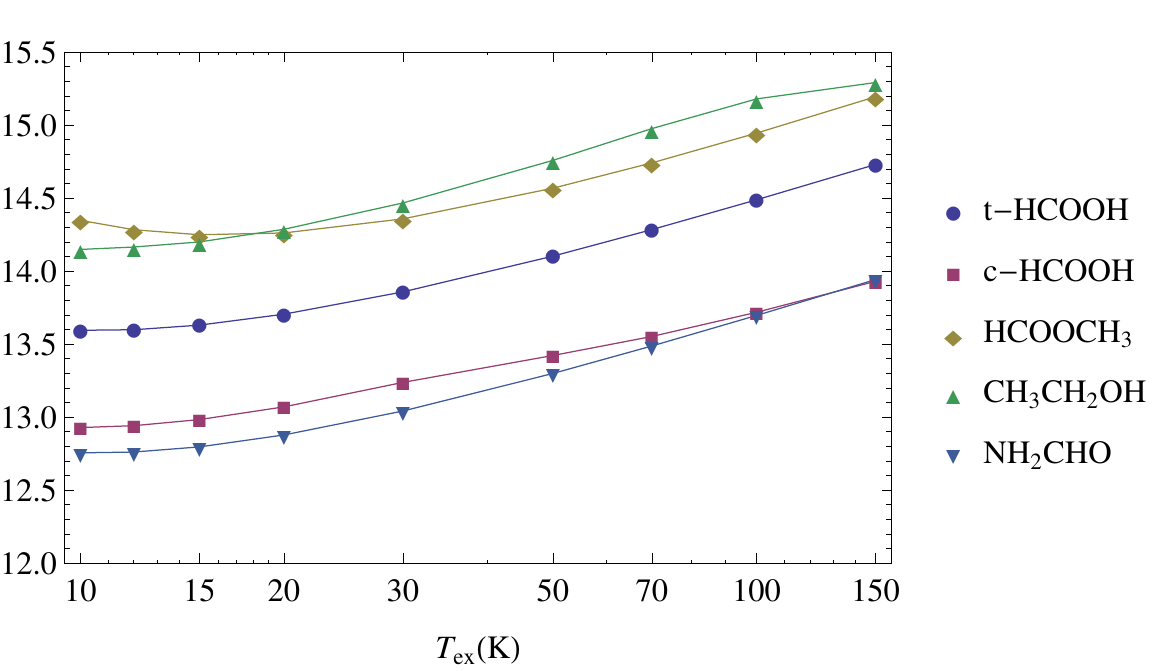}
\caption{Each curve shows the minimum  column density necessary to produce a Gaussian line peak
  of $5\sigma\approx0.3~\text{K}$ and $\text{FWHM}=4$ \kms, in at least one
  transition in our spectral coverage, versus the excitation temperature (assuming LTE). \label{fig-upp}}
\end{figure}

Finally, we emphasize that the lack of a molecule entry for a specific source in Table \ref{tab-cdt} is due to non-detection. The exception are the sulfur oxides for the CC core (SO, SO$_2$, and isotopologues), which are not present in the table because their spectra were analyzed in detail in \citetalias{Guzman2014ApJ} and their emission  cannot be  well fitted by a single temperature model. The general criterion to discard a detection is the absence of  any line  attributable to the molecule  over $2.5\sigma$ at a  $V_{\rm  LSR}$ between $-5$ and $-20$ \kms. However, in some cases detection of the main isotopologue of a species lends credibility to fainter spectral features located at the same  $V_{\rm LSR}$.  The non-detections are judged upon visual inspection of the spectra in the positions where the strongest lines are expected.  Figure \ref{fig-upp} shows the minimum column density needed to produce a Gaussian peak of $2.5\sigma\approx0.143$ K with $\Delta V=4.0$ \kms\ versus temperature, assuming LTE conditions for different molecules. This $\Delta V$ is typical of high-mass star formation regions but, in any case, the diagrams are easily scalable to other values because the peak of optically thin lines are proportional to the column density and inversely proportional to $\Delta V$. Figure \ref{fig-upp} includes the minimum column density of some species typical of hot-cores which are not (or only dubiously) detected in this work.

%This is related with the distinction we make in this work between LTE and SET models: the latter could  be a reasonable approximation for non-LTE 
%Note that in the case of optically thin emission the filling factor and the column density are degenerate parameters.

%Source 18

%%%%

%%%%%%%%%%%%%%%%%%%%%%%%%%%%%%%%%%%%%%%%%%%%%%%%%%%%%%%%%%%%
%%%%%%%%%%%%%%%%%%%%%%%%%%%%%%%%%%%%%%%%%%%%%%%%%%%%%%%%%%%%

{\section{FRACTIONATION AND ISOMERIZATION IN \clump}\label{sec-iso}}

%{\subsection{Isotopologue and Isotopomer Fractionation in \clump}\label{sec-iso}}

%Determining isotope relative abundances are an effective way to study the
%Galactic nucleosythesis and star formation history.  

The isotopic composition of molecular gas --- presumably preserved until the beginning of further stellar nucleosynthesis --- is a relevant initial condition of star formation.  Deriving the proportions from molecular \replaced{isotopomer}{isotopologue} ratios is not trivial, and entails determining what are the mechanisms that produce molecular fractionation.  Because these mechanisms are sensitive to the present and past conditions of the molecular gas, characterizing them could in turn provide us with important constrains about the physical parameters of the clump (like the gas temperature).

Using the data presented in this work, we can study the
fractionation of the following groups of isotopologues: 
SiO, $^{29}$SiO, and $^{30}$SiO; 
\hcqn\ and \htcn; 
\htcop\ and \hcdop;  
$^{34}$SO, SO, and $^{33}$SO; 
CS, and C$^{33}$S; OCS and O$^{13}$CS; and 
 \met\ and $^{13}$CH$_3$OH. In addition to these,
 we study the \htcn/\hntc\ \replaced{isomer}{isotopomer} fractionation.
 In principle, we can asses the silicon, sulfur, and carbon fractionations more directly because the  isotopic difference between the molecules involve  only one  atom. On the other hand, for hydrogen cyanide and formylium  the differences between the observed isotopologues involve two isotopes from different elements, complicating  somewhat the interpretation. 

The following sections present the analysis of the isotopic and isomeric
fractionation toward \clump. Table \ref{tab-iso} shows a summary of the
typical observed fractionations. Emission from the SiO, HCO$^+$, and HCN
\replaced{isotopomers}{isomers} is extended, hence, Table \ref{tab-iso} gives the average
fractionation ratios measured toward the clump. The rest of the ratios
characterize emission arising  from the central core.

The isotopic proportion of many atoms are found to vary with
Galactocentric radius \citep{Wilson1999RPPh}.\replaced{Throughout}{We } \deleted{this section we} use $6.9$ kpc as the
kinematic distance from \clump\ to the center of the Galaxy.
Throughout this section we use square parentheses to represent
abundances.

 %Throughout this work, we use the following solar isotopic abundance
%ratios \citep{Asplund2009ARA&A}: [$^{14}$N:$^{15}$N]$_\sun$=[98.89:0.23],
%[$^{28}$Si:$^{29}$Si:$^{30}$Si]$_\sun$=[92.23:4.68:3.09],

\subsection{Silicon Fractionation}

SiO and $^{29}$SiO have been observed toward \clump\ using the SEST telescope by \citet{Harju1998AA} and \cite{Miettinen2006AA}. The observed lines profiles are symmetric, with large wings \citep[FWZP$=27.3$ \kms,][]{Harju1998AA}. Excitation temperatures reported by \citet{Miettinen2006AA}
%toward the entire sample
are typically $4.3\pm0.1$ K, derived from simultaneous observations of the $3\shortrightarrow2$ transition.  

The velocity integrated emission from the $2\shortrightarrow1$ lines reported in the literature toward \clump\ are $5.3\pm0.1$ and $0.35\pm0.3$ K \kms\ for SiO and $^{29}$SiO, respectively. Their ratio, $15.1$, is lower than the solar abundance value of $19.7$ \citep{Asplund2009ARAA}. It is also lower than the
mean ratio found recently by \citet[]{Monson2017ApJ}  of \mbox{[SiO]/[$^{29}$SiO]}=$17.9\pm1.1$, who also argue for the lack of a Galactic gradient in the silicon isotopic distribution.
The isotopic ratio varies within an interquartile range of $[9.6,17.2]$ (Table \ref{tab-iso}),
but with no  evident systematic spatial trend. 
Line ratios lower than the expected isotopic abundance proportion are usually interpreted  as evidence of optically thick emission, which probably characterizes the  main isotopologue ($^{28}$SiO) lines \citep{Penzias1981ApJ}. Optical depth   introduces an additional complication, because its effects and abundance changes are in principle degenerate (the study by \citealp{Monson2017ApJ} models the opacity broadening of the line to estimate $\tau$ independently).

From our ALMA data, we find line ratios between SiO and $^{29}$SiO ranging between $9.2$ and $13.6$ (interquartile range), with a median value across the clump of $11.2$.  We integrated the line emission in the same velocity range for each line ($-24$ to $0$ \kms) to calculate these values, but we note that using the zero moment maps (obtained through moment masking) gives similar results.  The ratios we obtain are comparable to the ratios calculated by \citet{Monson2017ApJ} toward clouds in the central molecular zone of the Galaxy. \citeauthor{Monson2017ApJ} do not attribute these low values to fractionation, but explain them as due to optical depths  $\gtrsim1$.  This is likely only part of the explanation for the values obtained for \clump. We observe that the SiO/$^{29}$SiO line ratio increases to a median value of $13.7$ if we integrate the lines away from the central $V_{\rm   LSR}$. Presumably, emission from the line wings is more likely associated with optically thin column densities. However, the proportion is still low compared with the mean Galactic value. Because it is also possible that the main isotopologue line being affected by short spacing losses, we refrain from attributing the low [SiO]/[$^{29}$SiO] line ratios to fractionation. 

We derive $^{29}$SiO column densities toward the N-red cloud,  the S8 maser position,  the NW cloud, and  position (a) from the NW cloud, and found [SiO]/[$^{29}$SiO] values of, respectively, $14.8\pm1.0$, $19.5\pm2.0$, $18.6\pm2.3$,  and $12.3\pm1.7$, where the error bars represent formal uncertainties. These values are  somewhat higher that those obtained from the zero moment maps. It may be possible that, because they are associated with emission from rather compact sources, they could be less affected by short spacing losses. It is apparent that a fair assessment of the SiO fractionation toward \clump\ entails observing the likely optically thin line wings with an adequate short spacing coverage.

We expect the $^{29}$SiO and $^{30}$SiO emission to have opacities lower than those of the main isotopologue, and also being less affected by short baseline filtering. Indeed, the $^{29}$SiO/$^{30}$SiO velocity integrated line ratios are distributed within an interquartile range  of $[1.2,1.6]$ and a median of $1.4$. This value is the same as the median ratio obtained by \citet{Monson2017ApJ}, and it is very similar to the solar value  \citep[$1.5$,][]{Asplund2009ARAA}. We also derive $^{30}$SiO column densities toward the N-red cloud,  the S8 maser position,  and the NW cloud, observing little variation: we obtain [$^{29}$SiO]/[$^{30}$SiO] values of $1.5\pm0.1$, $1.2\pm0.1$, and $1.4\pm0.26$, respectively.

Finally, we emphasize that we do not find obvious spatial trend or pattern in the distribution of any of the SiO isotopologue ratios. This is in line with what is found by \citet{Monson2017ApJ}. Because the origin of SiO in the gas phase is likely dust sputtering in shocks, it is unlikely that this mechanism be sensitive to the rather small relative difference between the molecular masses of the SiO isotopologues. Furthermore, any chemical mechanisms affecting the SiO relative isotopic abundance in the clump \replaced{should}{probably} act on timescales larger than the depletion time of SiO onto dust grains, therefore, not generating detectable fractionation.
%Possible ways of testing this idea further would be to study the fractionation of other isotopologues like Si$^{18}$O.

%and other refractory silicon bearing molecules  like SiC. 
%% In our data, it is 
%% Nred        14.03\pm  0.02 12.86\pm  0.02 12.68\pm  0.02
%% S8maser     13.84\pm  0.03 12.55\pm  0.03 12.49\pm  0.04
%% NW cloud    13.52\pm  0.02 12.25\pm  0.05 12.09\pm  0.06
%% NW cloud(a) 13.81\pm  0.01 12.72\pm  0.06

\subsection{Sulfur Fractionation}

%In this section we analyze the relative abundances of sulfuretted isotopologues.  \citetalias{Guzman2014ApJ} already analyzed the SO/SO$^{34}$ fractionation toward the central HMC. Their results are consistent with the solar sulfur isotopic ratio of $22.5$.
%The rest of the sulfuretted isotopologue pairs are complicated to analyze because of either very high optical depths and self-absorption (like CS) or exceedingly faint detections, like $^{33}$SO.

In this section we analyze the relative abundances of the sulfur isotopologues $^{32}$S (main), $^{34}$S and $^{33}$S. 
The only source toward which we claim detection of $^{33}$SO is the CC core. Assuming an excitation temperature of $64$ K, the $^{34}$SO column density toward the CC core is $9.3\pm0.2\times10^{14}$ cm$^{-2}$. Combining this with the $^{33}$SO column density given in Table \ref{tab-cdt}, we obtain [$^{34}$SO]/[$^{33}$SO]$=5.4\pm0.6$. This value is remarkably close to the solar isotopic proportion [$^{34}$S]/[$^{33}$S]$=5.64$ \citep{Asplund2009ARAA}. The analysis of \citetalias{Guzman2014ApJ} also indicated that the $^{34}$SO abundance respect to the main isotopologue ($^{32}$SO) was consistent with solar abundance.  Note that the isotopic proportions of sulfur are predicted to vary little with Galactic radius \citep{Chin1996AA}: at $6.9$ kpc, the expected [S]/[$^{34}$S]$=27\pm5$ is still consistent with the solar abundance ratio.  We conclude that, toward the central HMC, the isotopic abundances of sulfur --- from SO emission lines --- are consistent with the solar value.
We refrain from analyzing the [CS]/[C$^{33}$S] ratio because in the directions where we detect C$^{33}$S, the CS line is optically thick.

\subsection{Carbon Fractionation}

We can compare the abundance of $^{13}$C-bearing molecules with their main isotopologue using lines from O$^{13}$CS, \tmet, and \hctccn. We avoid using the lines of H$^{13}$CCCN and HCC$^{13}$CN because they are blended with lines of HNCO and \acry, \added{respectively}.

O$^{13}$CS and \tmet\ are only detected toward  the CC core. 
Using the column densities of Table \ref{tab-cdt}, we obtain [OCS]/[O$^{13}$CS]$=25\pm2.4$, assuming the same excitation temperature for both molecules. This ratio is below the expected value for [C]/[$^{13}$C]$\sim60$ \citep{Wilson1999RPPh,Milam2005ApJ} and it is similar to the values found by \citet{Tercero2010AA} toward Orion-KL. The most likely cause for this difference is that the OCS line is optically thick. LTE fittings with derived column densities compatible with the solar isotopic abundances require peak optical depth ($\tau_p$) of $1.65$, filling factors $\lesssim0.3$, and excitation temperatures higher than $100$ K.
%However, we note that the OCS lines are symmetric and do not show self-absorption characteristics  which would suggest optically thick conditions. 

The isotopologue ratio calculated from the methanol column densities gives [\met]/[\tmet]$=26\pm4$. This number is comparable to the one obtained above using OCS, and it is a factor of $\sim2$ less than the values derived for Orion by \citet{Persson2007AA}. Similarly as in the case of OCS, LTE column densities derived for an optically thick  \met\ line ($\tau_p\approx2$ for the \maser\ transition) with a filling factor of $0.24$  give isotopic ratios $\sim50$, closer  to the expected [C]/[$^{13}$C] value.

Finally, we can calculate the [\hctn]/[HC$^{13}$CCN] quotient toward the CC core, the NEC-wall positions (a) and (b), the C8 maser position, and the C8 1\farcs5 radius source. The ratios obtained from Table \ref{tab-cdt} for these five sources are $32\pm4$, $41\pm4$, $37\pm5$, $60\pm6$, and $50\pm4$, respectively. The error bars give the formal uncertainty, but there is a likely more relevant systematic uncertainty associated with the excitation temperature. Note that because we only detect one line for each isotopologue, the assumed excitation temperature is critical. The ratio obtained  for the CC core emission is close to the values obtained for O$^{13}$CS and \tmet. Again, the column density quotient between the two isotopologues can accommodate a value closer to $50$ if the opacity near the peak of the \hctn\ line is $1.9$ and the filling factor $\lesssim0.5$. The ratios increase with distance from the center of \clump: from the CC core to the NEC-wall, reaching values consistent with the expected isotopic $^{13}$C abundance for the C8 sources. This is consistent with the view that quotients lower than the expected isotopic abundance ratio are caused by opacity.

\subsection{Formylium Cation: Oxygen and Carbon Fractionation}

The median value we calculate for the velocity integrated flux ratio between \htcop\ and
\hcdop, within the range $-15~\kms<V_{\rm LSR}<-9~\kms$,  is 6.9, with an interquartile range of [4.8, 8.5]. This abundance ratio depend on 
carbon and oxygen isotopic abundances, which at the clump's location  are [$^{12}$C]/[$^{13}$C]=$59.4\pm18$ and [$^{16}$O]/[$^{18}$O]=$442.8\pm116$, respectively \citep{Wilson1999RPPh}.  We infer that [\htcop]/[\hcdop]=$7.5\pm3.0$. Using instead [$^{12}$C]/[$^{13}$C]=$63.4$ \citep{Milam2005ApJ} we obtain [\htcop]/[\hcdop]=$7.0$, which is within the previous value uncertainty.
The median value found toward \clump\ is
remarkably close to expected ratio, considering the uncertainties.

However, in contrast to the SiO isotopologue, we do see a systematic trend
in the \htcop/\hcdop\ line ratios: the central parts of \clump\ (within
12\arcsec\ from \hmyso) are associated with ratios higher than average,
with a mean value of $\sim8.5$. Regions farther away from the
center of the clump, like the emission associated with Sources 7 , 3 , and
2 (the Continuum arc) and to sources in south east like 15 and 17; exhibit ratios $\sim5$. This behavior of the line ratios is rather
surprising because it is the opposite to what would be expected if the
lines from the central parts of \clump\ are optically thick. In principle,
short spacing losses could affect the line ratios, but they would not
explain the spatial differences.

Short spacing observations are needed to confirm whether the effect is real
or due to filtering. Additionally, HCO$^+$ observations would determine
whether the cause of this decrement is fractionation, and of which
isotopologue.  Finally, we mention a possible path to increase the \htcop\
abundance near the center of the clump respect to \hcdop: it may merely
reflect $^{13}$CO being better (self) shielded than C$^{18}$O from
radiation arising from \hmyso, simply because of the lower abundance of the
latter.  We note that the HMYSO has contracted and its high energy
radiation has already ionized a small region in the center of \clump.  If
HCO$^+$ is formed mainly by combining H$_3^+$ and CO, an increase of
$^{13}$CO respect to C$^{18}$O could produce a corresponding increase in
the formylium isotopologue.

\subsection{Hydrogen Cyanide: Nitrogen and Carbon Fractionation}

\begin{figure}
\includegraphics[angle=-0, ext=.pdf, width=\textwidth]{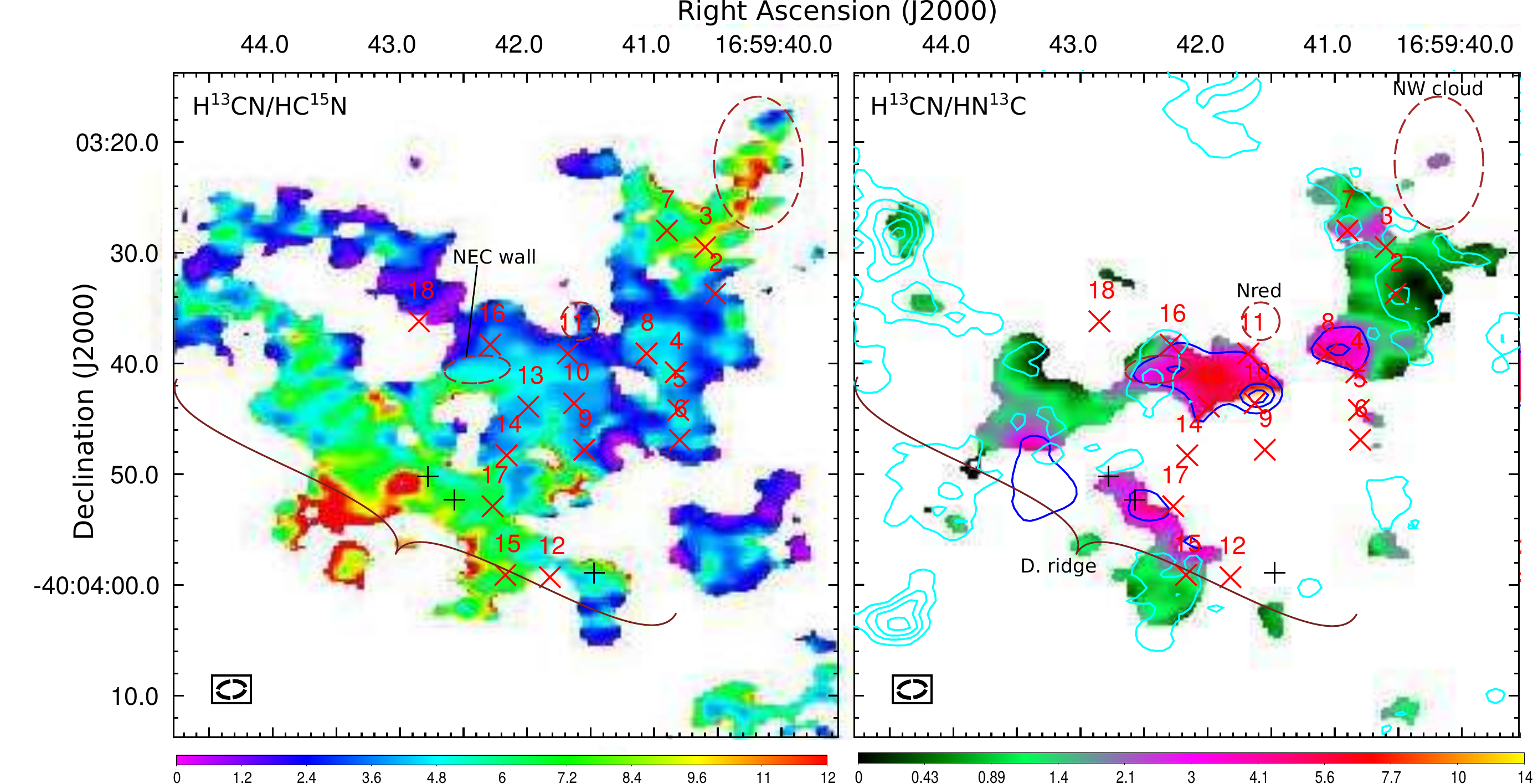}
\caption{\emph{Left panel.} Moment zero quotient between the \htcn\ and \hcqn, $J=0\shortrightarrow1$ transitions.  \emph{Right panel.} Moment zero quotient between the H$^{13}$CN and HN$^{13}$C, $J=0\shortrightarrow1$ transitions. Blue contours are drawn at the 25, 50, and 75\%\ of the peak H$^{13}$CN moment 0 emission. Cyan contours correspond to NH$_2$D emission as in Figure   \ref{fig-jhk}. In both panels we indicate the N-red, NW- and Diffuse ridge markers as in Figure \ref{fig-consp}.\label{fig-hcn}}
\end{figure}

Figure \ref{fig-hcn} shows in the left panel the quotient between the moment zero images of the \htcn\ and \hcqn, $J=0\shortrightarrow1$ transitions. These two molecules have different C and N isotopes, which means that their quotient depends on the fractionation of these two atoms. In practice, because of the better knowledge about the C vs.\ $^{13}$C proportion and because HCN transitions are usually optically thick, the [\htcn]/[\hcqn] ratio is commonly used to study the N vs.\ $^{15}$N fractionation. Single dish observations of \clump\ of the same isotopologue pair studied in this section  were performed by \citet{Dahmen1995AA}.

There has been some discrepancy in the literature respect to the
[N]/[$^{15}$N] behavior with Galactocentric radius. While
\citet{Wilson1999RPPh} --- based on \htcn\ and \hcqn\ observations
from \citet{Dahmen1995AA} and \citet{Wannier1981ApJ} --- suggests a value of $420$
at the distance of \clump,  \citet[]{Adande2012ApJ}, based on HNC and
  CN observations, proposes a  lower value of
$270$. The latter authors suggest that the difference is due to the
assumed [C]/[$^{13}$C] curves (illustrating the uncertainties
introduced by the double isotope corrections) but this is not clear
since there are already noticeable differences between their
[HN$^{13}$C]/[H$^{15}$NC] and the [H$^{13}$CN]/[HC$^{15}$N] values
given by \citet{Dahmen1995AA}. In view of these uncertainties,
we focus the quantitative analysis directly on the [\htcn]/[\hcqn]
ratio.

%Ideally, we have to calculate quotients between fluxes integrated within matching velocity ranges, as it was done in previously with SiO and HCO$^+$.  We note, however, that the conclusions of the previous sections  would not have been modified in any important way if we had used directly the ratio of the moment zero maps. Additionally, in the case of \htcn, selecting a velocity range is complicated by the blending of the  hyperfine components  due to the large linewidths. 

According to the image shown in Figure \ref{fig-hcn}, the mean \htcn/\hcqn\ line quotient (which is the same as [\htcn]/[\hcqn], assuming the same excitation temperature for both molecules) rises from values $\sim3.5$  near \hmyso, to typically  $4.0$ at 10\arcsec\ from the center of \clump. At these distances from the clump center, the  behavior of the quotient is dominated by the values obtained near the CC core, the NEC-wall and the C8. Farther away from the HMYSO, the mean ratio has a slightly steeper rise to reach values between $5.5$ and $6.0$ at 25\arcsec\ from the clump center. These values are dominated by the higher quotients measured toward the Diffuse ridge and the NW cloud. The values obtained for [\htcn]/[\hcqn] from the LTE fittings (Table \ref{tab-cdt}) for each source are similar to those read from Figure \ref{fig-hcn}. The mean [\htcn]/[\hcqn] is $5.7\pm2$, where the uncertainty represents the dispersion observed between different sources. This value is slightly lower, although within the uncertainties,  compared with that given  recently by  \citet{Colzi2018MNRAS} at 6.9 kpc.

Using single dish data, \citet{Dahmen1995AA} obtained [\htcn]/[\hcqn]$=6.32\pm0.16$ for \clump. It is not clear that the lower ratios obtained for the sources close to the center of \clump\ are compatible with this value using optically thick lines. For example, the ratio of 3.5 obtained for the CC core  is consistent with the value found by \citet{Dahmen1995AA} only if the \htcn\ line is associated with an optical depth at peak of $\sim4$. Such high opacity would affect the observed relative intensities of the hyperfine components, deteriorating noticeably the quality of the fits: we do not observe in general large hyperfine anomalies which could be evidence of an optically thick \htcn\ line. It is still not clear how much influence the short baseline filtering has on the line quotients, but at least for the CC core, it is not likely to dominate.

Hence, it is possible that there is real fractionation effect toward the densest cores of \clump. Measurements of the [\htcn]/[\hcqn] show that this proportion varies depending on the nature of the source.  For example, \citet{Wampfler2014AA} finds that the quotients associated with three low mass protostellar envelopes are $2.4$, $4.2$, and $5.3$. The causes for the different values are not entirely clear, but the lowest ratio (highest \hcqn\ fractionation) is associated with the hot corino IRAS 16293$-$2422 A. The latter source is somewhat colder and less affected by external illumination compared with the other two. \citet{Guzman2015ApJ814,Guzman2017ApJ} have also determined very low ratios ($\le3.0$) toward proto-planetary disks.  Of the two fractionation mechanisms suggested by \citet{Guzman2015ApJ814}, one of them only acts effectively in cold gas ($\le20$ K), which is not characteristic of \clump. The other, selective photo-dissociation (closely related with self-shielding, \citealp{Heays2014AA}), predicts that the [N]/[$^{15}$N] ratio should increase the more illuminated with dissociating radiation and diffuse a cloud is. In this case, the same self-shielding from the HMYSO radiation which may cause the \htcop-\hcdop\ spatial fractionation could also explain the observed [\htcn]/[\hcqn] ratios.

%The reason is that in dense clouds, N$_2$ can be optically thick (self-shielded) to dissociating radiation, whereas the rarer $^{15}$NN molecule is not. Dissociation of the latter injects equivalent amounts of the two isotopes, which could presumably will affect the isotopologue fractionation diminishing the [\htcn]/[\hcqn] ratio.

%IQR 2.5 3.6 5.2
%among all the sources the median is 5.5
{\subsection{H$^\textit{13}$\!CN/HN$^\textit{13}$\!C Isomer Ratio}\label{sec-isomer}}

We can explore the HCN to HNC isomer ratio in \clump\ using their $^{13}$C isotopologues, whose transitions have the advantage of being more likely optically thin compared to those of the main species. The HCN/HNC ratio has been found to depend on kinetic temperature \citep[e.g.,][]{Schilke1992AA}. Theory proposes that the  two molecules are produced in a ratio $\sim1$  mostly through dissociative recombination of HCNH$^+$ \citep{Herbst1978ApJ,Hirota1998ApJ}. Selective destruction of HNC through the neutral-neutral reaction $\text{HNC}+\text{H}\rightarrow\text{HCN}+\text{H}$ is expected to work at higher temperatures because of the presence of an activation barrier, depleting HNC in favor of HCN. The question of what is the energy of this activation barrier is still not completely resolved \citep{Graninger2014ApJ}.

The right panel of Figure \ref{fig-hcn} shows the quotient between the moment zero of the \htcn\ and \hntc, $1\shortrightarrow0$ lines. This figure shows a qualitative correspondence with what is expected for the HCN/HNC ratio. Away from the center of  \clump\ the temperatures are low and the \htcn/\hntc\ quotient is comparable to unity. 
%C8 maser, associated with a relatively low temperature ($$) and  
Close to the center of \clump\ and to Source 8  the temperature is higher,
\deleted{increasing} \added{and} the \htcn/\hntc\ quotient
\replaced{in an}{increases by one}
order of magnitude. It is also apparent that the molecular ratio is higher in the NE side of the Diffuse ridge, being presumably 
illuminated and heated more directly by \hmyso.
For the CC core, the column density given in Table \ref{tab-cdt} compared with the upper limits from  Figure \ref{fig-upp} suggests a ratio $\sim100$. Regions where we detect \damm\ in combination with \hntc, the \htcn/\hntc\ ratios have values close to unity, again consistent with the low temperatures expected to be traced by deuterated species \citep{Bergin2007ARA&A}. 

The right panel of Figure \ref{fig-hcn} shows the quotient between the moment zero of the \htcn\ and \hntc, $1\shortrightarrow0$ lines. There are significant differences between the \htcn\ and the \hntc\ emission, the most noticeable in the moment zero map is the lack of a CC core counterpart in \hntc, in contrast to  \htcn. In addition, it is not uncommon these lines having different $V_{\rm LSR}$, as shown for several sources in Table \ref{tab-cdt}. 

The differences can be understood as due to the depletion of HNC in favor of HCN in gas  at temperatures $>25$ K, which implies that  warmer gas along a given line of sight will weight more into the \htcn\ emission than \hntc. Arising from different locations, warm and cold gas do not necessarily share the same $V_{\rm LSR}$. This interpretation implies also that  the effective temperature of the HCN emission should be higher than that of HNC \citep{Jin2015ApJ}.
Conversely, low temperatures should be associated with HCN/HNC column density ratios close  to unity,  the same effective temperature for both species, and consistent $V_{\rm LSR}$.

%These caveats prevent us from interpreting directly the \htcn/\hntc\ quotient shown in Figure \ref{fig-hcn} as the abundance ratio between these two molecules.

%%%%%%%%%%%%%%%%%%%%%%%%%%%%%%%%%%%%%%%%%%%%%%%%%%%%%%%%%%%%
%%%%%%%%%%%%%%%%%%%%%%%%%%%%%%%%%%%%%%%%%%%%%%%%%%%%%%%%%%%%

{\section{DISCUSSION}\label{sec-discussion}}

We can gain some insight on the chemistry and physics of the clump by
comparing ours with independent observations. In Section \ref{sec-nir} we
compare our ALMA with NIR data taken from the literature. The latter are
some of the few data on \clump\ with comparable angular resolution and
coverage.  Finally, Section \ref{sec-chem} analyzes in more detail and draw
some conclusions about the possible chemical processes at work in \clump.

{\subsection{Near-Infrared Counterparts of Molecular Features}\label{sec-nir}}

\begin{figure}
\includegraphics[angle=-0, ext=.pdf, width=\textwidth]{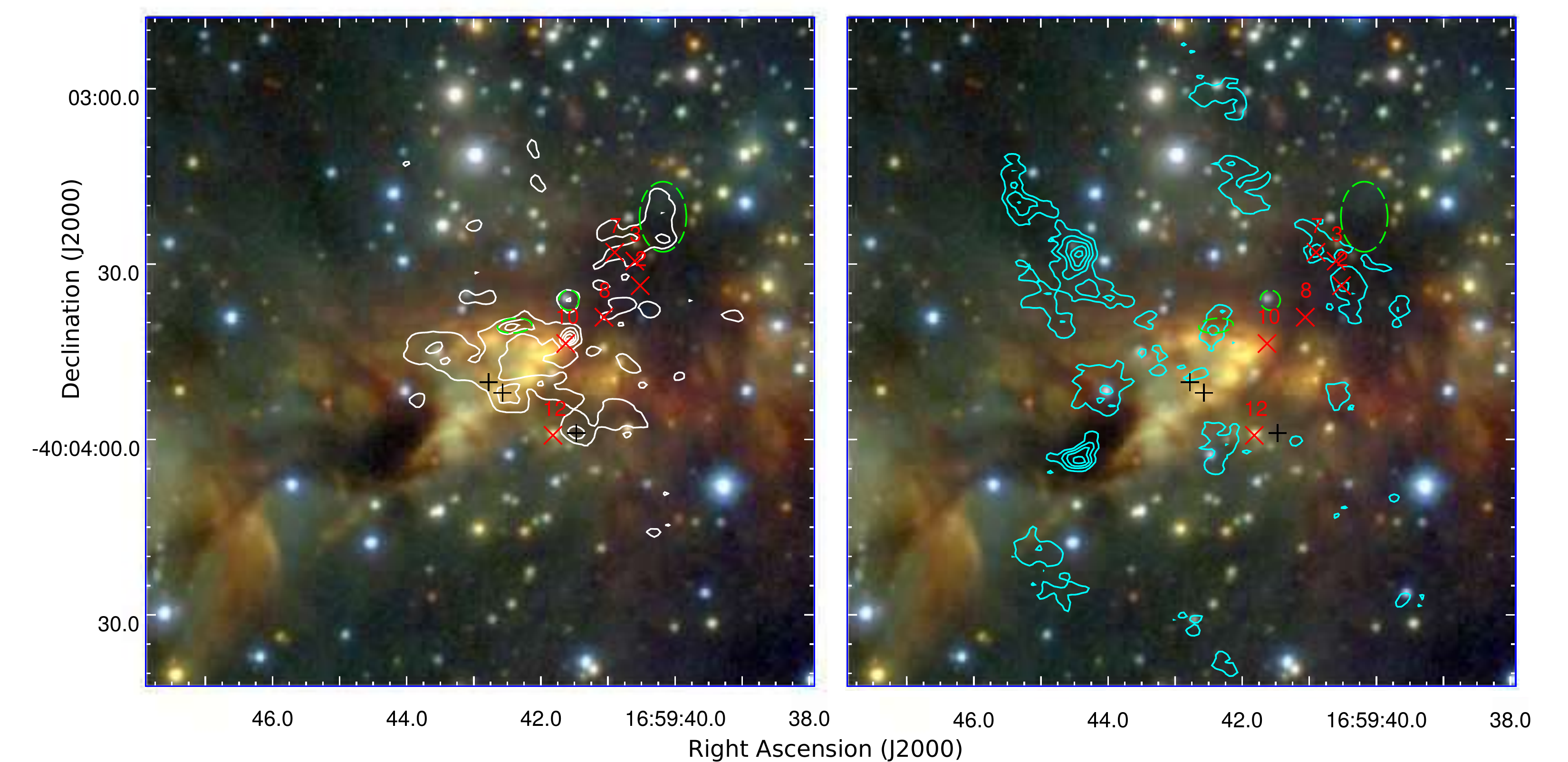}
\caption{Both panels show in the background three color NIR images (red,  green, and blue for K$_s$, H, and J band filters, respectively) obtained  from the VVV survey. \emph{Left panel.} White contours show  \met, $2_{1,1}\shortrightarrow1_{1,0}$ zero moment emission. Levels: 10, 30,  50, 70, and 90\% of the peak ($0.51$ Jy beam$^{-1}$ \kms). %Levels: 0.5, 1.5, 4.5, and 15 Jy beam$^{-1}$  \citep[$\sigma\approx60$ mJy,][]{Contreras2013AA}.
\emph{Right panel.}  Cyan contours show deuterated ammonia $(J,K)=(1,1)$ zero moment  emission. Levels: 10, 30, 50, 70, and 90\% of the peak ($0.10$ Jy  beam$^{-1}$ \kms). Red crosses  mark the continuum sources 2, 3, 7, 8, 10, and 12. Green-dashed ellipses show the location of the NW and N-red clouds, and the NEC-wall. Black crosses mark the DR(a), (b), and (c) positions.\label{fig-jhk}}
\end{figure}

Figure \ref{fig-jhk} shows a three color NIR image (JHK$_S$ filters)
of \clump\ using data from the VVV survey \citep{Minniti2010NewA}. We
have superimposed in the left and right panels contours of the zero moment \met\ and
\damm\ maps, respectively.
%The left
%panel also shows contours of the 870 \um\ continuum emission from
%ATLASGAL \citep[$19\farcs2$ beam size,][]{Schuller2009AA}.
\citet{Guzman2016ApJ} analyzed similar NIR images toward \hmyso,
describing among other features the illuminated blueshifted outflow
cavity and some continuum sources with NIR counterparts. In this
section we focus on the molecular emission counterparts.

There are three NIR conspicuous features which correlate with the
\met\ zero moment emission. The most evident is the CC core, which is
coincident with a NIR source associated with the HMYSO \hmyso.  As
already noted before, the CC core emission extends a few arcseconds in
the  north and north-east directions respect to \hmyso.  The second
feature is coincident with the
northern wall of the blueshifted illuminated cavity, particularly,
the strong \met\ emission associated with the NEC-wall.
Additional methanol emission seems to follow the location of the
southern wall of this cavity, although in a less clear way compared
to the NEC-wall.  The third feature is evident in the left panel of
Figure \ref{fig-jhk}, which shows that the Diffuse Ridge seen in the
\met\ contours coincides with a dark filament seen across
the brightly illuminated blueshifted outflow cavity.  From the
brightness decrement produced by the filament in the K$_s$ image,
which amounts to $A_{K_s}\approx0.3$ magnitudes, we derive an H$_2$
column density of $1.8\times10^{21}$ cm$^{-2}$
\citep{Rieke1985ApJ,Heiderman2010ApJ}. The median $V_{\rm LSR}$ of the
Diffuse Ridge is $-11.7$ \kms\ (Section \ref{sec-cdt}), which is
redshifted respect to the CC core.  A likely interpretation is that the
Diffuse Ridge material is in front of the outflow cavity, obscuring the
illuminated cavity, and possibly
falling toward the clump.

The right panel of Figure \ref{fig-jhk} shows \damm\ contours against the
NIR background. Also indicated in this panel are the positions of continuum
Sources 1, 2, 3, 7, 15, and 16. We can see that Sources 2, 3, and 7
delineate what we called in Section \ref{sec-results} the Continuum Arc,
which is also seen in \damm. The NIR image shows that the Continuum Arc is
embedded within a large IR-dark absorption region. The absence of stars
observed toward this area suggests large column densities.
%, which is also
%consistent with a slight distortion of the 870 \micron\ contours.
There is
a less evident NIR counterpart to the  \damm\ peak located in the
northeast part of \clump. The \damm\ emission corresponds with an
absorption patch where we see only most likely foreground stars, evidenced
by their blueish hue. The deuterated ammonia line peaks at $-15.1$
\kms\ and has a FWHM of $\sim4$ \kms. Note that the true linewidth is smaller and closer to $2$ \kms\ due to the instrumental broadening (see Section \ref{sec-obs}).

The most clear NIR counterpart to the \damm\ emission is located to the
southeast, coincident with a very clear IR-dark cloud $\sim15\arcsec$
size. This IR-dark cloud was already noted by \citet{Guzman2016ApJ} because
it is located nearby the outer-eastern ionized lobe of the \hmyso\ jet. The
spatial coincidence and the kinematics of the jet's lobe suggest that the
jet may be interacting with this cloud. Its \damm\ emission peaks at
$-11.7$ \kms\ with a FWHM of $4.5$ \kms, which implies that the dark cloud
is indeed part of \clump, supporting the previous interpretation.

IR-dark regions are usually associated with dense and cold starless gas
\citep{Rathborne2006ApJ}. In the case of \clump, the cold nature of the gas
associated with the dark IR regions is corroborated by their association
with \damm\ \citep{Bergin2007ARA&A} and by all the rest of the detected
spectral features being in absorption. Lines of CS, \propyne, CCH, \htcn,
\hntc, and \met\ appear in absorption toward both the south and northeast
\damm\ peak positions.

{\subsection{Chemistry in \clump}\label{sec-chem}}

Studying the chemistry of a single high-mass protostellar molecular clump is a complicated task since it  encompass many of the mechanisms expected to be at work in the ISM. Until the recent advent of instruments such as ALMA the chemistry of high-mass protostellar clumps had been studied using single-dish telescopes. These  mix different environments in their beams, having to rely on velocity differences to discern them. With high resolution and sensitive observations we  are now able to study in detail these regions, and spatially separate the emission of each molecule. Such studies are very useful to advance our chemical knowledge.  

%% In a broad sense, chemistry follows two paths: one associated with dust
%% grains, specially dust grain mantles, and the other associated with gas
%% phase chemistry.  The chemical products of these two paths gradually mix
%% together in the gas phase as the clump's increases its temperature, forming
%%  new molecules as well.

Near the end of the prestellar stage, a high-mass molecular clump is
characterized by temperatures $<15$ K, hydrogen column densities
$\gtrsim10^{22}$ cm$^{-2}$, and densities $\gtrsim10^4$ cm$^{-3}$. These
are the characteristics of the so-called quiescent infrared dark clouds
\citep{Guzman2015ApJ,Rathborne2006ApJ}. In this stage, dust grains are
covered by (mostly water) icy mantles together with other adsorbed
atoms  and molecules like H, CO, N$_2$, and possibly CS and H$_2$S
\citep{Viti2004MNRAS}. At these low temperatures, only H is mobile
on the dust's surface and it reacts with other molecules (like CO)
using the dust grain as a third body which helps dissipating the excess
energy.
Chemical products formed in the dust grain icy
layer during this stage are typically  hydrogenated
species like H$_2$CO and \met\ \citep{Vasyunina2014ApJ}. 
Meanwhile, the cold gas in the clump also evolves chemically,
specially through ion-neutral and barrier-less neutral-neutral reactions. A
permanent, small amount of cosmic-ray ionization facilitates the
gaseous ion-neutral reactions and formation  of simple unsaturated molecules.

At these early evolutionary stages, most of the molecules formed on  dust grains
 remain there: the temperature is too low to sublimate them from the
surface. About all of the ices will co-desorb with water  when the temperature reaches $\approx100$ K. However, even at low temperature several non-thermal
mechanisms can help desorbing some of these products, e.g., shock
sputtering, chemical desorption, and photo-desorption. Some of these
mechanisms --- not being entirely clear \emph{which one(s)} --- are
observed to be efficient at releasing large quantities of 
\met\ during the prestellar phase (e.g., \citealp{Sanhueza2013ApJ}; \added{\citealp{Vastel2014ApJ,Cosentino2018MNRAS}}).

As the clump evolves, it contracts, and eventually young stars are
born. These (HM)YSOs increase the clump's temperature and introduce
turbulence through outflows and winds.  Temperature rising to $20$--$30$ K
reduces the time spent by H in the surface of the grains, hampering further
hydrogenation reactions. The temperature rise increases
the mobility of heavier radicals in the mantles, allowing further reactions
and promoting the formation of more complex molecules. This phase of
chemical evolution receives the name of warm-up \citep{Garrod2008ApJ}.  The
radicals may be produced by photolysis of other molecules induced by secondary cosmic rays UV-photons. Some other species (CO, N$_2$,
O$_2$, CH$_4$) desorb with the increasing  temperature
\citep{Viti2004MNRAS}, potentially opening new \added{gaseous} chemical routes.

The third phase of chemical evolution comes only near the regions where the
temperature increases to $\ge100$ K, that is, very near the young stars and inside the 
HMCs. Here, the water ices completely evaporate and liberate all the
molecules formed in the dust mantles to the gas phase.  The temperature
rise also allows reactions with barriers and endothermic reactions to occur.
Characteristic chemical products generated during the three stages
described above --- cold prestellar, warm-up, and HMC --- are called by
\citet{Herbst2009ARA&A} zeroth-, first-, and second-generation species,
respectively.

Note that the dominating chemical mechanism in each stage 
is mainly determined by the temperature of the gas, which ultimately depends on the distance to
the young forming stars.  Protostellar clumps, like \clump, are expected
 to roughly follow the chemical timeline presented above. They
should likewise display all these stages at the same time depending on the
location of the material: cool pockets of starless ``pristine'' gas
remain present until the clump is ultimately dispersed; warm, less dense
gas presumably occupies most of the volume of the clump; and finally, hot
gas in the immediate surrounding of YSOs displays the chemical signatures
of HMCs. Following the processes and classification 
described above, we analyze in the next sections
some aspects about the chemistry of \clump. 

\subsubsection{HMCs and embedded YSOs}

% what is a HMC. Remember definition
HMCs are defined observationally as dense ($>10^7$ cm$^{-3}$), warm
(100--500 K), and compact ($<0.05$ pc) molecular cores \citep{Allen2017AA}.
The definition is not strict and it has changed at the same time that
telescopes resolve more compact and hotter components
\citep[]{vanDishoeck1998ARA&A}.  A rich spectrum including molecular
transitions of hot ammonia, water, and COMs like \met\ and CH$_3$CN
signpost the presence of a HMC.  In general, it is assumed that the heating
source of a HMC is a HMYSO located near its center.

% 4 possible HMCs in clump. Talk about the two less important
There are four compact continuum sources within \clump\ which are clearly
associated with high excitation molecular emission:  Sources
3, 8, 10, and 18. All of these are detected  in \met\ lines. Source 3 is 
a continuum source associated with hot methanol  ($T_\text{ex}\approx200$ K)
% and a column density of $1.7\pm0.3\times10^{16}$ cm$^{-2}$.
but no IR counterpart, which  prompts us concluding that
it correspond to a HMC associated with a very embedded YSO.  Source
18, on the other hand, is conspicuous in IR (2MASS and \emph{Spitzer}/IRAC bands) and it
is associated with the high-to-intermediate mass YSO GLIMPSE
G345.4977+01.4668 \citep[$L_\text{bol}\approx10^4\Lsun$,][]{Benjamin2003PASP}.
It is not clear that
Source 18 is associated with a HMC because --- being bright at NIR --- the YSO
cannot be very embedded. In addition,  the \met\  temperature ($64$ K) and  column density ($5.7\pm0.5\times10^{15}$ cm$^{-2}$) are significantly lower than those of the rest of the HMCs.

%Sources 8 and 10 are associated with HMCs. 
Source 8 and 10 are associated with two molecular cores, respectively, C8 and the
CC core.  These cores are the richest sources of molecular lines in
\clump.  The CC core is associated with temperatures $>100$ K and hydrogen
saturated COMs, which are characteristic of HMCs.  The heating source is
the HMYSO \hmyso, which ionizes an associated HC \hii\ region. It is
natural to assume that this ionizing radiation will also dissociate a large
fraction of molecules in the immediate surroundings of the HMYSO, thus, the
molecular emission does not need to coincide exactly with the HMYSO
position \citep[e.g.,][]{Mookerjea2007ApJ}.  C8, on the other hand, is
embedded in a dense molecular envelope which extends to the northwest.
This envelope
apparently engulfs the position of Maser \texttt{a},  which 
is associated with a previously unreported continuum
source of $2.1$ mJy at 3 mm.
Seemingly, Maser \texttt{a} pinpoints the location of another embedded YSO.
C8 is also characterized by \met\ lines with
$T_\text{ex}>100$ K and other COMs with $T_\text{ex}>70$ K revealing  the
presence of a HMC. In
contrast to the CC core, no evidence of free-free emission is observed
toward C8.

%We will chemically compare these two HMCs. We start/organize separating
%N-O molecules
In order to contrast the chemical composition between the CC core and C8,
we compare the abundances of several molecules respect to methanol.  The
main reason to derive methanol-normalized abundances is that the hydrogen
column density toward the CC core is very difficult to estimate due to most
of the 3 mm continuum corresponding to free-free emission arising from the
HC \hii\ region.  We focus first on the abundance of O- and N-bearing
molecules, which is a rather common approach in the literature 
\citep[e.g.,][]{Widicus-Weaver2012ApJ,Allen2017AA}.  This
strategy is also justified because the only hydrocarbon COM (that is, without N or O), \propyne, 
seems to trace the clump material on a large scale rather than 
gas associated with the molecular cores. 
Identifying the O- and N-bearing groups was originally motivated by
observations of Orion-KL, which show that nitrogenated and oxygenated
molecules are better associated with two distinct structures, respectively, the
HMC and the compact ridge  \citep[and references
  therein]{Feng2015AA}.  A thermal segregation is also observed, with
N-bearing molecules tracing hotter gas compared to the O-bearing species
\citep{Crockett2015ApJ}.  N-O segregation has been found toward other
high-mass star forming regions as well \citep{Su2005prpl,Oberg2013ApJ,Allen2017AA},
but it does not seem to be universal \citep{Fontani2007AA}.
%% Another very clear
%% example, which bear some resemblances with \clump, is G9.62+0.19. This
%% high-mass star forming region is associated with several \hii\ regions, among them a
%% ultracompact and a HC denoted `E' and `F', respectively
%% \citep{Garay1993ApJ,Testi2000AA}. Both sources `E' and `F' are associated
%% with HMCs and bright \met\ emission, with source `E' displaying strong
%% \methoxy\ and HCOOCH$_3$ lines. Source `F' is associated with strong
%% molecular outflows and it is more intense in \prop, \hcqn, HNCO, and
%% sulfuretted species \citep{Liu2005IAUS,Su2005prpl,Liu2017arXiv}.

\begin{figure}
\plotone{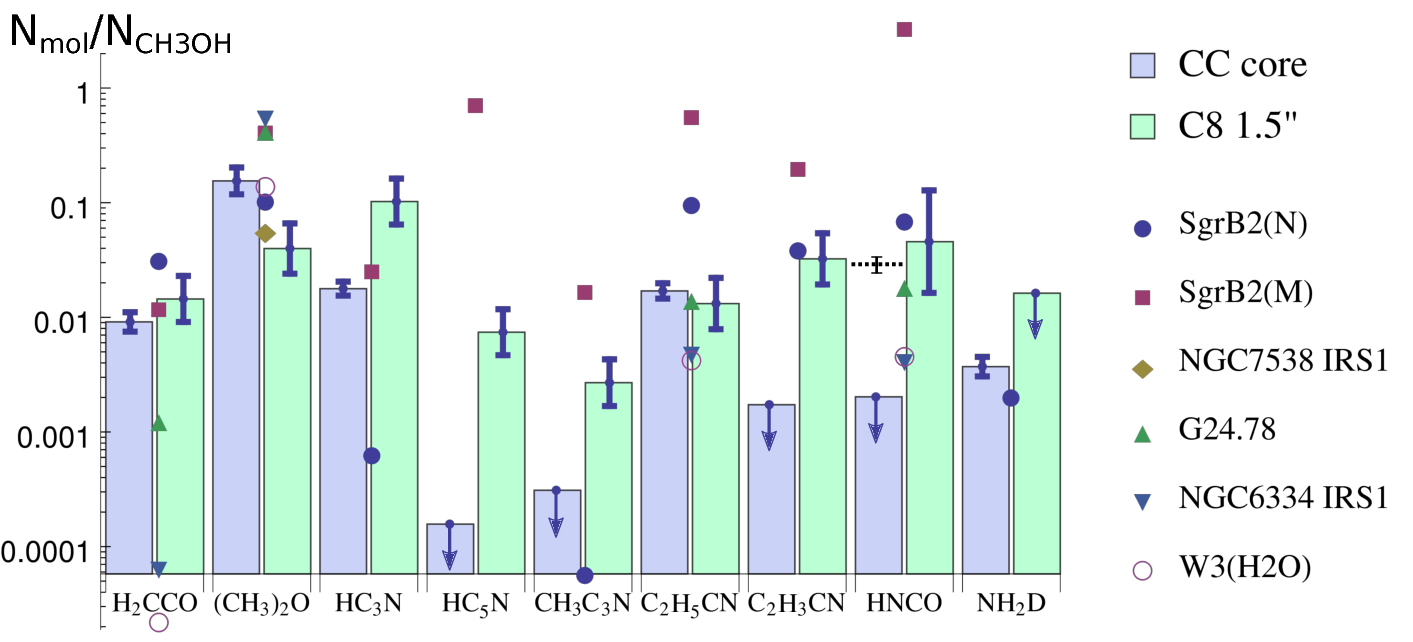}
\caption{Blue and green bars show column density ratios for O- and
  N-bearing molecules more complex than four atoms toward the CC core and
  C8 (C8 1\farcs5), respectively.  We use the values given in Table
  \ref{tab-cdt} for C8 1\farcs5 and the first velocity component toward the
  CC core.  Error bars indicate formal uncertainties and arrows indicate
  lower bounds.  We also show abundance ratios measured toward six HMCs
  from the literature
  (Sgr B2 sources from \citealp{Belloche2013AA}, NGC 7538 from \citealp{Oberg2014FaDi}, and the rest of the sources from \citealp{Bisschop2007AA}). The black dashed
  line and error bar indicate the [HNCO]/[\met] ratio obtained toward the
  CC core using the column density of HNCO given in Table \ref{fig-ccvsc8}
  associated with the \emph{second} velocity component.
\label{fig-ccvsc8}}
\end{figure}

% Overall results of the abundance comparison.
Figure \ref{fig-ccvsc8} compares the
methanol-normalized abundance of O- and N-bearing molecules with four atoms
or more, detected  toward the CC core or C8 (thus excluding \acet).
Figure \ref{fig-ccvsc8} also shows  abundance ratios associated with
well-known HMC sources such as Sgr B2 (N) and (M) \citep{Belloche2013AA},
NGC 7538 IRS1 \citep{Oberg2014FaDi}, G24.78, NGC6334 IRS1, and W3(H$_2$O) \citep{Bisschop2007AA}.
We note  that the relative abundance ratio respect to \met\
for different sources can vary within several orders of magnitude, depending on the molecule.
Part of this variation may be due to most  values reported in Figure \ref{fig-ccvsc8}
are obtained from   single dish telescope observations, which combines emission from several  regions in the beam.
Therefore, these ratios  are useful only as rough estimations.
In any case,  the abundance ratios toward the CC core and C8 are comparable to those of other sources,
within the observed variations.

The only molecule in Figure \ref{fig-ccvsc8} more abundant in the CC core
than in C8 respect to \met\ is \methoxy. We do not detect \damm\ toward
C8, but its abundance upper limit is above the [\damm]/[\met] ratio
measured toward the CC core.
We measure the same abundances respect to
methanol of \prop\ and \ethe\ for both cores. For the rest of the molecules
(\hctn, \hccn, HNCO, \acry, and \cyan), we observe larger
abundances toward C8 than toward the CC core.  We add a note of caution
about HNCO.  Lines from this molecule toward the CC core are characterized
by a $T_\text{ex}$ and $V_{\rm LSR}$ similar to those of \propyne, hence
they are not directly associated with the HMC.  However, there seems to be
in addition a hotter component, prompting us to conclude that the observed
HNCO lines blend emission from the HMC and its colder envelope.  We
indicate in Figure \ref{fig-ccvsc8} both the [\met]/[HNCO] lower limit and
the ratio using the column density of Table \ref{tab-cdt}.

%Focus on two important COMs: methoxymethane and propanenitrile
There are two COMs more complex than \met\ detected toward both the CC core and
C8: methoxymethane  and propanenitrile (\prop).
These two molecules are representative of
O-bearing and N-bearing COMs, respectively.
The \methoxy\ and \prop\ lines
toward C8 peak at the location of Source 8.  Toward the CC core, they peak
together with the highest energy \met\ transitions despite \methoxy\ and
\prop\ lines coming from relatively low upper energy ($E_{\rm up}<40$ K)
transitions.  Similar to \met, a hotter \methoxy\ component is blended
in the CC core emission: the LTE fitting ($T_\text{ex}=138$ K)
underpredicts the $16_{3,14}\shortrightarrow15_{4,11}$ transition with
$E_{\rm up}=136.6$ K.

%Focus on methoxymethane
Methoxymethane to methanol ratios of $1.7\pm0.03\times10^{-1}$ and
$4\pm2\times10^{-2}$ are measured toward the CC core and C8, respectively. These
ratios are within the values found toward most high-mass star forming
clumps.  The [\methoxy]/[\met] ratio does not vary
much among different sources, which \citet{Oberg2014FaDi} interprets as
evidence of \methoxy\ forming on the ice mantles of dust grains and
co-desorbing with \met.  Compared with methanol, methoxymethane is slightly less abundant
toward C8 compared to the CC core. In this regard, C8 is somewhat more
consistent with colder ($<50$ K) sources \citep{Oberg2014FaDi}.
 
% Focus on propanenitrile
The [\prop]/[\met] ratio  shows a range wider than that of \methoxy\ between different
sources, from $\sim10^{-3}$ in G35.03A \citep[]{Allen2017AA} to
$0.2$ in Orion-KL `D' \citep{Wright1996ApJ}. Toward the CC core and C8 we
measure the same ratio $[\prop]/[\met]\approx1.6\times10^{-2}$, which is consistent with
other hot cores like G34.26+0.15SE \citep{Mookerjea2007ApJ} and the sample
of \citet{Bisschop2007AA} (see Figure \ref{fig-ccvsc8}).  We do not detect clear evidence of thermal or
spatial N-O COM segregation in either core.  However, we note that at least
in Orion-KL, CH$_3$CN shows the N-O segregation much more clearly than
\prop\ \citep{Crockett2015ApJ}.

%% Ch3Ch2CN
%%  ra:    16:59:41.59236 +/- 0.00022 s 
%% dec: -040.03.42.95525 +/- 0.00104 arcsec
%% major axis FWHM:     937.4 +/- 16.5 marcsec
%% minor axis FWHM:     553.3 +/- 9.3 marcsec
%% met vt=1
%% --- ra:    16:59:41.59328 +/- 0.00022 
%% --- dec: -040.03.42.91104 +/- 0.00088 
%% unresolved
%% met7-6
%% ra:    16:59:41.59247 +/- 0.00061
%% dec: -040.03.42.90255 +/- 0.00323 arcsec
%% met 211-110
%% ra:    16:59:41.5840 +/- 0.0094 s (0
%% dec: -040.03.42.5587 +/- 0.1040 arcsec

% difference in abundance between both cores. C8 more evolved than CCC
We can explain some  of the abundance differences shown in Figure \ref{fig-ccvsc8}
if C8 is chemically more evolved than the CC core.
There are two facts which support this view:
 (i) the [\acry]/[\prop] abundance is larger in the C8 core 
than in the CC core  --- acrylonitrile is not detected in the latter --- and (ii)
the relative abundance of cyanopolyynes \hctn\ and \hccn\ in both cores.

% acry is higher in more evolved HMCs.
We assume that emission from  \acry\ and \prop\ 
associated with the CC core and C8 traces the hot
molecular gas  because of their high
temperatures, which is also usually the case toward other HMCs \citep{Fontani2007AA}.
Using HMC chemical models \citet{Caselli1993ApJ} find that, following grain mantle
evaporation, the \acry\ abundance increases respect to \prop\ because of
the activation of several gas destruction routes of \prop\ which produce
\acry. The [\acry]/[\prop] ratio, therefore, increases monotonically with time.    An underlying assumption
which justifies this conclusion and that C8 is chemically more evolved than the CC core is that, right after grain mantle evaporation, the [\acry]/[\prop]
ratio of the C8 core was similar (or even perhaps lower) than that of the CC
core.
%% Because of the
%% detection of both \acry\ and \prop\ toward hot cores, they were originally thought to
%% be produced on grain surfaces by succesive hydrogenation of
%% \hctn\ \citep[e.g.,][]{Blake1987ApJ}. However, it is possible that .
%% cold gas formation routes for \acry\ 
%% %$\mathrm{C_2H_4}+\mathrm{CN}\rightarrow\mathrm{\acry}+\mathrm{H}$
%% \citep{Herbst1990AA} do not seem to play an important role in high-mass
%% star formation regions (including \clump) because \acry\ is usually
%% detected at high temperatures and in places where there is also
%% \prop\ \citep{Fontani2007AA}.  Probably, in the timescales associated with
%% high-mass star formation the cold gas formation route is not too relevant.
%% Another potential complication to the picture presented above is that the
%% results of \citet{Belloche2009AA} disfavor the \hctn\ hydrogenation route
%% in grains for \prop\ formation. They suggest that \prop\ is formed by
%% another routes which do not include \acry\ in the path, which has the
%% effect that the final abundance

% HC5N/HC3N higher in C8 than in CC
The [\hccn]/[\hctn] (cyanodiacetylene to cyanoacetylene)
ratios of both cores also supports C8 being more
evolved than the CC core.
These ratios are
$\approx7\text{--}8\times10^{-2}$ (assuming $T_\text{ex}$ of cyanoacetylene ranging
between 17 and 77 K) and $\le9\times10^{-3}$ for C8 and the CC core,
respectively. This is a difference of  more than an order of magnitude in the
relative abundances of these two cyanopolyynes.
Because toward the CC core the $V_{\rm LSR}$ of \hctn\ and \met\ are very
similar we assume that these two molecules trace gas at the same
temperature (107 K). We use this  temperature to calculate the CC core upper limits
on the \hccn\ column density.  Toward C8 we are able to derive an
excitation temperature of $77\pm10$ K for \hccn, which is very similar to that of \prop\ but somewhat  lower than that of \met.
%In any case, we expect subthermal excitation is affecting the
%cyanopolyynes' column density estimations, we expect a similar effect
%between to be slightly lower than those of
%\hctn\ \citep{Snell1981ApJ}. For the \hctn\ $11\shortrightarrow10$
%transition, $n_{\rm crit.}\approx5\times10^{-5}$ cm$^{-3}$.  Therefore, if
%some of the emission from the cyanopolyyines comes from sub-thermally
%excited gas, we expect this to affect both molecules in a similar way.
Toward C8, \hctn\ and \hccn\ are detected at a consistent $V_{\rm LSR}$.

%% which means CC core younger than C8
Cyanopolyynes abundance models in HMCs \citep{Chapman2009MNRAS} indicate
that important gas formation routes for \hctn\ and \hccn\ start with
acetylene (also ethyne, C$_2$H$_2$) and CN. \citet{Chapman2009MNRAS} find
that \hctn\ and \hccn\ are formed in succession. The [\hccn]/[\hctn]
abundance ratio is consistent with \hccn\ not having formed yet in the CC core whereas in C8,
being older, \hccn\ and \hctn\ have already formed in large quantities.  In
addition and in consistency with this picture, \hctn\ is more abundant respect
to \met\ in C8 respect to the CC core. In the literature,
cyanopolyynes more complex than \hctn\ are not commonly detected toward HMCs. Figure \ref{fig-ccvsc8} shows that
the [\hctn]/[\met] ratios of both cores are comparable to  that of Sgr B2 (M). On the other hand, \hccn\ is not detected toward Sgr B2 (N), while it is very abundant respect to \met\ in Sgr B2 (M).

%% CC core is younger despite being associated with ionized gas
We propose that the  CC core is younger than C8 despite the former is associated with a HC
\hii\ region whereas C8 is not.  This is consistent with the HMYSO associated
with the CC core, namely \hmyso, being more massive than the YSO heating the C8
HMC.  The typical timescale for protostars to contract and evolve to the
main sequence is given by the Kelvin-Helmholtz contraction, which is 
faster for high-mass protostars \citep{Tan2014prpl}.  Furthermore, we also
expect that pre-stellar contraction --- characterized roughly by the
free-fall timescale --- was faster  for the denser CC core compared to that of C8.

%$\text{C}_{2}N+\text{CN}\rightarrow\text{HC}_3\text{N}+\text{H}$ and 
%$\text{C}_{4}\text{H}_2+\text{CN}\rightarrow\text{HC}_5\text{N}+\text{H}$.

% CH3CC-CN difference.
Another molecule with a different abundance in both cores is \cyan.
It is not clear, however, that cyanopropyne
is tracing the HMCs because of its  low temperature and $V_{\rm LSR}$
similar to that of propyne. More likely, \cyan\ abundances are representative of the envelope
material of the YSOs associated with the CC core and C8.
The cyanopropyne physical
parameters are consistent with it forming from propyne
through the gas reaction  
$\propyne+\text{CN}\rightarrow\cyan+\text{H}$ \citep{Balucani2000ApJ}.
Higher amounts of \cyan\ associated with C8 could reflect an older
age for this clump respect to the CC core. Note that the previous reaction
also generates cyanoallene (CH$_2$CHCCN) with a theoretical branching ratio
of 1.  Our data is not sensitive enough to detect a column density of
cyanoallene similar to that of cyanopropyne, but in principle, this would
be a way to test this theoretical chemical formation path.

%sulfuretted molecules also lots differences
While some differences in the abundance of N- and O-bearing
molecules between both cores  are noticeable, the chemical differences are most conspicuous in
sulfuretted species, and specially in the sulfur oxides.  Figures
\ref{fig-cen1} and \ref{fig-cen2} show that most of the SO and SO$_2$ is
associated with the central source, with no SO$_2$ and little SO emission
detection toward C8.
\citetalias{Guzman2014ApJ} already shows that sulfur oxides are good tracers of the
central HMC and, as the complex COMs, their emission indicate the presence
of more than one thermal component. 

%sulfur oxides, owever, seem to trace yet another structure
However, a careful analysis of the sulfur oxides compared with other HMC
tracers indicates that the SO and SO$_2$ emission does not originate from 
what we have called the CC core.  This is evidenced by a kinematic feature
which is rather unique to the SO and SO$_2$ emission: the rotating core
morphology around \hmyso.  Among our detected species \citep[and CH$_3$CN,
  see][]{Cesaroni2017AA}, there is no other molecule which displays this
feature. Toward the CC core direction --- $\sim0\farcs7$ northwest of
\hmyso --- the amount of SO$_2$ and methanol are comparable, that is, their
abundance ratio is on the higher end of the range of observed [SO$_2$]/[\met]
values toward other sources, which fluctuate between 1 and $10^{-2}$
\citep{Wright1996ApJ,Mookerjea2007ApJ,Allen2017AA}. Closer to \hmyso, however, the SO$_2$ and SO column densities
are approximately one order of magnitude larger than that of \met. 

% can we explain the chemistry of this third source (the rotating core)?
What special characteristics does the gas in the rotating core have? What
chemical processes are responsible for the enhanced abundance of sulfur
oxides? Definitive answer to these questions cannot be given, but it is
possible that part of it is related with the illumination of the core by
UV-photons from \hmyso. Based on the results of \citet{Ferrante2008ApJ},
\citetalias{Guzman2014ApJ} already argued for prolonged UV-illumination of
ices to explain the dominance of sulfur oxides in the rotating core versus
other sulfuretted molecules like OCS or CS, which are not formed efficiently in the gaseous phase \citep{Charnley1997ApJ}. 
However,  explaining the lack of the rest of the
molecules is more difficult.

Other systems which may bear some resemblance with the rotating core are
the circumstellar envelopes of M-type (O-rich) asymptotic giant branch
(AGB) stars. They consist of dense and mostly molecular gas, heavily
influenced by the UV radiation from the central degenerate core. They do
not show signs of methanol or the type of hydrogenated COMs commonly seen
toward hot cores \citep{Olofsson2005IAUS}. Interestingly enough, SO and
SO$_2$ are  common toward O-rich AGB envelopes, with these
molecules sometimes located very near the degenerate core \citep{Danilovich2016AA}. Formation of sulfur oxides may proceed through gas reactions involving OH such as $\text{S}+\text{OH}\rightarrow\text{SO}+\text{H}$ and
$\text{SO}+\text{OH}\leftrightarrow\text{SO}_2+\text{H}$. Consistently, large quantities of OH toward \hmyso\ are attested by the presence of OH masers \citep{Caswell1998MNRAS297215}. Detection of the
rotating core in molecules beside the sulfur oxides will help supporting
or rejecting these speculations.
Other sulfuretted molecules like \hcsp, OCS, and C$^{33}$S show no rotation
and are not even clearly associated with the  CC core because  their  emission peak farther in the northwest direction  and with a different
$V_{\rm LSR}$.

%% Other molecules
%% like \htcop, \hcdop, CS, and CCH in Table \ref{tab-cdt} have no obvious CC
%% core counterpart but they were detected in the line of sight. Their
%% emission is more consistent with the ambient cloud ($-11$ to $-13$ \kms)
%% velocity than that of the CC core. Molecules other than the sulfur oxides
%% not listed in the CC core section of Table \ref{tab-cdt} were not detected.

%Other observations are consistent with what we see in sources 8 and 10
As mentioned in Section \ref{sec-morph}, recent high spatial resolution
continuum and methyl cyanide observations of the central source in
\clump\ \citep{Cesaroni2017AA} show that it is actually composed of two
distinct sources: one associated with \hmyso\ and another one displaced
$0\farcs8$ northwest. The latter dominates the CH$_3$CN emission toward the
center of \clump.  It is possible that, along with that of
methyl cyanide, most of the COMs emission actually arise from this second
source, which we identify with the CC core.
Our data cannot resolve the two sources, but it is reasonable to
assume that the northwest displacements respect to \hmyso\ observed in
several species (Section \ref{sec-morph}) are due to these molecules
peaking at the position of this second component.  Sulfur oxides, on the
other hand, apparently dominate the ``rotating core,'' which is
a component much more closely associated with the HMYSO.

Finally, we mention two possible caveats of using the \met\ column
densities to normalize the abundances.  The first caveat is the use of a filling
factor of 1 for the LTE models. While this is certainly a simplification, we
think it is in part justified because \met\ emission is rarely
point-like, being commonly associated with extended envelope
material which covers the beam. It is possible, however, that
lines with the highest upper
energy levels are tracing a more compact and unresolved component.
The second caveat
is that, at least for the CC core, the SET model does not account for the
observed intensities of the two lines with upper energy levels $E_{\rm up}=340.1$
and $536.8$ K.  This is evidence that emission from another,
hotter component, is blended with the observed intensities.  Modeling the
\met\ intensities using two LTE models allows us to better fit the lines
with temperatures of $90$ and $350$ K, and a $\sim50\%$ higher total
methanol column density. However, this fit is very uncertain because we do
not detect enough lines to adequately constrain all the free parameters. Hence,
we refrain from complicating the model and we use the values given in Table
\ref{tab-met} on the understanding that these correspond to average values.
Note that toward the CC core the peak position of the \met\ lines varies
within 0\farcs4, with the lower energy lines peaking farther from
\hmyso\ compared to the higher energy transitions, consistent with the
hottest material being closer to the
HMYSO.  A similar thermal gradient is
observed toward C8, with $>50$ K energy transitions peaking closer to
Source 8 compared to lower energy lines, which  peak  0\farcs7 farther to the west.

\subsubsection{Extended emission in \clump}

In Section \ref{sec-corr} we performed a systematic analysis of the
extended emission in the clump. Our main interest was to determine
morphological similarities between the emission of different molecules.
%% We
%% focused on molecules with significant extended emission, among them, all of
%% the previously detected toward \clump\ using single dish telescopes like
%% CS, SiO, and \propyne.
We found that the extended molecular emission patterns can be collected in two groups
which we call the Shock and the Continuum groups.
Since temperature variations
away from the central HMYSO are rather smooth (Section \ref{sec-prop}),
differences in the extended emission morphology are mainly
related with different column density and abundance.  In the following,
we propose some physical and chemical ideas which help interpreting the
observed correlations.

% Shock group
\emph{\underline{The Shock group.}}
Our results  indicate that species in the Shock group giving rise to the extended emission
can be related with shocks in two ways: (i) as directly sputtered from the grain mantle or
the refractory core, or (ii) as being formed in recently
shocked gas.

% SiO via sputtering
Perhaps the most common and specific shock tracer molecule is SiO, which is
assumed to be produced by sputtering of Si/SiO from dust grains due to 
shocks (%$v_s>10$ \kms,
\citealp{Jimenez-Serra2008AA}) usually attributed to
YSO outflows.  Significant amounts of SiO exists in \clump\ and, consistent
with the shock interpretation, SiO lines are the broadest among all the
molecules.
There are several elongated features resembling outflows or
cavity walls in the SiO maps. Some of these also reveal line wings,
like the feature extending in the $\text{P.A.}=-135\arcdeg$ from the
approximate location of Source 6. However, in general it was not possible
to identify collimated molecular outflows unambiguously. Part of the reason is likely
confusion: less massive embedded clusters are observed to be
associated with a plethora of outflows \citep[e.g.\ NGC 1333, see][]{Plunkett2013ApJ} and we
 expect this  to be the case in high-mass protostellar clumps as well.

% SO and shocks
The remarkable correlation between the SO \emph{extended} emission (that
is, away from \hmyso) and SiO also indicates the  association of SO
with shocks.  This association has been observed previously \citep[e.g.,][]{Jimenez-Serra2005ApJ,Podio2015AA}. The SO zero moment map
illustrates clearly the double origin of SO: on the one hand, associated
with hot molecular gas near the HMYSO and no SiO emission, and on the other, extended emission
tracing the shocked gas.  The origin of sulfuretted species can be
explained (at least in part) by gas reactions facilitated by the special
physical conditions associated with shocked gas. In particular, sequential
reactions of S with OH and O$_2$ will enhance the formation of sulfur
oxides. In \clump, we observe SO correlated with SiO, but no
SO$_2$ whatsoever.
This  is
consistent with models of low velocity shocks 
which predict large amounts of SO and little SO$_2$  \citep{Pineau1993MNRAS}.
% grain mantles. ice sputtering
Shocks and sputtering are ways of releasing molecules from the dust grains
to the gaseous phase.  While other mechanisms like photo- and chemical
desorption are likely to be at work in \clump\ as well, shocks relate
naturally with  tracers like SiO and SO and with the strong turbulence
which characterizes high-mass clumps.  Shocks can produce ice sputtering
but they can also destroy molecules like methanol if the velocities become
$\gtrsim15$ \kms\ \citep{Suutarinen2014MNRAS}.

%tension existing between the requirements on
%the shock velocity for SiO sputtering (high velocity shock) versus the [SO]/[SO$_2$] abundance
%ratio (low velocity shock).

% CS and HCS+ production
The production of secondary UV photons associated with the shocks can also
increase the amount of S$^+$ and accelerate the rate of ion neutral
reactions like $\text{S}^++\text{CH}_2\rightarrow\hcsp+\text{H}$
\citep{Yamamoto2017Springer}. \hcsp\ may later form CS through dissociative
recombination. The good correlation between C$^{33}$S and \hcsp\ indicates
that this route is maybe important in the formation of these two
molecules, although there are also several other relevant formation routes
for CS, for example, from $\text{SO}+\text{C}\rightarrow\text{CS}+\text{O}$
\citep{Pineau1993MNRAS}. The latter reaction path is consistent with the  observed correlation between SO and CS.
Additionally,  ion-neutral
formation routes for \hcsp\ and CS are also efficient, which
helps explaining the relatively good correlation of the former with CCH.
In
fact, \hcsp\ is somewhat in a middle ground between both the Shock and the Continuum group.

% mantle hydrogenation products. partly and completely hydrogenated C+O and C+C+O products
Molecules like \met, OCS, \acet, \ethe, and HNCO show very similar spatial
features. Because methanol is the archetypal dust mantle molecule,    we interpret the morphological similarities as these species being formed in grain mantles
and subsequently removed from the dust, probably by ice sputtering.
\met, \acet, and  \ethe\ are part of two
hydrogenation paths which were  active when the clump's temperature was lower, as it is now too high for H to remain long in the dust grain surface. One of these paths starts with CO, leading to H$_2$CO and
finally \met. The other includes an additional carbon atom addition and
progresses through \ethe, \acet, and finally CH$_3$CH$_2$OH (ethanol)
\citep[][]{Charnley1999ASIC}.
The latter path is probably less common
than the first one, judging from the usually lower amounts of ethanol
compared to methanol.
In \clump, we do not detect ethanol but we note that
our observations are only sensitive to rather large column densities
($>2\times10^{15}$ cm$^{-2}$ at 100 K).

Because hydrogenation in the dust mantles drops once
the grain reaches $T_d\gtrsim20$ K due to fast H evaporation, some of the
intermediate unsaturated products like \ethe\ and \acet\ remain in the ices
and they are prone to be removed afterwards by shocks (or other forms of
ice erosion).
This interpretation for the origin of \met,  \acet, and \ethe\ 
is supported by, for example, the  detection \added{and correlation} of these species toward 
\added{embedded low-mass protostars \citep{Bergner2017ApJ}} and in shocks associated with low-mass star formation outflows in L1157-B1
\citep{Lefloch2017MNRAS}.

Despite the overall good correlation of \ethe\ and \acet\ in
the diffuse gas, contrary to \ethe, \acet\ is not detected toward either
the CC core or C8. Consistently, \citet{Oberg2014FaDi}  reports
acetaldehyde depletion with increasing  temperature
in HMCs.  A possible explanation for the lack of
\acet\ in the HMCs of \clump\ is the activation of  gas destruction routes for \acet,
possibly back into \ethe. \deleted{We note that}
\ethe\ and \acet\ have \added{both} been detected \deleted{both} toward relatively cold gas \citep[$<50$ K,][]{Bisschop2007AA} and hot
HMCs at temperatures above $\ge100$ K \citep[][]{Belloche2013AA}.

%OCS and HNCO
The good correlation observed between the OCS and methanol zero moment maps
can be explained by most of the OCS coming from dust mantles as well. This picture
is confirmed by OCS being one of the few molecules which has been detected
directly in solid state in the ISM \citep{Gibb2004ApJS}.
Note that this  proposed formation mechanisms opposes to that of other sulfuretted species, such as the  sulfur oxides, CS, and \hcsp, \added{at least in the diffuse clump  gas}.
There is no difference between the CC core and C8 in their [OCS]/[\met]
abundance ratios, consistent with most of the OCS actually codesorbing
(thermally and non-thermally) with \met\ to the gas phase.  Isocyanic acid
has also been identified as a (slow) shock tracer at galactic scales
\citep{Watanabe2016ApJ,Ueda2017PASJ,Kelly2017AA} and in combination with
\met. The HNCO abundance is seemingly enhanced in shocks due to ice erosion and
direct formation in the heated post-shock gas
\citep{Rodriguez-Fernandez2010AA}.

% intro about the cont goup
\emph{\underline{The Continuum group.}}
Emission from molecules in this group display good  correlation with the 3 mm continuum map. Extended emission from this map is thought to be dominated by thermal dust emission, with free-free being relevant  toward specific sources like 10 and 18 \citepalias{Guzman2014ApJ}.  While the amounts of all species are expected to increase with increasing material, some molecules (like those of the Shock group) increment their abundances in a higher proportion due to specific physical phenomena.  The abundance of molecules in the Continuum group, in contrast, seems to be more stable, suggesting  ion-neutral chemical formation paths. For example,   paths dominated by the barrier-less Langevin rates  are indeed independent of temperature \citep{Yamamoto2017Springer}. 

% characteristics: hydrocarbons and hydrocarbon+CN molecules, absence of O-bearing and S-bearing except HCO+
There are at least two characteristics of the abundance of molecules in the Continuum group worth noticing. The first is the presence of  carbon-chain related species, either in just hydrocarbon form or in combination with the cyanide group (C{\tbond}N). The second is the absence of oxygenated and sulfuretted species. All sulfuretted and oxygenated species are in the Shock group, except the HCO$^+$ isotopologues. 
%The third characteristic is that all molecules in this group, except CCH, have a strong C8 counterpart. 

% hydrocarbons form in gas phase..
Gas phase formation routes dominated by  ion-neutral and barrierless reactions  characterize the formation of hydrocarbons  such as CCH, \cyc, and \propyne. The three molecules display a good correlation between each other as expected for a common formation pathway.
\replaced{These}{However, reactions involving these} molecules
\replaced{represent but a}{are only a}
small \replaced{fraction}{part} of the  complex  chemical networks which characterize
hydrocarbon formation\replaced{, which}{. This}   complicates reaching solid conclusions about their chemistry.
In general, there are three main formation paths for hydrocarbons and carbon chains: (i) ion-neutral reactions, which dominate at high densities and are expected to  correlate with the column density, (ii) warm carbon chain chemistry (WCCC), which could dominate at high temperatures, and (iii)  top-down chemistry, which is  efficient in UV-illuminated, diffuse regions.

% methane not likely forming a relevan role in \clump\
From the correlations observed  \replaced{between}{involving} \explain{It is not only the correlation between them, but also the lack of correlation with molecules formed through other processes (like methanol).} CCH, \cyc, and \propyne, we can conclude that their production  has not been likely affected by WCCC processes.
The defining characteristic of WCCC is the expansion of the hydrocarbon's chemical networks allowed by the injection of methane (CH$_4$) into the gas phase from dust mantles \citep{Sakai2013ChRv}. In \clump, all hydrocarbons are in the Continuum group, which means that their abundances do not seem to be  enhanced by any type of dust-released species  (either thermally or non-thermally) including methane. In addition, WCCC suggests that the formation of methane in the ice mantles would follow from a fast gravitational contraction.
%The latter would  lock carbon in atomic form into the mantles.
Contraction of high-mass clumps including \clump, however, is rather slow \citep{Guzman2011ApJ}.
In consequence, hydrocarbon formation in the clump is probably dominated by ion-neutral chemistry. It may be  possible that top-down chemistry could also contribute through secondary UV-photons from shocks or in regions near the outskirts of \clump. 
%Future confirmation of these ideas will be hindered  because of relevant molecules in the formation routes like acetylene, ethylene (C$_2$H$_4$), and methane are undetectable in mm/sub-mm rotational transitions.

% hydrocarbons+CN 
Besides hydrocarbons, another set of species characteristic of the Continuum group  are CN-bearing species.
The good correlation between the cyanopolyynes, the HCN \replaced{isotopomers}{isomers} (\htcn, \hntc, and \hcqn), and the hydrocarbons  indicate that these species likely share common formation (or destruction) paths. 
The most natural molecule acting as the bridge between all these species is CN, which can react with several hydrocarbons to form larger N-bearing molecules. 
We already described formation paths for \hctn\ and \hccn\ involving the combination  of, respectively, acetylene and diacetylene with CN. These neutral-neutral reactions are barrier-less, therefore, they are also important in the clump away from heating sources. As remarked in the previous subsection, \cyan\ can be formed from \propyne\ through the addition of CN. Consistently, \propyne\ is the MCP of \cyan. 
Formation of HCN, on the other hand, naturally starts with CN which reacts with H$_3^+$ (ion-neutral reaction, presumably close to the Langevin rate) to create HCNH$^+$, which will form HCN by dissociative recombination \citep[e.g.,][]{Prasad1980ApJ}. Of course, H$_3^+$ is not a hydrocarbon but it plays a crucial role in starting the hydrocarbon reaction chains.
In the same vein, \acry\ can be formed in the gas phase through a $\text{CN}+\text{'hydrocarbon'}$ reaction \citep[specifically, ethylene][]{Herbst1990AA}. We caution, however, that \acry\ is only well detected near C8 (see Figure \ref{fig-cont}), hence,  it is unclear  how well is \acry\ related with the rest of the molecules in the Continuum group.

Summarizing, molecules in the Continuum group are hydrocarbons, cyanides, cyanopolyynes, and HCO$^+$.  All these species have efficient gas phase formation routes which are chemically connected through key molecules like H$_3^+$ and CN.  Hydrocarbons have relevant formation routes starting from C and H$_3^+$, the latter being important as well in the creation of HCO$^+$ and HCN. In addition, a fraction of the hydrocarbons combines with CN producing the cyanopolyynes, \cyan, and possibly \acry.  Gas phase chemistry is efficacious in producing all these species even at low temperatures, which helps explaining why their abundances are relatively unaffected by special circumstances like shocks or photo illumination. In fact, emission from species in this group correlates well with the total mass column density as traced by the optically thin dust continuum. 

%Another molecule which seems to avoid \hmyso\ is \cyc. This molecule shows a strong absorption feature associated with the CC core.

\subsubsection{Cold, starless gas}
%Absorption dark cloud SE:
%CS, \propyne, CCH, \htcn, \hntc, \met, 
% Emission
% NH2D & $15$ & $  13.88\pm  0.03$ & $ -11.94\pm   0.2$ & $  7.367\pm   0.5$ & 
% deuteration of 0.02 en promedio Lackington 2016

One remarkable feature found in \clump\ --- a protostellar clump in the
brink of developing an \hii\ region --- is the presence of cold and dense
gas clearly differentiated spatially and chemically from the rest of the molecular emission.
There are two starless clouds seen only in \damm\ located in the
northeast and southeast regions of the clump. We
refer to these two clouds as the NE and SE clouds, respectively. As noted in Section \ref{sec-nir},
the SE
cloud is clearly evident as an IR-dark feature against the brightly
illuminated outflow cavity associated with \hmyso.

Emission from \damm\ is most readily explained by a high deuterated ammonia
fractionation. Deuterium fractionation is expected in cold and dense
molecular environments for all species whose formation path involves
H$_3^+$ \citep{Bergin2007ARA&A}.  Gas phase reactions which produce ammonia
in this way also include N$_2$H$^+$ (N$_2$D$^+$ for \damm) in the formation
route \citep{Yamamoto2017Springer}. Therefore, because CO readily destroys
N$_2$H$^+$, this chemical path is most effective combined with strong CO
depletion \citep{Roueff2005AA}. Confirming large fractions of \damm\ and of
N$_2$D$^+$ respect to their main isotopologues, and also measuring CO
depletion factors, will help solidifying these theoretical expectations.
Alternatively, deuterium enrichment may occur directly in the dust grains
\citep{Fedoseev2015MNRAS} but this requires an efficient non-thermal
desorption mechanism to release deuterated ammonia into the gas phase.

Determining whether these starless clouds --- namely, without evidence of
embedded YSOs in any of our observations --- could collapse and form stars
in the future entails deriving their masses. We derive a \damm\ mean column
density of $3.8$ and $5.7\times10^{13}$ cm$^{-2}$ toward the SE and NE
clouds, respectively, assuming an excitation temperature of 15 K.  Assuming
a mean deuterated ammonia fractionation of 0.02
\citep[]{Lackington2016MNRAS} characteristic of IRDCs, an ammonia abundance
respect to H$_2$ ranging between $5\text{--}30\times10^{-8}$
\citep{Wienen2012AA}, and a size given by the 30\% of the peak contour of
the \damm\ zero moment map in Figure \ref{fig-jhk}, we derive mass ranges
for the SE and NE clouds of 0.25--1.5 and 0.4--2.3 \Msun,
respectively. Column densities of H$_2$ range between 10$^{22\text{--}23}$
cm$^{-2}$. Whereas these column densities are large, the cloud masses are
too low to virialize these clouds of radius $\sim0.02$ pc and $\Delta
V\ge2$ \kms. We conclude that the clouds are the remnants of the cold
prestellar clump which gave birth to \clump\ and that they will likely
disperse and sustain no further star formation.

%%%%%%%%%%%%%%%%%%%%%%%%%%%%%%%%%%%%%%%%%%%%%%%%%%%%%%%%%%%%
%%%%%%%%%%%%%%%%%%%%%%%%%%%%%%%%%%%%%%%%%%%%%%%%%%%%%%%%%%%%
{\section{SUMMARY}\label{sec-summ}}
%%%%%%%%%%%%%%%%%%%%%%%%%%%%%%%%%%%%%%%%%%%%%%%%%%%%%%%%%%%%
%%%%%%%%%%%%%%%%%%%%%%%%%%%%%%%%%%%%%%%%%%%%%%%%%%%%%%%%%%%%

We performed 3 mm ALMA observations  of the protostellar high-mass  molecular clump \clump. We investigated the chemistry of the clump by analyzing  its  associated molecular line emission detected between   84.5--88.0 and 96.7--100.2 GHz.  We summarize our main results as follows:
\begin{enumerate}
\item{We detect emission from 22 molecular species which encompass 34 \replaced{isotopomers}{isomers}. In addition to these, the spectra shows emission lines of methyl formate and formamide at the $2\sigma$ level with consistent V$_{\rm LSR}$. }
\item{We derive physical parameters --- column densities and excitation temperatures --- based on the SET modeling of the spectra associated with the most prominent features in the clump. We use the Radex non-LTE model to fit the methanol emission. Typical temperatures for the protostellar cores are $70$--$120$ K. Most of the extended emission is characterized by temperatures 20--40 K, depending on the distance to the central, dominating HMYSO \hmyso.} 
\item{The morphological characteristics of the extended emission allow us to collect the molecules in two groups.
  The Shock group gathers molecules whose emission is more similar to that of SiO, $J=2\shortrightarrow1$.
  It collects  molecules formed in shocked gas (like SO) and in dust grains (like \met, \ethe, and \acet) later released into the gas phase.
  The Continuum group collects molecules  whose emission morphology is more similar to that of the 3 mm dust continuum and seem to depend first on the column density. This group includes hydrocarbons, cyanopolyynes, and other cyanides (HCN and \cyan) which have effective gas phase formation routes. There is no clear evidence of efficient WCCC processes in forming hydrocarbons.}
%\item{We detected molecular emission clearly associated with previously detected continuum Sources 8, 10 (the central source), and  18 \citepalias[see][]{Guzman2014ApJ}. There is a less clear molecular emission association with Sources 2, 3, 7, 11, and 16, and no clear counterparts for the rest of the Sources.}
\item{The HMYSO \hmyso\ is associated with two different structures: a HMC conspicuous in hydrogen saturated COMs and a previously detected rotating core, conspicuous in sulfur oxides. This rotating core is not detected in any other species.}
\item{Source 8 is associated with a second  HMC within \clump, linked with a HMYSO less massive than \hmyso. Its chemical characteristics --- \acrylonitrile/\propanenitrile\  and cyanopolyyne abundances ---  suggest that the core associated with Source 8 is more evolved than the one associated with \hmyso.}
\item{We detect \damm\ emission arising from cold, IR-dark, starless clouds of $\sim1$ \Msun. These clouds are chemically and spatially differentiated from the rest of the gas in the clump. They are remnants of the prestellar stage of the clump, and they will not likely sustain further star formation activity.}
\item{We observe a strong  spatial segregation in \clump\ between \damm, Shock, and Continuum group molecules. This segregation  illustrates the need to separate the physical origin of different (groups of) molecules when modeling the chemistry of high-mass clumps  using unresolved observations \added{from, for example, single dish telescopes}.}
\end{enumerate}

\acknowledgements{A.E.G.\ thanks support from Fondecyt 3150570. V.V.G.\ acknowledges support from the National Aeronautics and Space Administration under grant No. 15XRP15\_20140 issued through the Exoplanets Research Program. G.G.\ and L.B.\ acknowledge support from CONICYT project PFB-06. This paper makes use of the following  ALMA data: ADS/JAO.ALMA\#2011.0.00351.S. ALMA is a partnership of ESO (representing its member states), NSF (USA) and NINS (Japan), together with NRC (Canada), NSC and ASIAA (Taiwan), and KASI (Republic of Korea), in cooperation with the Republic of Chile. The Joint ALMA Observatory is operated by ESO, AUI/NRAO and NAOJ. Portions of the analysis presented here made use of the Perl Data Language (PDL) developed by K.~Glazebrook, J.~Brinchmann, J.~Cerney, C.~DeForest, D.~Hunt, T.~Jenness, T.~Lukka, R.~Schwebel, and C.~Soeller. Analysis of this paper made use of CASSIS, developed by IRAP-UPS/CNRS (\url{http://cassis.irap.omp.eu}).}
\facility{Atacama Large Millimeter/submillimeter Array (ALMA).}
\software{CASA v4.7.2 \citep{Petry2012ASPC}.  Radex \citep{vanderTak2007AA}. CASSIS \citep[]{Caux2011IAUS}. Perl Data Language (\url{http://pdl.perl.org}).}
  %.  Radex \citep{vanderTak2007AA}. CASSIS, developed by IRAP-UPS/CNRS (\url{http://cassis.irap.omp.eu}). Portions of the analysis presented here made use of the Perl Data Language (PDL) developed by K.~Glazebrook, J.~Brinchmann, J.~Cerney, C.~DeForest, D.~Hunt, T.~Jenness, T.~Lukka, R.~Schwebel, and C.~Soeller and can be obtained from \url{http://pdl.perl.org}. PDL provides a high-level numerical functionality for the Perl scripting language (Glazebrook \& Economou, The Perl Journal, 5, 5).}

%%%%%%%%%%%%%%%%%%%%%%%%%%%%%%%%%%%%%%%%%%%%%%%%%%%%%%%%%%%%
%%%%%%%%%%%%%%%%%%%%%%%%%%%%%%%%%%%%%%%%%%%%%%%%%%%%%%%%%%%%
%%%   TABLES 
%%%%%%%%%%%%%%%%%%%%%%%%%%%%%%%%%%%%%%%%%%%%%%%%%%%%%%%%%%%%
%%%%%%%%%%%%%%%%%%%%%%%%%%%%%%%%%%%%%%%%%%%%%%%%%%%%%%%%%%%%

\startlongtable
\begin{deluxetable}{lllrr}
\tablewidth{0pc} \tablecolumns{5} \tabletypesize{\small}
\tablecaption{Observed frequencies and upper energies from identified lines\label{tab-lines}}
\tablehead{\colhead{\#}&
\colhead{Molecule %(Sym)
} & \colhead{Transition} & \colhead{LSRK}&\colhead{\mbox{$E_{\rm up}$}}\\\colhead{}&\colhead{}&\colhead{Frequency}&\colhead{}\\
\colhead{}&\colhead{~}&\colhead{~}&\colhead{(GHz)}&\colhead{(K)}
}
\startdata
\sidehead{Carbon monosulfide}\\%m1
1&CS       & $J=2\rightarrow1$ & 97.9854 & 7.1 \\
2&C$^{33}$S & $J=2\rightarrow1$ & 97.1759 & 7.0\\
\sidehead{Carbon oxide sulfide (carbonyl sulfide)}\\%m1
3&OCS & $J=7\rightarrow6$ & 85.1432 & 16.3\\
 &   & $J=8\rightarrow7$ & 97.3058 & 20.9 \\
4&O$^{13}$CS & $J=7\rightarrow6$ & 84.8690 & 16.3\\
 &         & $J=8\rightarrow7$ & 96.9929 & 21.0 \\          
\sidehead{Thioformylium}\\%m1
5& \hcsp & $J=2\rightarrow1$ & 85.3511 & 6.1 \\
\sidehead{Sulfur monoxide}\\%m1
6&SO & $J,N=2,2\rightarrow1,1$ & 86.0978 & 19.3\\
 && $J,N=3,2\rightarrow2,1$ & 99.3045 & 9.2\\
 && $J,N=4,5\rightarrow4,4$& 100.0343 & 38.6 \\
7&$^{34}$SO & $J,N=4,5\rightarrow4,4$ & 96.7856 & 38.1\\
  &       & $J,N=3,2\rightarrow2,1$ & 97.7195& 9.1 \\
8&$^{33}$SO & $J,N,F=3,2,\frac{9}{2}\rightarrow 2,1,\frac{7}{2}$ & 98.4976 & 9.2\\
\sidehead{Sulfur dioxide }\\
9&SO$_2$ & $J_{K_aK_c}=8_{3,5}\rightarrow9_{2,8}$     & 86.6430 & 55.2 \\
  &     & $J_{K_aK_c}=20_{2,18}\rightarrow21_{1,21}$ & 86.8329 & 207.8 \\
  &     & $J_{K_aK_c}=7_{3,5}\rightarrow8_{2,6}$     & 97.7067 & 47.8 \\
  &     & $J_{K_aK_c}=28_{7,21}\rightarrow29_{6,24}$ & 98.9817 & 493.7 \\
  &     & $J_{K_aK_c}=29_{4,26}\rightarrow28_{5,23}$ & 99.3978 & 440.7 \\
\sidehead{Ammonia (azane)}\\%m1
10&NH$_2$D &$J_{K_aK_c}=1_{1, 1},0^s\rightarrow 1_{0, 1},0^a$ & 85.9223 & 20.7 \\
\sidehead{Silicon monoxide}\\%m1
11&SiO & $J=2\rightarrow1$ & 86.8499 & 6.3 \\
12&$^{29}$SiO & $J=2\rightarrow1$ & 85.7612 & 6.2\\
13&$^{30}$SiO & $J=2\rightarrow1$ & 84.7489$^{b11}$ & 6.1\\
\sidehead{Propyne (methylacetylene)}\\%m1
13&\propyne & $J,K=5,3\rightarrow4,3$ & 85.4467 & 77.3 \\
&& $J,K=5,2\rightarrow4,2$ & 85.4548 & 41.2 \\
&& $J,K=5,1\rightarrow4,1$ & 85.4605$^{b1}$ & 19.5 \\
&& $J,K=5,0\rightarrow4,0$ & 85.4605$^{b1}$ & 12.3 \\
\sidehead{Formylium}\\%m1
14&\htcop  &$J=1\rightarrow0$  & 86.7582$^{b6}$ & 4.2\\
15&\hcdop &$J=1\rightarrow0$  & 85.1660 & 4.1\\
\sidehead{Cyclopropenylidene }\\%m1
16&c-C$_3$H$_2$ & $J_{K_aK_c}=2_{1,2}\rightarrow1_{0,1}$ & 85.3421 & 6.5 \\
   &       & $J_{K_aK_c}=4_{3,2}\rightarrow4_{2,3}$ & 85.6597 & 29.1\\
\sidehead{Ethynyl}\\%m1
17&CCH    &$J,N,F=\frac{3}{2},1,1 \rightarrow \frac{1}{2},0,1$ & 87.2876 & 4.2\\
  &       &$J,N,F=\frac{3}{2},1,2 \rightarrow \frac{1}{2},0,1$ & 87.3201 & 4.2\\
  &       &$J,N,F=\frac{3}{2},1,1 \rightarrow \frac{1}{2},0,0$ & 87.3319 & 4.2\\
  &       &$J,N,F=\frac{1}{2},1,1 \rightarrow \frac{1}{2},0,1$ & 87.4054 & 4.2\\
  &       &$J,N,F=\frac{1}{2},1,0 \rightarrow \frac{1}{2},0,1$ & 87.4106 & 4.2\\
  &       &$J,N,F=\frac{1}{2},1,1 \rightarrow \frac{1}{2},0,0$ & 87.4499 & 4.2\\
\sidehead{Acetaldehyde}\\%m1
18 & CH$_3$CHO  & $({\rm E})~J_{K_aK_c}=5_{1,4}\rightarrow4_{1,3}$ & 98.8673 & 16.6\\
 &              & $({\rm A})~J_{K_aK_c}=5_{1,4}^{-}\rightarrow4_{1,3}^{-}$ & 98.9052 & 16.5\\
\sidehead{Propanenitrile (ethyl cyanide)}\\
19 &CH$_3$CH$_2$CN & $J_{K_aK_c}=10_{1,10}\rightarrow9_{1,9}$ & 86.8242 & 24.1 \\
   &            & $J_{K_aK_c}=11_{0,11}\rightarrow10_{0,10}$ & 96.9245 & 28.1 \\
   &            & $J_{K_aK_c}=11_{2,10}\rightarrow10_{2,9}$ & 98.1824 & 32.8 \\
   &            & $J_{K_aK_c}=11_{6,6}\rightarrow10_{6,5}$ & 98.5293$^{b2}$ & 68.4 \\
   &            & $J_{K_aK_c}=11_{6,5}\rightarrow10_{6,4}$ & 98.5293$^{b2}$ & 68.4 \\
   &            & $J_{K_aK_c}=11_{7,4}\rightarrow10_{7,3}$ & 98.5293$^{b2}$ & 82.8 \\
   &            & $J_{K_aK_c}=11_{7,5}\rightarrow10_{7,4}$ & 98.5293$^{b2}$ & 82.8 \\
   &            & $J_{K_aK_c}=11_{8,3}\rightarrow10_{8,2}$ & 98.5385$^{b3}$ & 99.5 \\
   &            & $J_{K_aK_c}=11_{8,4}\rightarrow10_{8,3}$ & 98.5385$^{b3}$ & 99.5 \\
   &            & $J_{K_aK_c}=11_{5,7}\rightarrow10_{5,6}$ & 98.5385$^{b3}$ & 56.2 \\
   &            & $J_{K_aK_c}=11_{5,6}\rightarrow10_{5,5}$ & 98.5385$^{b3}$ & 56.2 \\
   &            & $J_{K_aK_c}=11_{9,2}\rightarrow10_{9,1}$ & 98.5493$^{b4}$ & 118.4 \\
   &            & $J_{K_aK_c}=11_{9,3}\rightarrow10_{9,2}$ & 98.5493$^{b4}$ & 118.4 \\
   &            & $J_{K_aK_c}=11_{4,8}\rightarrow10_{4,7}$ & 98.5701$^{b5}$ & 46.2 \\
   &            & $J_{K_aK_c}=11_{4,7}\rightarrow10_{4,6}$ & 98.5712$^{b5}$ & 46.2 \\
   &            & $J_{K_aK_c}=11_{3,9}\rightarrow10_{3,8}$ & 98.6154       & 38.4 \\
   &            & $J_{K_aK_c}=11_{3,8}\rightarrow10_{3,7}$ & 98.7062       & 38.4 \\
   &            & $J_{K_aK_c}=11_{2,9}\rightarrow10_{2,8}$ & 99.6867       & 33.0 \\
\sidehead{Cyanopropyne (2-butynenitrile)}\\
20& CH$_3$C$_3$N & $J,K=21,1\rightarrow20,1$ & 86.7596$^{b6}$ & 53.3 \\
&& $J,K=24,3\rightarrow23,3$ & 99.1467 & 126.9 \\
&& $J,K=24,2\rightarrow23,2$ & 99.1505 & 89.4 \\
&& $J,K=24,1\rightarrow23,1$ & 99.1537$^{b7}$ & 70.0 \\
&& $J,K=24,0\rightarrow23,0$ & 99.1537$^{b7}$ & 59.5 \\
\sidehead{Cyanodiacetylene (2-4 Pentadiynenitrile)}\\%m1% alkanes, alkenes, alkynes
21& HC$_5$N & $J=32\rightarrow31$ & 85.2057 & 67.5 \\
&& $J=33\rightarrow32$ & 87.8675 & 71.7 \\
&& $J=39\rightarrow38$ & 98.5172 & 89.8 \\
\sidehead{Cyanoacetylene (propynenitrile)}\\%m1
22& \hctn & $J=11\rightarrow10$ & 100.0815 & 28.8 \\
23& HCC$^{13}$CN & $J=11\rightarrow10$ &  99.6665$^{b8}$ & 28.7 \\
24& HC$^{13}$CCN & $J=11\rightarrow10$ &  99.6568 & 28.7 \\
25& H$^{13}$CCCN & $J=11\rightarrow10$ &  96.9873$^{b9}$ & 27.9 \\
\sidehead{Hydrogen cyanide and isocyanide}\\%m1
26& HC$^{15}$N & $J=1\rightarrow0$ & 86.0591 & 4.1 \\
27& H$^{13}$CN & $J=1\rightarrow0$ & 86.3440 & 4.1 \\
28& HN$^{13}$C &  $J=1\rightarrow0$ & 87.0942 & 4.2 \\
\sidehead{Isocyanic acid}\\
29& HNCO & $J_{K_aK_c}=4_{1,4}\rightarrow3_{1,3}$ & 87.6016 & 53.8 \\
&& $J_{K_aK_c}=4_{2,3}\rightarrow3_{2,2}$ & 87.9026$^{b10}$ & 180.8 \\
&& $J_{K_aK_c}=4_{2,2}\rightarrow3_{2,1}$ & 87.9026$^{b10}$ & 180.8 \\
&& $J_{K_aK_c}=4_{0,4}\rightarrow3_{0,3}$ & 87.9295 & 10.6 \\
\sidehead{Methanol}\\%m1 %revisar vastel si incluye energia de formacion?
30 &\met\  
 & $({\rm E_2})$ $J_{K_aK_c}=5_{-1,0}\rightarrow4_{0,0},\,v_t=0$ & 84.5246 & 32.5 \\
&& $({\rm E_1})$ $J_{K_aK_c}=19_{4,0}\rightarrow18_{5,0},\,v_t=0$ & 84.7486$^{b11}$ & 528.9 \\
&& $({\rm E_2})$ $J_{K_aK_c}=6_{-2,5}\rightarrow7_{-1,7},\,v_t=0$ & 85.5725 & 66.8 \\
&& $({\rm E_2})$ $J_{K_aK_c}=2_{-1,2}\rightarrow1_{-1,1},\,v_t=0$ & 96.7442 & 4.6\\
&& $({\rm E_1})$ $J_{K_aK_c}=2_{0,2}\rightarrow1_{0,1},\,v_t=0$ & 96.7495  &12.2\\
&& $({\rm E_1})$ $J_{K_aK_c}=2_{1,1}\rightarrow1_{1,0},\,v_t=0$ & 96.7604 & 28.0\\
&& $({\rm E_1})$ $J_{K_aK_c}=6_{1,6}\rightarrow5_{0,5},\,v_t=1$ & 99.7363  &340.1\\
&CH$_3$OH (A)
 & $({\rm A^-})$ $J_{K_aK_c}P=7_{2,6}^-\rightarrow6_{3,3}^-,\,v_t=0$ & 86.6201 & 102.7 \\
&& $({\rm A^+})$ $J_{K_aK_c}P=7_{2,5}^+\rightarrow6_{3,4}^+,\,v_t=0$ & 86.9074 & 102.7 \\
&& $({\rm A^+})$ $J_{K_aK_c}P=2_{0,2}^+\rightarrow1_{0,1}^+,\,v_t=0$ & 96.7463  &7.0\\
&& $({\rm A^-})$ $J_{K_aK_c}P=2_{1,1}^-\rightarrow1_{1,0}^-,\,v_t=0$ & 97.5878  &21.6\\
31&$^{13}$CH$_3$OH 
 & $({\rm A^+})$ $J_{K_aK_c}P=8_{0,8}^+\rightarrow7_{1,7}^+,\,v_t=0$ & 84.9745 & 81.5 \\
\sidehead{Ethenone (ketene)}\\
32 & \ethe & $J_{K_aK_c}=5_{1,5}\rightarrow4_{1,4}$ &100.0992 & 27.5 \\
\sidehead{Methoxymethane (dimethyl ether)}\\
33 &CH$_3$OCH$_3$ 
& $J_{K_aK_c}=5_{2,4}\rightarrow5_{1,5}\,({\rm EE})$ & 96.8554 & 19.3 \\
&& $J_{K_aK_c}=16_{3,14}\rightarrow15_{4,11}\,({\rm EE})$ & 97.9983 & 136.6 \\ 
&& $J_{K_aK_c}=4_{1,4}\rightarrow3_{0,2}$\tablenotemark{$\dagger$} &99.3303 & 10.2 \\
\sidehead{Acrylonitrile (vinyl cyanide, propenenitrile)}
34& CH$_2$CHCN
& $J_{K_aK_c}=9_{0,9}\rightarrow8_{0,8}$ & 84.9494 & 20.4 \\
&& $J_{K_aK_c}=9_{2,8}\rightarrow8_{2,7}$ & 85.3062 & 29.1 \\
&& $J_{K_aK_c}=9_{6,3}\rightarrow8_{6,2}$ & 85.4354$^{b12}$ & 98.3 \\ 
&& $J_{K_aK_c}=9_{6,4}\rightarrow8_{6,3}$ & 85.4354$^{b12}$ & 98.3 \\ 
&& $J_{K_aK_c}=9_{3,6}\rightarrow8_{3,5}$ & 85.4390 & 40.0 \\
&& $J_{K_aK_c}=9_{2,7}\rightarrow8_{2,6}$ & 85.7191 & 29.2 \\
&& $J_{K_aK_c}=9_{1,8}\rightarrow8_{1,7}$ & 87.3163 & 23.1 \\
&& $J_{K_aK_c}=10_{1,9}\rightarrow9_{1,8}$ & 96.9873$^{b9}$ & 27.8 \\
&& $J_{K_aK_c}=3_{2,1}\rightarrow4_{1,4}$ & 99.6650$^{b8}$ & 11.4 \\
\enddata
\tablenotetext{\mbox{$bn$}}{\phn\,Blending between the lines marked with the same $^{bn}$, where $n$ is a number.}
\tablenotetext{\dagger}{Blending of the four (AA, EE, EA, and AE) transitions.}
\tablenotetext{b\dagger}{\phn Possible blending with CH$_3$OCHO.}
\end{deluxetable}

\begin{deluxetable}{llll}
\tablewidth{0pc} \tablecolumns{4} \tabletypesize{\normalsize}
\tablecaption{Isotopic and Isomeric Ratios\label{tab-iso}}
\tablehead{\colhead{Species}&\colhead{Observed} & \colhead{Solar} & \colhead{ISM at}\\
\colhead{}&\colhead{}&\colhead{}&\colhead{$R_{\rm Gal}=6.9$ kpc}}
\startdata
$[\text{SiO}]/[^{29}\text{SiO}]$        & $15.1\pm0.4$ &  $19.7$ & $17.9\pm1.1$\\
$[^{29}\text{SiO}]/[^{30}\text{SiO}]$    & $1.4\pm0.2$  &  $1.5$  & $1.42\pm0.05$\\
$[\text{SO}]/[^{34}\text{SO}]$          & $22.5$\tablenotemark{a}& $22.5$ & $24.4\pm5.0$\\
$[\text{SO}_2]/[^{34}\text{SO}_2]$      & $22.5$\tablenotemark{a}& $22.5$ & $24.4\pm5.0$\\
$[^{34}\text{SO}]/[^{33}\text{SO}]$      & $5.4\pm0.6$  & $5.6$   & $6.3\pm1$ \\
$[\text{OCS}]/[\text{O}^{13}\text{CS}]$ & $25\pm2.4$   &  $89.4$ & $59\pm18$\\
$[\met]/[\tmet]$                       & $26\pm4$     &  $89.4$ & $59\pm18$\\
$[\hctn]/[\text{HC}^{13}\text{CCN}]$    & $32\pm4$     &  $89.4$ & $59\pm18$\\
$[\htcop]/[\hcdop]$                    & $6.9\pm2.7$  &  $5.6$  & $7.5\pm3.0$ \\
$[\htcn]/[\hcqn]$                      & $5.7\pm2$    &  $4.9$  & $4.5\pm1.6$ \\
$[\htcn]/[\hntc]$                      & $1.8\pm1.6$  &  $1$    & \nodata\\ 
\enddata
\tablerefs{Solar abundances from \citet{Asplund2009ARAA} Silicon fractionation from \citet{Monson2017ApJ}. Sulfur and carbon's from \citet{Wilson1999RPPh}. Nitrogen fractionation from \citet{Adande2012ApJ}.}
\tablenotetext{a}{Assumed in \citetalias{Guzman2014ApJ} and found consistent with the data.}
\end{deluxetable}

\clearpage
% 2017-07-05 15:35:09 INFO imfit	       --- ra:    16:59:41.71641 +/- 0.00106 s (0.01221 arcsec along great circle)
%2017-07-05 15:35:09 INFO imfit	       --- dec: -040.03.45.12872 +/- 0.00491 arcsec
%2017-07-05 15:35:09 INFO imfit	       --- Integrated:   8.63 +/- 0.12 Jy
%-13.762 +/- 0.035 km/s     2.93 +/- 0.12 km/s

% 2017-07-05 16:25:32 INFO imfit	       --- ra:    16:59:40.76506 +/- 0.00072 s (0.00821 arcsec along great circle)
%2017-07-05 16:25:32 INFO imfit	       --- dec: -040.03.36.75892 +/- 0.00294 arcsec
%2017-07-05 16:25:32 INFO imfit	       --- Integrated:   7.931 +/- 0.075 Jy
%-10.008 +/- 0.041 3.074 +/- 0.099

%2017-07-05 16:03:13 INFO imfit	       --- ra:    16:59:42.2831 +/- 0.0061 s (0.0695 arcsec along great circle)
%2017-07-05 16:03:13 INFO imfit	       --- dec: -040.03.39.6308 +/- 0.0497 arcsec
%2017-07-05 16:03:13 INFO imfit	       --- Integrated:   4.15 +/- 0.40 Jy
% -10.321 +/- 0.01 3.001 +/- 0.053 km/s
\begin{deluxetable}{lllccc}
\tablewidth{0pc} \tablecolumns{6} \tabletypesize{\normalsize}
\tablecaption{Parameters of \met, $5_{-1,0}\shortrightarrow4_{0,0}$ Maser Spots\label{tab-mas}}
\tablehead{\colhead{Maser}&\colhead{R.A.} & \colhead{Decl.} & \colhead{Flux density\tablenotemark{a}}&\colhead{$V_\text{LSR}$}&\colhead{FWHM}\\\colhead{~}&\colhead{(J2000)}&\colhead{(J2000)}&\colhead{(Jy)}&\colhead{(\kms)}&\colhead{(\kms)}\\
\colhead{~}&\colhead{16:59:\ldots}&\colhead{$-40$:03:\ldots}&\colhead{~}&\colhead{~}}
\startdata
\texttt{a} & $40.765\pm0.001$ & $36.759\pm0.003$ & $8.46\pm0.09$ & $-10.01\pm0.04$ & $3.07\pm0.1$\\
\texttt{b} & $41.716\pm0.001$ & $45.129\pm0.005$ & $8.63\pm0.1$  & $-13.76\pm0.04$ & $2.93\pm0.1$\\
\texttt{c} & $42.283\pm0.006$ & $39.63\pm0.05$   & $4.27\pm0.4$  & $-10.32\pm0.01$ & $3.00\pm0.1$ \\
\enddata
\tablenotetext{a}{Flux density values are primary beam corrected.}
\end{deluxetable}

\begin{deluxetable}{lclllcp{0.2\linewidth}}
\tablewidth{0pc} \tablecolumns{7} \tabletypesize{\small}
\tablecaption{Methanol (non-)LTE Model Parameters\label{tab-met}}
\tablehead{\colhead{Source}&\colhead{$T$} & \colhead{Column} & \colhead{$V_{\rm LSR}$} & \colhead{$\Delta V$} & \colhead{Density}&\colhead{Comment}\\
\colhead{~} & \colhead{(K)} & \colhead{$\log\left(\frac{N}{\text{cm}^{-2}}\right)$}  & \colhead{(\kms)} & \colhead{(\kms)} & 
\colhead{$\log\left(\frac{n}{\text{cm}^{-3}}\right)$}&\colhead{}
}
\startdata
CC core     & $107\pm 10$ & $16.57\pm 0.06$ & $-15.33\pm 0.05$  & $4.65\pm 0.1$   & $\ge 10$      & +Hotter component.\\
N-red cloud & $28\pm 4$   & $15.40\pm 0.05$ & $-7.5\pm0.5$      & $4.65\pm0.3$    & $6.9\pm 0.2$  & Blueshift absorption.\\
NEC-wall (a)& $57\pm10$   & $15.9\pm0.4$    & $-9.60\pm0.04$    & $3.7\pm 0.1$    & $7.2\pm0.3$   &  Exclude \maser.\\
NEC-wall (b)& $56\pm 5$   & $15.61\pm  0.06$& $-9.99\pm  0.06$ & $3.14\pm   0.1$ & $7.5\text{-}10.0$ & Exclude \maser. \\
%perl ../fitRad.pl consp/lsb_237-268_NECW.dat consp/usb_237-268_NECW.dat e-ch3oh.dat,a-ch3oh.dat 84000-85500 56,15.6,-9.95,2.97,0,7.5,0.5 -fix 1,5 -norun
Diffuse Ridge (a) & $33\pm2$    & $15.52\pm0.1$   & $-11.5\pm0.2$     & $3.6\pm0.1$     & $6.42\pm0.3$  & ~\\
Diffuse Ridge (b) & $38\pm 4$   & $15.6\pm0.1 $   & $-12.0$           & $3.5$           & $6.7\pm0.3$   & ~\\
Diffuse Ridge (c) & $25\pm15$   & $15.71\pm0.1$   & $-11.5\pm 0.3$    & $4.7\pm 0.5$    & $6.34\pm0.1$  &  Exclude \maser. \\
%perl ../fitRad.pl consp/lsb_263-206_DR.dat consp/usb_263-206_DR.dat e-ch3oh.dat,a-ch3oh.dat 84000-85500 25,15.71,-11.5,4.7,0.0,6.4,0.5 -fix 1,3,4,5,7 -norun
C8 maser   & $20\pm 2$   & $14.8\pm  0.09$ & $-10.19\pm  0.05$ & $  2.66\pm 0.1$ & $7.0\text{-}10.0$      & Exclude \maser.\\
C8 1\farcs5& $115\pm 30$ & $15.39\pm  0.2$ & $-10.93\pm   0.1$ & $  3.265\pm0.2$ & $7.2\text{-}10.0$      & Strong absorption in $E_{\rm up}\le40$ K lines.\\% 74.7 in emission
NW cloud    & $20\pm  10$ & $15.25\pm 0.3$  & $-10.2\pm  0.1$   & $   5.2\pm 0.3$ &$6.73\pm0.3$   &  Exclude \maser.\\
NW cloud (a)& $50\pm10$   & $15.95\pm0.1$   & $-10.2\pm  0.1$   & $   5.2\pm 0.3$ &$6.8\pm0.2$    &  ~\\
Source 3 &$  221\pm 30$ & $ 16.22\pm  0.08$ & $ -11.15\pm   0.2$ & $   5.34\pm   0.3$ & $7.0\text{-}10.0$ & Exclude \maser.\\
Source 18   & $64\pm4$    & $15.76\pm 0.04$ & $-14.08\pm  0.06$ & $   4.25\pm0.1$ & $\ge 10$  & Absorption in  $E_{\rm up}\le40$ K lines.\\
\enddata
\end{deluxetable}

\startlongtable
\begin{deluxetable}{lclllcp{0.25\linewidth}}
\tablewidth{0pc} \tablecolumns{7} \tabletypesize{\small}
\tablecaption{Physical Parameters of SET Emission Models\tablenotemark{$\star$} \label{tab-cdt}}
\tablehead{ \colhead{Molecule} & \colhead{$T$\tablenotemark{a}}     & \colhead{Column}  & \colhead{$V_{\rm LSR}$} & \colhead{$\Delta V$} & \colhead{Vel.} & \colhead{Comment}\\ \colhead{~}    & \colhead{~}       & \colhead{Density} & \colhead{}          & \colhead{}        & \colhead{Cmp.} & \colhead{} \\ \colhead{~}    & \colhead{(K)}     & \colhead{$\log\left(\frac{N}{\text{cm}^{-2}}\right)$} & \colhead{(\kms)}    & \colhead{(\kms)}  & \colhead{~} & \colhead{~}}
\startdata
% perl ../fitLTE_1mol.pl consp/lsb_s10.dat consp/usb_s10.dat SO - 64,13.,-15.,5.4,0.0 -fix 1,5
\sidehead{\textbf{CC core}}% ($\text{R.A.}=16^\text{h}59^\text{m}41\fs45$, $\text{decl.}=-40\arcdeg03\arcmin41\farcs2$)% 258-259 16:59:41.447 -40:03:41.21 0.056192 Jy/b
\damm          & $108\pm 10$       & $14.14\pm 0.06$   & $-15.7\pm 0.3$      & $6.59\pm 0.9$     & 1   & ~\\% ncrti 5.0 $>n_c\approx5.0$
% perl fitLTE_1mol.pl pixelSpecs/lsb_258-259.dat pixelSpecs/usb_258-259.dat NH2D - 50,15.33,-15.4,4.6,0.0  -fix 5 
%
\methoxy       & $138\pm20$ & $15.76\pm0.1$ & $-15.65\pm0.4$ & $5.54\pm 0.8$ & 1   & +Hotter component.\\
%& $107^\dag$        & $15.61\pm 0.05$   & $-15.69\pm 0.4$     & $5.4\pm 1$        & 1   & ~\\
% perl ../fitLTE_1mol.pl ../pixelSpecs/lsb_258-259.dat ../pixelSpecs/usb_258-259.dat CH3OCH3 97000-97990,84000-84600,96920-96930 105,15.6,-15.4,4.5,0.0  -fix 5 -device pdfs/spectra/CCcore_CH3OCH3/cps
%
\ethe          & $107^\dag$        & $14.53\pm 0.06$   & $-15.56\pm 0.4$     & $5.7\pm 1$        & 1   & ~\\
% perl fitLTE_1mol.pl pixelSpecs/lsb_258-259.dat pixelSpecs/usb_258-259.dat H2CCO - 105,15,-15.,4.6,0.0 -fix 1,5
%
\hctn          & $107^\dag$        & $14.82\pm 0.01$   & $-15.55\pm 0.07$    & $4.63\pm 0.2$     & 1   & ~\\%$>5.7$
%perl fitLTE_1mol.pl pixelSpecs/lsb_258-259.dat pixelSpecs/usb_258-259.dat HC3N - 105,15,-15.,4.6,0.0 -fix 1,5
%
HC$^{13}$CCN   & $107^\dag$        & $13.32\pm 0.05$   & $-15.2\pm 0.3$      & $4.7\pm 0.6$      & 1   & ~\\
% perl fitLTE_1mol.pl pixelSpecs/lsb_258-259.dat pixelSpecs/usb_258-259.dat HC13CCN - 105,14,-15.,4.6,0.0 -fix 1,5
\met           & $107\pm 10$       & $16.57\pm 0.06$   & $-15.33\pm 0.05$    & $4.65\pm 0.1$     & 1   & \dag, +hotter component.\\
% perl fitLTE_1mol.pl pixelSpecs/lsb_258-259.dat pixelSpecs/usb_258-259.dat CH3OH - 100,15.6,-15.4,4.5,0.0  -fix 5 
%
$^{13}$\met    & $107^\dag$        & $15.16\pm 0.05$   & $-15.33$   & $4.65$            & 1   & $V_{\rm LSR}$ and $\Delta V$ from \met.\\
% perl fitLTE_1mol.pl pixelSpecs/lsb_258-259.dat pixelSpecs/usb_258-259.dat 13CH3OH - 107,15,-15.33,4.6,0.0  -fix 1,3,4,5
\propanenitrile& $123\pm 9$        & $14.80\pm 0.03$   & $-15.27\pm 0.04$    & $4.9\pm 0.09$     & 1   & ~\\
%
%
% perl fitLTE_1mol.pl pixelSpecs/lsb_258-259.dat pixelSpecs/usb_258-259.dat CH3CH2CN 86755-86760,97000-97200,99600-99675,99695-2e5 123,14.7,-15.,4.6,0.0  -fix 5
\htcn          & $64^\ddag$        & $14.83\pm  0.01$  & $-14.78\pm   0.1$   & $5.145\pm   0.4$  & 2   & ~\\
%perl fitLTE_1mol.pl pixelSpecs/lsb_258-259.dat pixelSpecs/usb_258-259.dat H13CN_jpl - 64,14.5,-15.,4.6,0.0 -fix 1,5
HNCO           & $65.1\pm 5$       & $15.01\pm 0.03$   & $-14.64\pm 0.09$    & $6.7\pm 0.2$      & 2   & +Hotter component.\\%ncri 6.0
% perl fitLTE_1mol.pl pixelSpecs/lsb_258-259.dat pixelSpecs/usb_258-259.dat HNCO - 74,15.33,-15.4,4.6,0.0  -fix 5 
\hcqn          & $64^\ddag$        & $14.28\pm 0.009$  & $-14.51\pm  0.06$   & $5.3\pm   0.1$    & 2   & ~\\%$>6.5$
%perl fitLTE_1mol.pl pixelSpecs/lsb_258-259.dat pixelSpecs/usb_258-259.dat HC15N - 64,14.5,-15.,4.6,0.0 -fix 1,5
OCS            & $64^\ddag$        & $15.81\pm 0.01$   & $-14.41\pm  0.05$   & $5.4\pm   0.1$    & 2   & ~\\
% perl fitLTE_1mol.pl pixelSpecs/lsb_258-259.dat pixelSpecs/usb_258-259.dat O13CS 96985-96990 64,14.4,-15.,5.4,0.0 -fix 1,5
O$^{13}$CS     & $64^\ddag$        & $14.41\pm  0.04$  & $-14.21\pm   0.3$   & $5.1\pm   0.6$    & 2   & ~\\
\propyne       & $64.0\pm 4$       & $15.82\pm 0.03$   & $-14.23\pm 0.06$    & $4.67\pm 0.1$     & 2   & \ddag\\ %ncrit 4.1 $>n_c\approx4.1$
% perl fitLTE_1mol.pl pixelSpecs/lsb_258-259.dat pixelSpecs/usb_258-259.dat CH3CCH - 50,15.33,-15.4,4.6,0.0  -fix 5 
%
$^{33}$SO      & $64^\ddag$        & $14.23\pm  0.05$  & $-14.3\pm   0.4$    & $5.4$             & 2   & Faint.\\% nc 5.5
% perl fitLTE_1mol.pl pixelSpecs/lsb_258-259.dat pixelSpecs/usb_258-259.dat 33SO - 64,14.3,-15.,5.4,0.0 -fix 1,4,5
% CCH associated with a much broader component and centered at -9.5 kms
%CCH & $64\pm     0$ & $15.52\pm  0.02$ & $-9.487\pm   0.2$ & $8.389\pm   0.4$ &
% perl fitLTE_1mol.pl pixelSpecs/lsb_258-259.dat pixelSpecs/usb_258-259.dat CCH - 64,15.33,-11.4,5.6,0.0  -fix 1,5
\htcop   &  $64^\ddag$ & $  13.83\pm  0.02$ & $-13.38\pm   0.2$ & $  6.4\pm   0.4$ & & \propyne\ $T_\text{ex}$ from group 2.\\
\hcdop   &  $64^\ddag$ & $  12.63\pm   0.1$ & $-13.4\pm   0.6$ & $  4.0\pm     1$ & & \propyne\ $T_\text{ex}$ from group 2.\\
C$^{33}$S &  $64^\ddag$ & $  13.96\pm  0.04$ & $-12.84\pm   0.3$ & $  7.2\pm     1$ & & Self-absorption? \propyne\ $T_\text{ex}$ from group 2.\\
\hcsp    &  $64^\ddag$ & $  13.74\pm  0.07$ & $-11.43\pm   0.7$ & $  9.4\pm     2$ & & \propyne\ $T_\text{ex}$ from group 2.\\
% perl fitLTE_1mol.pl pixelSpecs/lsb_258-259.dat pixelSpecs/usb_258-259.dat HCS+ - 64,13.,-15.,5.4,0.0 -fix 1,5 -device m/pdfs/spectra/CCcore_HCSp/cp
CCH      &  $64^\ddag$ & $  15.52\pm  0.02$ & $-9.487\pm   0.2$ & $   8.39\pm   0.4$ & & \propyne\ $T_\text{ex}$ from group 2.\\
%%%%%%%%%%%%%%%%%%%%%%%%%%%%%%-------------------------------------
%
\sidehead{\textbf{N-red cloud}}% continuum is negative.... -Jy/b
\met           & $28\pm 4$         & $15.40\pm 0.05$   & $-7.5\pm0.5$        & $4.65\pm0.3$      & 1   & \dag, blueshift absorption.\\
%perl ../fitRad.pl consp/lsb_Nred.dat consp/usb_Nred.dat  e-ch3oh.dat,a-ch3oh.dat -  28.01,15.38,-7.5,5.0,0.0,6.9,0.5 -fix 1,3,4,5,7 -norun
%perl ../fitRad.pl consp/lsb_Nred.dat consp/usb_Nred.dat  e-ch3oh.dat,a-ch3oh.dat -  24.01,15.38,-7.5,5.0,0.0,6.9,0.5 -fix 1,3,4,5,7 -norun
%perl ../fitRad.pl consp/lsb_Nred.dat consp/usb_Nred.dat  e-ch3oh.dat,a-ch3oh.dat -  32.01,15.38,-7.5,5.0,0.0,6.9,0.5 -fix 1,3,4,5,7 -norun
OCS            & $28^\dag$         & $14.41\pm  0.02$  & $-7.4\pm  0.1$      & $5.2\pm  0.3$     & 1   & ~\\
%perl ../fitLTE_1mol.pl consp/lsb_Nred.dat consp/usb_Nred.dat OCS - 25,13.00,-7.5,4.6,0.0  -fix 1,5
SO             & $8.5\pm1$         & $14.65\pm  0.01$  & $-7.22\pm  0.07$    & $5.408\pm   0.2$  & 1   & \ddag, subthermal.\\
%perl ../fitLTE_1mol.pl consp/lsb_Nred.dat consp/usb_Nred.dat SO - 8.5,13.,-6.5,4.6,0.0  -fix 1,5
$^{34}$SO      & $8.5^\ddag$       & $13.27\pm  0.03$  & $-7.158\pm   0.2$   & $5.042\pm   0.4$  & 1   & ~\\
%perl ../fitLTE_1mol.pl consp/lsb_Nred.dat consp/usb_Nred.dat 34SO - 8.5,13.00,-7.5,4.6,0.0  -fix 1,5
SiO            & $8.5^\ddag $      & $14.03\pm  0.02$  & $-6.91\pm   0.1$    & $6.59\pm   0.2$   & 1   & ~\\
$^{29}$SiO     & $8.5^\ddag $      & $12.86\pm  0.02$  & $-6.87\pm   0.2$    & $6.86\pm   0.3$   & 1   & ~\\
$^{30}$SiO     & $8.5^\ddag $      & $12.68\pm  0.02$  & $-6.87\pm   0.2$    & $7.16\pm   0.5$   & 1   & ~\\
%perl ../fitLTE_1mol.pl consp/lsb_Nred.dat consp/usb_Nred.dat SiO   - 8.5,14.50,-7.5,4.6,0.0  -fix 1,5
%perl ../fitLTE_1mol.pl consp/lsb_Nred.dat consp/usb_Nred.dat 29SiO - 8.5,13.00,-7.5,4.6,0.0  -fix 1,5
%perl ../fitLTE_1mol.pl consp/lsb_Nred.dat consp/usb_Nred.dat 30SiO - 8.5,13.00,-7.5,4.6,0.0  -fix 1,5
% some detected emission at  CCH at -9. kms but not evident in mom0 either
\acet          & $28^\dag$         & $13.73\pm   0.2$  & $-7.68\pm   0.8$    & $4.238\pm  0.6$   & 1   & ~\\
%perl ../fitLTE_1mol.pl consp/lsb_Nred.dat consp/usb_Nred.dat CH3CHO 84500-84600 28,12.9,-7.5,4.6,0.0  -fix 1,5
\hctn          & $8.5^\ddag$       & $13.94\pm  0.03$  & $-6.93\pm   0.2$    & $4.15\pm   0.4$   & 1   & Strong blueshift absorption.\\
% perl ../fitLTE_1mol.pl consp/lsb_Nred.dat consp/usb_Nred.dat HC3N - 8.5,13,-7.5,4.6,0.0  -fix 1,5
HNCO           & $8.5^\ddag$       & $13.64\pm  0.02$  & $-6.70\pm   0.1$    & $6.10\pm 0.3$     & 1   & ~\\
% perl ../fitLTE_1mol.pl consp/lsb_Nred.dat consp/usb_Nred.dat HNCO - 8.5,13,-7.5,4.6,0.0  -fix 1,5 
\ethe          & $28^\dag$         & $13.28\pm  0.08$  & $-6.97\pm   0.6$    & $5.8\pm 1$        & 1   & ~\\
%perl ../fitLTE_1mol.pl consp/lsb_Nred.dat consp/usb_Nred.dat H2CCO - 8.5,12.8,-7.5,4.6,0.0  -fix 1,5
%
%
CS             & $28^\dag$         & $14.44\pm  0.01$  & $-5.93\pm0.3$       & $5.6\pm 0.2$      & 2   & Blueshift absorption. Radiative trapping. \\
%perl ../fitLTE_1mol.pl consp/lsb_Nred.dat consp/usb_Nred.dat CS - 28,14.7,-6.5,4.6,0.0  -fix 1,5
C$^{33}$S      & $28^\dag$         & $12.33\pm0.2$      & $-5.93$             & $5.6$             & 2   & CS line profile.\\%[C$^{32}$S]/[C$^{33}$S]=127.\\
%
%perl ../lte.pl consp/lsb_Nred.dat consp/usb_Nred.dat CS,C33S 126.7,1 28,14.44,-5.93,5.6,0.0
\hcqn          & $8.5^\ddag$       & $12.15\pm  0.05$  & $-5.59\pm   0.2$    & $4.75\pm   0.6$   & 2   & Strong blueshift absorption.\\
%perl ../fitLTE_1mol.pl consp/lsb_Nred.dat consp/usb_Nred.dat HC15N - 8.5,14.50,-7.5,4.6,0.0  -fix 1,5
% H13CN there is emission logNcol ~12.7 but lots of absorption and complication
%
%%%%%%%%%%%%%%%%%%%%%%%%%%%%%%-------------------------------------
%
\sidehead{\textbf{NEC-wall}}
\sidehead{NEC-wall (a): Methanol Peak\tablenotemark{b}}%225-266 16:59:42.466 -40:03:40.91 0.001990Jy/b
CS             & $57^\dag$         & $14.76\pm  0.07$  & $-9.13$             & $3.7$             &     & Complex line profile.\\
SO             & $57^\dag$         & $13.59\pm  0.08$  & $-9.13$             & $3.7$             &     & Faint emission.\\
\met           & $57\pm10$         & $15.9\pm0.4$      & $-9.60\pm0.04$      & $3.7\pm 0.1$      &     & \dag, exclude \maser.\\
%perl ../fitRad.pl consp/lsb_225-266_NECW.dat consp/usb_225-266_NECW.dat e-ch3oh.dat,a-ch3oh.dat 84000-85500  57,15.90,-9.6,3.7,0.0,7.2,0.5 -fix 3,4,5,6,7 -norun
\acet          & $57^\dag$         & $14.28\pm  0.06$  & $-10.09\pm   0.2$   & $3.318\pm 0.5$    &     & ~\\
$^{13}$\met    & $57^\dag$         & $14.66\pm   0.1$  & $-10.23\pm0.2$      & $3.553\pm   0.9$  &     & Tentative.\\
CCH            & $57^\dag$         & $15.68\pm 0.005$  & $-10.46\pm  0.02$   & $3.878\pm  0.06$  &     & ~\\
\hcqn          & $57^\dag$         & $13.64\pm  0.03$  & $-10.86\pm   0.2$   & $6.31\pm   0.4$   &     & ~\\
%perl ../fitLTE_1mol.pl consp/lsb_225-266_NECW.dat consp/usb_225-266_NECW.dat HC15N -  57,13.50,-9.6,3.7,0.0 -fix 1,5 -device pdfs/spectra/NEC-wall/NECw_MP_HC15N/cps
\htcn          & $57^\dag$         & $14.33\pm 0.006$  & $-10.86\pm  0.06$   & $5.49\pm   0.1$   &     & ~\\
%perl ../fitLTE_1mol.pl consp/lsb_225-266_NECW.dat consp/usb_225-266_NECW.dat H13CN_jpl -  42,14,-10,5.,0.0 -fix 1,5
HC$^{13}$CCN   & $42^\ddag$        & $12.85\pm  0.04$  & $-11.22\pm   0.2$   & $4.532\pm   0.5$  &     & ~\\
\hntc          & $42^\ddag$        & $13.55\pm  0.02$  & $-11.55\pm  0.08$   & $4.32\pm   0.2$   &     & ~\\
\htcop         & $42^\ddag$        & $13.66\pm 0.01$   & $-11.46\pm  0.07$   & $4.146\pm   0.1$  &     & ~\\
\hcdop         & $42^\ddag$        & $12.7\pm 0.05$    & $-12.04\pm   0.2$   & $4.23\pm   0.5$   &     & ~\\
\propyne       & $42.1\pm 2$       & $15.45\pm  0.02$  & $-12.13\pm  0.07$   & $5.504\pm   0.1$  &     & \ddag\\
\hctn          & $42^\ddag$        & $14.46\pm  0.01$  & $-12.31\pm  0.08$   & $5.864\pm   0.2$  &     & ~\\
\damm          & $42^\ddag$        & $13.81\pm  0.05$  & $-14.61\pm   0.5$   & $8.539\pm     1$  &     & ~\\
\sidehead{NEC-wall (b): Sulfur Monoxide Peak}%237 268  16:59:42.153 -40:03:40.31 0.001131Jy/b
\met           & $56\pm 5$         & $15.61\pm  0.06$  & $-9.993\pm  0.06$   & $3.14\pm   0.1$   & 1   & \ddag, exclude \maser. \\
%perl ../fitLTE_1mol.pl consp/lsb_237-268_NECW.dat consp/usb_237-268_NECW.dat CH3OH 84000-85500 56,15.6,-9.99,3.14,0 -fix 5
%
CS             & $56^\ddag$        & $15.05\pm  0.04$  & $-12.36\pm  0.3$    & $6.64\pm 0.8$     &     & Complex line profile. Optically thick?\\
C$^{33}$S      & $56^\ddag$        & $13.76\pm  0.03$  & $-10.11\pm   0.2$   & $4.23\pm   0.4$   & 1   & ~\\
%perl ../fitLTE_1mol.pl consp/lsb_237-268_NECW.dat consp/usb_237-268_NECW.dat CS - 80,15.6,-9.99,3.14,0 -fix 5
OCS            & $56^\ddag$        & $14.49\pm  0.02$  & $-10.08\pm  0.08$   & $3.52\pm   0.2$   & 1   & ~\\
% no O13CS
\hcsp          & $56^\ddag$        & $13.35\pm  0.07$  & $-9.586\pm   0.3$   & $4.19\pm   0.8$   & 1   & ~\\
CCH            & $56^\ddag$        & $15.55\pm  0.01$  & $-10.32\pm  0.06$   & $5.48\pm   0.1$   & 1   & ~\\
\htcop         & $80^\dag$         & $13.73\pm  0.01$  & $-11.51\pm  0.07$   & $4.25\pm   0.2$   & 2   & ~\\
%perl ../fitLTE_1mol.pl consp/lsb_237-268_NECW.dat consp/usb_237-268_NECW.dat H13CO+ - 80,13.6,-9.99,3.14,0 -fix 1,5
\propyne       & $43.2\pm     3$   & $15.37\pm  0.03$  & $-12.46\pm  0.07$   & $4.00\pm   0.1$   & 2   & ~\\
SO             & $80\pm 40$        & $15.35\pm   0.2$  & $-11.89\pm  0.08$   & $5.93\pm   0.2$   & 2   & \dag, poorly constrained temperature.\\
%perl ../fitLTE_1mol.pl consp/lsb_237-268_NECW.dat consp/usb_237-268_NECW.dat SO - 72,15,-13.,3.5,0.0 -fix 5
$^{34}$SO      & $80^\dag$         & $14.12\pm  0.04$  & $-12.23\pm   0.4$   & $7.36\pm   0.7$   & 2   & ~\\
\hccn          & $32.89\pm7$       & $13.28\pm   0.1$  & $-12.27\pm   0.1$   & $3.58\pm   0.3$   & 2   & ~\\
\hctn          & $80^\dag$         & $14.35\pm 0.01$   & $-12.53\pm  0.04$   & $4.34\pm   0.1$   & 2   & ~\\
HC$^{13}$CCN   & $80^\dag$         & $12.78\pm  0.06$  & $-12.4\pm   0.2$    & $2.94\pm   0.4$   & 2   & ~\\
\hcqn          & $80^\dag$         & $13.85\pm  0.01$  & $-11.76\pm  0.07$   & $4.30\pm   0.2$   & 2   & ~\\
\htcn          & $80^\dag$         & $14.46\pm 0.007$  & $-12.31\pm  0.06$   & $4.69\pm   0.1$   & 2   & ~\\
\hntc          & $80^\dag$         & $13.57\pm  0.02$  & $-11.92\pm   0.1$   & $3.722\pm  0.2$   & 2   & ~\\
SiO            & $80^\dag$         & $13.29\pm  0.04$  & $-13.67\pm   0.2$   & $3.42\pm   0.4$   &     & ~\\
%
%%%%%%%%%%%%%%%%%%%%%%%%%%%%%%-------------------------------------
%
\sidehead{\textbf{Diffuse Ridge}}% 
\sidehead{DR(a)\tablenotemark{b}}% 213-235 16:59:42.780 -40:03:50.21, diffuse part v separation? 0.000170Jy/b
%CS            & $8.5\pm0.4$       & $15.21$           & $-10.87\pm   0.1$   & $1.3\pm   0.2$    &     & Optically thick. [C$^{32}$S]/[C$^{33}$S]=127.\\
CS             & $33^\dag$         & $14.25\pm  0.01$  & $-10.5\pm  0.04$    & $2.957\pm  0.08$  &     & Optically thick. Radiative trapping.\\
%perl ../fitLTE_1mol.pl consp/lsb_213-235_DR.dat consp/usb_213-235_DR.dat CS - 33,14,-10.7,4,0 -fix 1,5
\hcqn          & $6^\ddag$         & $12.13\pm   0.1$  & $-10.97\pm   0.4$   & $3.00\pm   0.7$   &     & ~\\
%perl ../fitLTE_1mol.pl consp/lsb_213-235_DR.dat consp/usb_213-235_DR.dat HC15N - 6.0,13,-11.7,3,0 -fix 1,5
C$^{33}$S      & $6^\ddag$         & $13.11\pm  0.07$  & $-11.4\pm   0.3$    & $4.04\pm   0.9$   &     & ~\\
\met           & $33\pm2$          & $15.52\pm0.1$     & $-11.5\pm0.2$       & $3.6\pm0.1$       &     & \dag\\
% perl ../fitRad.pl consp/lsb_213-235_DR.dat consp/usb_213-235_DR.dat e-ch3oh.dat,a-ch3oh.dat -  32.2,15.52,-11.5,3.6,0.0,6.42,0.5 -fix 1,3,4,5,7  -norun
HNCO           & $6^\ddag$         & $13.69\pm  0.05$  & $-11.67\pm   0.2$   & $3.59\pm   0.4$   &     & ~\\
OCS            & $33^\dag$         & $14.25\pm  0.03$  & $-11.70\pm   0.1$   & $3.36\pm   0.2$   &     & ~\\
\htcn          & $6^\ddag$         & $13.20\pm  0.03$  & $-11.98\pm   0.1$   & $3.00\pm   0.2$   &     & ~\\
%perl ../fitLTE_1mol.pl consp/lsb_213-235_DR.dat consp/usb_213-235_DR.dat H13CN_jpl - 6.0,13,-11.7,3,0 -fix 1,5
\acet          & $33^\dag$         & $14.24\pm   0.04$ & $-12.0\pm   0.2$    & $4.20\pm     0.5$ &     & ~\\
%perl ../fitLTE_1mol.pl consp/lsb_213-235_DR.dat consp/usb_213-235_DR.dat CH3CHO 84500-84600 33,14,-11.7,4,0 -fix 1,5
\ethe          & $6^\ddag$         & $14.5\pm  0.06$   & $-12.30\pm   0.3$   & $4.04\pm   0.5$   &     & ~\\
\htcop         & $6^\ddag$         & $12.88\pm  0.03$  & $-12.32\pm  0.08$   & $2.715\pm   0.2$  &     & ~\\
\hntc          & $6^\ddag$         & $12.87\pm  0.03$  & $-12.44\pm  0.09$   & $2.932\pm   0.2$  &     & ~\\
%perl ../fitLTE_1mol.pl consp/lsb_213-235_DR.dat consp/usb_213-235_DR.dat HN13C - 6.0,13,-11.7,3,0 -fix 1,5
SiO            & $6^\ddag$         & $12.87\pm  0.09$  & $-12.70\pm   0.3$   & $3.14\pm   0.7$   &     & Redshifted absorption.\\
\hcdop         & $6^\ddag$         & $11.94\pm   0.1$  & $-12.70\pm   0.5$   & $3.207\pm   0.9$  &     & ~\\
SO             & $6.0\pm0.8$       & $14.3\pm    0.1$  & $-12.83\pm  0.09$   & $2.89\pm   0.2$   &     & \ddag\\
\hcsp          & $36.4^{\dagg}$ & $13.13\pm  0.07$  & $-13.00\pm     0.3$ & $3.30\pm   0.6$   &     & \\
\propyne       & $36.4\pm 4$       & $14.8\pm   0.06$  & $-13.11\pm  0.08$   & $3.181\pm   0.2$  &     & $\dagg$\\
CCH            & $6^\ddag$         & $14.38\pm  0.03$  & $-13.20\pm   0.1$   & $2.567\pm   0.2$  &     & ~\\
%
%\sidehead{DR 2}%188-246, eastern core
%
\sidehead{DR(b)}%221-228 16:59:42.571 -40:03:52.31 0.000276Jy/b
\met           & $38\pm 4$         & $15.6\pm0.1 $     & $-12.0$             & $3.5$             & 1   & \\
%perl ../fitRad.pl consp/lsb_221-228_DR.dat consp/usb_221-228_DR.dat e-ch3oh.dat,a-ch3oh.dat - 38,15.6,-12.,3.5,0.0,6.7,0.5 -fix 1,3,4,5,7  -norun
\propyne       & $37.7\pm 4$       & $14.92\pm  0.05$  & $-12.23\pm 0.1$     & $3.63\pm 0.2$     & 1   & \dag\\
\htcop         & $38^\dag$         & $13.44\pm  0.01$  & $-12.25\pm  0.05$   & $3.21\pm   0.1$   & 1   & ~\\
\hcdop    & $38^\dag$         & $12.68\pm  0.02$  & $-12.74\pm   0.1$   & $3.57\pm   0.2$   & 1   & ~\\
CCH            & $11^\ddag$        & $14.74\pm  0.01$  & $-11.52\pm  0.05$   & $3.6\pm   0.1$    & 1   & SO $T_{ex}$ from group 2.\\
\acet          & $38^\dag$         & $14.29\pm0.05$    & $-12.47\pm0.2$      & $3.7\pm     0.4$  & 1   & ~\\
%perl ../fitLTE_1mol.pl consp/lsb_221-228_DR.dat consp/usb_221-228_DR.dat CH3CHO 84500-84600 38,14.6,-12.,3.5,0.0 -fix 1,5
\hctn          & $11^\ddag$        & $14.32\pm  0.01$  & $-12.17\pm  0.03$   & $2.98\pm  0.07$   & 1   & SO $T_{ex}$ from group 2.\\
\hntc          & $11^\ddag$        & $12.96\pm  0.03$  & $-11.78\pm   0.1$   & $3.3\pm   0.2$    & 1   & SO $T_{ex}$ from group 2.\\
\hcqn          & $11^\ddag$        & $12.68\pm  0.04$  & $-10.81\pm   0.2$   & $3.8\pm   0.4$    & 2   & ~\\
\htcn          & $11^\ddag$        & $13.5\pm  0.03$   & $-10.74\pm   0.2$   & $3.84\pm   0.4$   & 2   & ~\\
HNCO           & $11^\ddag$        & $13.52\pm  0.05$  & $-10.59\pm   0.2$   & $4.16\pm   0.5$   & 2   & ~\\
SO             & $10.7\pm 2$       & $14.66\pm 0.04$   & $-10.97\pm   0.1$   & $5.241\pm   0.3$  & 2   & \ddag\\
$^{34}$SO      & $11^\ddag$        & $13.21\pm  0.09$  & $-10.78\pm   0.6$   & $4.7\pm     1$    & 2   & ~\\
CS             & $38^\dag$        & $14.71\pm0.03$    & $-10.32\pm  0.1$    & $4.17\pm   0.3$   & 2   & Optically thick. Radiative trapping. \\%[C$^{32}$S]/[C$^{33}$S]=127.\\
%perl ../fitLTE_1mol.pl consp/lsb_221-228_DR.dat consp/usb_221-228_DR.dat CS  11,14.97,-12.,3.5,0.0 -fix 1,5
C$^{33}$S      & $11^\ddag$        & $12.87\pm  0.03$  & $-10.21\pm   0.1$   & $3.8\pm   0.3$    & 2   & ~\\
\hcsp          & $38^\dag$         & $13.28\pm  0.04$  & $-11.16\pm   0.2$   & $4.22\pm   0.5$   & 2   & \propyne\ $T_{ex}$ from group 1.\\
SiO            & $11^\ddag$        & $13.69\pm  0.02$  & $-8.334\pm   0.2$   & $7.49\pm   0.3$   &     & SO $T_{ex}$ from group 2.\\%
\sidehead{DR(c)\tablenotemark{b}}%263-206 16:59:41.473 -40:03:58.91 0.000158Jy/b
CS             & $7^\dag$          & $14.39\pm  0.03$  & $-7.876\pm   0.1$   & $5.851\pm   0.3$  &     & ~\\
C$^{33}$S      & $7^\dag$          & $12.43\pm 0.1$    & $-7.88$             & $4.264\pm 1$      &     & ~\\
\htcn          & $7^\dag$          & $13.31\pm  0.02$  & $-10.12\pm   0.3$   & $7.552\pm   0.8$  &     & ~\\%
SiO            & $7^\dag$          & $13.73\pm  0.02$  & $-10.46\pm   0.1$   & $6.16\pm   0.3$   &     & ~\\
\hcqn          & $7^\dag$          & $12.57\pm  0.05$  & $-10.46\pm   0.3$   & $5.372\pm   0.8$  &     & ~\\
SO             & $6.7\pm0.9$       & $14.55\pm0.1$     & $-10.85\pm  0.08$   & $5.515\pm   0.3$  &     & \dag\\
OCS            & $19.8\pm 5$       & $14.49\pm0.02$    & $-11.01\pm   0.1$   & $5.96\pm   0.2$   &     & ~\\
%perl ../fitLTE_1mol.pl consp/lsb_263-206_DR.dat consp/usb_263-206_DR.dat OCS - 10,14.5,-11,3.9,0.0 -fix 5 -device pdfs/spectra/DR/DR_c_OCS/cps
\met           & $25\pm15$         & $15.71\pm0.1$     & $-11.5\pm 0.3$      & $4.7\pm 0.5$      &     & \ddag, exclude \maser. \\
%perl ../fitRad.pl consp/lsb_263-206_DR.dat consp/usb_263-206_DR.dat e-ch3oh.dat,a-ch3oh.dat 84000-85500 25,15.71,-11.5,4.7,0.0,6.4,0.5 -fix 1,3,4,5,7 -norun
HNCO           & $7^\dag$          & $13.92\pm  0.03$  & $-11.59\pm   0.2$   & $6.436\pm   0.5$  &     & ~\\
\hctn          & $7^\dag$          & $14.35\pm  0.04$  & $-11.72\pm   0.4$   & $8.55\pm   0.9$   &     & ~\\
\acet          & $25^\ddag$        & $14.18\pm  0.04$  & $-12.48\pm   0.1$   & $4.77\pm   0.3$   &     & ~\\
%perl ../fitLTE_1mol.pl consp/lsb_263-206_DR.dat consp/usb_263-206_DR.dat CH3CHO 84500-84600 25,14.,-12.,3.9,0.0 -fix 1,5
\hntc          & $7^\dag$          & $12.56\pm  0.05$  & $-13.09\pm   0.2$   & $3.471\pm   0.4$  &     & ~\\%lN 12.75 si 20 K
%perl ../fitLTE_1mol.pl consp/lsb_263-206_DR.dat consp/usb_263-206_DR.dat H13CN_jpl - 7,12.5,-12.,3.9,0.0 -fix 1,5
%
%%%%%%%%%%%%%%%%%%%%%%%%%%%%%%-------------------------------------
%
\sidehead{\textbf{C8}}
\sidehead{C8 maser}% 0.003847Jy/b -> uncatalogued continuum source!
CS             & $17^\dag$         & $13.93\pm   0.1$  & $-8.094\pm   0.8$   & $5.0\pm     1$  &     & Strong blueshifted absorption.\\
SiO            & $17^\dag$         & $13.84\pm  0.03$  & $-8.683\pm   0.4$   & $10.8\pm     1$  &     & ~\\
$^{29}$SiO     & $17^\dag$         & $12.55\pm  0.03$  & $-8.764\pm   0.3$   & $9.30\pm   0.7$   &     & ~\\
$^{30}$SiO     & $17^\dag$         & $12.49\pm  0.04$  & $-9.32\pm     0.4$  & $9.30\pm   0.9$   &     & ~\\
%perl ../fitLTE_1mol.pl consp/lsb_S8_maser.dat consp/usb_S8_maser.dat 29SiO -  17,13.3,-10.7,2.7,0 -fix 1,5
C$^{33}$S      & $17^\dag$         & $12.68\pm  0.06$  & $-9.479\pm   0.2$   & $2.63\pm   0.5$  &     & ~\\
\met           & $20\pm     2$     & $14.8\pm  0.09$   & $-10.19\pm  0.05$   & $2.66\pm   0.1$   &     & \dagg, exclude \maser.\\
% methanol maser very strong
%perl ../fitLTE_1mol.pl consp/lsb_S8_maser.dat consp/usb_S8_maser.dat CH3OH 84000-85500 20,14.1,-10.2,2.7,0 -fix 1,5
%perl ../fitRad.pl consp/lsb_S8_maser.dat consp/usb_S8_maser.dat e-ch3oh.dat,a-ch3oh.dat 84000-85500 20,14.8,-10.2,2.7,0.0,8.433,0.5 -fix 1,3,4,5,7 -norun
\cyc           & $10.2\pm   1$     & $13.16\pm 0.03$   & $-10.29\pm  0.07$   & $3.60\pm   0.1$  &     & ~\\
\htcn          & $17^\dag$         & $13.43\pm  0.02$  & $-10.35\pm   0.2$   & $4.49\pm   0.4$  &     & ~\\
%perl ../fitLTE_1mol.pl consp/lsb_S8_maser.dat consp/usb_S8_maser.dat H13CN_jpl -  17,13.3,-10.7,2.7,0 -fix 1,5
\hntc          & $17^\dag$         & $13.15\pm  0.02$  & $-10.38\pm  0.07$   & $3.11\pm   0.2$  &     & ~\\
\hcsp          & $20^{\dagg}$        & $12.89\pm  0.06$  & $-10.39\pm   0.2$   & $3.38\pm   0.3$  &     & ~\\
CCH            & $17^\dag$         & $15.19\pm 0.01$   & $-10.57\pm  0.03$   & $3.25\pm  0.06$   &     & ~\\
SO             & $16.9\pm     7$   & $14.21\pm  0.08$  & $-10.6\pm   0.1$    & $3.69\pm   0.3$  &     & \dag~\\
\acet          & $17^\dag$         & $13.52\pm 0.06$   & $-10.71\pm   0.3$   & $4.04\pm    0.6$  &     & ~\\
%perl ../fitLTE_1mol.pl consp/lsb_S8_maser.dat consp/usb_S8_maser.dat CH3CHO 84500-84600 17,13.3,-10.7,2.7,0 -fix 1,5
\hcdop         & $40^\ddag$        & $12.61\pm  0.03$  & $-10.77\pm   0.1$   & $3.57\pm   0.3$  &     & ~\\
\hccn          & $28.57\pm     5$  & $13.09\pm 0.1$    & $-10.77\pm   0.1$   & $3.81\pm   0.3$  &     & ~\\
\hcqn          & $17^\dag$         & $12.76\pm  0.03$  & $-10.82\pm   0.1$   & $3.63\pm   0.3$  &     & ~\\
OCS            & $40^\ddag$        & $13.93\pm  0.04$  & $-10.83\pm   0.1$   & $3.10\pm   0.3$  &     & ~\\
\htcop         & $40^\ddag$        & $13.6\pm  0.02$   & $-10.89\pm  0.06$   & $3.49\pm   0.1$   &     & ~\\
HC$^{13}$CCN   & $17^\dag$         & $12.48\pm 0.04$   & $-11.05\pm   0.2$   & $3.27\pm   0.3$  &     & ~\\
\propyne       & $40.34\pm     2$  & $15.44\pm 0.02$   & $-11.08\pm  0.03$   & $3.41\pm  0.06$   &     & \ddag\\
%perl ../fitLTE_1mol.pl consp/lsb_S8_maser.dat consp/usb_S8_maser.dat CH3CCH 40,15.3,-11.,2.7,0 -fix 5
HNCO           & $17^\dag$         & $12.95\pm  0.07$  & $-11.58\pm   0.3$   & $3.05\pm   0.4$   &     & ~\\
\hctn          & $17^\dag$         & $14.26\pm  0.02$  & $-11.89\pm  0.06$   & $2.97\pm   0.1$  &     & ~\\
%
%% %
\sidehead{C8 1\farcs5 radius\tablenotemark{b}} % usar datos del paper mean=0.00463Jy/b
SiO            & $17^\dag$         & $13.3\pm  0.01$   & $-10.02\pm  0.08$   & $5.09\pm   0.2$  &     & ~\\
%perl ../fitLTE_1mol.pl consp/lsb_s8_1.5.dat consp/usb_s8_1.5.dat SiO - 17,14.,-12.3,3.7,0.0  -fix 1,5
SO             & $16.74\pm     4$  & $13.8\pm  0.05$   & $-10.15\pm  0.07$   & $3.80\pm   0.2$  &     & \dag\\
%perl ../fitLTE_1mol.pl consp/lsb_s8_1.5.dat consp/usb_s8_1.5.dat SO - 60,14.,-12.3,3.7,0.0  -fix 5
\hntc          & $17^\dag$         & $13.04\pm  0.02$  & $-10.57\pm  0.06$   & $3.35\pm   0.2$   &     & ~\\
HNCO           & $100\pm    70$    & $14.05\pm   0.4$  & $-10.74\pm   0.8$   & $5.1\pm     2$  &     & ~\\
%perl ../fitLTE_1mol.pl consp/lsb_s8_1.5.dat consp/usb_s8_1.5.dat HNCO - 100,13.,-12.3,3.7,0.0 -fix 5
CCH            & $17^\dag$         & $14.81\pm  0.01$  & $-10.8\pm  0.03$    & $3.18\pm  0.07$   &     & ~\\
\met           & $114.5\pm    30$  & $15.39\pm   0.2$  & $-10.93\pm   0.1$   & $3.27\pm   0.2$  &     & Strong absorption in low $E_{\rm up}$ transitions.\\
%perl ../fitLTE_1mol.pl consp/lsb_s8_1.5.dat consp/usb_s8_1.5.dat CH3OH 84000-85500,96700-96755 60,14.3,-10.2,2.7,0 -fix 5
\cyc           & $34.5\pm     2$  & $14.11\pm  0.02$  & $-11.26\pm  0.03$   & $4.02\pm  0.08$  &     & ~\\
%perl ../fitLTE_1mol.pl consp/lsb_s8_1.5.dat consp/usb_s8_1.5.dat c-C3H2 - 10,14.,-12.3,3.7,0.0  -fix 5 
\htcop         & $60^\ddag$        & $13.87\pm 0.007$  & $-11.3\pm  0.04$    & $4.46\pm  0.08$  &     & ~\\
\cyan          & $48.1\pm 9$      & $12.82\pm  0.04$  & $-11.37\pm   0.2$   & $2.95\pm   0.4$  &     & ~\\
%perl ../fitLTE_1mol.pl consp/lsb_s8_1.5.dat consp/usb_s8_1.5.dat CH3C3N 86740-88000 50,12.8,-12.3,3.7,0.0  -fix 5 -device pdfs/spectra/S8E/S8E_15_CH3C3N/cps
OCS            & $60^\ddag$        & $14.51\pm  0.02$  & $-11.41\pm  0.07$   & $4.30\pm   0.2$  &     & ~\\
%perl ../fitLTE_1mol.pl consp/lsb_s8_1.5.dat consp/usb_s8_1.5.dat OCS - 60,14.,-11.3,3.7,0.0  -fix 1,5 
\propyne       & $59.7\pm  1$     & $15.77\pm  0.01$  & $-11.46\pm  0.02$   & $3.71\pm  0.04$  &     & \ddag\\
%perl ../fitLTE_1mol.pl consp/lsb_s8_1.5.dat consp/usb_s8_1.5.dat CH3CCH - 60,15.5,-11.3,3.7,0.0  -fix 5 
\hcdop         & $60^\ddag$        & $12.72\pm  0.04$  & $-11.52\pm   0.2$   & $4.18\pm   0.4$  &     & ~\\
\propanenitrile& $74.8\pm  20$    & $13.51\pm0.1$     & $-11.57\pm   0.3$   & $5.56\pm   0.7$  &     & ~\\
%perl ../fitLTE_1mol.pl consp/lsb_s8_1.5.dat consp/usb_s8_1.5.dat CH3CH2CN 98605-98620 60,13.3,-12.3,3.7,0.0  -fix 5
\hcqn          & $17^\dag$         & $13.06\pm  0.02$  & $-11.62\pm  0.09$   & $4.30\pm   0.2$  &     & ~\\
\hccn          & $77.2\pm    10$  & $13.26\pm  0.01$  & $-11.64\pm  0.05$   & $3.69\pm   0.1$   &     & ~\\
%perl ../fitLTE_1mol.pl consp/lsb_s8_1.5.dat consp/usb_s8_1.5.dat HC5N - 60,13.,-12.3,3.7,0.0 -fix 5
C$^{33}$S      & $17^\dag$         & $12.45\pm  0.05$  & $-11.72\pm   0.2$   & $3.19\pm   0.4$  &     & ~\\
%perl ../fitLTE_1mol.pl consp/lsb_s8_1.5.dat consp/usb_s8_1.5.dat C33S - 17,12.,-11.5,3.7,0.0  -fix 1,5 -device pdfs/spectra/S8E/S8E_15_C33S/cps
\htcn          & $17^\dag$         & $13.78\pm  0.01$  & $-11.74\pm  0.07$   & $4.09\pm   0.1$  &     & ~\\
% perl ../fitLTE_1mol.pl consp/lsb_s8_1.5.dat consp/usb_s8_1.5.dat H13CN_jpl - 17,13.5,-11.5,3.7,0.0  -fix 1,5 -device pdfs/spectra/S8E/S8E_15_H13CN/cps 
%strangely, the H13CN model has no quantum anomalies
\acrylonitrile & $81.7\pm    20$   & $13.90\pm   0.1$   & $-11.96\pm   0.2$   & $5.00\pm   0.4$  &     & H$^{13}$CCCN blend at 99986 MHz.\\
%perl ../fitLTE_1mol.pl consp/lsb_s8_1.5.dat consp/usb_s8_1.5.dat CH2CHCN  85440-85448,85452.5-85500,87318-87400,96980-96995 30,13.5,-12.3,3.7,0.0  -fix 5 -device pdfs/spectra/S8E/S8E_15_CH2CHCN/cps
HC$^{13}$CCN   & $17^\dag$         & $12.7\pm0.03$     & $-11.96\pm0.09$     & $3.08\pm   0.2$  &     & ~\\
\methoxy         & $17^\dag$         & $13.99\pm  0.09$  & $-12.11\pm   0.4$   & $2.70\pm     1$  &     & Faint.\\
CS             & $17^\dag$         & $13.9\pm  0.01$   & $-12.29\pm  0.04$   & $2.63\pm  0.08$  &     & ~\\
\ethe          & $17^\dag$         & $13.55\pm  0.02$  & $-12.33\pm  0.08$   & $3.22\pm   0.2$  &     & ~\\
%perl ../fitLTE_1mol.pl consp/lsb_s8_1.5.dat consp/usb_s8_1.5.dat H2CCO - 17,13.5,-12.3,3.7,0.0 -fix 1,5
\hctn          & $17^\dag$         & $14.4\pm  0.01$   & $-12.4\pm  0.05$    & $3.52\pm   0.1$   &     & ~\\
%perl ../fitLTE_1mol.pl consp/lsb_s8_1.5.dat consp/usb_s8_1.5.dat HC3N - 17,13.,-12.3,3.7,0.0 -fix 1,5
% 14.36 HC3N a 80 K
%perl ../fitLTE_1mol.pl consp/lsb_s8_1.5.dat consp/usb_s8_1.5.dat HC3N - 77,13.,-12.3,3.7,0.0 -fix 1,5
% HC3N & $     77\pm     0$ & $  14.35\pm  0.01$ & $  -12.4\pm  0.04$ & $  3.927\pm   0.1$ &
%perl ../fitLTE_1mol.pl consp/lsb_s8_1.5.dat consp/usb_s8_1.5.dat CH3CHO - 50,14.5,-12.3,3.7,0.0 -fix 1,5 -norun
%%-------------------------------------
%%
\sidehead{\textbf{NW cloud}}% mean 6.30779234507612e-05Jy/b, 786 pixels 
\ethe          & $20^{\dagg}$         & $13.3\pm  0.03$   & $-12.38\pm   0.2$   & $4.62\pm   0.4$  &     & \met\ $T_{ex}$ from group 1.\\
%perl ../fitLTE_1mol.pl consp/lsb_WNCloud.dat consp/usb_WNCloud.dat H2CCO - 20,13.,-12.,3.5,0.0 -fix 1,5
\hctn          & $16^\dag$         & $13.65\pm  0.01$  & $-11.92\pm  0.08$   & $5.43\pm   0.2$  &     & SO $T_{ex}$ from group 1.\\
\met           & $20\pm     10$    & $15.25\pm 0.3$    & $-10.2\pm  0.1$     & $5.20\pm   0.3$    & 1   & \dagg\\
%perl ../fitRad.pl consp/lsb_WNCloud.dat consp/usb_WNCloud.dat e-ch3oh.dat,a-ch3oh.dat 84000-85500 20,15.25,-10.2,5.2,0.0,6.73,0.5  -fix 1,3,4,5,7 -norun  Final chi2 red=41.0233191310942
%perl ../fitRad.pl consp/lsb_WNCloud.dat consp/usb_WNCloud.dat e-ch3oh.dat,a-ch3oh.dat 84000-85500 10,15.19,-10.2,5.2,0.0,7.048,0.5  -fix 1,3,4,5,7 -norun Final chi2 red=36.3793483940624
%perl ../fitRad.pl consp/lsb_WNCloud.dat consp/usb_WNCloud.dat e-ch3oh.dat,a-ch3oh.dat 84000-85500 30,15.4,-10.2,5.2,0.0,6.93,0.5  -fix 1,3,4,5,7   -norun Final chi2 red=48.7183104992527
\propyne       & $44.26\pm    10$  & $14.13\pm   0.1$  & $-10.41\pm   0.1$    & $1.77\pm   0.2$   & 1   & \ddag\\
SO             & $16^\dag$         & $14.07\pm  0.02$  & $-9.91\pm   0.1$    & $6.90\pm   0.3$   & 1   & Poor fit. Blueshift absorption. \\
$^{34}$SO      & $16^\dag$         & $12.71\pm  0.06$  & $-10.46\pm   0.4$   & $5.86\pm   0.8$   & 1   & ~\\
OCS            & $16.29\pm     3$  & $13.96\pm  0.03$  & $-10.48\pm  0.07$   & $4.51\pm   0.2$   & 1   & \dag, possibly subthermal\\
\htcop         & $44^\ddag$        & $12.93\pm  0.01$  & $-10.23\pm  0.04$   & $2.79\pm  0.09$   & 1   & ~\\
\hcdop         & $44^\ddag$        & $12.04\pm  0.07$  & $-10.49\pm   0.2$   & $2.78\pm   0.6$   & 1   & ~\\
%perl ../fitLTE_1mol.pl consp/lsb_WNCloud.dat consp/usb_WNCloud.dat CH3CHO 84500-84600 16,13.7,-10.,3.5,0.0 -fix 1,5
CCH            & $16^\dag$         & $14.22\pm  0.02$  & $-10.00\pm  0.05$   & $2.67\pm   0.1$   & 1   & ~\\
\acet          & $16^\dag$         & $13.66\pm  0.03$  & $-11.0\pm   0.2$    & $4.90\pm   0.4$    & 1   & ~\\
%perl ../fitLTE_1mol.pl consp/lsb_WNCloud.dat consp/usb_WNCloud.dat CH3CHO 84500-84600 16,13.7,-10.,3.5,0.0 -fix 1,5
\hcqn          & $16^\dag$         & $12.13\pm  0.04$  & $-10.40\pm   0.2$    & $3.87\pm   0.4$   & 1   & ~\\
\htcn          & $16^\dag$         & $13.1\pm  0.02$   & $-10.10\pm   0.1$    & $4.36\pm   0.3$   & 1   & ~\\
\hntc          & $16^\dag$         & $12.76\pm  0.01$  & $-10.02\pm  0.04$   & $3.13\pm  0.09$   & 1   & ~\\
HNCO           & $16^\dag$         & $13.37\pm  0.01$  & $-10.53\pm  0.06$   & $4.67\pm   0.1$   & 1   & ~\\
CS             & $16^\dag$         & $14.00\pm  0.01$  & $-8.53\pm  0.08$    & $6.32\pm   0.2$   & 2   & SO $T_{ex}$ from group 1.\\
SiO            & $16^\dag$         & $13.52\pm  0.02$  & $-8.92\pm   0.1$    & $7.77\pm   0.3$   & 2   & SO $T_{ex}$ from group 1, blueshift absorption.\\
$^{29}$SiO     & $16^\dag$         & $12.25\pm  0.05$  & $-10.00\pm  0.3$     & $5.90\pm   0.7$    &     & SO $T_{ex}$ from group 1, blueshift absorption.\\
$^{30}$SiO     & $16^\dag$         & $12.09\pm  0.06$  & $-8.40\pm 0.6$       & $8.2\pm     1$    & 2   & SO $T_{ex}$ from group 1.\\
%debil CH3OCH3 & $16$& $13.6\pm  0.09$& $-10.36\pm   0.6$& $3.761\pm     1$& 
\sidehead{NW cloud (a)} % 313 316   16:59:40.167 -40:03:25.9 negative -0.000043Jy/b
\met           & $50\pm10$         & $15.95\pm0.1$     & $-10.20\pm  0.1$     & $5.20\pm   0.3$    & 1   & \dag\\
%perl ../fitRad.pl consp/lsb_313-316_NWC.dat consp/usb_313-316_NWC.dat e-ch3oh.dat,a-ch3oh.dat - 50,15.95,-10.2,5.2,0.0,6.8,0.5  -fix 1,3,4,5,7 -norun
%perl ../fitRad.pl consp/lsb_313-316_NWC.dat consp/usb_313-316_NWC.dat e-ch3oh.dat,a-ch3oh.dat - 40,15.9,-10.2,5.2,0.0,6.6,0.5  -fix 1,3,4,5,7 -norun
% perl ../fitRad.pl consp/lsb_313-316_NWC.dat consp/usb_313-316_NWC.dat e-ch3oh.dat,a-ch3oh.dat - 60,16.03,-10.2,5.2,0.0,6.97,0.5  -fix 1,3,4,5,7 -norun
OCS            & $16$              & $14.12\pm  0.05$  & $-11.73\pm   0.2$   & $3.63\pm   0.4$  & 1   & $T_{ex}$ from NW cloud.\\
\acet          & $50^\dag$         & $14.74\pm  0.04$  & $-11.43\pm   0.2$   & $5.57\pm   0.5$  & 1   &~\\
%perl ../fitLTE_1mol.pl consp/lsb_313-316_NWC.dat consp/usb_313-316_NWC.dat CH3CHO 84500-84600  50,14.43,-11.2,5.6,0.0  -fix 1,5
HNCO           & $16$              & $13.69\pm  0.04$  & $-11.22\pm   0.2$   & $4.51\pm   0.5$  & 1   & $T_{ex}$ from NW cloud.\\
SiO            & $16$              & $13.81\pm  0.01$  & $-8.38\pm   0.1$   & $8.22\pm   0.3$  & 2   & $T_{ex}$ from NW cloud.\\
$^{29}$SiO     & $16$              & $12.72\pm  0.06$  & $-8.4$              & $7.96$            & 2   & $T_{ex}$ from NW cloud. Faint.\\
\sidehead{\textbf{Source 3}}% 
\met  & $  221\pm 30$ & $ 16.22\pm  0.08$ & $ -11.15\pm   0.2$ & $   5.34\pm   0.3$ & & Peculiar \maser\ profile.\\
% perl ../fitLTE_1mol.pl consp/lsb_s3.dat consp/usb_s3.dat  CH3OH 84000-84550 150,15.6,-12.4,4.5,0.0  -fix 5 -device pdfs/spectra/S3_CH3OH/cps
\sidehead{\textbf{Source 18}}% 0.002067 usar paper surly
\met           & $64\pm     5$     & $15.76\pm  0.04$  & $-14.08\pm  0.06$   & $4.25\pm   0.1$   &     & Strong absorption in low $E_{\rm up}$ transitions.\\
%% % perl ../fitLTE_1mol.pl consp/lsb_s18.dat consp/usb_s18.dat CH3OH 84000-85500,96700-96755 64,15.76,-14.09,4.22,0 -fix 5 -norun
%% %lots of absorption from SiO and CS -> continuum source?
\enddata
\tablenotetext{\star}{Except for the \met\ non-LTE modeling described by the parameter in Table \ref{tab-met}. }
\tablenotetext{\rm a}{For each source, temperatures marked with $^\dag$, $^\ddag$, and $^\dagg$ are taken from the molecule with the same symbol in column (7).}
\tablenotetext{\rm b}{This source has a continuous distribution of velocities. No discrete $V_{\rm LSR}$ components are identified.}
%\tablenotetext{\rm b}{}
\end{deluxetable}

%%%%%%%%%%%%%%%%%%%%%%%%%%%%%%%%%%%%%%%%%%%%%%%%%%%%%%%%%%%%
%%%%%%%%%%%%%%%%%%%%%%%%%%%%%%%%%%%%%%%%%%%%%%%%%%%%%%%%%%%%
%%%   FIGURES 
%%%%%%%%%%%%%%%%%%%%%%%%%%%%%%%%%%%%%%%%%%%%%%%%%%%%%%%%%%%%
%%%%%%%%%%%%%%%%%%%%%%%%%%%%%%%%%%%%%%%%%%%%%%%%%%%%%%%%%%%%

%%%%%%%%%%%%%%%%%%%%%%%%%%%%%%%%%%%%%%%%%%%%%%%%%%%%%%%%%%%%
%%%%%%%%%%%%%%%%%%%%%%%%%%%%%%%%%%%%%%%%%%%%%%%%%%%%%%%%%%%%
\clearpage
\appendix
\counterwithin{figure}{section}
%%%%%%%%%%%%%%%%%%%%%%%%%%%%%%%%%%%%%%%%%%%%%%%%%%%%%%%%%%%%
%%%%%%%%%%%%%%%%%%%%%%%%%%%%%%%%%%%%%%%%%%%%%%%%%%%%%%%%%%%%

{\section{Short Spatial Frequencies Filtering}\label{sec-ssf}}
The data presented in this work was taken using the ALMA with only the
12 m antennas with baselines between 453 and 21 m. This coverage does
not guarantee good recovery of structures with angular scales larger
than
$\sim19\arcsec$.\footnote{\url{http://almascience.nrao.edu/about-alma/alma-basics}}
The most noticeable effect produced by this lack of short spacing on
the images are regions of negative emission. The filtering of emission
on large scales has two effect: first, it filters out diffuse,
extended components associated with compact sources, and second, it
leaves local decrements in intensity below the zero level of the
synthesized images.
Thus, negative emission is not necessarily consequence  of negative sidelobes or spurious data: they could be merely the reflection of diminished intensity compared with nearby positions, either because of absorption, or just because of the specific morphology of the source.    

In order to evaluate how the observations presented in this work recover the  short spacing, we compare the flux of the CS, $J=2\shortrightarrow1$ emission with that obtained by \citet{Bronfman1996AAS} using the \emph{Swedish-ESO Submillimetre Telescope} (SEST). This 15 m single dish antenna has a beam only slightly smaller than  ALMA's primary beam.
%\footnote{The \emph{Atacama Pathfinder Experiment} antenna is more similar to ALMA's, but it does not observe at 3 mm.} 
The emission from CS toward \clump\ has an extended morphology, which makes it adequate to estimate how much short spacing flux is missing in our observations.

\begin{figure}
\plottwo{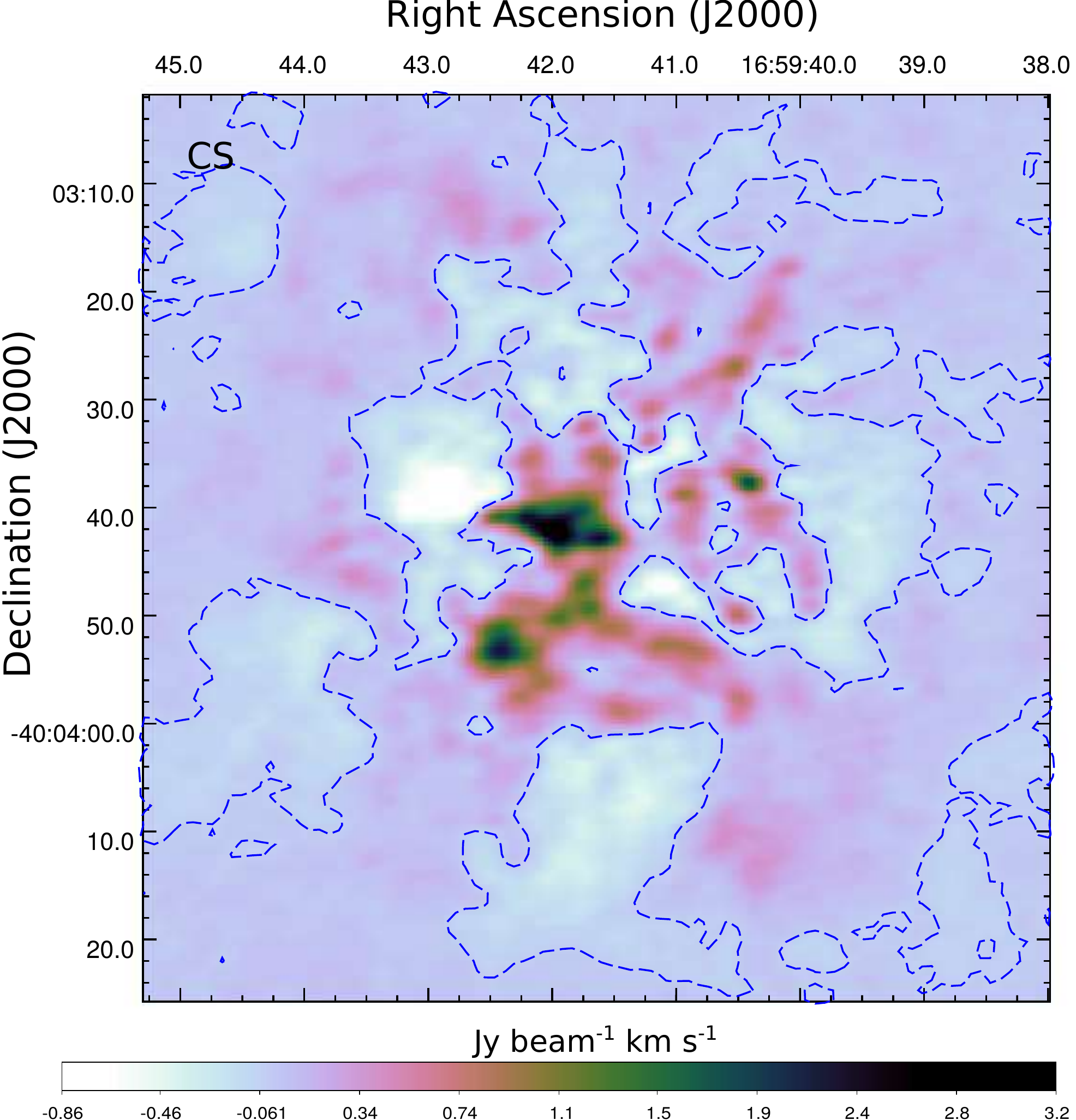}{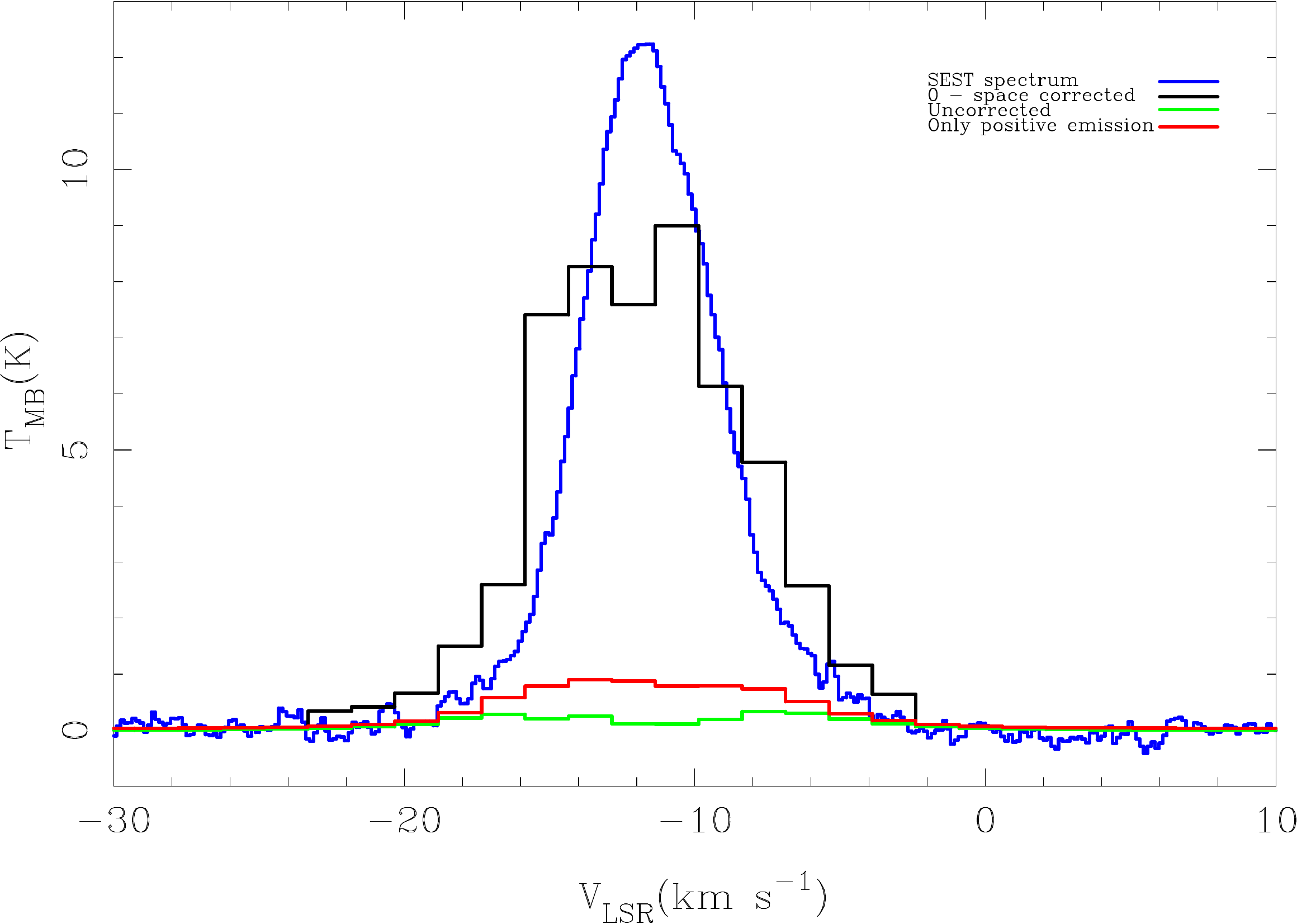}
\caption{\emph{Left panel:} zero moment of the CS map. Dashed contours
  indicate the $-0.03$ Jy beam$^{-1}$ \kms\ level. \emph{Right panel:} in   blue, we show the SEST CS, $J=2\shortrightarrow1$ spectrum of  \clump\ \citep{Bronfman1996AAS}. Black and green lines show,  respectively, the zero-space corrected and uncorrected CS spectra from  the ALMA data cubes weighted by the SEST beam. The spectrum in red  considers only the positive part of the emission from the ALMA  cubes.\label{fig-cs}}
\end{figure}

We test the hypothesis that the negative CS emission in our maps is due to short baseline filtering by adding an uniform constant level, per channel, to all channels with negatives below $-7\sigma$. This constant is  \replaced{minus}{the negative of} the minimum intensity of each channel, ensuring that each channel corrected has only positive emission. This procedure can be thought as a crude way to correct the lack of zero spacing from the data cube. We integrate spatially this corrected CS cube using the SEST beam.  Figure \ref{fig-cs} show the SEST data and the corrected  CS spectrum in blue and black lines, respectively.
We regard the agreement as reasonable, considering the systematic uncertainties involved in the single dish and interferometer flux calibrations of these two instruments. Our rough correction procedure  appears to overestimate the emission in the channels with velocities $<-15$ \kms.
This is expected if, for example,  the zero-space emission does not cover the entire field of view. On the other hand, at the peak of the line, our zero space correction accounts for about 60\% of the single dish flux.

Therefore, we conclude that most of the negative emission seen in the maps presented in this study are produced by zero-space filtering. In a way, the interferometer has recovered spatial information indicating  us that in these negative regions there is a decrement of the emission compared to the surrounding clump. Figure \ref{fig-cs} also shows in green and red the ALMA CS spectra of the uncorrected data cube and of the positive part of the emission, respectively. The uncorrected and positive part  spectra reach only $8$ and $\sim1\%$ of the line peak flux. 

\clearpage
\section{Uncertainties of the Propyne Column density and temperature fitting}

\begin{figure}[h]
\includegraphics[angle=0, ext=.pdf, width=\textwidth]{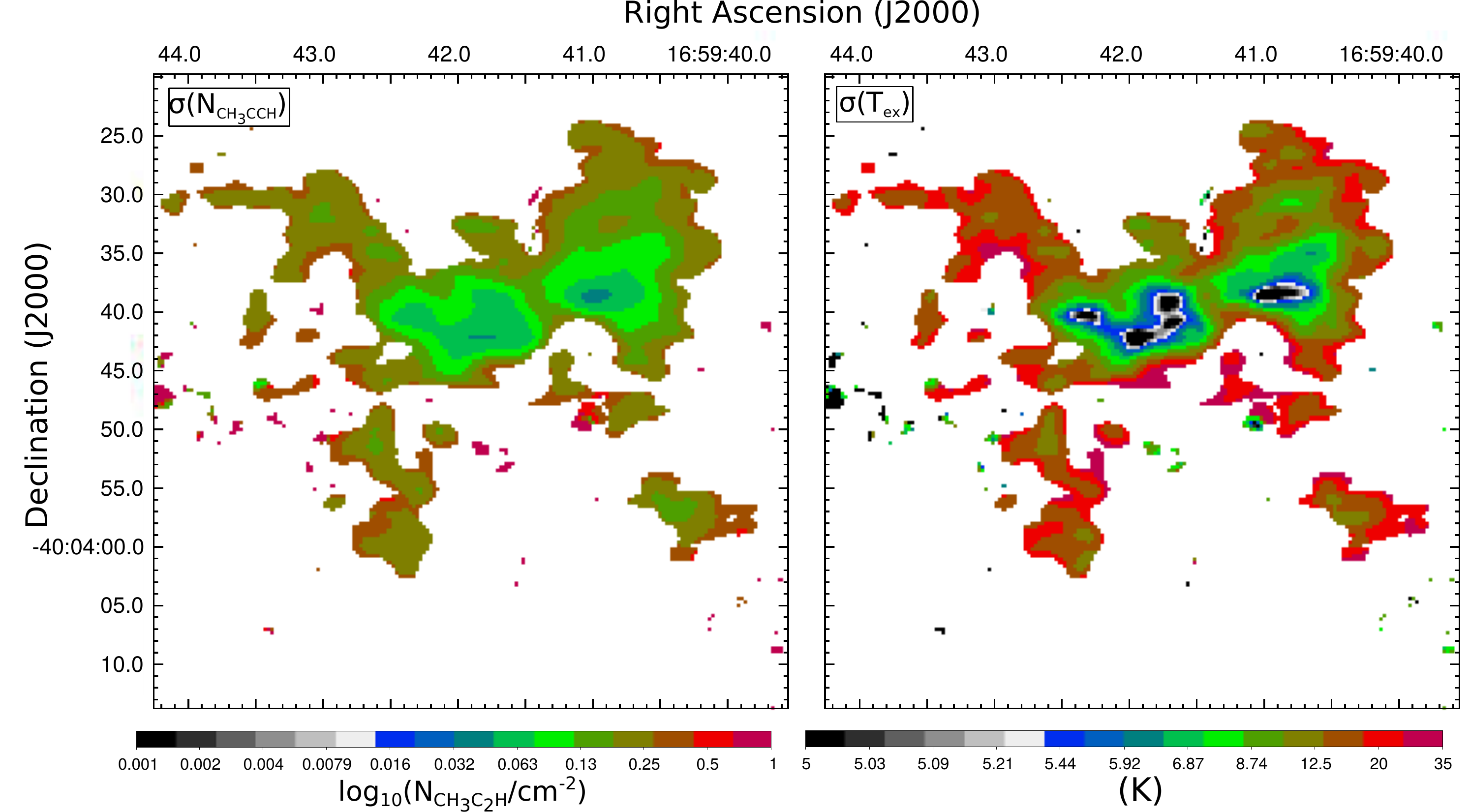}
\caption{Left and right panels show the formal uncertainties in the  \added{\propyne}\ column density and excitation temperature fitting, respectively (Section \ref{sec-cdt}). The best-fit parameters are shown in Figure \ref{fig-prop}.\label{fig-errorPropyne}}
\end{figure}

\clearpage
{\section{Spectra From Conspicuous Sources}\label{sec-figSpec}}
%%%%

\begin{figure}[b]
\subfloat[][]{\includegraphics[angle=-90,ext=.pdf,width= 1.00 \textwidth]{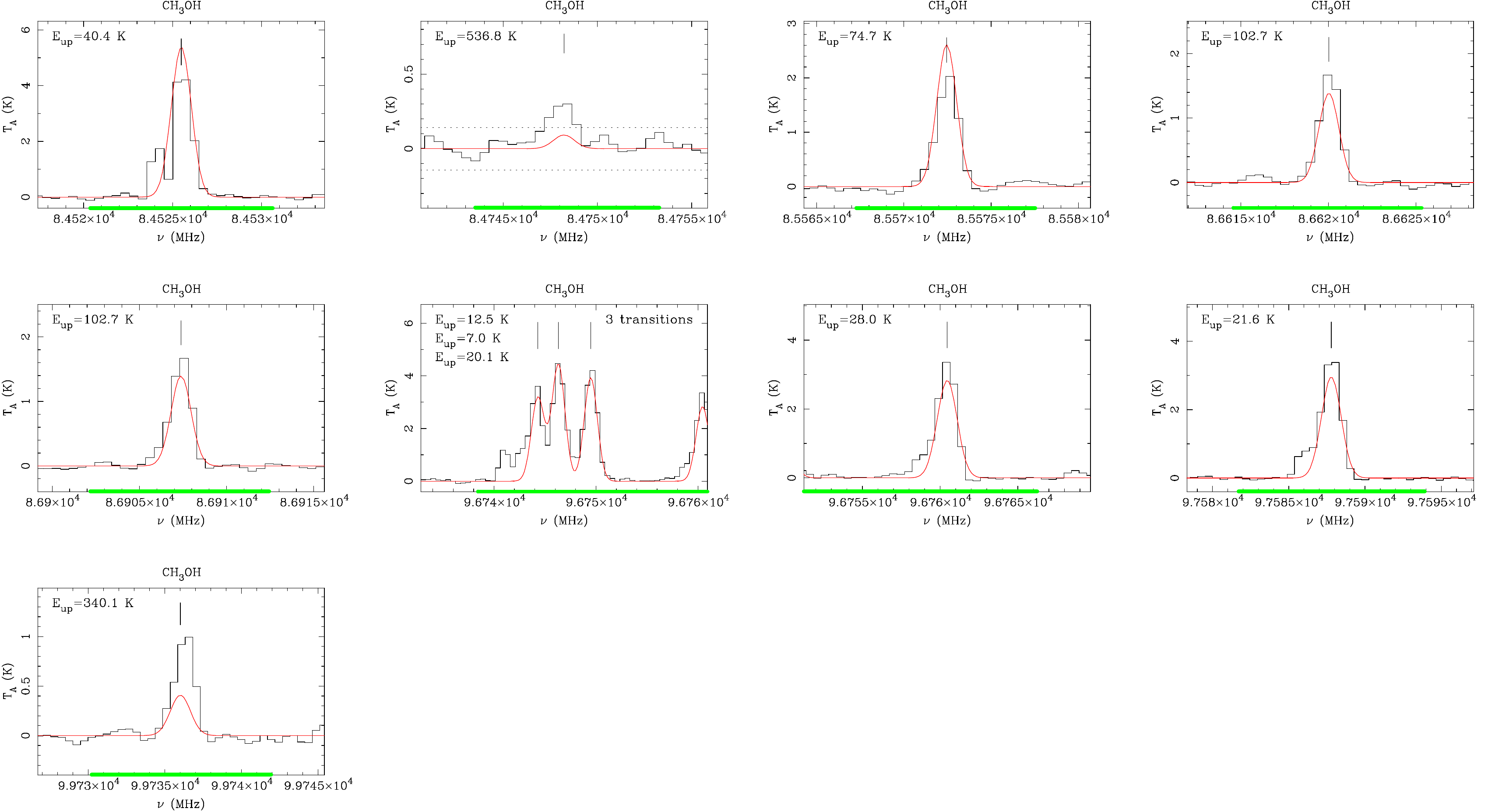}}\\
\subfloat[][]{\includegraphics[angle=-90,ext=.pdf,width= 0.50 \textwidth]{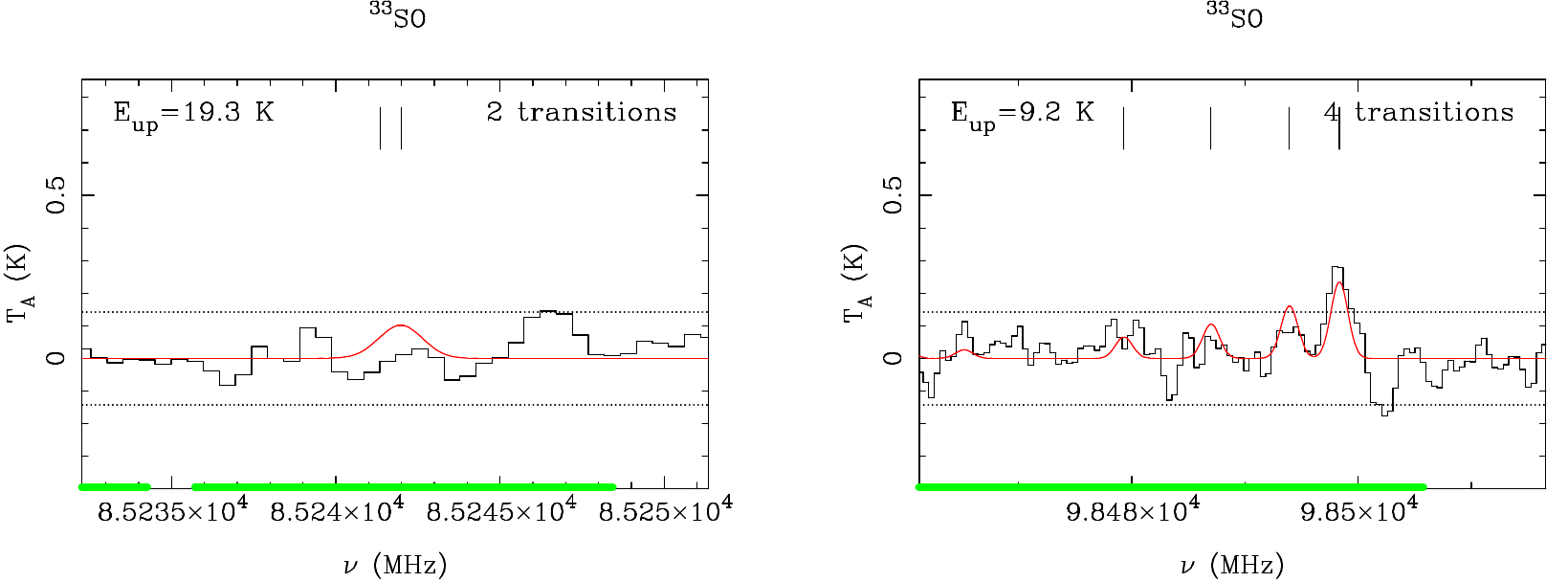}}
\subfloat[][]{\includegraphics[angle=-90,ext=.pdf,width= 0.25 \textwidth]{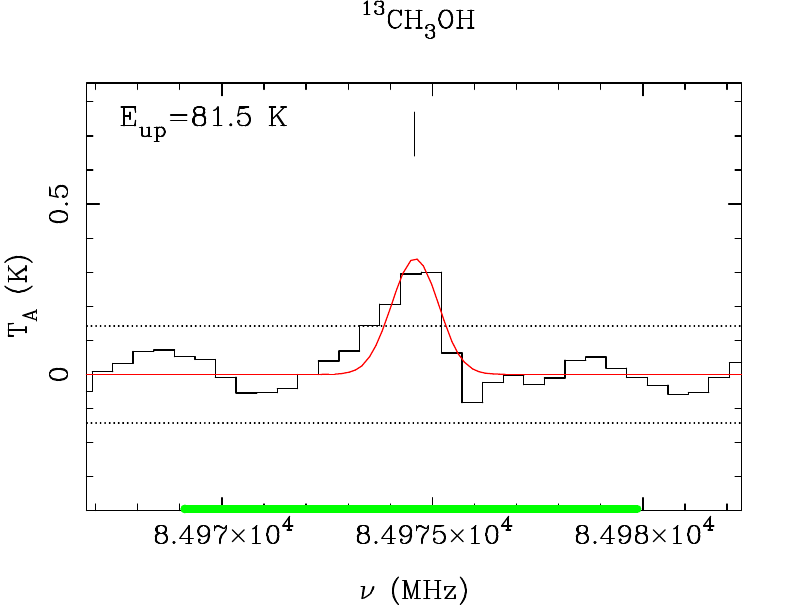}}
\subfloat[][]{\includegraphics[angle=-90,ext=.pdf,width= 0.25 \textwidth]{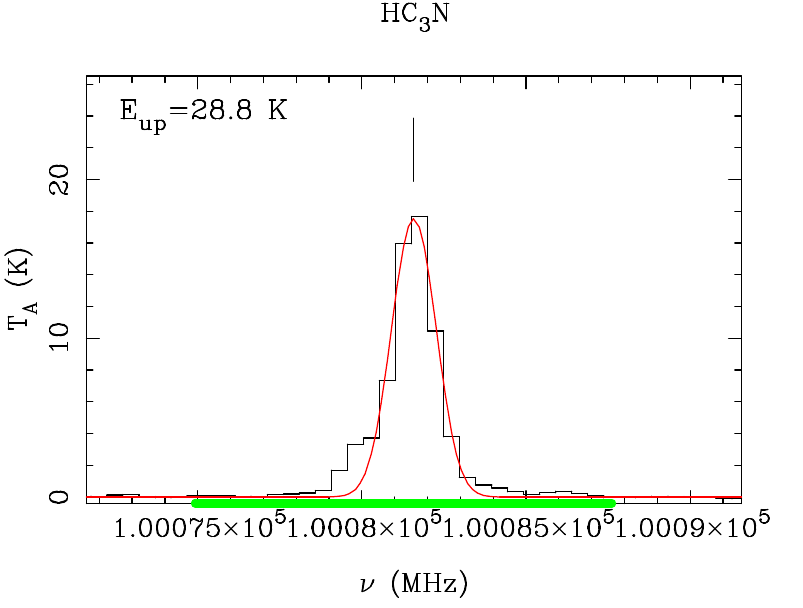}}\\
\caption{Emission lines from several molecules toward the CC core, in K
  (primary beam corrected antenna temperatures) vs.\ frequency (MHz).
  Black and red continuous lines show the data and the model whose
  parameters are given in Table \ref{tab-cdt}. In each panel we mark the
  $V_{\rm LSR}$ of the model with a vertical black line, and the upper
  energy level of each transition. Thick green lines indicate the section
  of the spectra used to get the best-fit values. In panels where the peak
  intensity does not reach $5\sigma$, $\sigma=0.06$ K we show the
  $\pm2.5\sigma$ levels using dotted black lines. Panels (a) to (d) show
  \met, $^{33}$SO, $^{13}$CH$_3$OH, and \hctn, respectively. \added{The
    \met\ spectrum suggests the presence of a additional and hotter
    component.}\label{fig-SpecCCcore1}}
\end{figure}
\begin{figure}
\ContinuedFloat
\subfloat[][]{\includegraphics[angle=-90,ext=.pdf,width= 0.25 \textwidth]{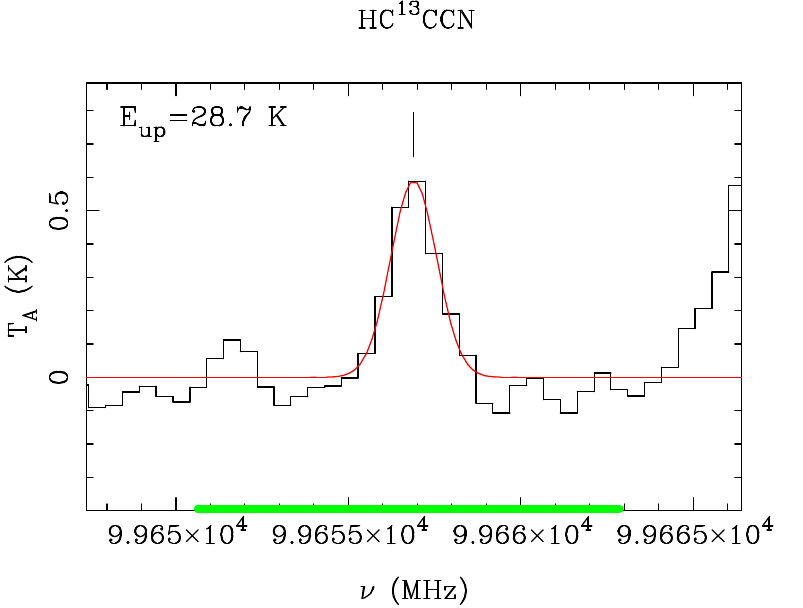}}
\subfloat[][]{\includegraphics[angle=-90,ext=.pdf,width= 0.25 \textwidth]{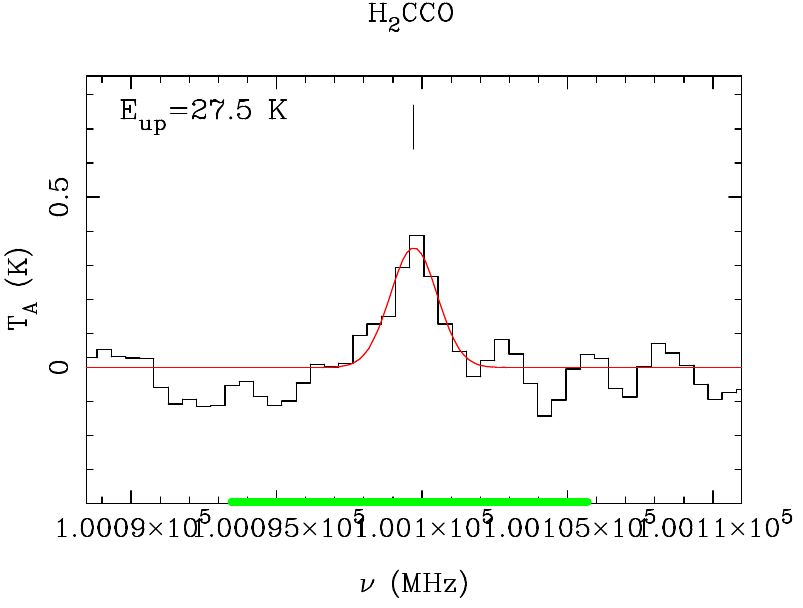}}
\subfloat[][]{\includegraphics[angle=-90,ext=.pdf,width= 0.25 \textwidth]{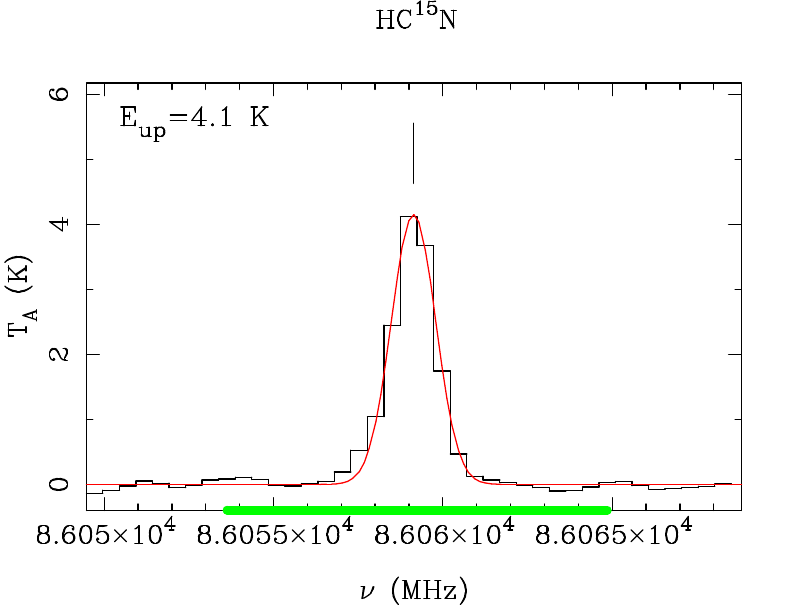}}
\subfloat[][]{\includegraphics[angle=-90,ext=.pdf,width= 0.25 \textwidth]{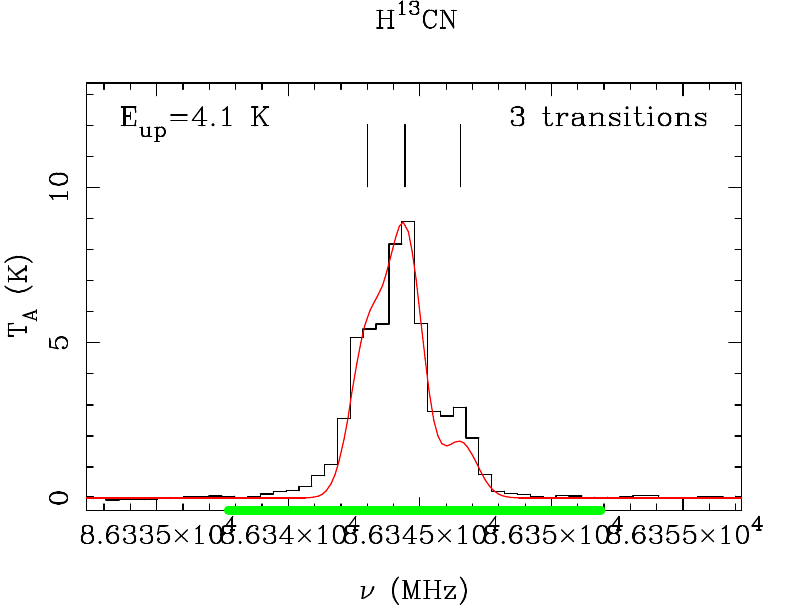}}
\caption{\textbf{\hspace{-0.2 em}(cont.)}\ Same as the previous
  plot. Panels (e) to (h) show HC$^{13}$CCN, \ethe, \hcqn, and \htcn,
  respectively.}
\end{figure}

\begin{figure}
\ContinuedFloat
\subfloat[][]{\includegraphics[angle=0,ext=.pdf,width=1.0 \textwidth]{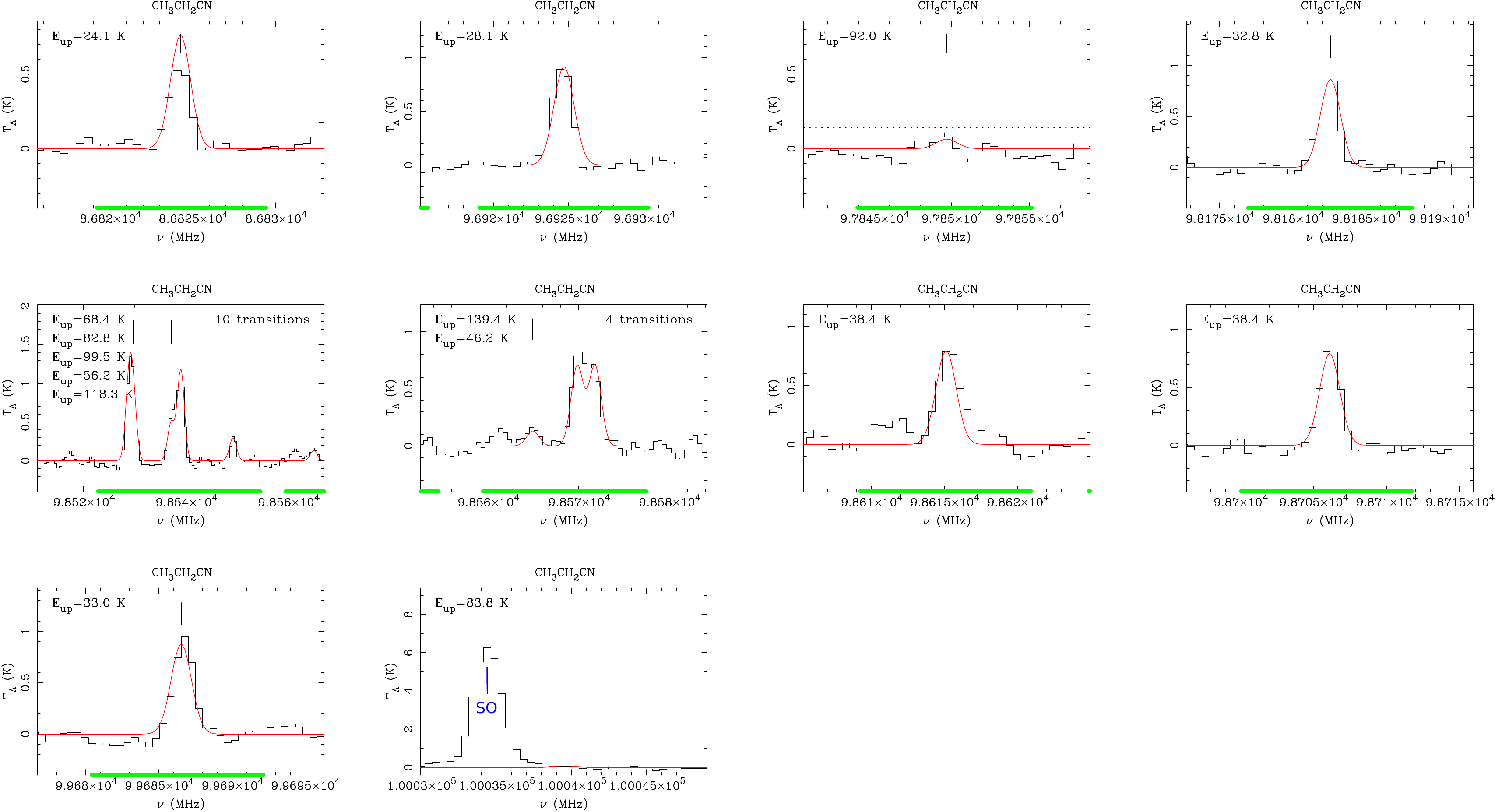}}\\
\caption{\textbf{\hspace{-0.2 em}(cont.)}\ Same as the previous plot. Panels (i)  show the \prop\ lines. \added{Strong lines from other species within the displayed frequency window are marked in blue}}
\end{figure}
\begin{figure}
\ContinuedFloat
\subfloat[][]{\includegraphics[angle=-90,ext=.pdf,width=0.5 \textwidth]{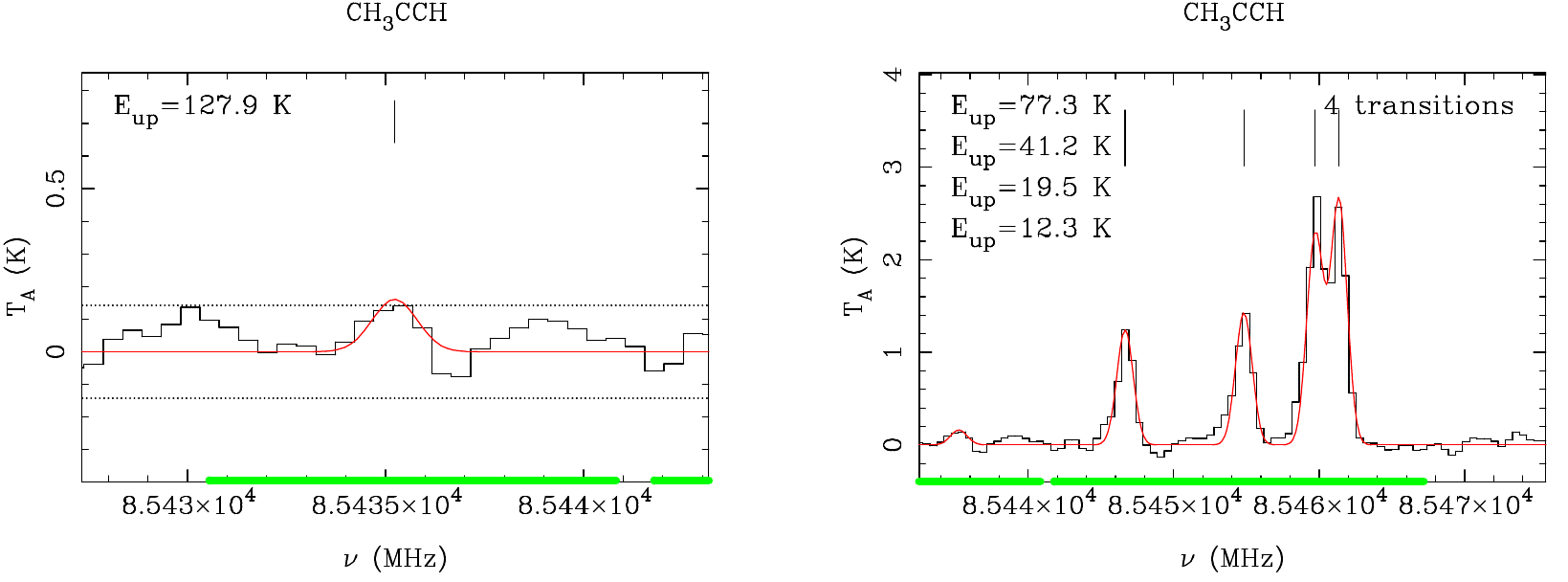}}
\subfloat[][]{\includegraphics[angle=-90,ext=.pdf,width=0.5 \textwidth]{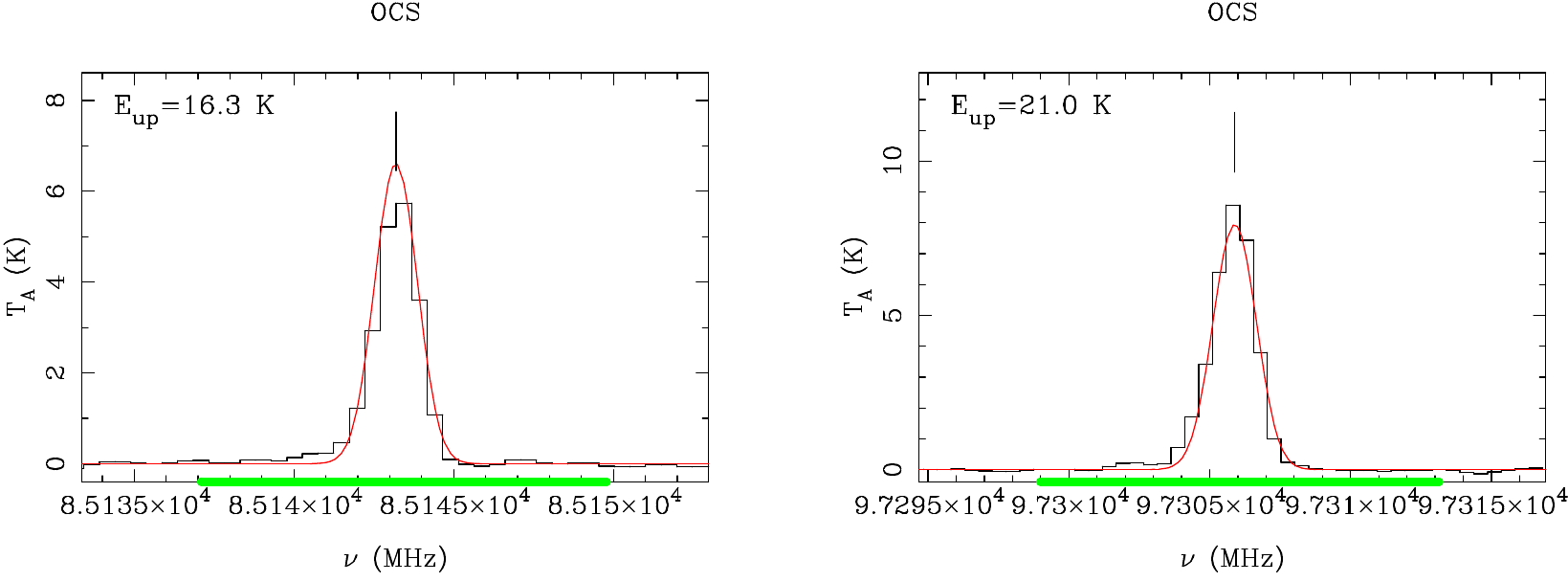}}\\
\subfloat[][]{\includegraphics[angle=-90,ext=.pdf,width=0.5 \textwidth]{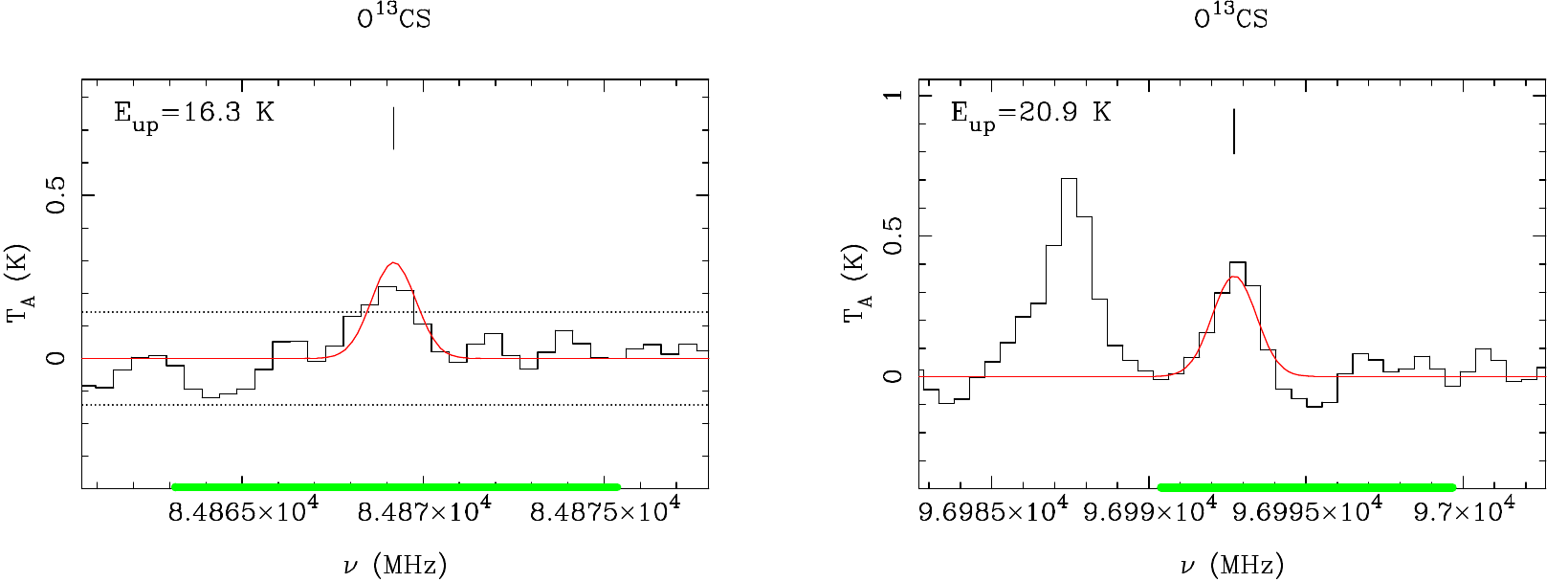}}
\subfloat[][]{\includegraphics[angle=-90,ext=.pdf,width=0.5 \textwidth]{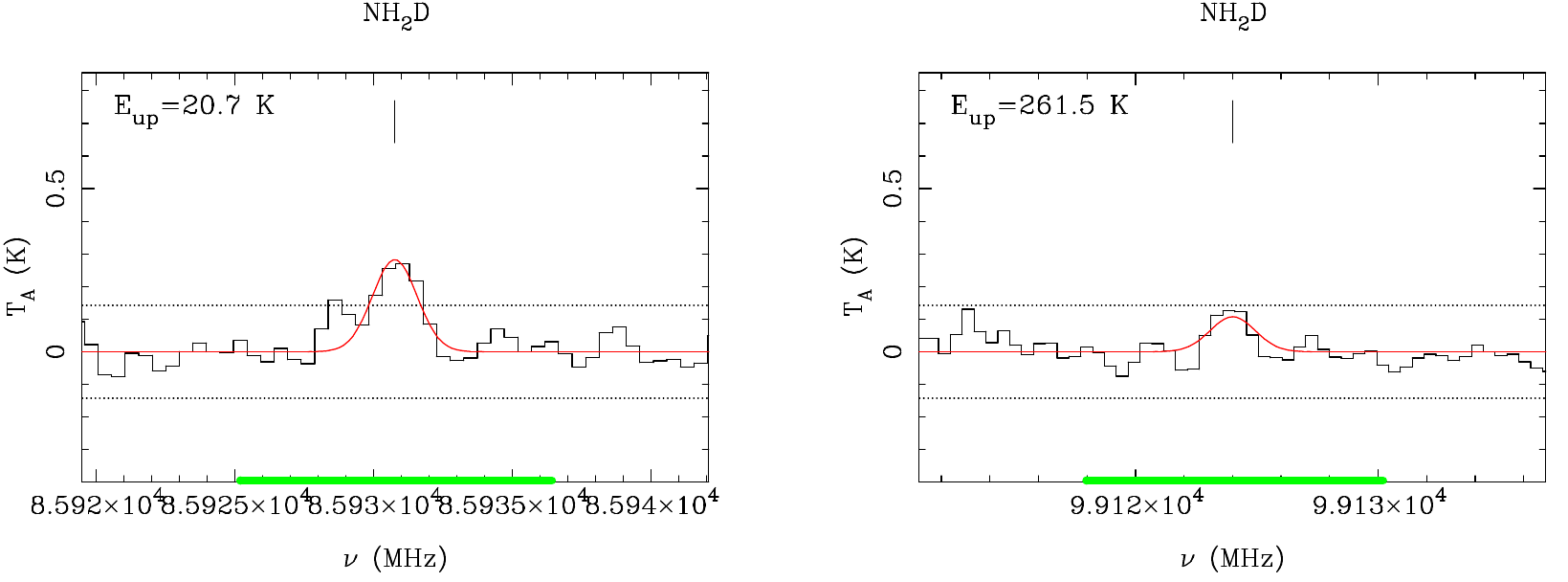}}\\
\subfloat[][]{\includegraphics[angle=-90,ext=.pdf,width=0.75 \textwidth]{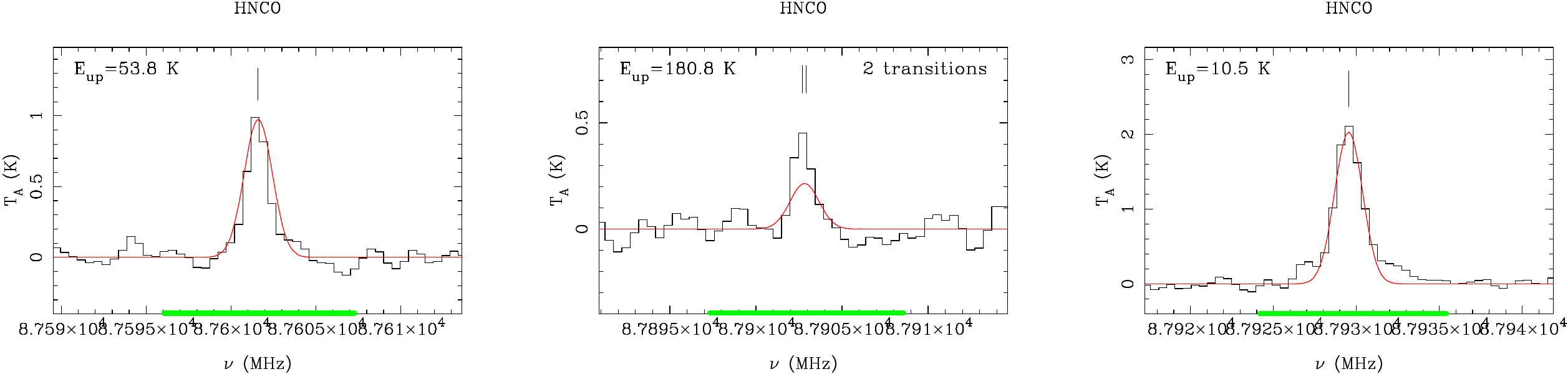}}
\caption{\textbf{\hspace{-0.2 em}(cont.)}\ Same as the previous plot. Panels (j) to (n) show, respectively, \propyne,  OCS, O$^{13}$CS, NH$_2$D, and HNCO.}
\end{figure}

\begin{figure}
\ContinuedFloat
\subfloat[][]{\includegraphics[angle=-0,ext=.pdf,width=1.0 \textwidth]{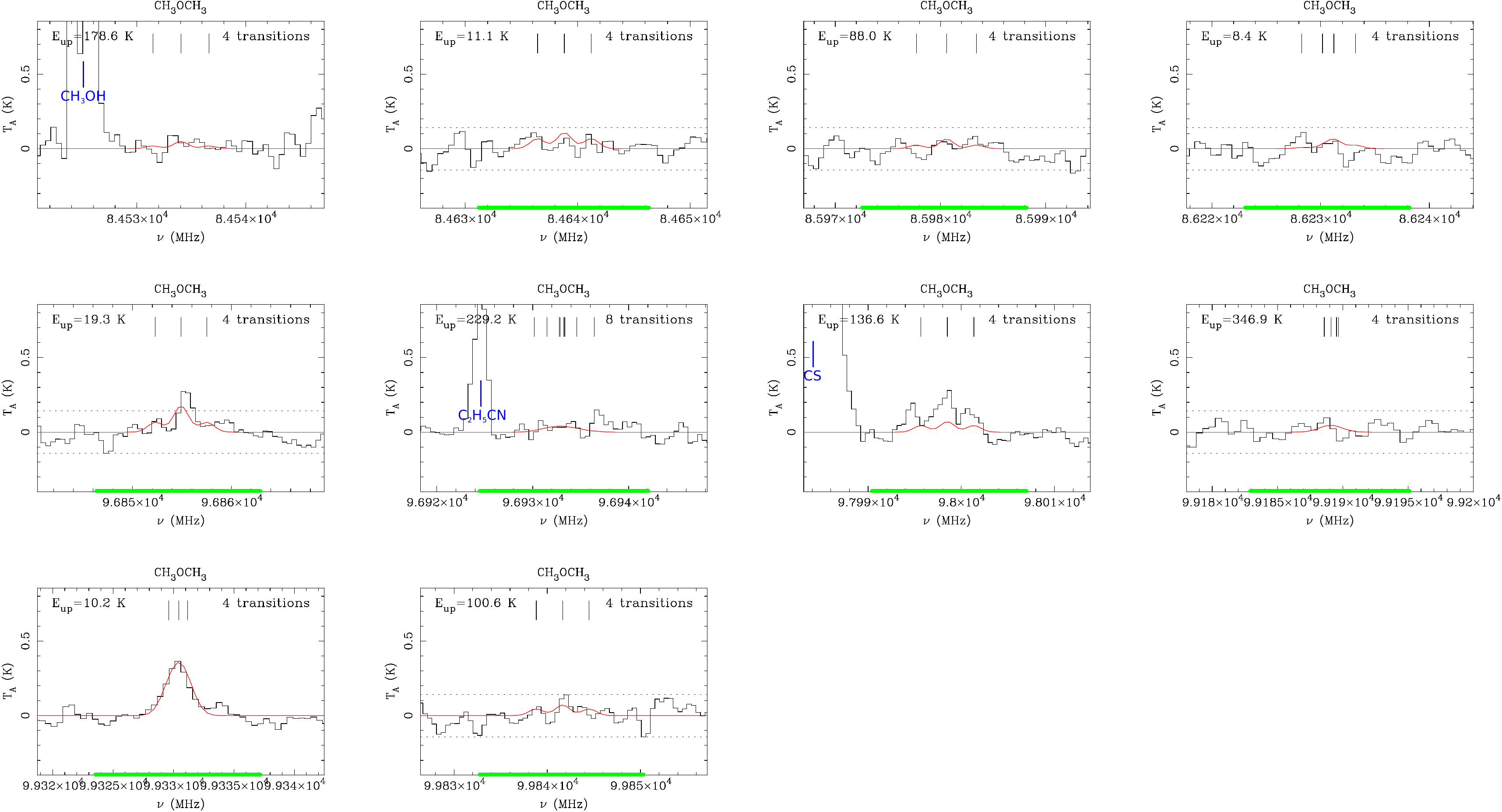}}
\caption{\textbf{\hspace{-0.2 em}(cont.)}\ Same as the previous plot. Panel (o) shows the CH$_3$OCH$_3$ spectra of the CC core. \added{Strong lines from other species within the displayed frequency windows are marked in blue.}}
\end{figure}

\begin{figure}
\ContinuedFloat
\subfloat[][]{\includegraphics[angle=-90,ext=.pdf,width=0.25 \textwidth]{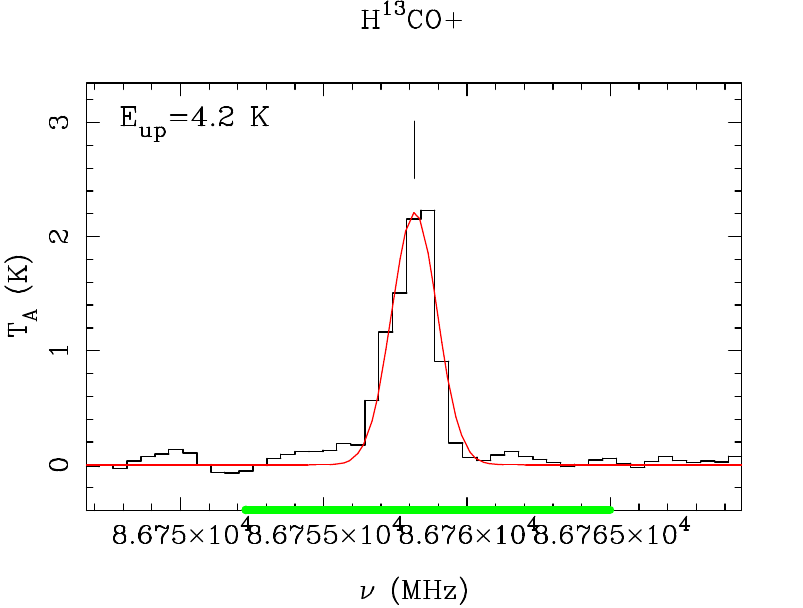}}
\subfloat[][]{\includegraphics[angle=-90,ext=.pdf,width=0.25 \textwidth]{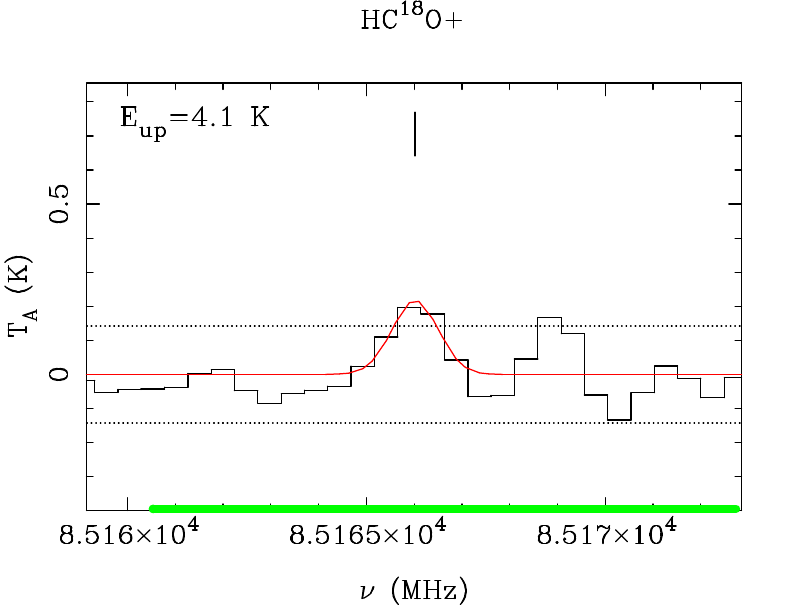}}
\subfloat[][]{\includegraphics[angle=-90,ext=.pdf,width=0.25 \textwidth]{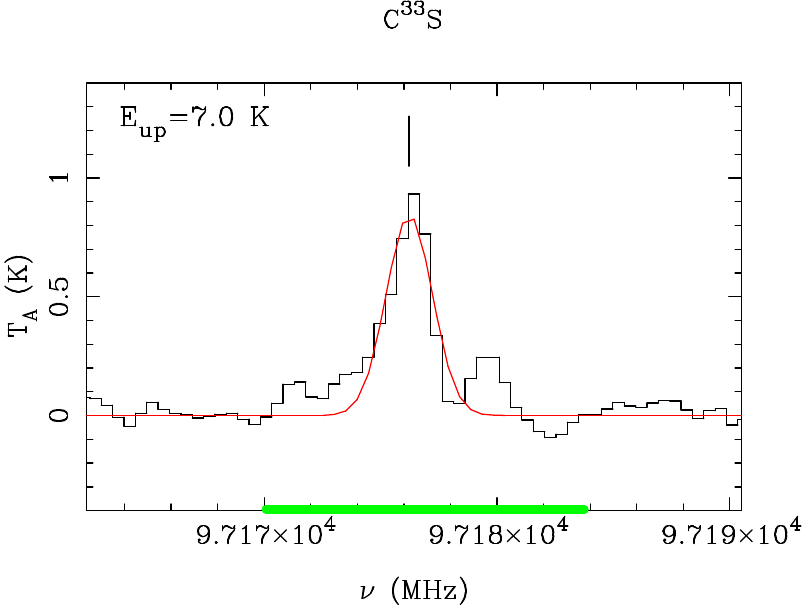}}
\subfloat[][]{\includegraphics[angle=-90,ext=.pdf,width=0.25 \textwidth]{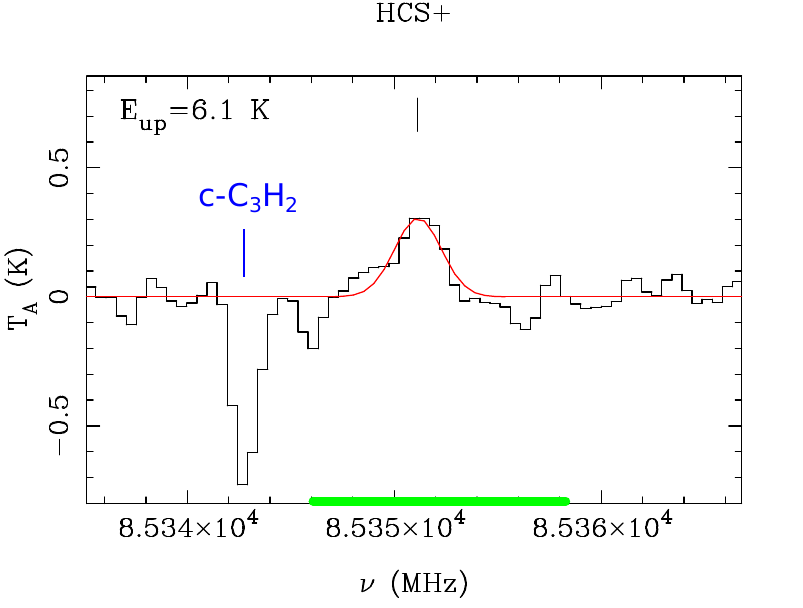}}\\
\subfloat[][]{\includegraphics[angle=-90,ext=.pdf,width=1.00 \textwidth]{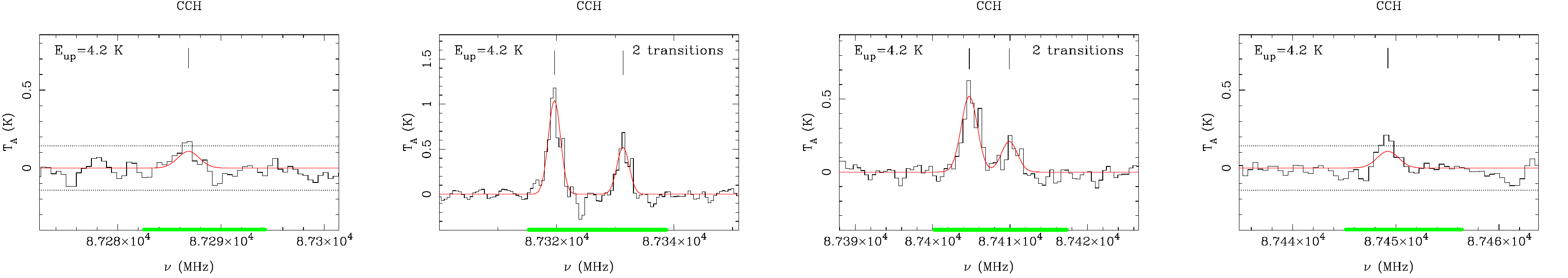}}
\caption{\textbf{\hspace{-0.2 em}(cont.)}\ Same as the previous plot. Panels (q) to (r) show, respectively, the \htcop, \hcdop, C$^{33}$S, \hcsp, and CCH spectra of the CC core. \added{A strong absorption feature due to  \cyc\ within the  displayed frequency window of the \hcsp\ \!(1-0) line is  marked in blue.}}
\end{figure}

%%%%%%%%%%%%%%%%%%%%%%%%%%%%%%%%%%%%%%%%%%%%%%%%%%

\begin{figure}
  \subfloat[][]{\includegraphics[angle=-90,ext=.pdf,width= 1.0   \textwidth]{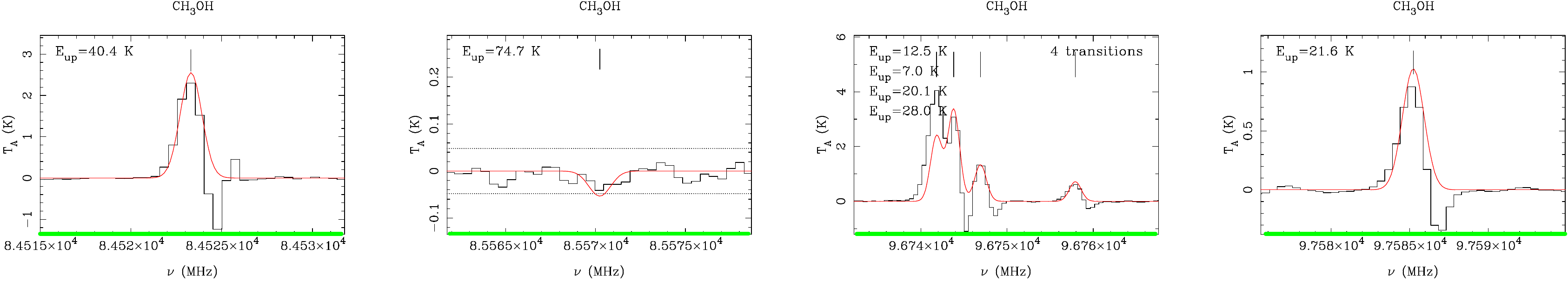}}\\
  \subfloat[][]{\includegraphics[angle=-90,ext=.pdf,width= 0.25   \textwidth]{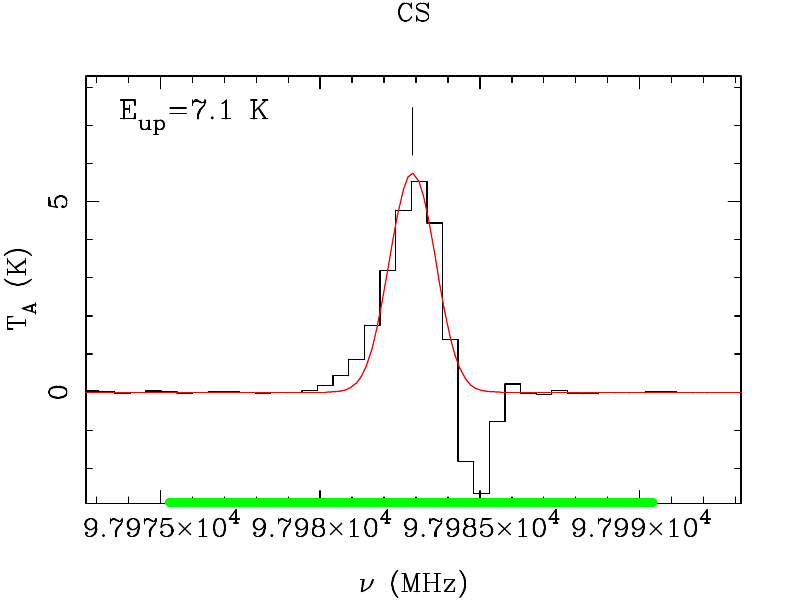}}
  \subfloat[][]{\includegraphics[angle=-90,ext=.pdf,width= 0.25   \textwidth]{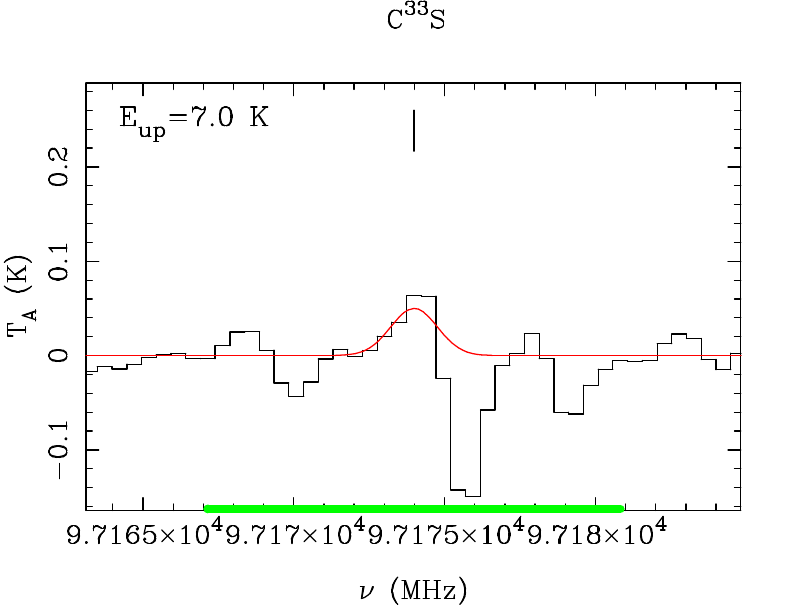}}
  \subfloat[][]{\includegraphics[angle=-90,ext=.pdf,width= 0.5   \textwidth]{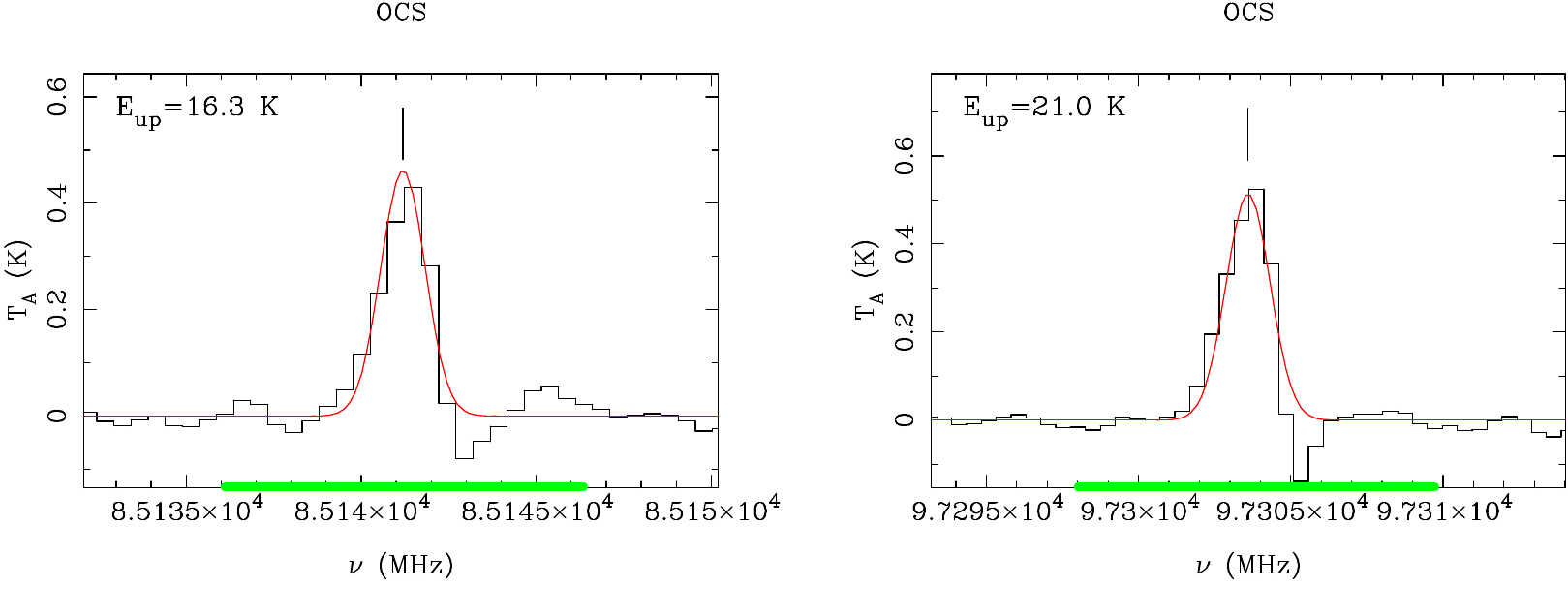}}\\
  \subfloat[][]{\includegraphics[angle=-90,ext=.pdf,width= 0.75   \textwidth]{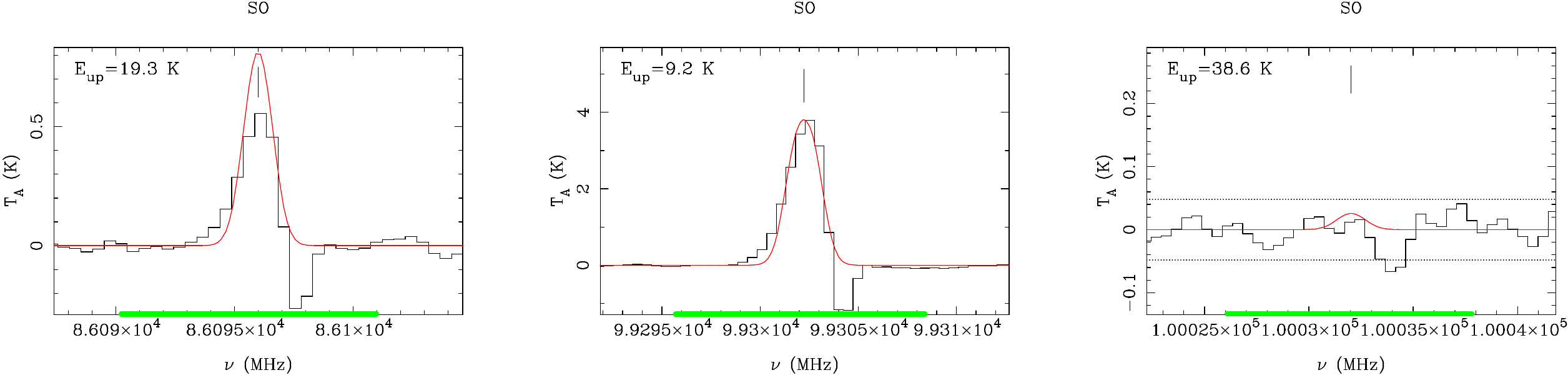}}
  \subfloat[][]{\includegraphics[angle=-90,ext=.pdf,width= 0.25   \textwidth]{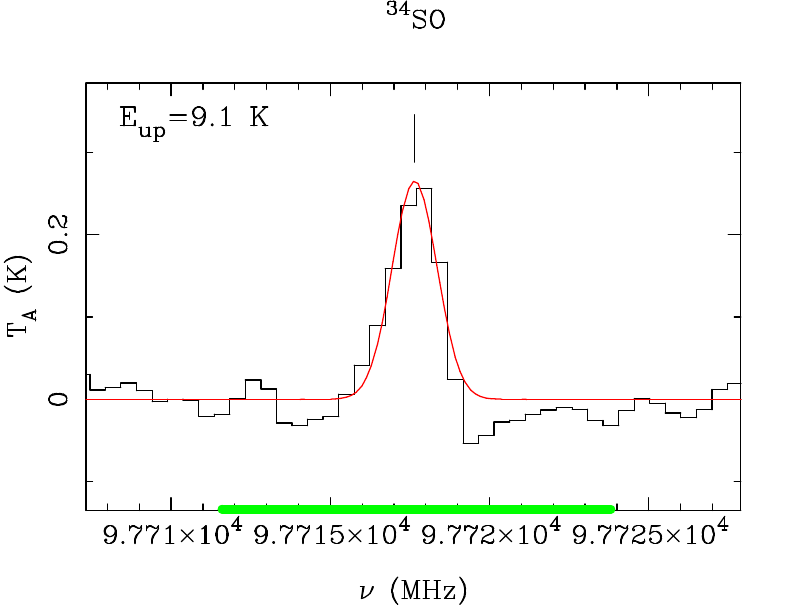}}
  \caption{Emission lines from several molecules toward the N-red
    cloud. Line types and colors as described in Figure
    \ref{fig-SpecCCcore1}, with $\sigma=0.02$ K. Panels (a) to (f) show lines of \met, CS,
    C$^{33}$S, OCS, SO, and $^{34}$SO, respectively. \label{fig-SpecNredC}}
\end{figure}
\begin{figure}
  \ContinuedFloat
  \subfloat[][]{\includegraphics[angle=-90,ext=.pdf,width= 0.25 \textwidth]{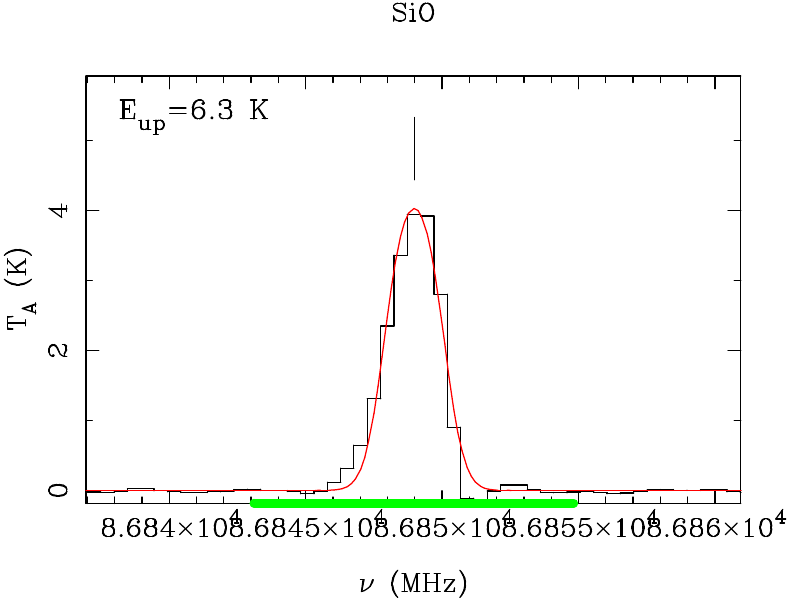}}
  \subfloat[][]{\includegraphics[angle=-90,ext=.pdf,width= 0.25 \textwidth]{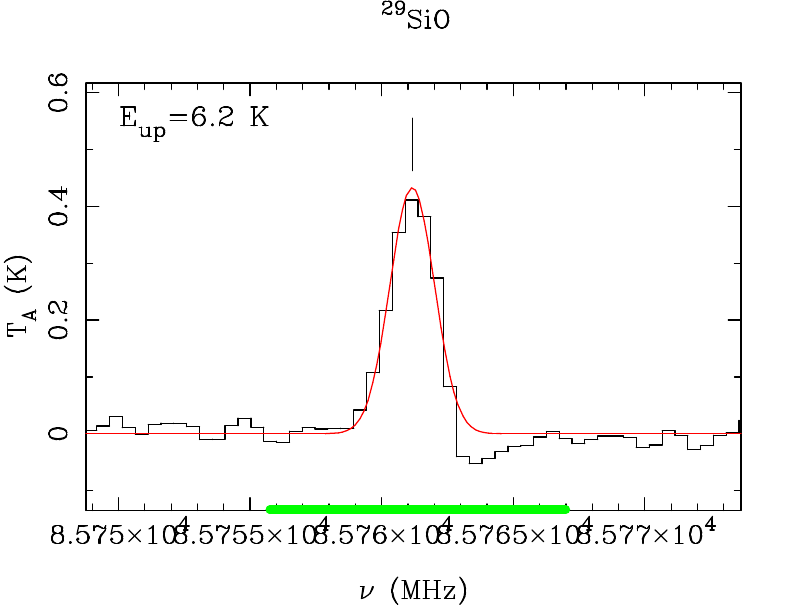}}
  \subfloat[][]{\includegraphics[angle=-90,ext=.pdf,width= 0.25 \textwidth]{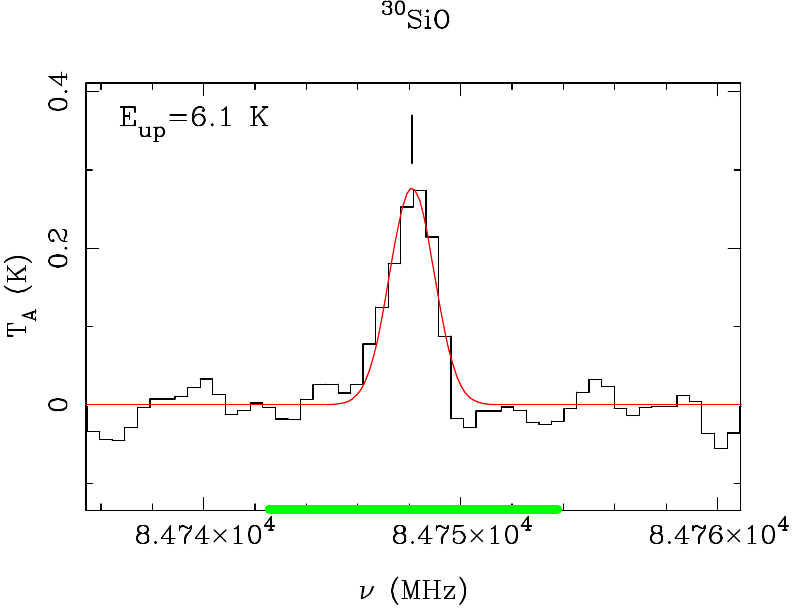}}
  \subfloat[][]{\includegraphics[angle=-90,ext=.pdf,width= 0.25 \textwidth]{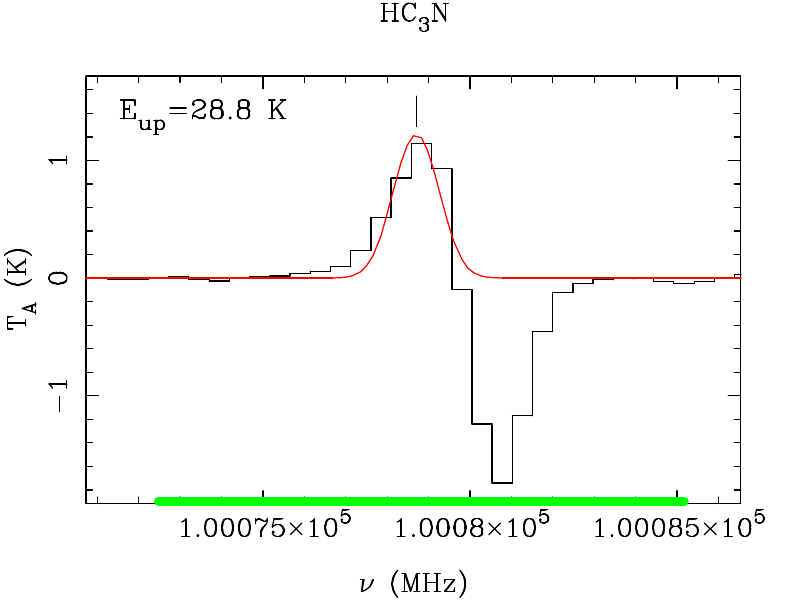}}\\
  \subfloat[][]{\includegraphics[angle=-90,ext=.pdf,width= 0.50 \textwidth]{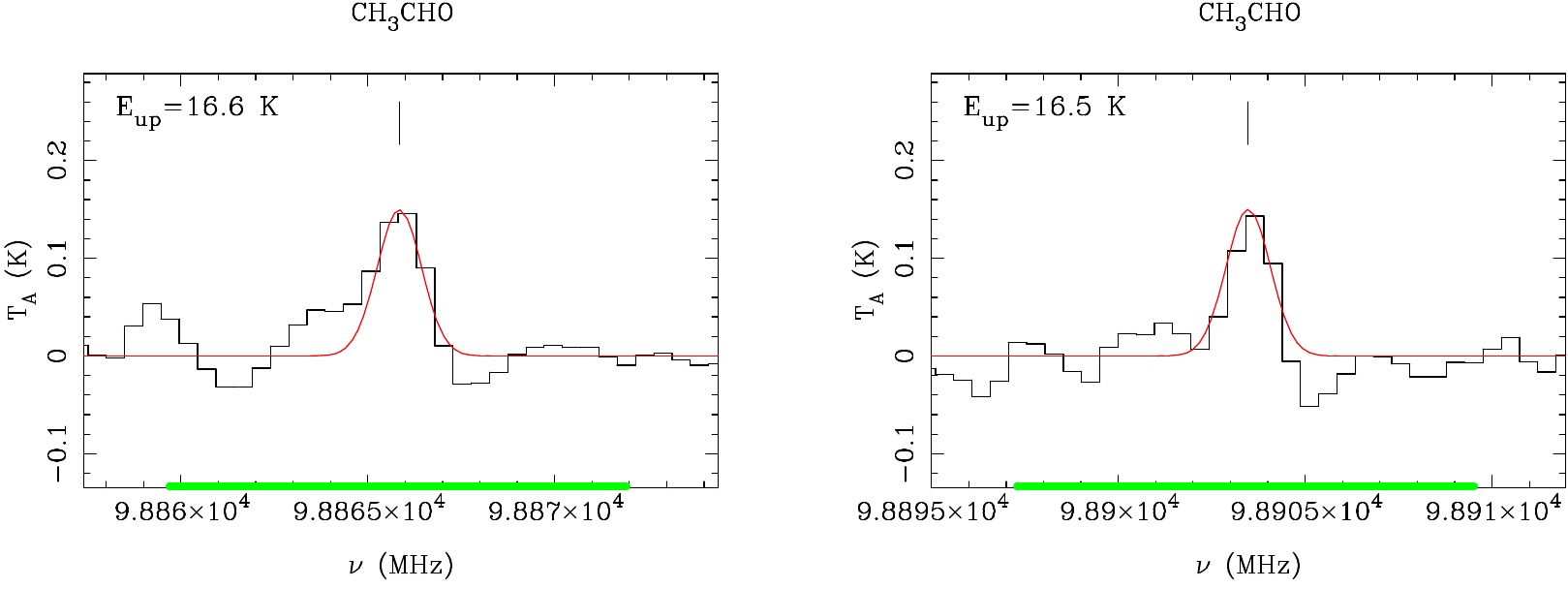}}
  \subfloat[][]{\includegraphics[angle=-90,ext=.pdf,width= 0.25 \textwidth]{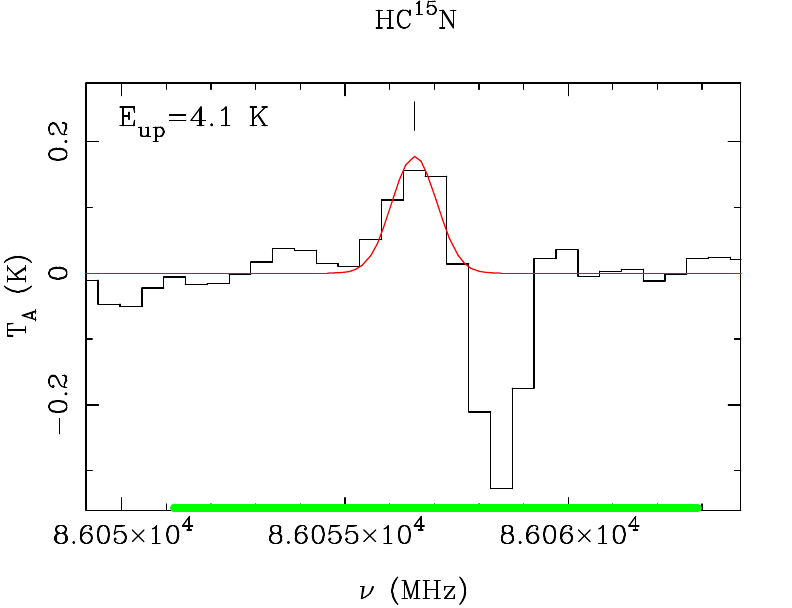}}
  \subfloat[][]{\includegraphics[angle=-90,ext=.pdf,width= 0.25 \textwidth]{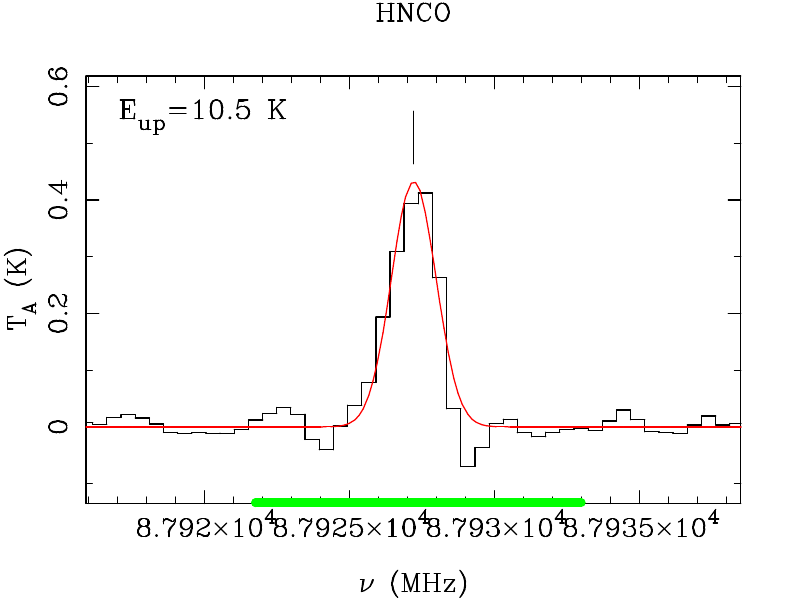}}\\
  \subfloat[][]{\includegraphics[angle=-90,ext=.pdf,width= 0.25 \textwidth]{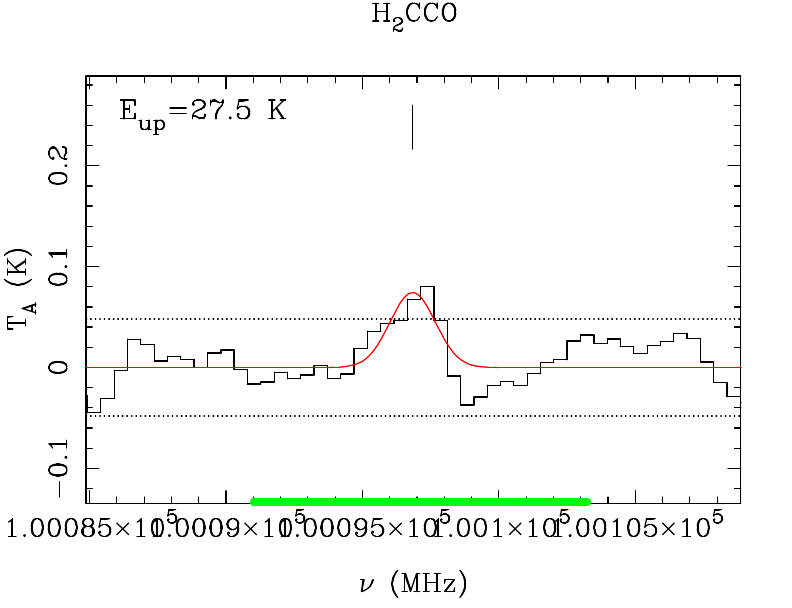}}
  \caption{\textbf{\hspace{-0.2 em}(cont.)}\ Same as in Figure \ref{fig-SpecNredC}. Panels (g) to (n) 
    show lines of SiO, $^{29}$SiO, $^{30}$SiO, HC$_3$N, CH$_3$CHO, HC$^{15}$N, HNCO, and H$_2$CCO, respectively.}
\end{figure}
%%%%%%%%%%%%%%%%%%%%%%%%%%%%%%%%%%%%%%%%%%%%%%%%%%
\begin{figure}
  \subfloat[][]{\includegraphics[angle=-90,ext=.pdf,width= 1.00 \textwidth]{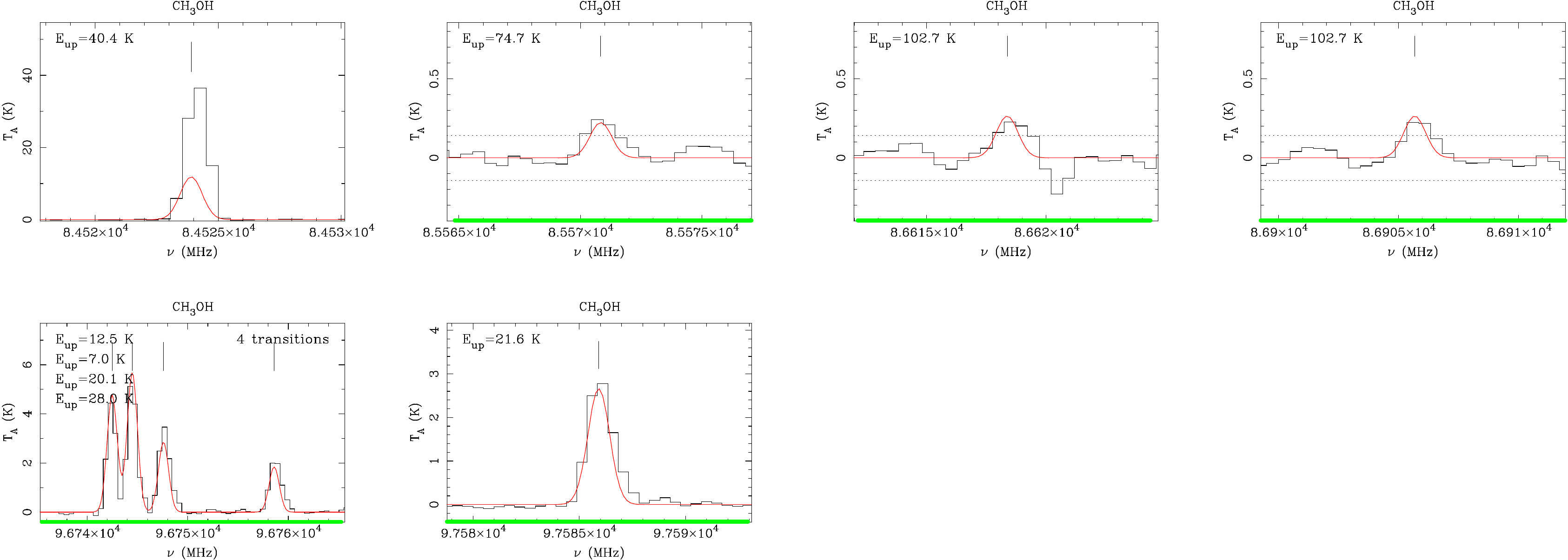}}\\
  \subfloat[][]{\includegraphics[angle=-90,ext=.pdf,width= 0.25 \textwidth]{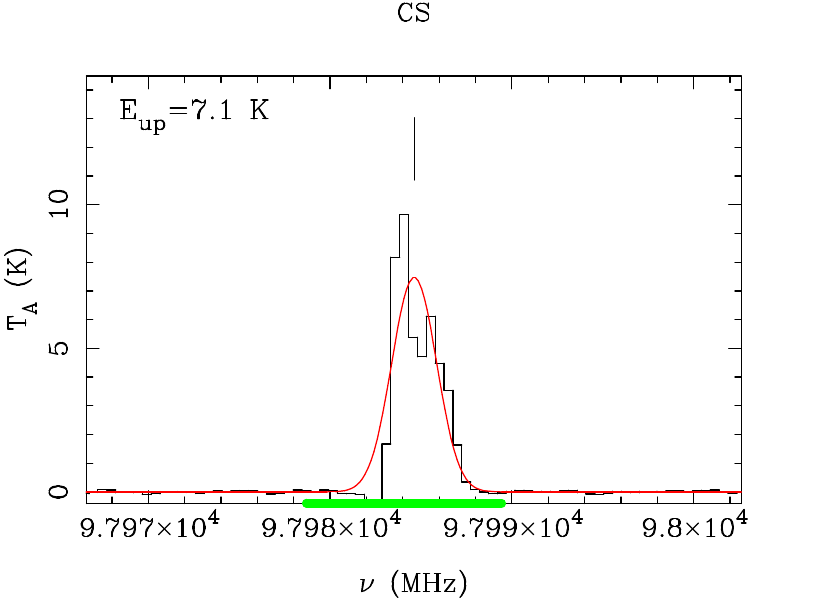}}
  \subfloat[][]{\includegraphics[angle=-90,ext=.pdf,width= 0.50 \textwidth]{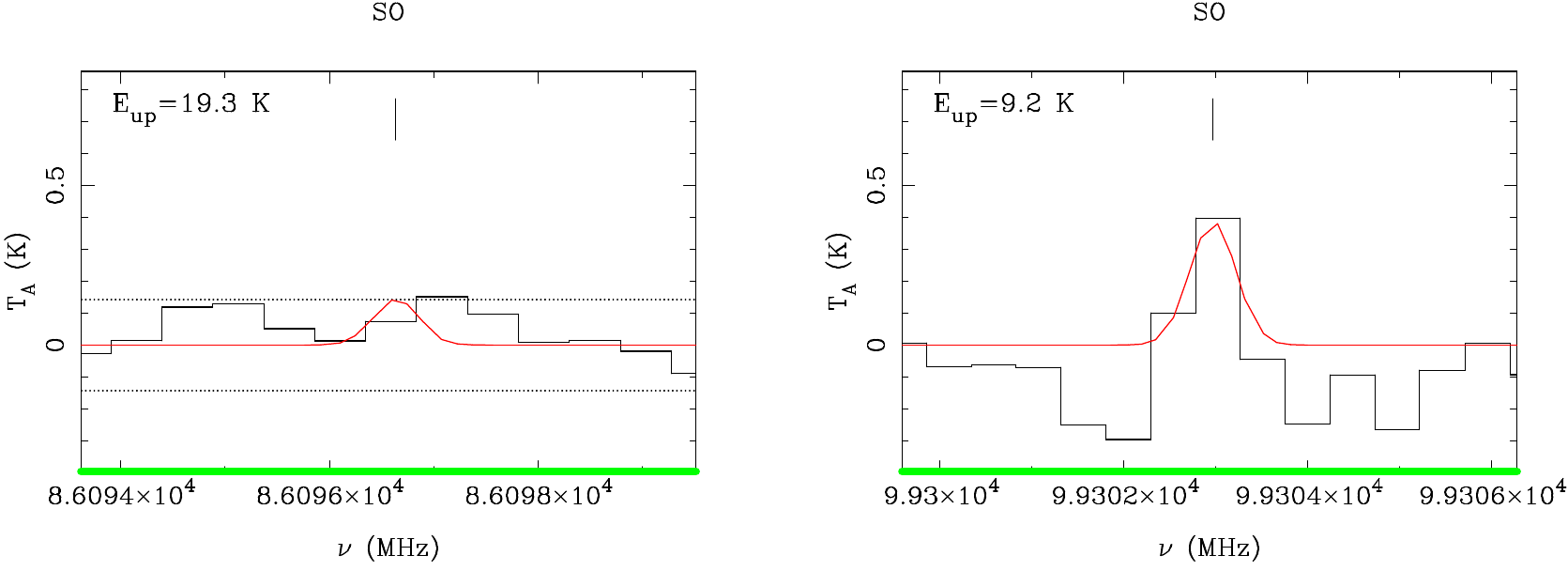}}
  \subfloat[][]{\includegraphics[angle=-90,ext=.pdf,width= 0.25 \textwidth]{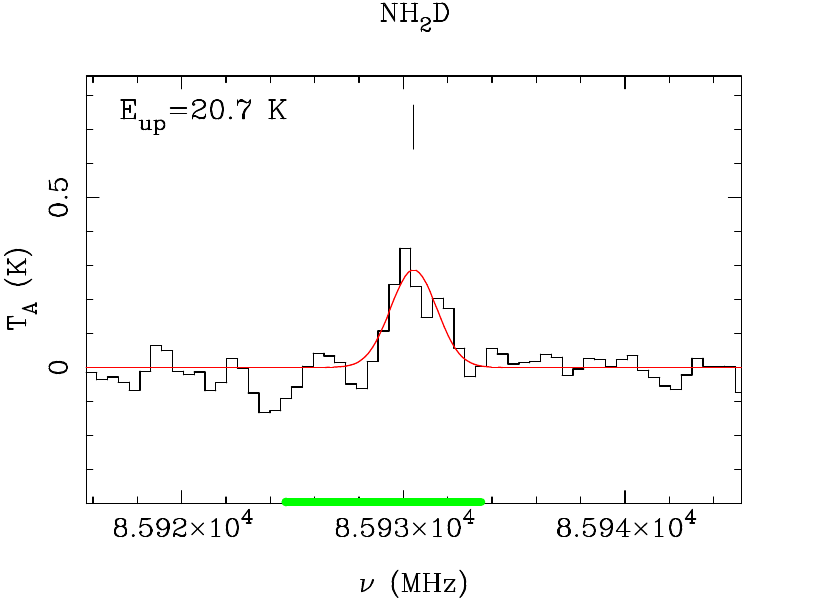}}\\
  \subfloat[][]{\includegraphics[angle=-90,ext=.pdf,width= 0.25 \textwidth]{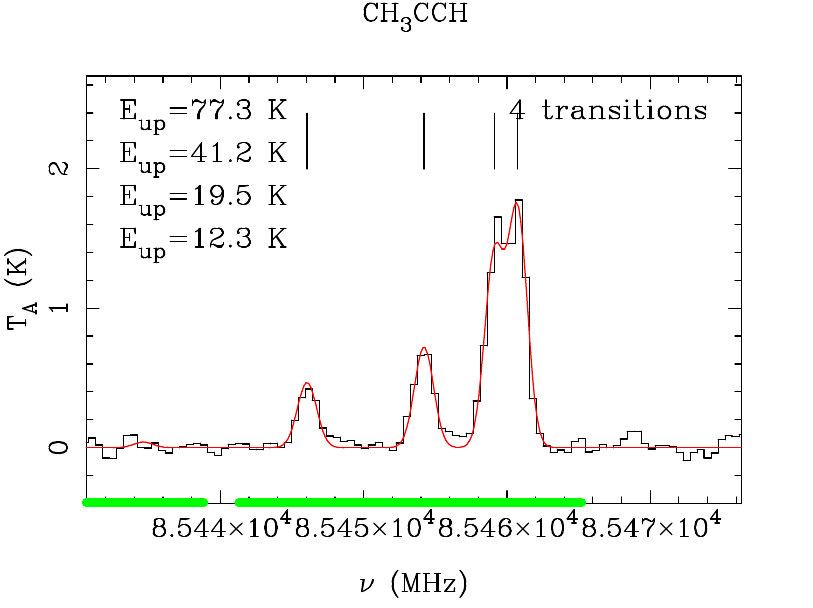}}
  \subfloat[][]{\includegraphics[angle=-90,ext=.pdf,width= 0.25 \textwidth]{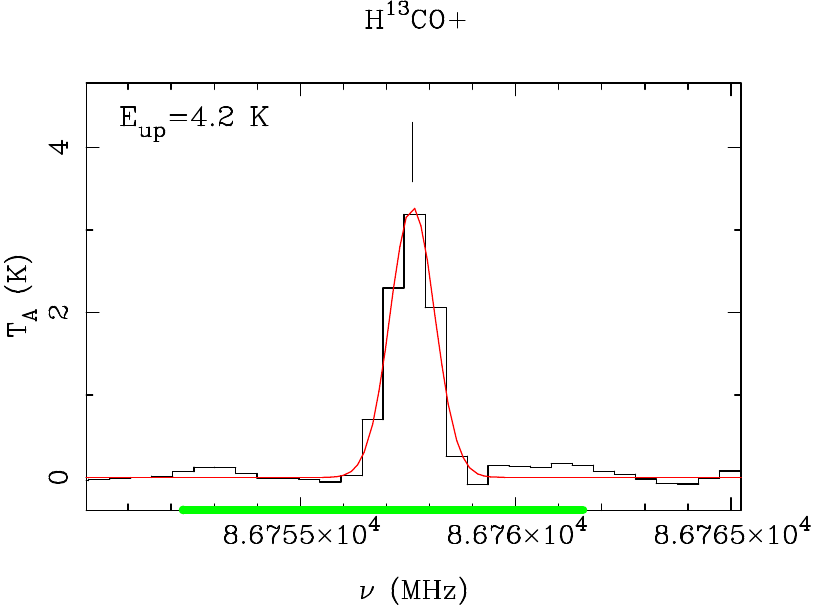}}
  \subfloat[][]{\includegraphics[angle=-90,ext=.pdf,width= 0.25 \textwidth]{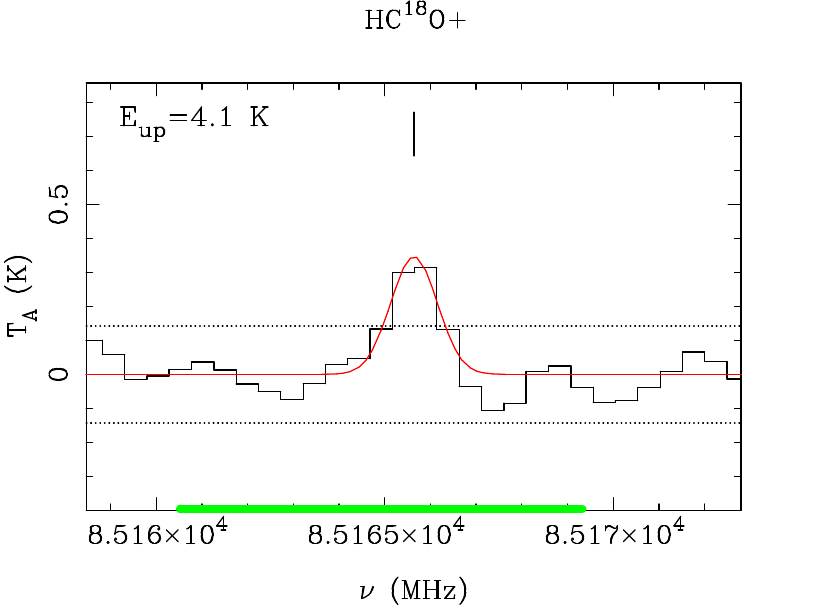}}
  \subfloat[][]{\includegraphics[angle=-90,ext=.pdf,width= 0.25 \textwidth]{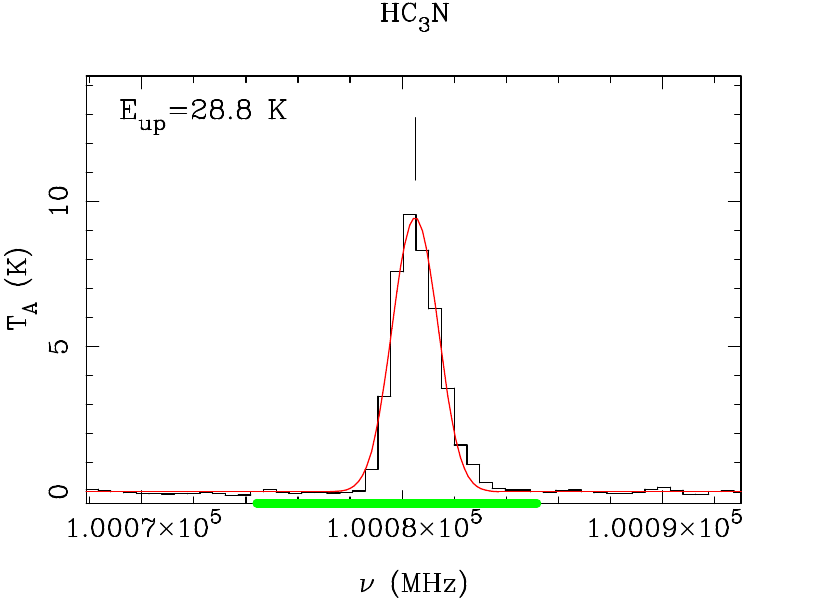}}\\
  \caption{Emission lines from several molecules toward the methanol peak
    of the NEC-wall. Line types and colors as described in Figure
    \ref{fig-SpecCCcore1}, with $\sigma=0.06$ K. Panels (a) to (h) show lines of \met, CS, SO,
    NH$_2$D, CH$_3$CCH, H$^{13}$CO$^+$, HC$^{18}$O$^+$, and HC$_3$N,
    respectively. \added{We note in (a) that the strong non-LTE \met, \maser\ emission cannot be reproduced with the models used in this work.} \label{fig-SpecNECwNP}}
\end{figure}

\begin{figure}
  \ContinuedFloat
  \subfloat[][]{\includegraphics[angle=-90,ext=.pdf,width= 1.00 \textwidth]{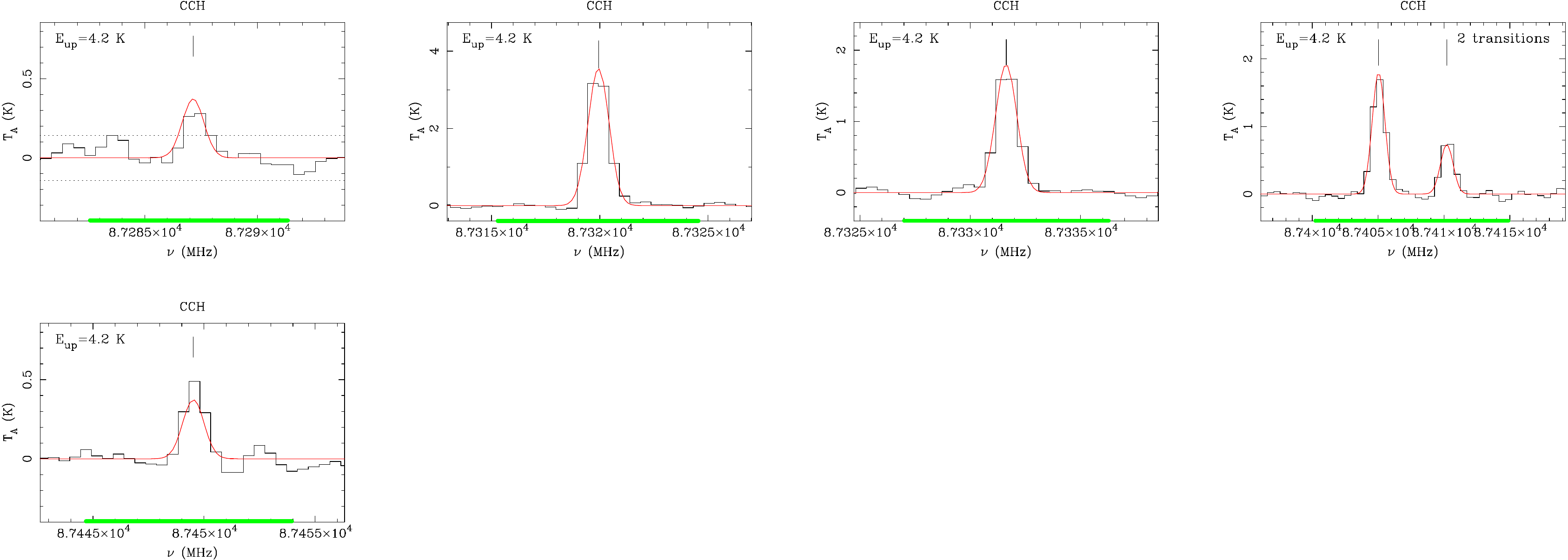}}\\
  \subfloat[][]{\includegraphics[angle=-90,ext=.pdf,width= 0.50 \textwidth]{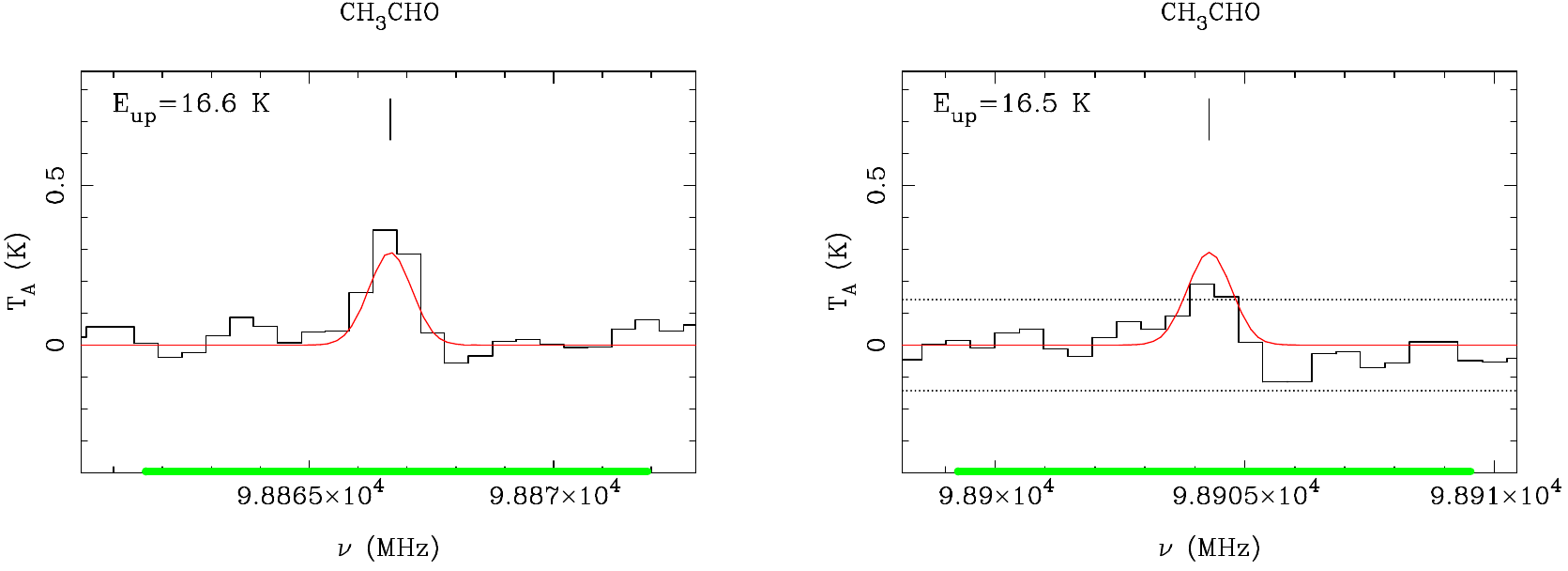}}
  \subfloat[][]{\includegraphics[angle=-90,ext=.pdf,width= 0.25 \textwidth]{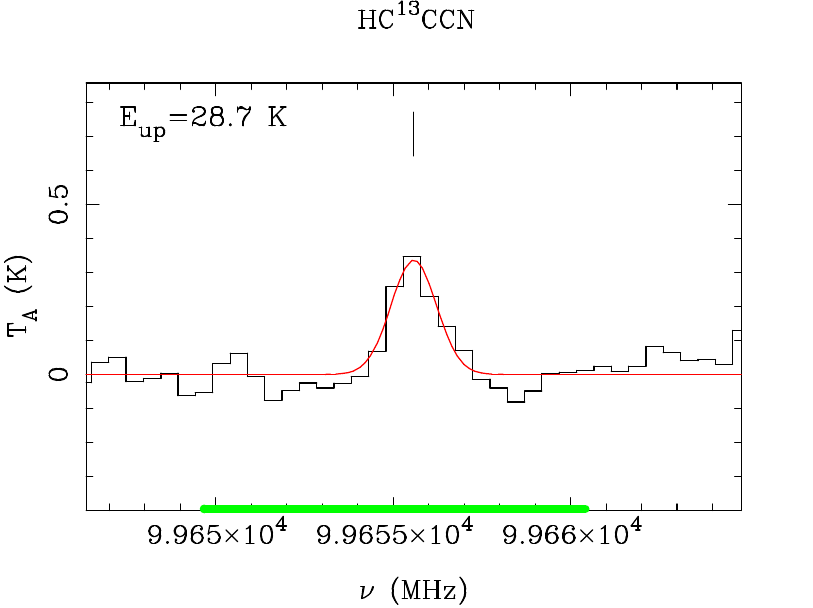}}
  \subfloat[][]{\includegraphics[angle=-90,ext=.pdf,width= 0.25 \textwidth]{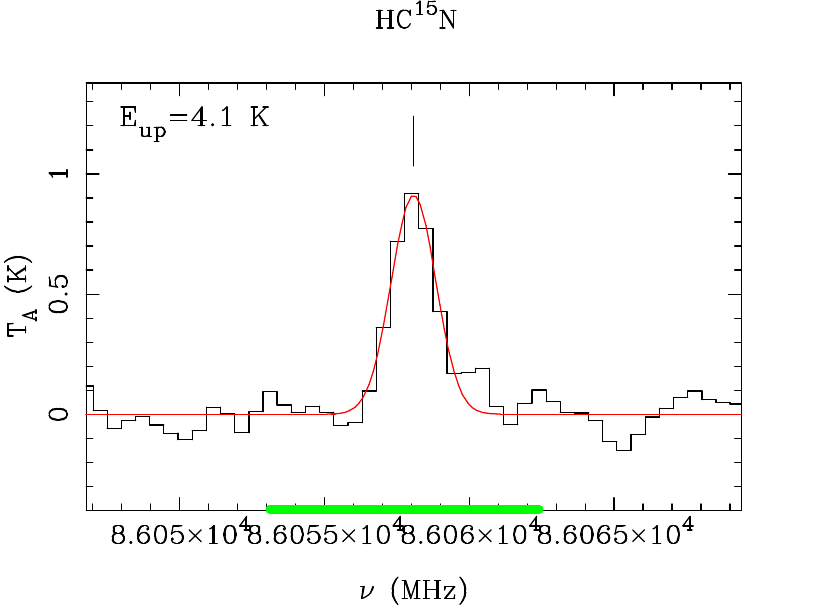}}\\
  \subfloat[][]{\includegraphics[angle=-90,ext=.pdf,width= 0.25 \textwidth]{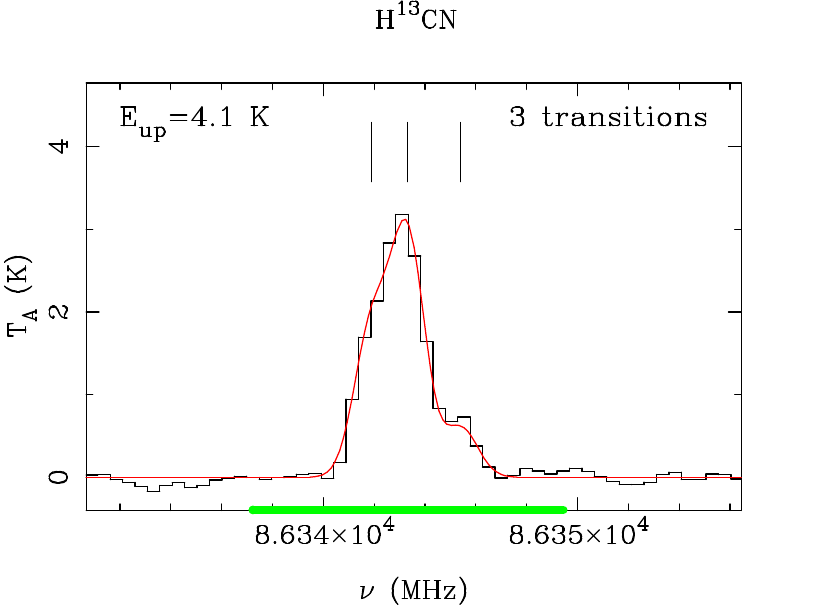}}
  \subfloat[][]{\includegraphics[angle=-90,ext=.pdf,width= 0.25 \textwidth]{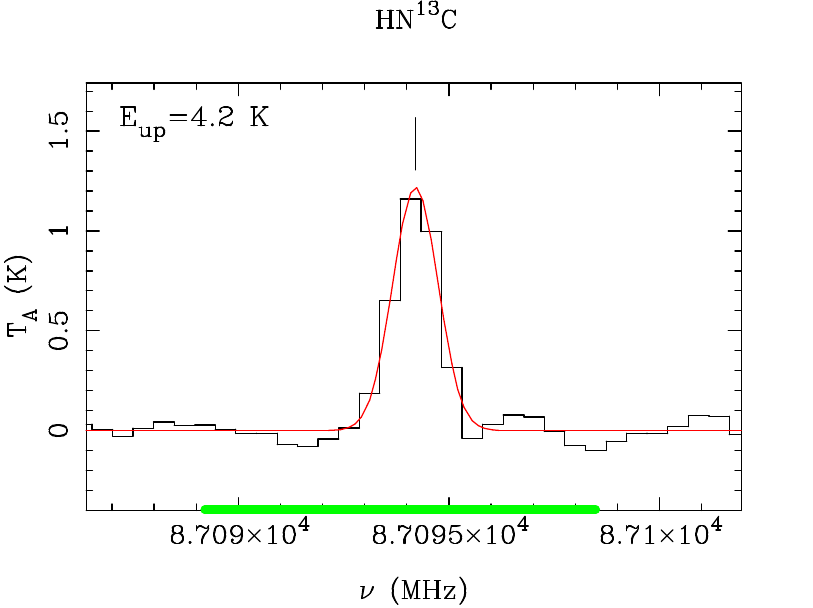}}
  \subfloat[][]{\includegraphics[angle=-90,ext=.pdf,width= 0.25 \textwidth]{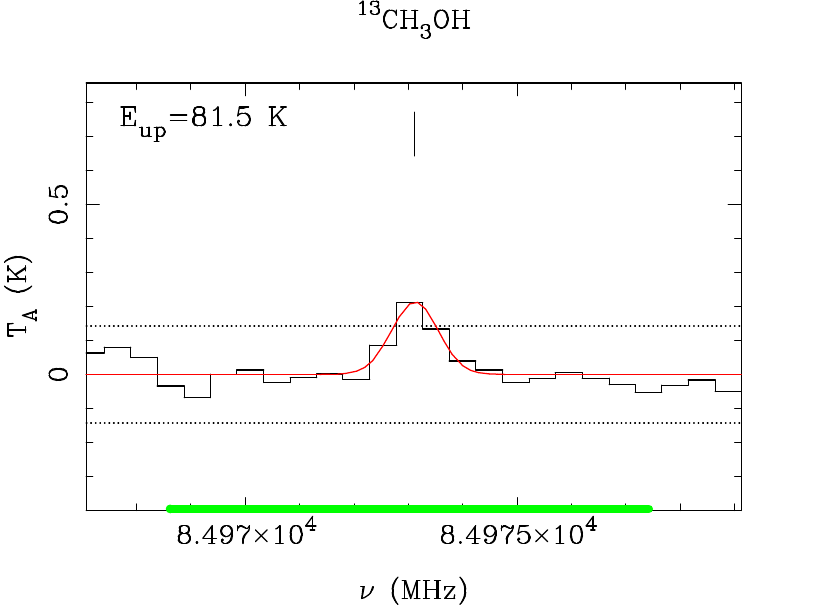}}
  \caption{\textbf{\hspace{-0.2 em}(cont.)}\ Same as Figure \ref{fig-SpecNECwNP}. Panels (i) to (o) show lines of CCH, CH$_3$CHO, HC$^{13}$CCN, HC$^{15}$N, H$^{13}$CN, and $^{13}$CH$_3$OH, respectively.}
\end{figure}
%%%%%%%%%%%%%%%%%%%%%%%%%%%%%%%%%%%%%%%%%%%%%%%%%%
\begin{figure}
   \subfloat[][]{\includegraphics[angle=-90,ext=.pdf,width= 1.00 \textwidth]{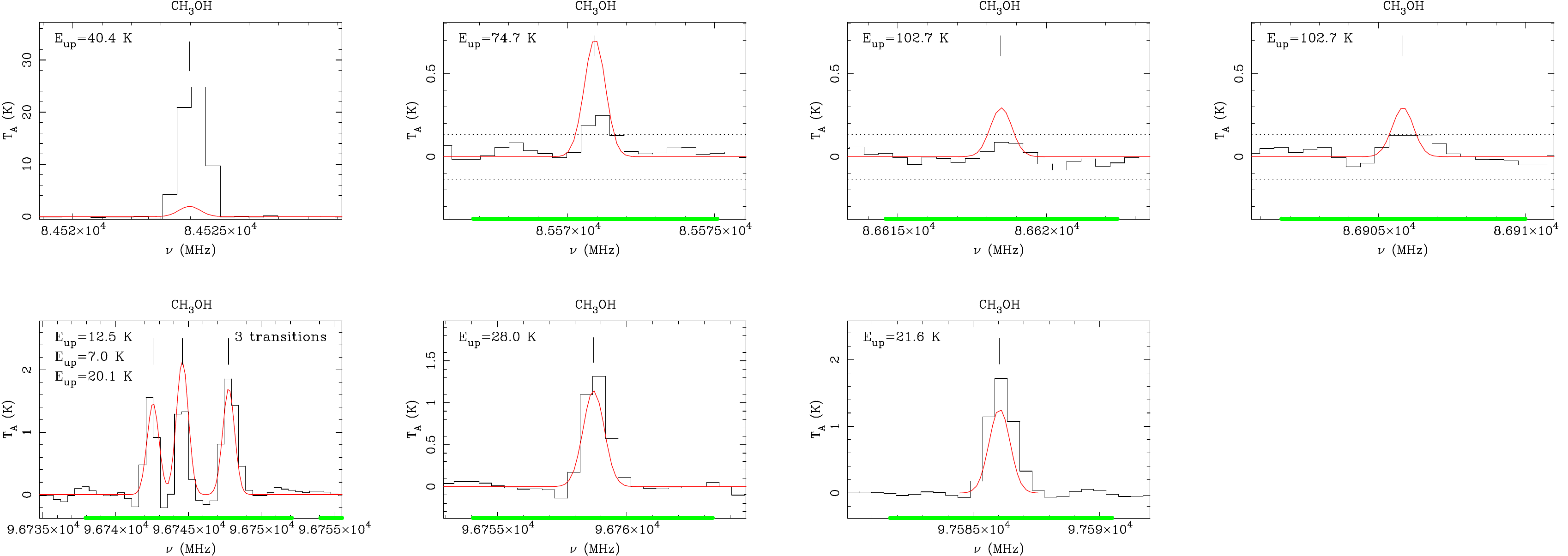}}\\
   \subfloat[][]{\includegraphics[angle=-90,ext=.pdf,width= 0.25 \textwidth]{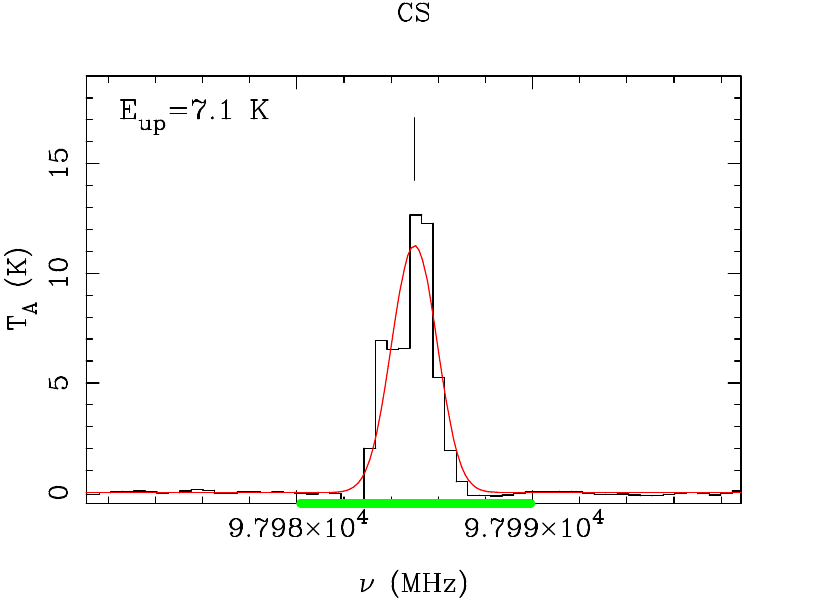}}
   \subfloat[][]{\includegraphics[angle=-90,ext=.pdf,width= 0.25 \textwidth]{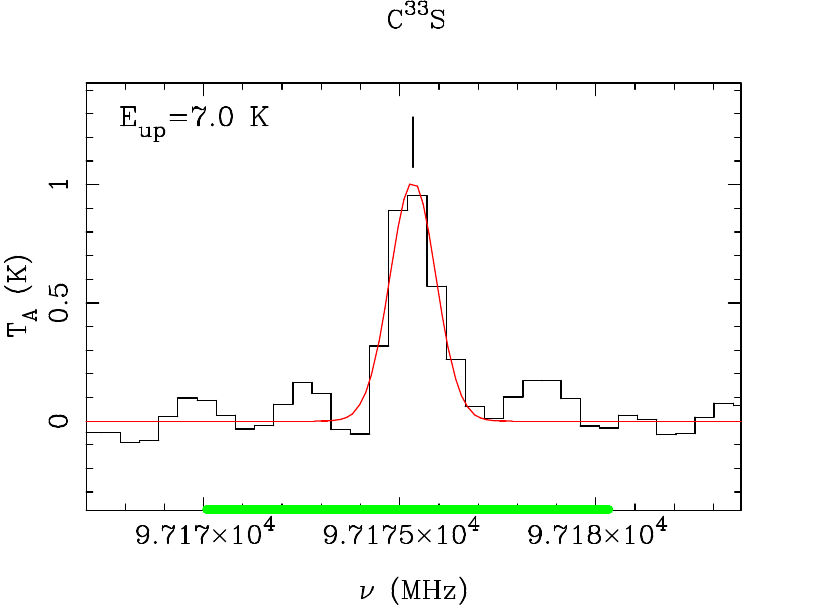}}
   \subfloat[][]{\includegraphics[angle=-90,ext=.pdf,width= 0.50 \textwidth]{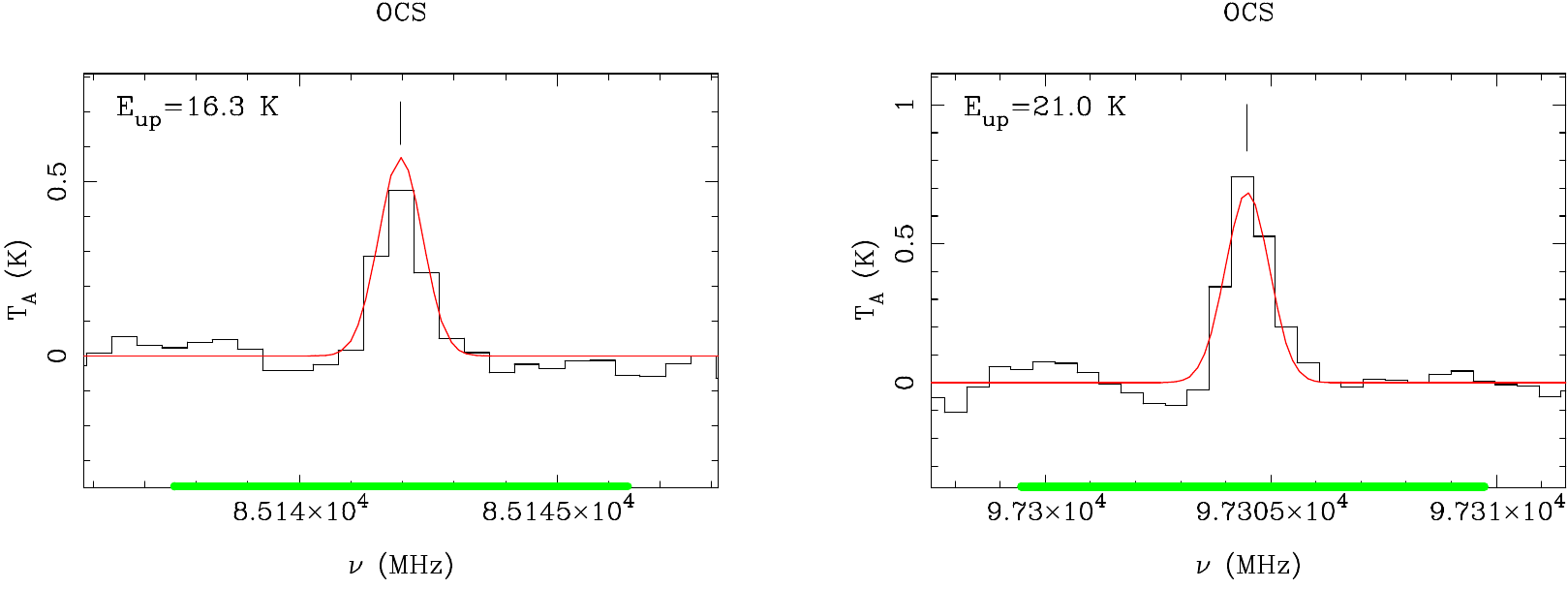}}\\
   \subfloat[][]{\includegraphics[angle=-90,ext=.pdf,width= 0.75 \textwidth]{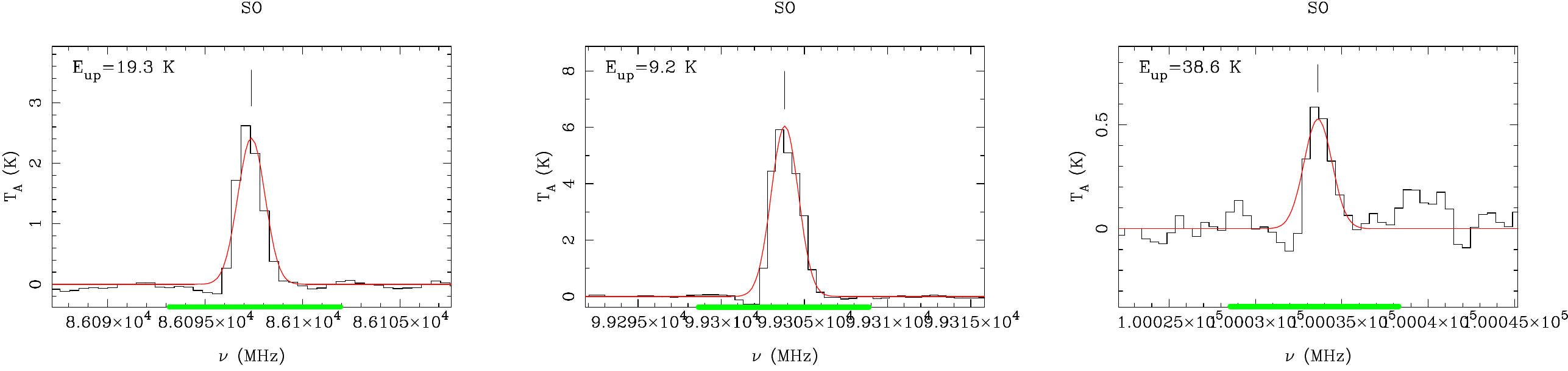}}
   \subfloat[][]{\includegraphics[angle=-90,ext=.pdf,width= 0.25 \textwidth]{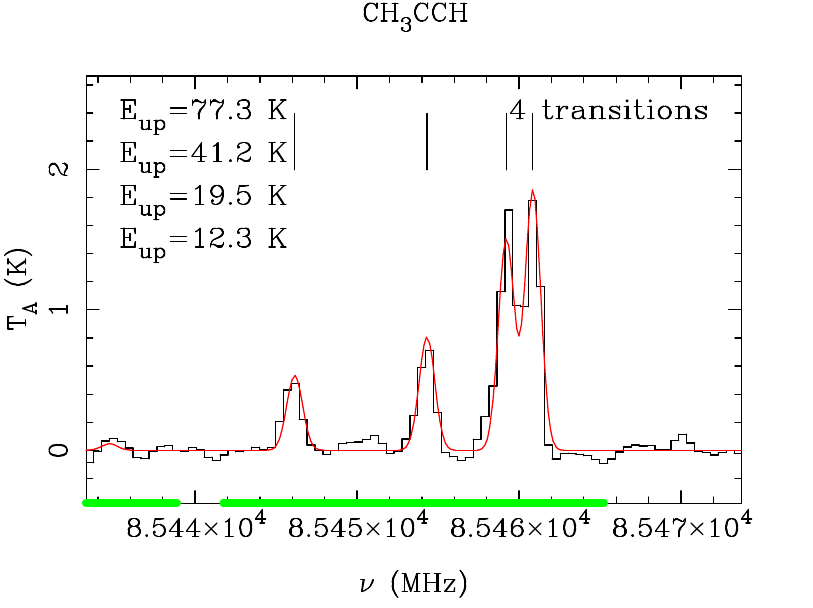}}\\
   \subfloat[][]{\includegraphics[angle=-90,ext=.pdf,width= 0.25 \textwidth]{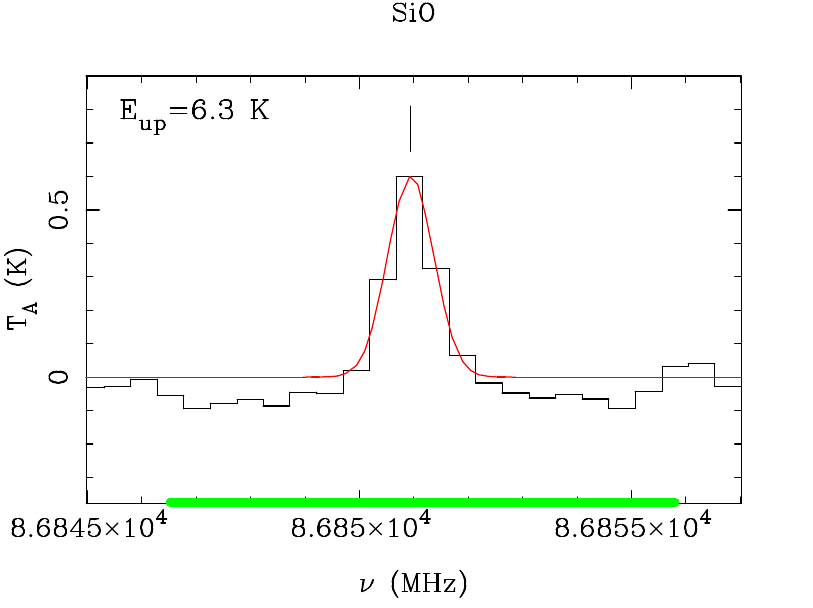}}
   \subfloat[][]{\includegraphics[angle=-90,ext=.pdf,width= 0.25 \textwidth]{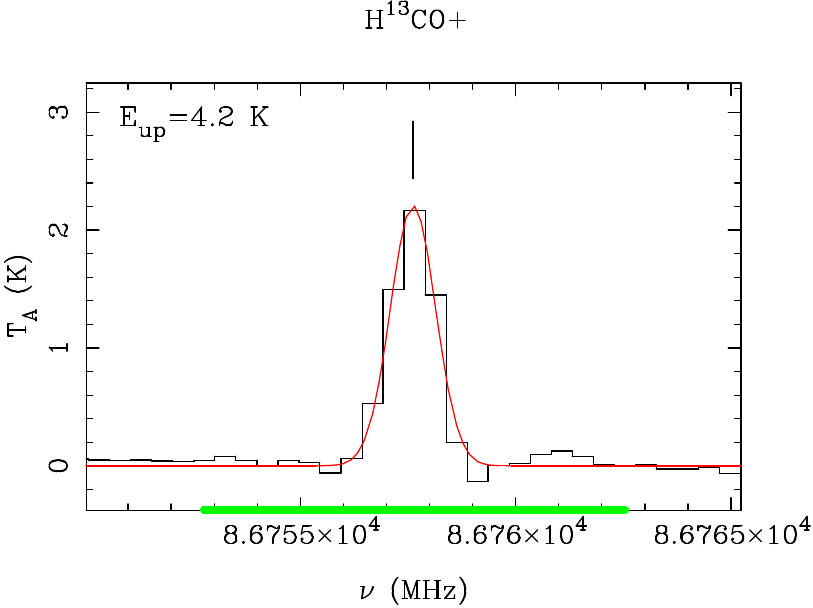}}
   \subfloat[][]{\includegraphics[angle=-90,ext=.pdf,width= 0.25 \textwidth]{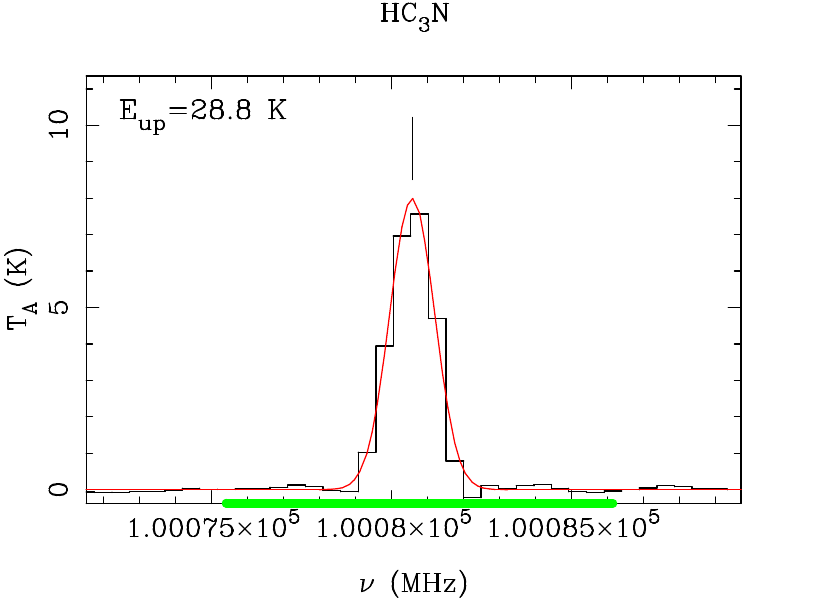}}
   \subfloat[][]{\includegraphics[angle=-90,ext=.pdf,width= 0.25 \textwidth]{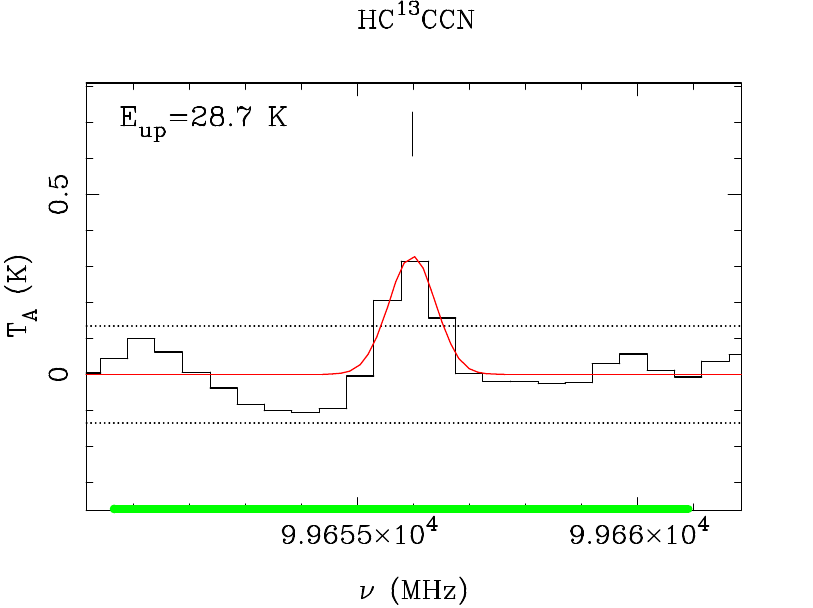}}
   \caption{Emission lines from several molecules toward the sulfur monoxide peak
    of the NEC-wall. Line types and colors as described in Figure
    \ref{fig-SpecCCcore1} with $\sigma=0.05$ K. Panels (a) to (j) show lines of \met, CS, C$^{33}$S, OCS, SO, \propyne, SiO, H$^{13}$CO$^+$, 
\hctn, and HC$^{13}$CCN, respectively. \added{The strong non-LTE \met, \maser\ emission shown in (a) cannot be reproduced by the models used in this work.}\label{fig-SpecNECwSP}}
\end{figure}
\begin{figure}
  \ContinuedFloat
  \subfloat[][]{\includegraphics[angle=-90,ext=.pdf,width= 1.00 \textwidth]{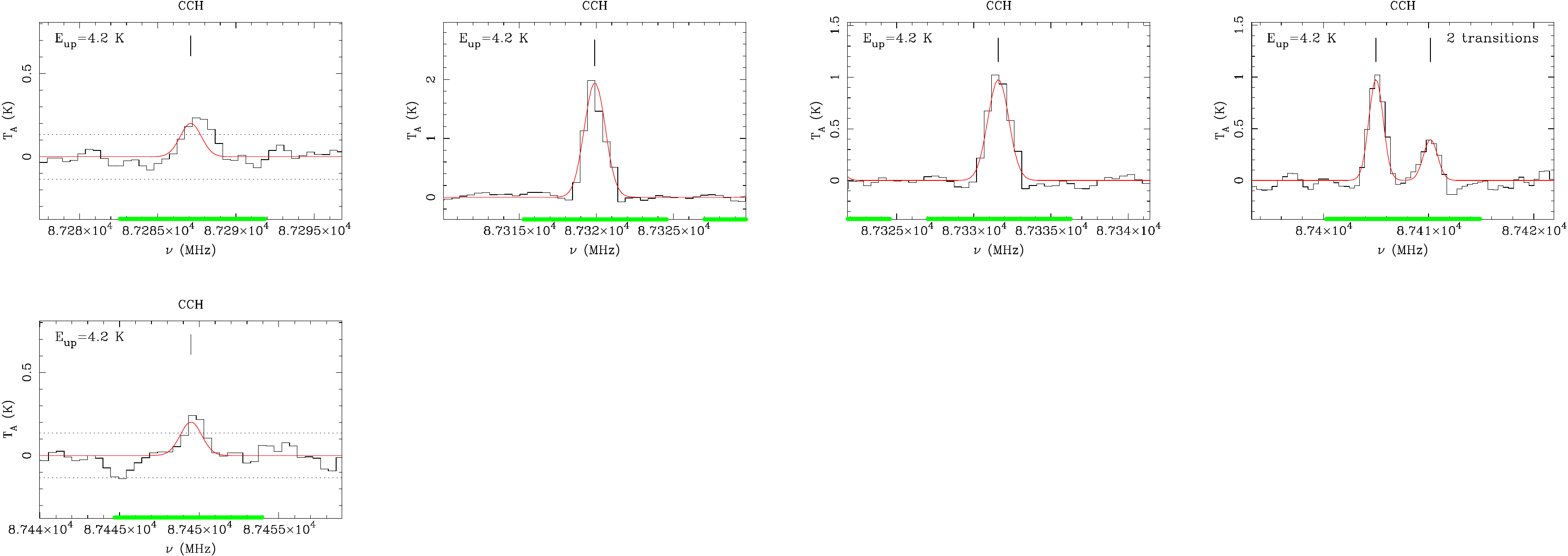}}\\
  \subfloat[][]{\includegraphics[angle=-90,ext=.pdf,width= 0.75 \textwidth]{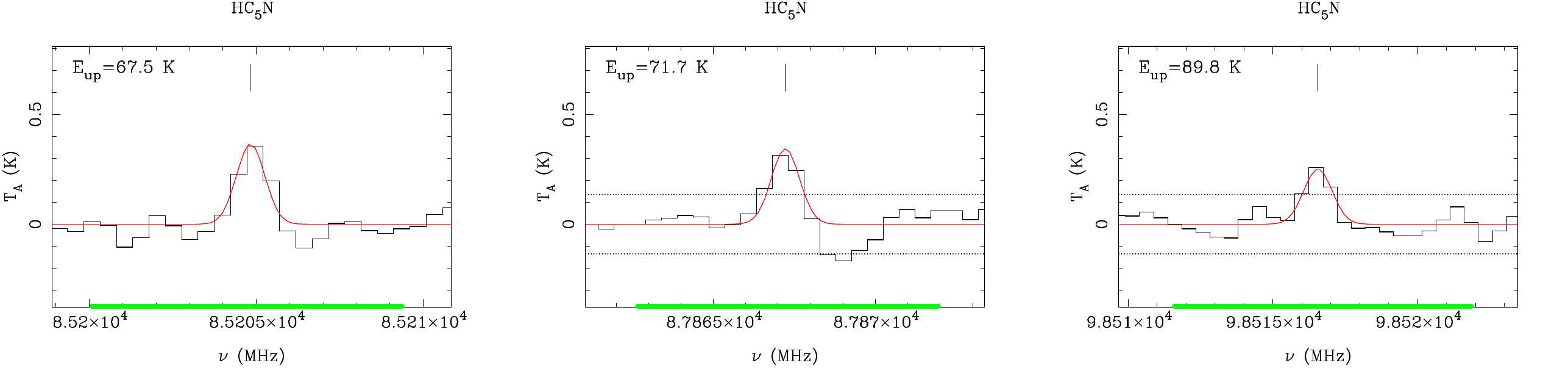}}
  \subfloat[][]{\includegraphics[angle=-90,ext=.pdf,width= 0.25 \textwidth]{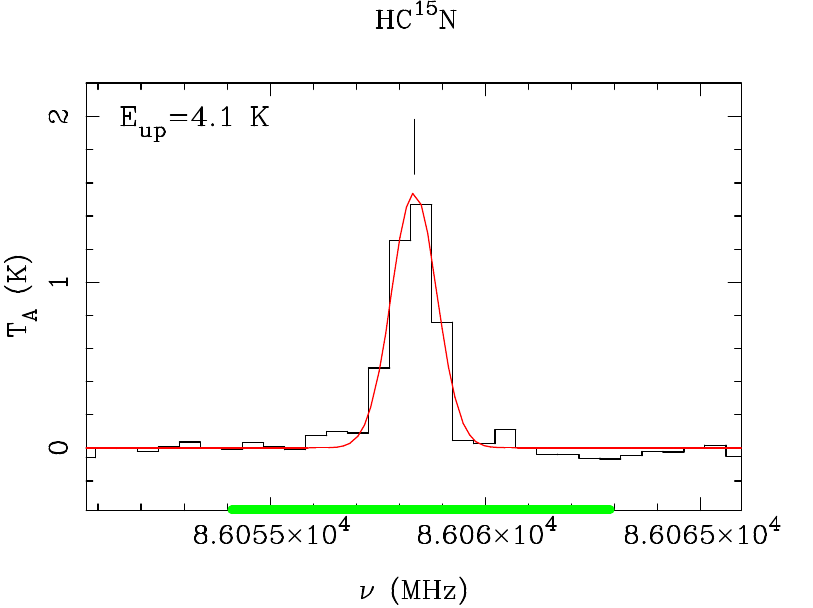}}\\
  \subfloat[][]{\includegraphics[angle=-90,ext=.pdf,width= 0.25 \textwidth]{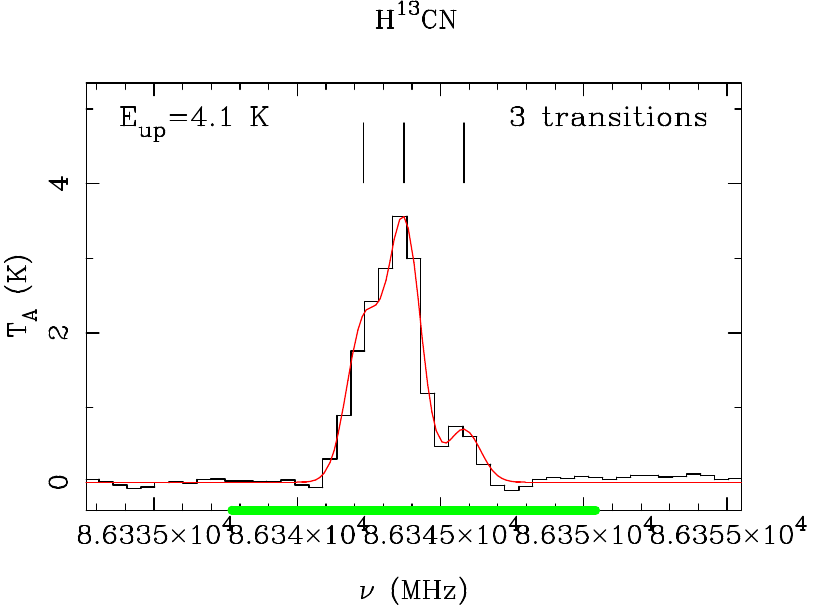}}
  \subfloat[][]{\includegraphics[angle=-90,ext=.pdf,width= 0.25 \textwidth]{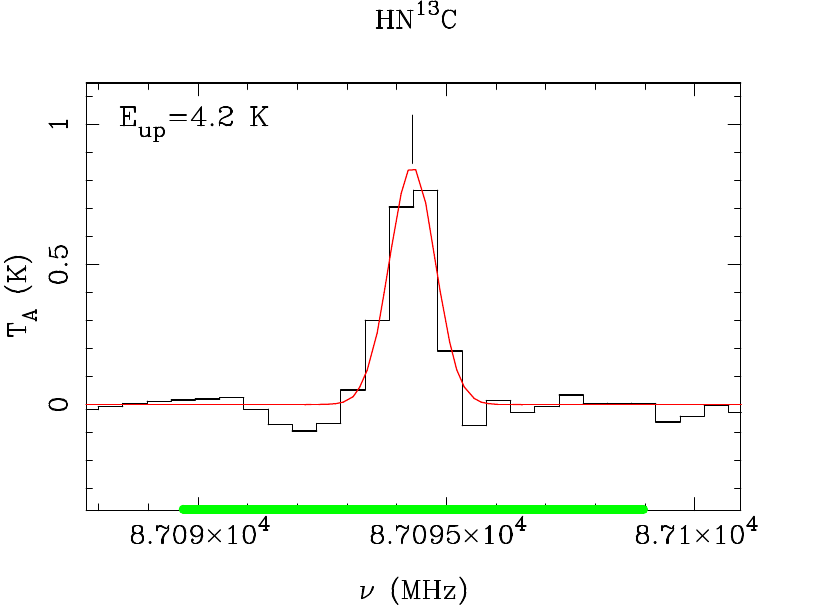}}
  \caption{\textbf{\hspace{-0.2 em}(cont.)}\ Same as Figure \ref{fig-SpecNECwSP}. Panels (k) to (n) show lines of CCH, \hccn, \hcqn, \htcn, and \hntc, respectively.}
\end{figure}
%%%%%%%%%%%%%%%%%%%%%%%%%%%%%%%%%%%%%%%%%%%%%%%%%%
\begin{figure}
\subfloat[][]{\includegraphics[angle=-90,ext=.pdf,width= 0.25 \textwidth]{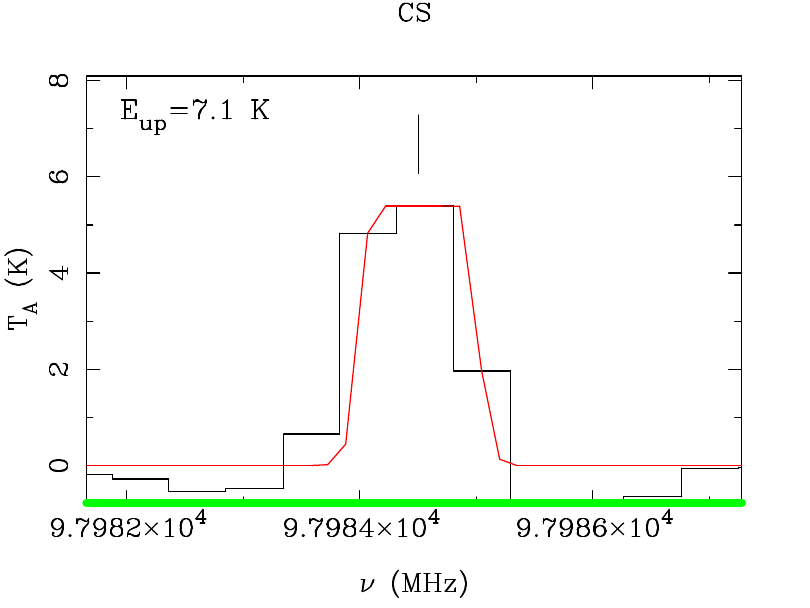}}
\subfloat[][]{\includegraphics[angle=-90,ext=.pdf,width= 0.25 \textwidth]{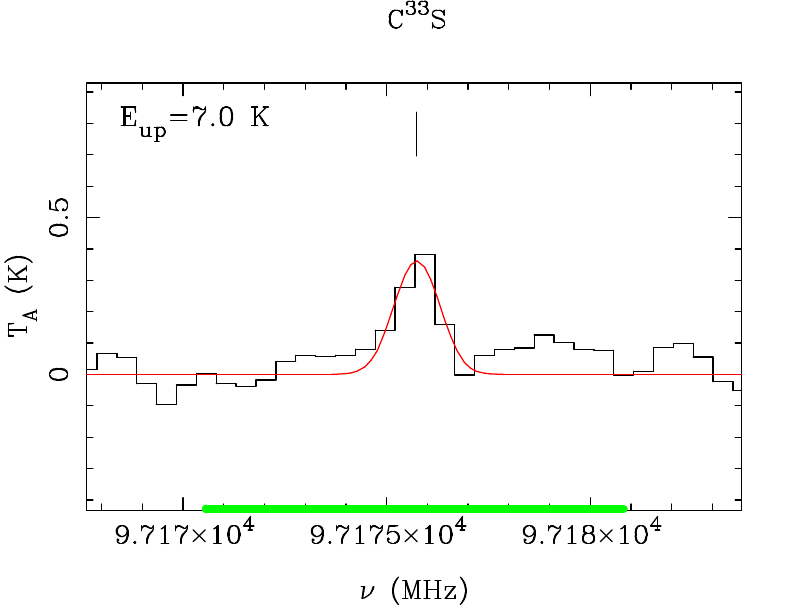}}
\subfloat[][]{\includegraphics[angle=-90,ext=.pdf,width= 0.50 \textwidth]{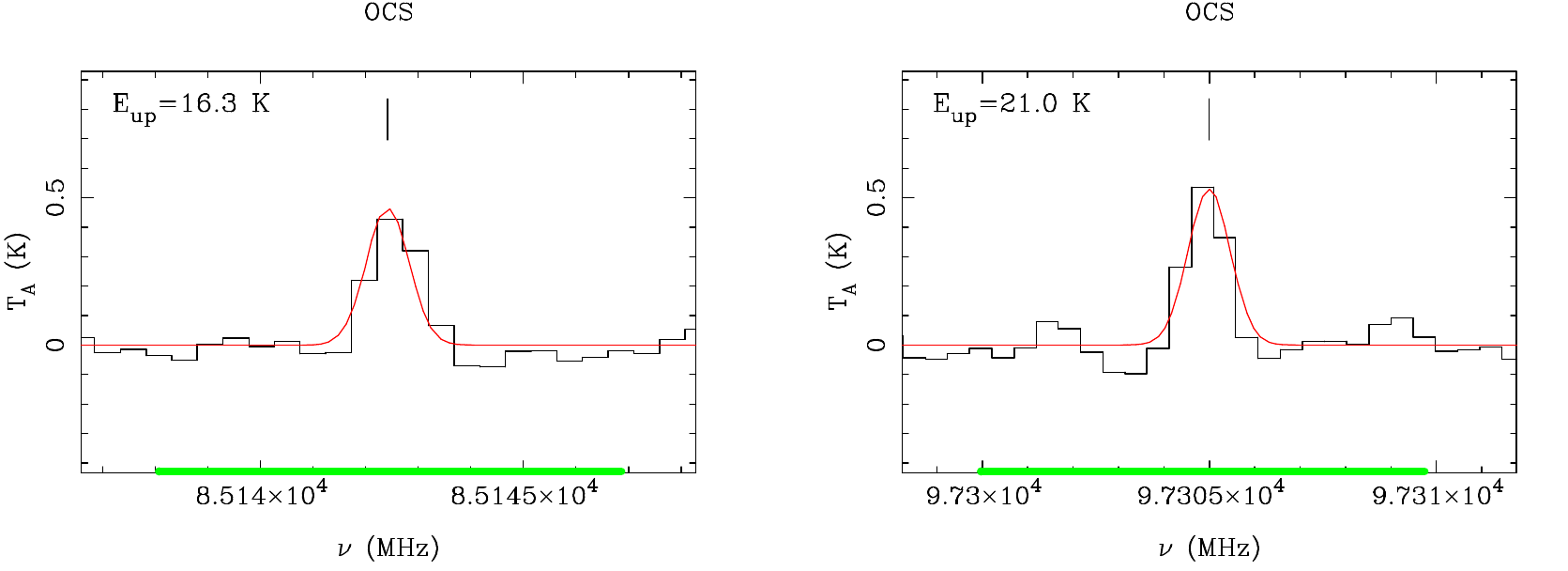}}
\caption{Emission lines from several molecules toward the point (a) in the Diffuse ridge (DR (a)). 
  Line types and colors as described in Figure
    \ref{fig-SpecCCcore1} with $\sigma=0.06$ K. Panels (a) to (c) show lines of CS, C$^{33}$S, and OCS, respectively.\label{fig-SpecDRa}}
\end{figure}
\begin{figure}
\ContinuedFloat
\subfloat[][]{\includegraphics[angle=-90,ext=.pdf,width= 1.00 \textwidth]{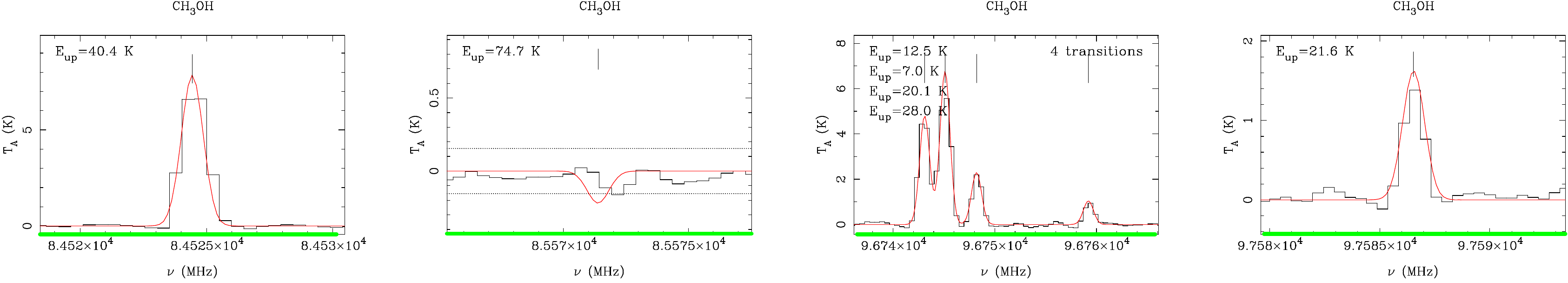}}\\
\subfloat[][]{\includegraphics[angle=-90,ext=.pdf,width= 0.50 \textwidth]{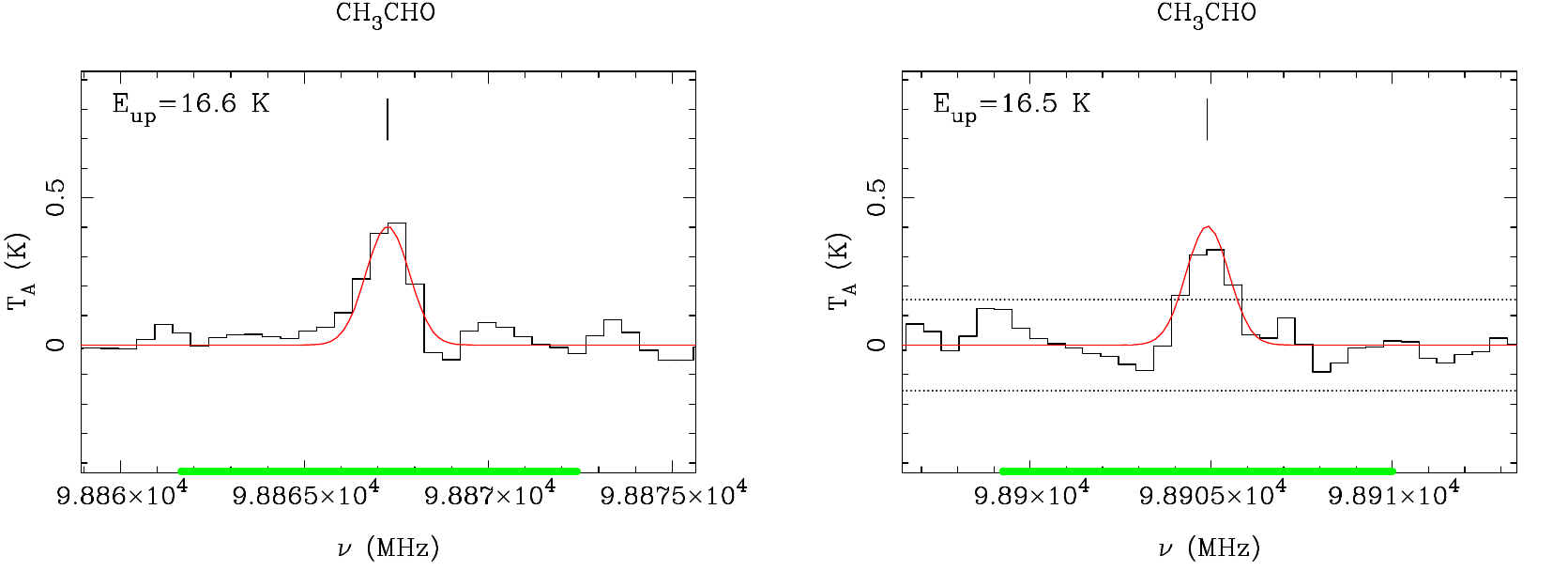}}
\subfloat[][]{\includegraphics[angle=-90,ext=.pdf,width= 0.25 \textwidth]{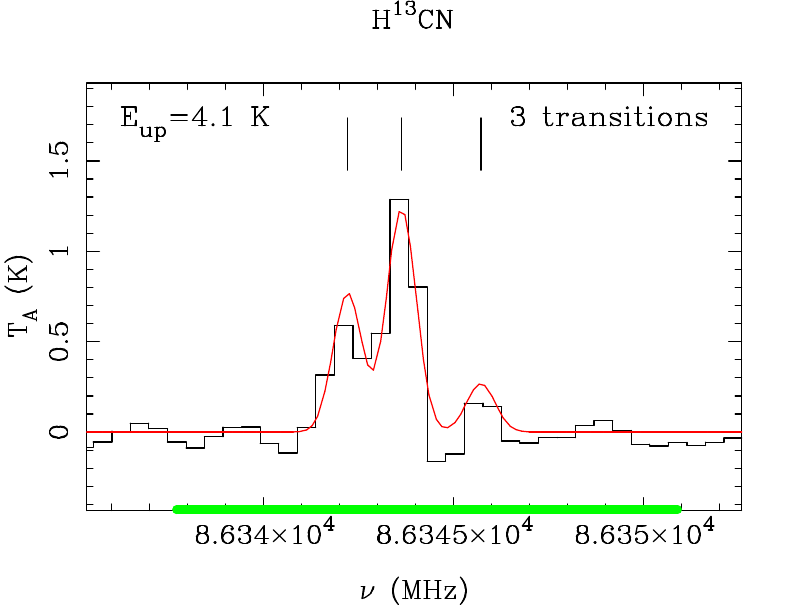}}
\subfloat[][]{\includegraphics[angle=-90,ext=.pdf,width= 0.25 \textwidth]{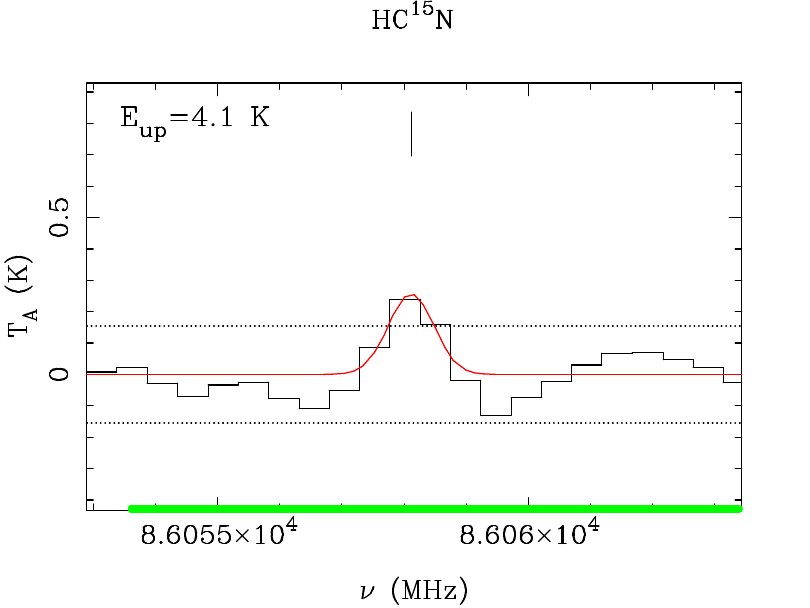}}\\
\subfloat[][]{\includegraphics[angle=-90,ext=.pdf,width= 0.25 \textwidth]{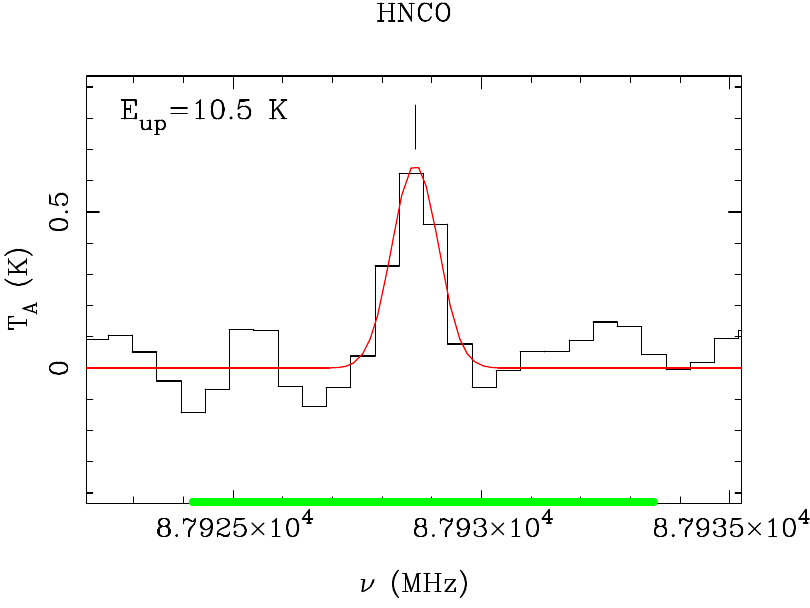}}
\subfloat[][]{\includegraphics[angle=-90,ext=.pdf,width= 0.25 \textwidth]{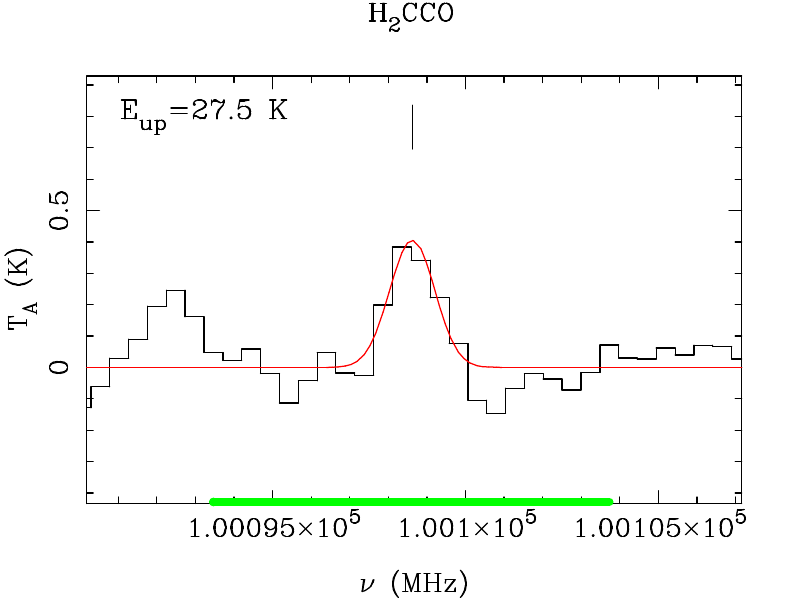}}
\subfloat[][]{\includegraphics[angle=-90,ext=.pdf,width= 0.25 \textwidth]{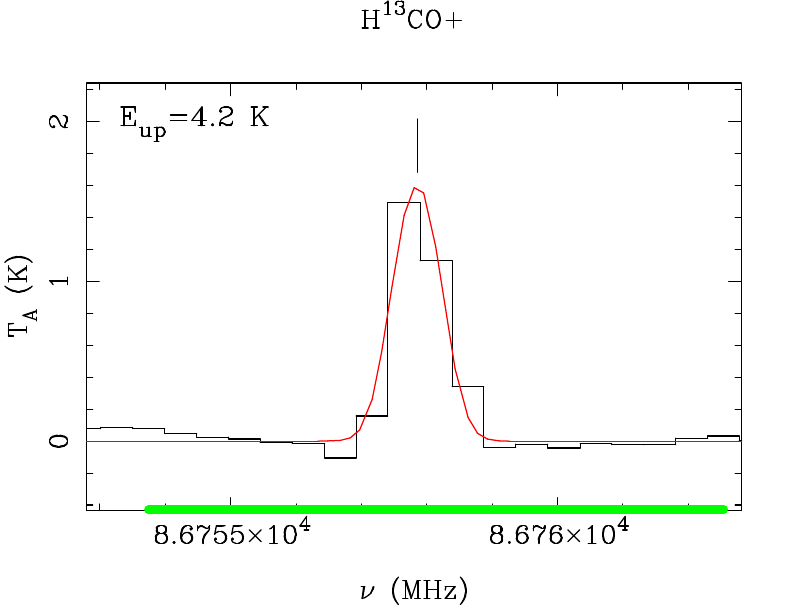}}
\subfloat[][]{\includegraphics[angle=-90,ext=.pdf,width= 0.25 \textwidth]{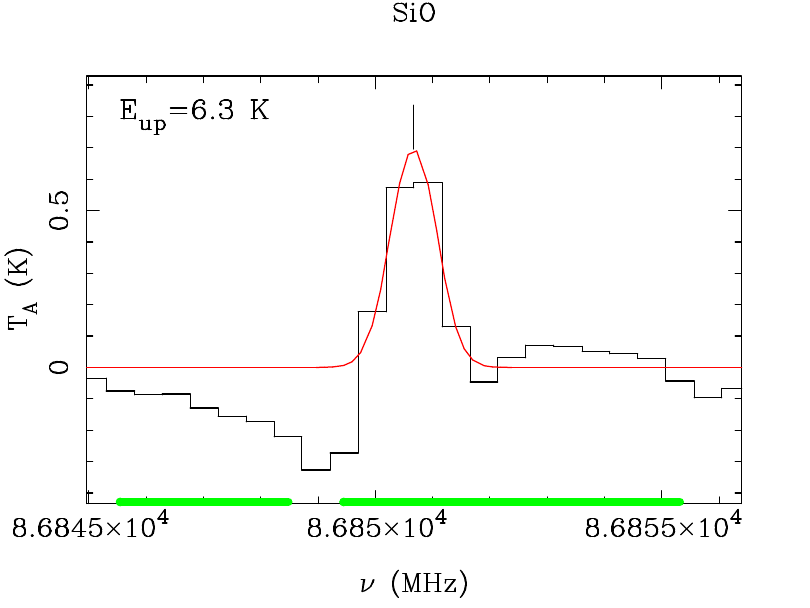}}\\
\subfloat[][]{\includegraphics[angle=-90,ext=.pdf,width= 0.25 \textwidth]{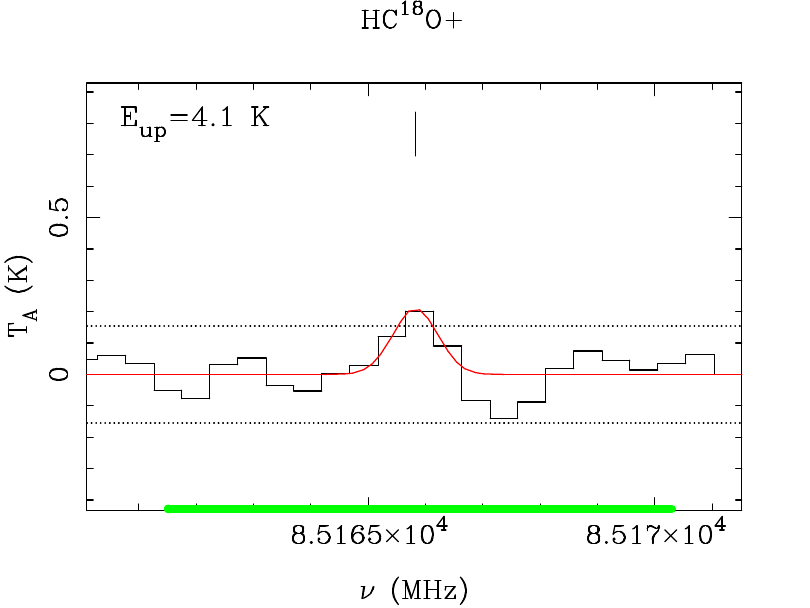}}
\subfloat[][]{\includegraphics[angle=-90,ext=.pdf,width= 0.50 \textwidth]{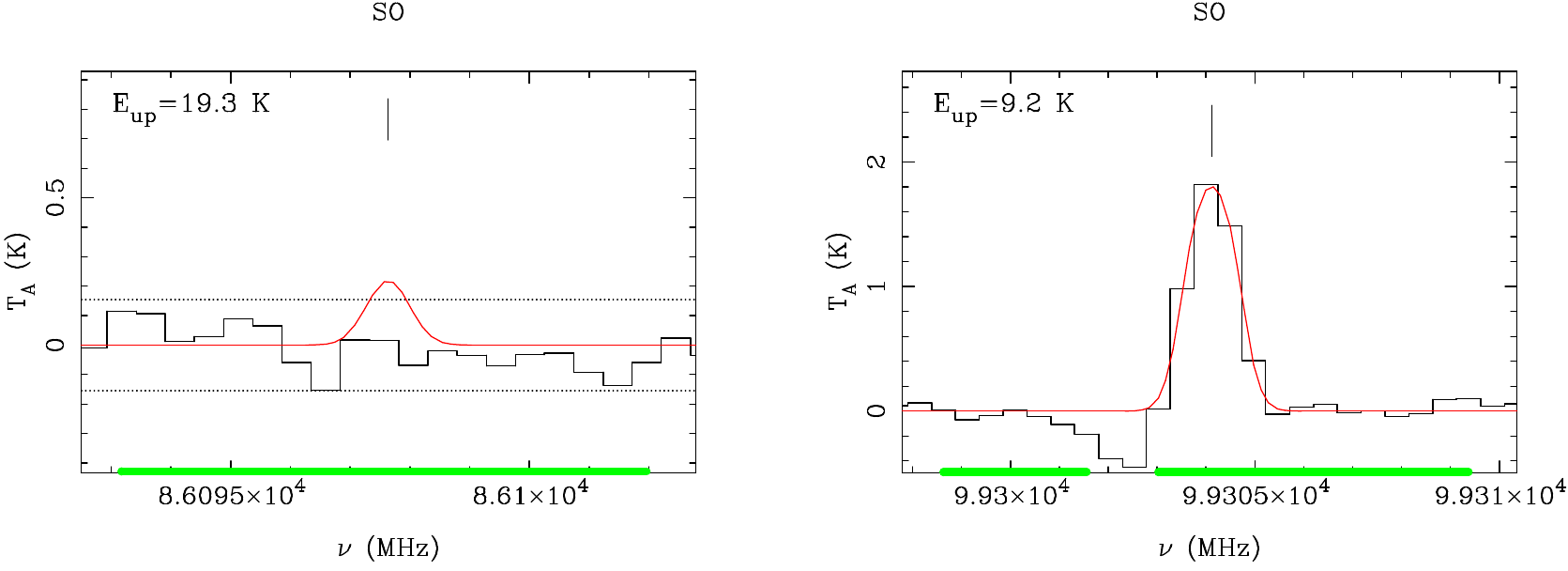}}
\subfloat[][]{\includegraphics[angle=-90,ext=.pdf,width= 0.25 \textwidth]{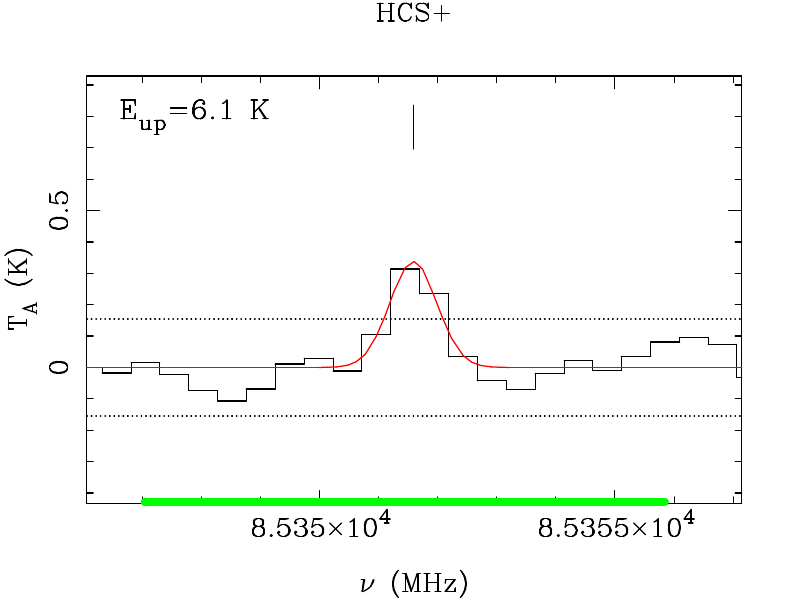}}\\
\subfloat[][]{\includegraphics[angle=-90,ext=.pdf,width= 0.50 \textwidth]{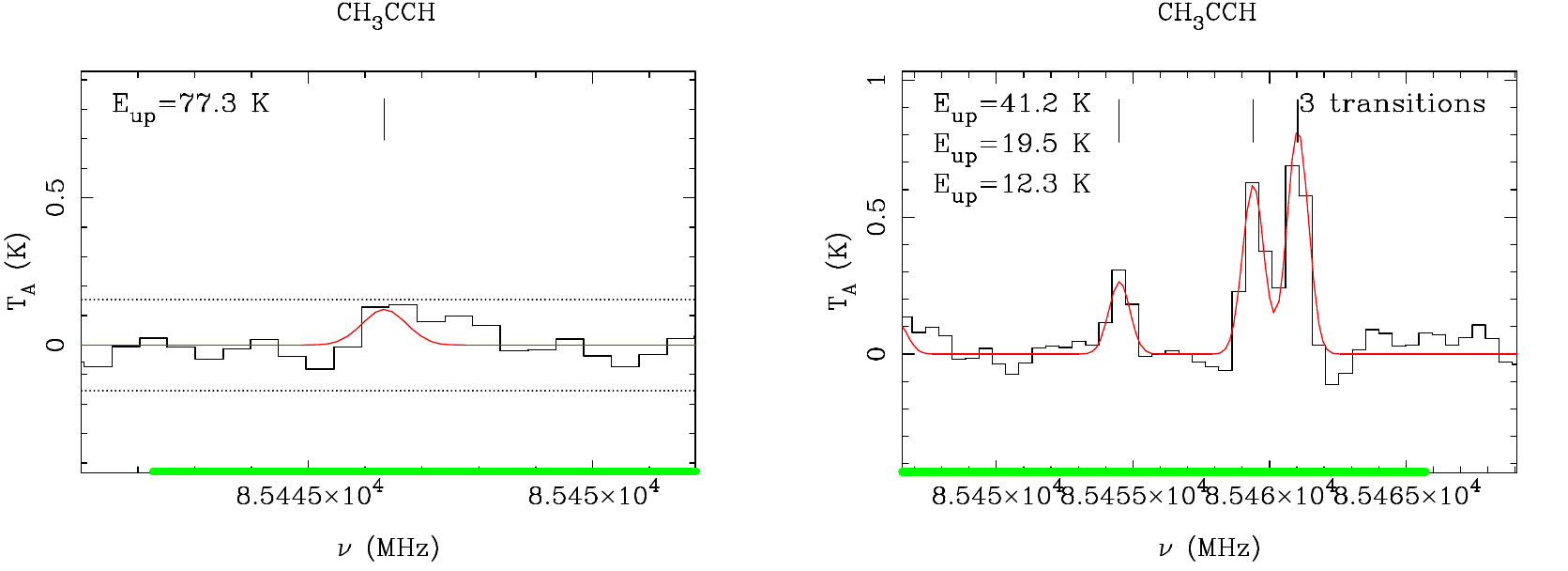}}\\
\caption{\textbf{\hspace{-0.2 em}(cont.)}\ Same as Figure \ref{fig-SpecDRa}. Panels (d) to (o) show lines of \met, \acet, \htcn, \hcqn, HNCO, \ethe, \htcop, SiO, \hcdop, SO, \hcsp, and \propyne, respectively.}
\end{figure}
\begin{figure}
\ContinuedFloat
\subfloat[][]{\includegraphics[angle=-90,ext=.pdf,width= 1.0 \textwidth]{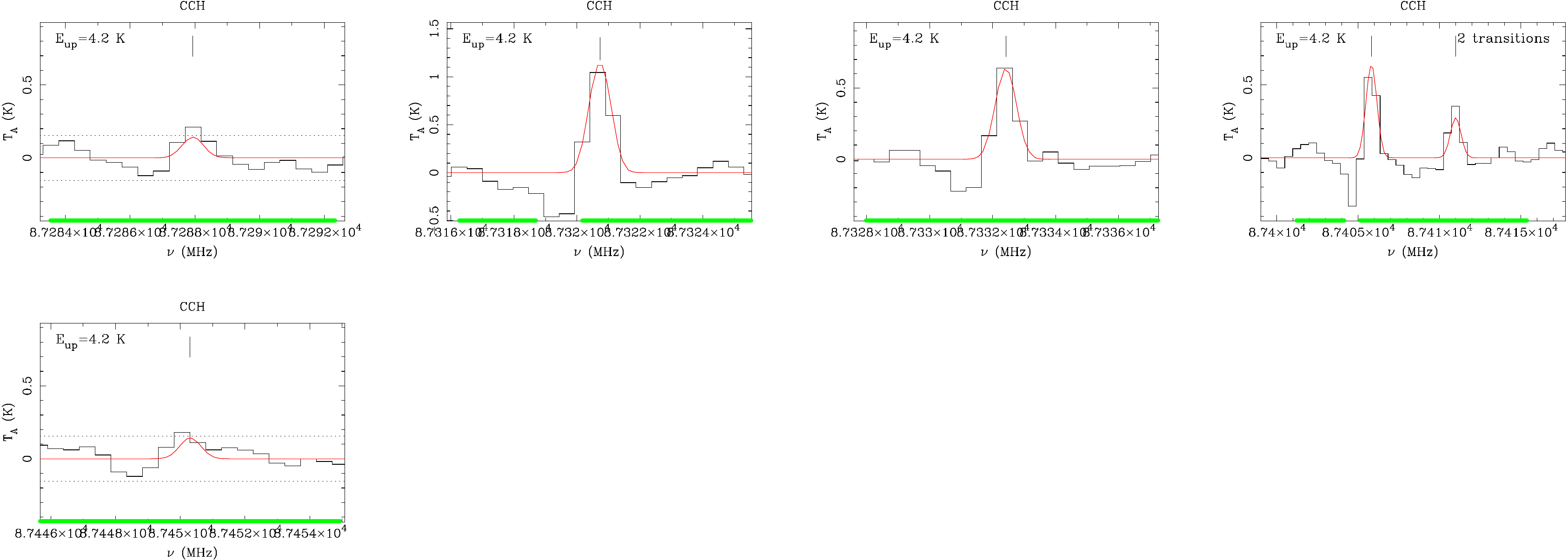}}
\caption{\textbf{\hspace{-0.2 em}(cont.)}\ Same as Figure
  \ref{fig-SpecDRa}. Panel (p) shows the CCH lines toward DR (a).}
\end{figure}

\begin{figure}
\subfloat[][]{\includegraphics[angle=-90,ext=.pdf,width= 1.00 \textwidth]{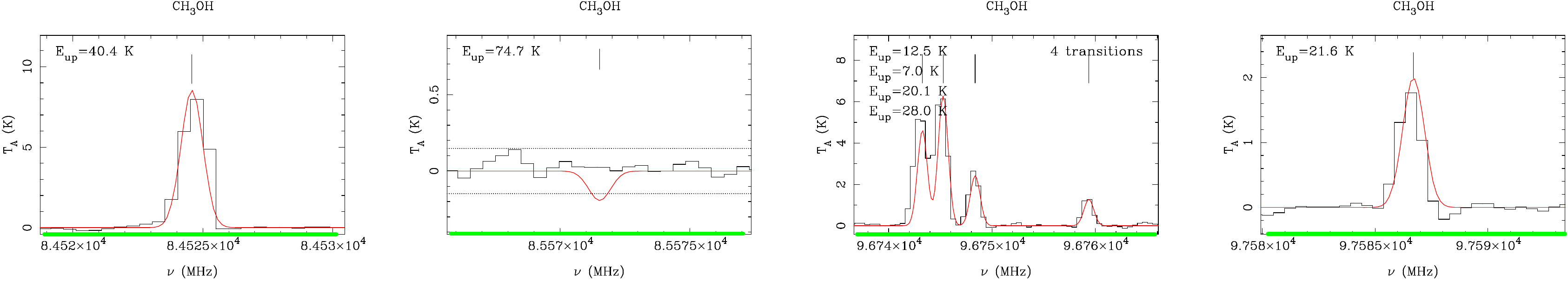}}\\
\subfloat[][]{\includegraphics[angle=-90,ext=.pdf,width= 0.50 \textwidth]{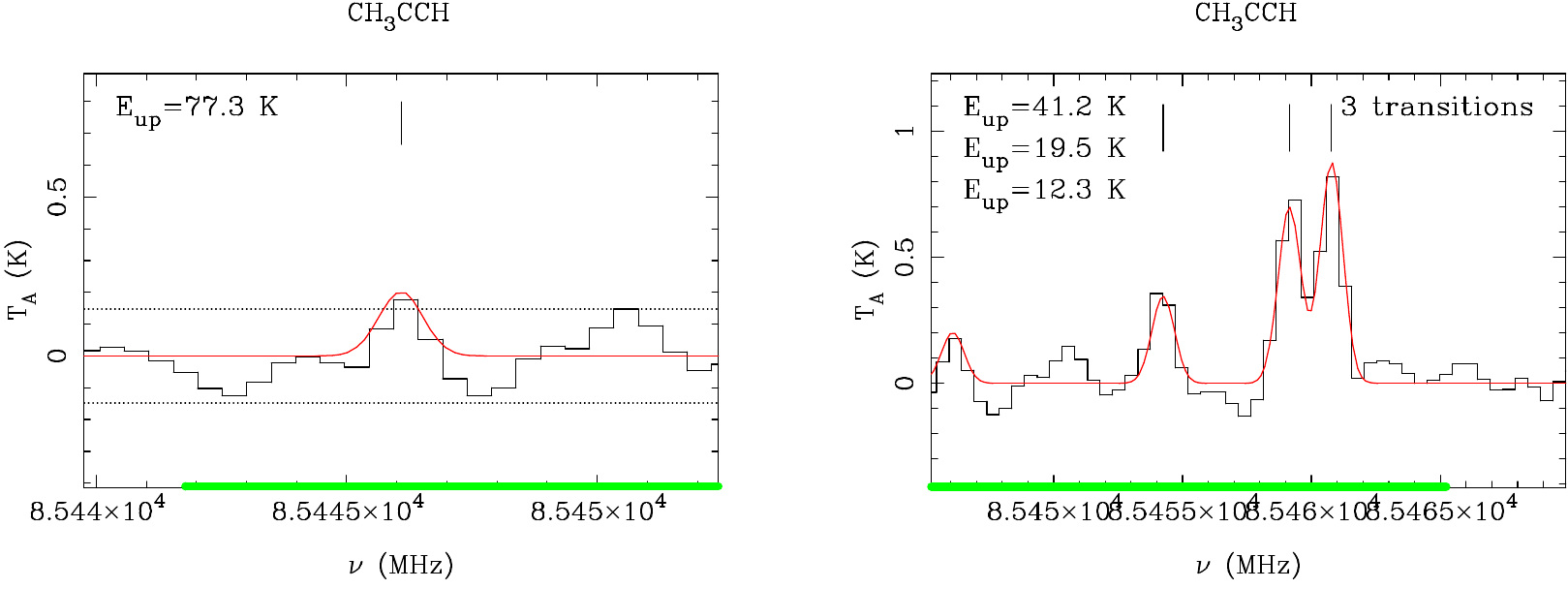}}
\subfloat[][]{\includegraphics[angle=-90,ext=.pdf,width= 0.25 \textwidth]{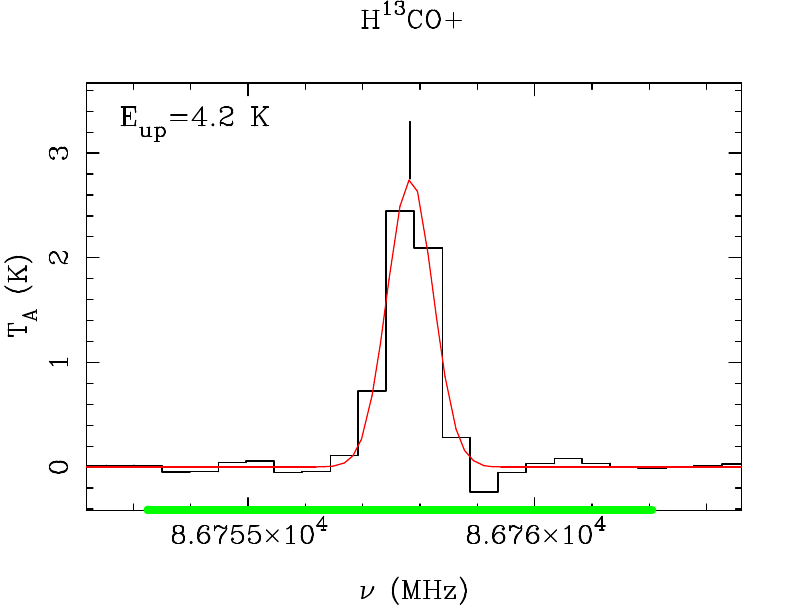}}
\subfloat[][]{\includegraphics[angle=-90,ext=.pdf,width= 0.25 \textwidth]{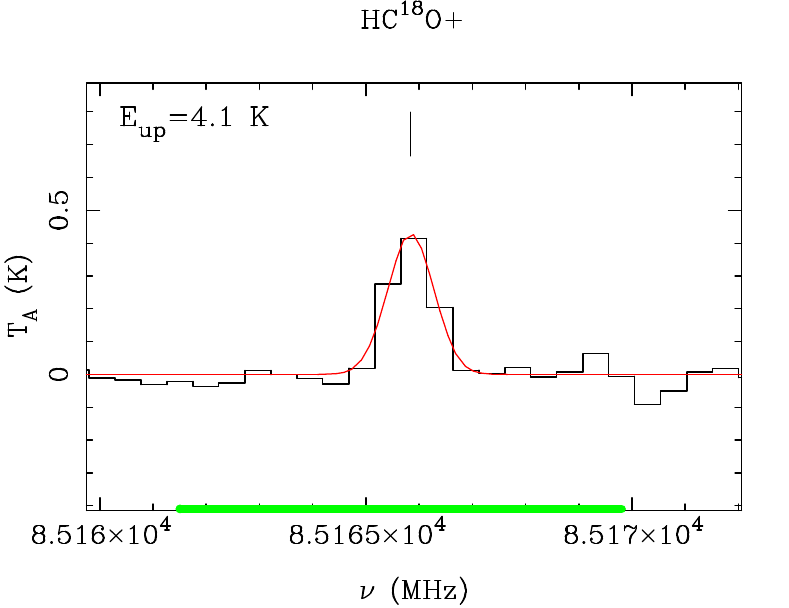}}\\
\subfloat[][]{\includegraphics[angle=-90,ext=.pdf,width= 0.50 \textwidth]{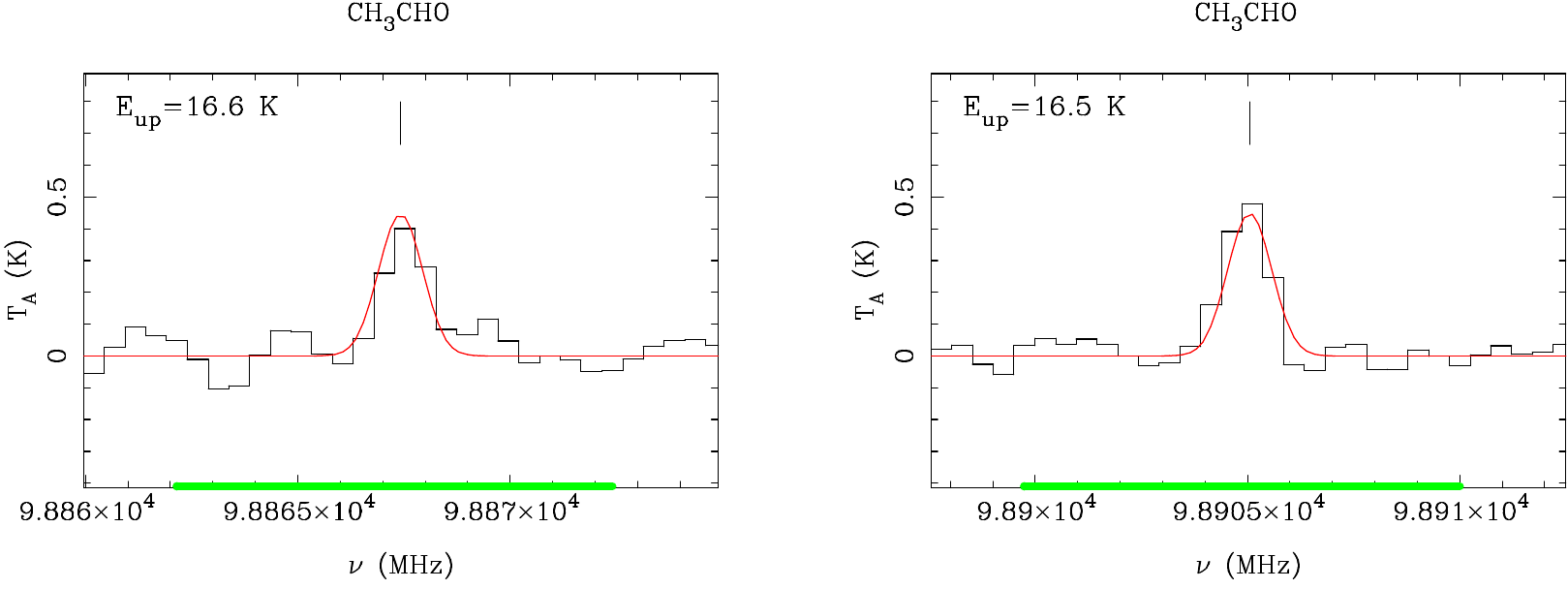}}
\subfloat[][]{\includegraphics[angle=-90,ext=.pdf,width= 0.25 \textwidth]{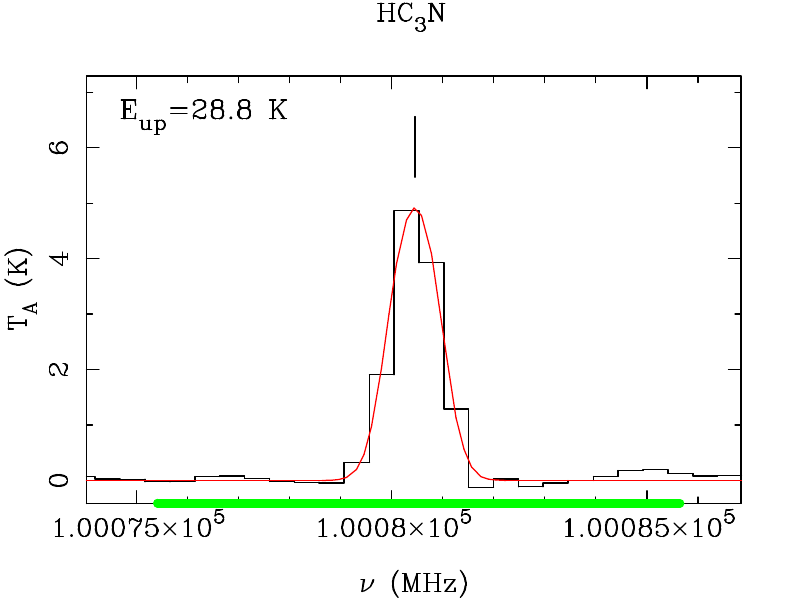}}
\subfloat[][]{\includegraphics[angle=-90,ext=.pdf,width= 0.25 \textwidth]{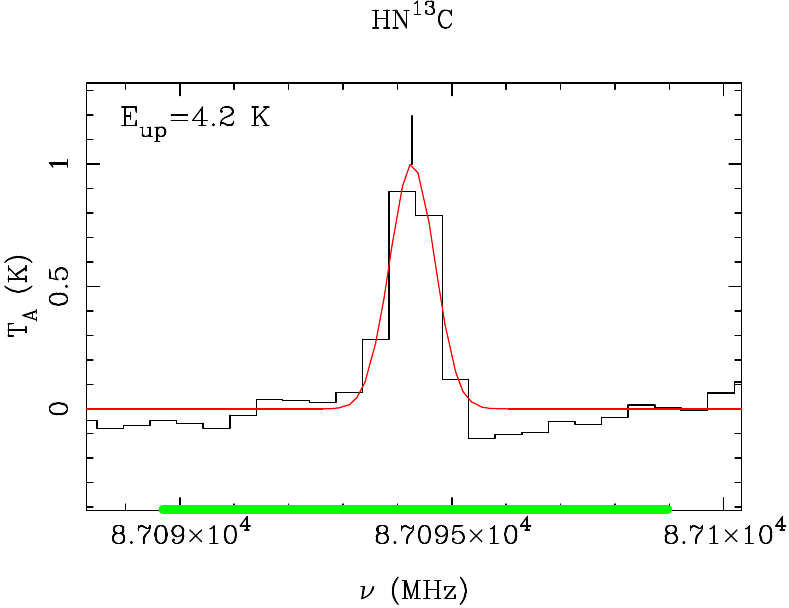}}
\caption{Emission lines from several molecules toward the point (b) in the Diffuse ridge (DR (b)). 
  Line types and colors as described in Figure
    \ref{fig-SpecCCcore1} with $\sigma=0.06$ K. Panels (a) to (g) show lines of \met, \propyne, \htcop, \hcdop, \acet, \hctn, and \hntc, respectively.\label{fig-SpecDRb}}
\end{figure}

\begin{figure}
\ContinuedFloat
\subfloat[][]{\includegraphics[angle=-90,ext=.pdf,width= 1.00 \textwidth]{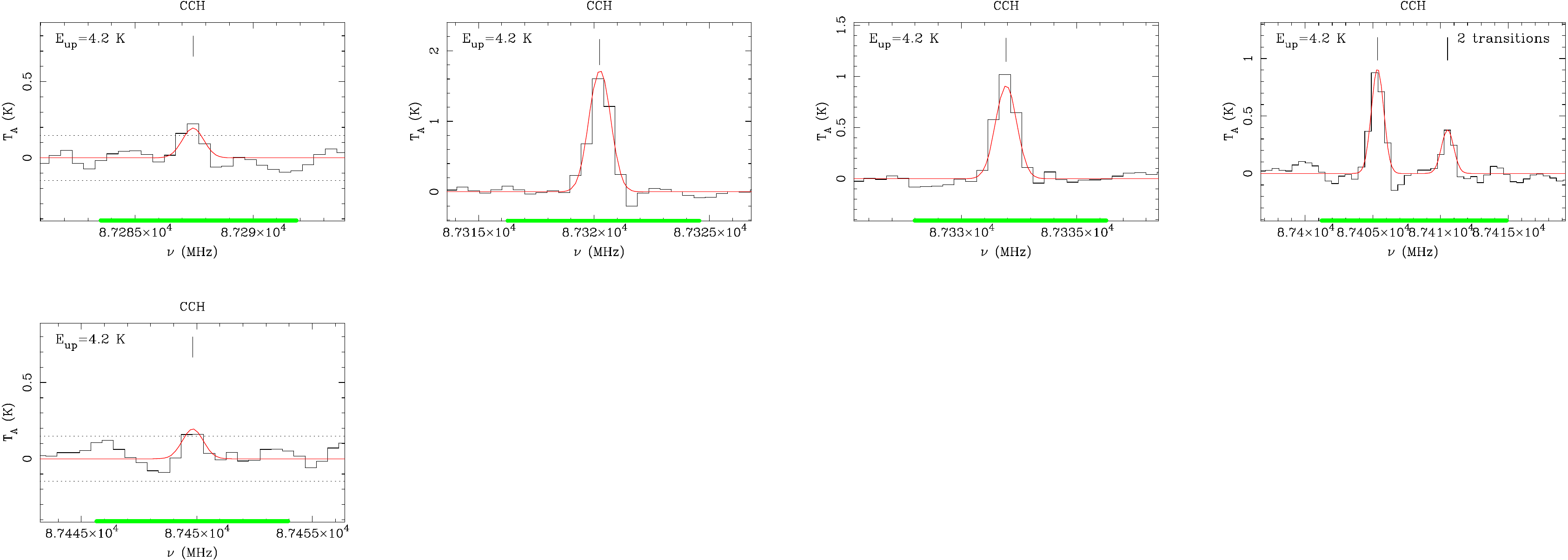}}\\
\subfloat[][]{\includegraphics[angle=-90,ext=.pdf,width= 0.25 \textwidth]{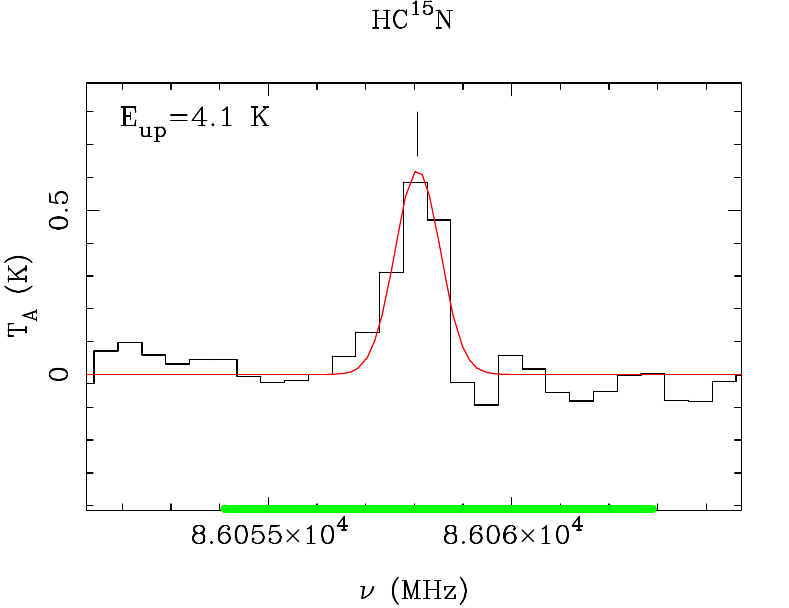}}
\subfloat[][]{\includegraphics[angle=-90,ext=.pdf,width= 0.25 \textwidth]{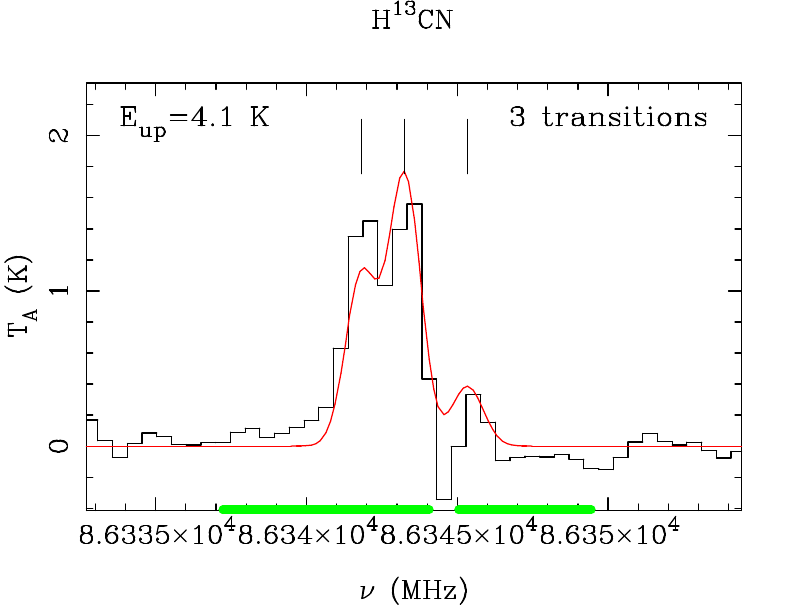}}
\subfloat[][]{\includegraphics[angle=-90,ext=.pdf,width= 0.25 \textwidth]{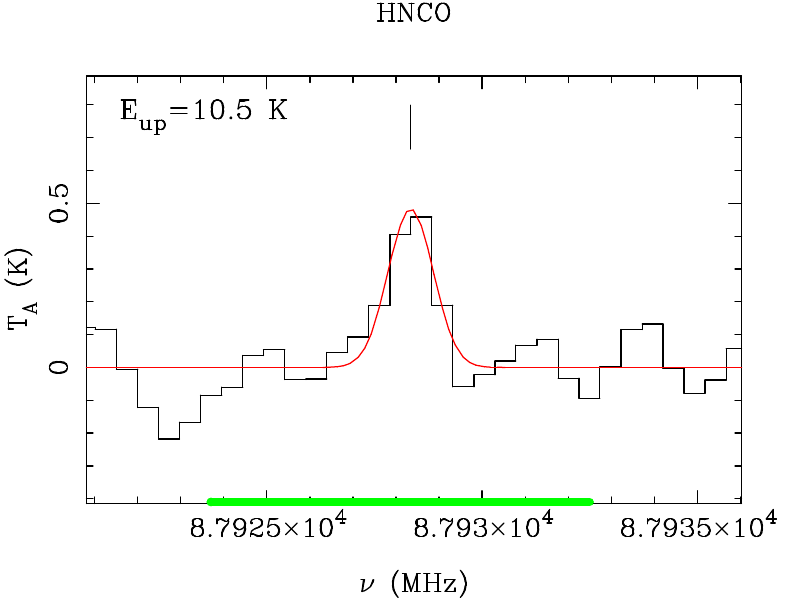}}
\subfloat[][]{\includegraphics[angle=-90,ext=.pdf,width= 0.25 \textwidth]{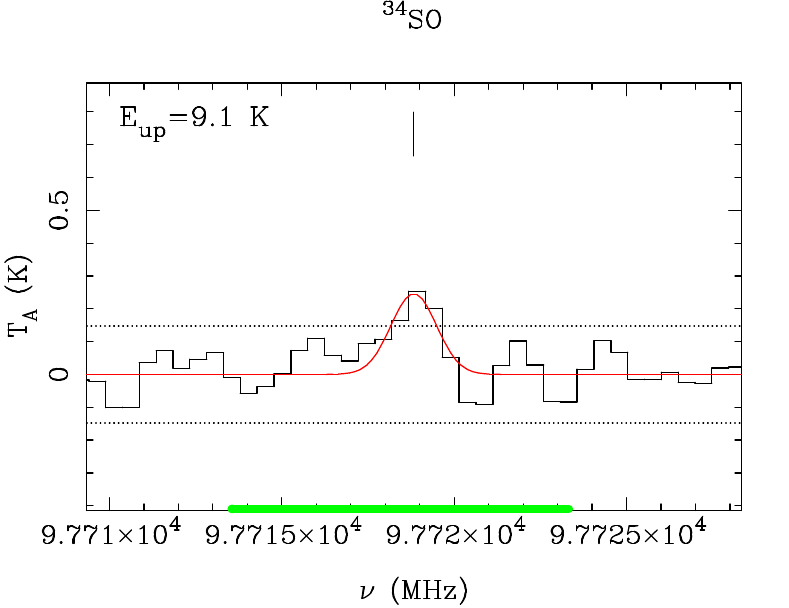}}\\
\subfloat[][]{\includegraphics[angle=-90,ext=.pdf,width= 0.50 \textwidth]{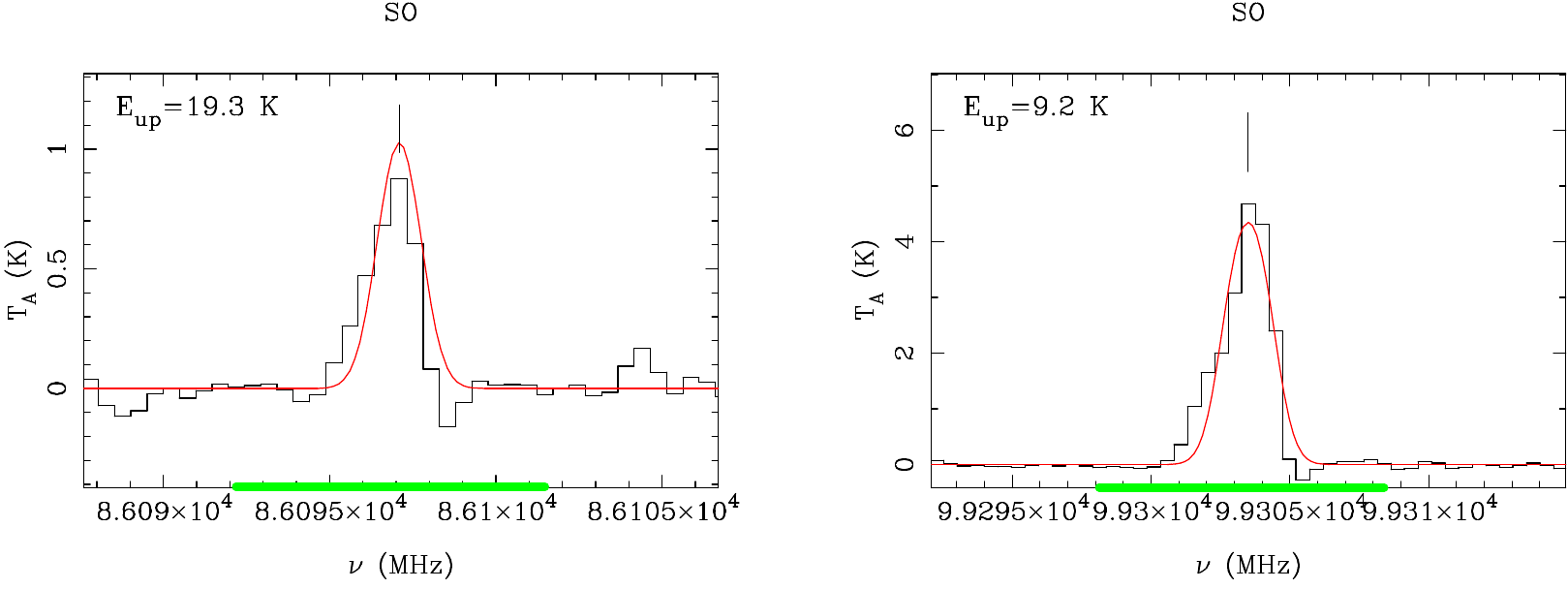}}
\subfloat[][]{\includegraphics[angle=-90,ext=.pdf,width= 0.25 \textwidth]{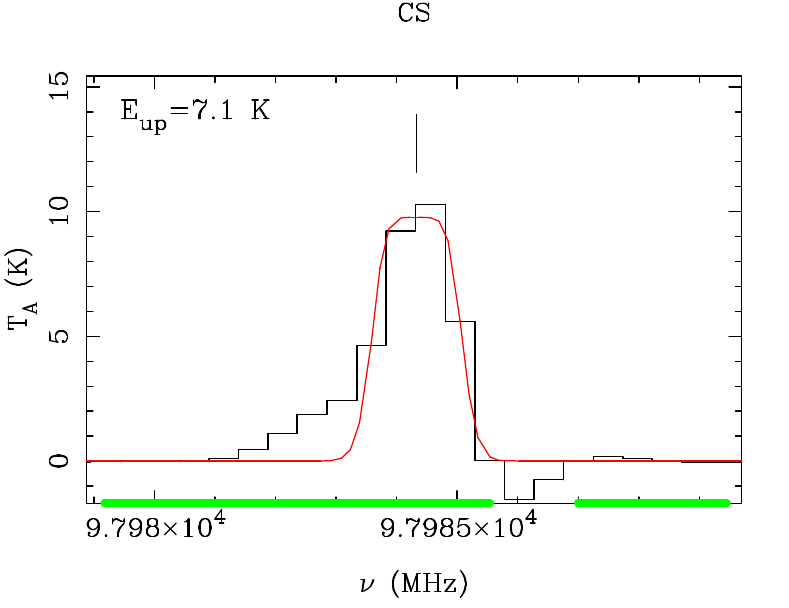}}
\subfloat[][]{\includegraphics[angle=-90,ext=.pdf,width= 0.25 \textwidth]{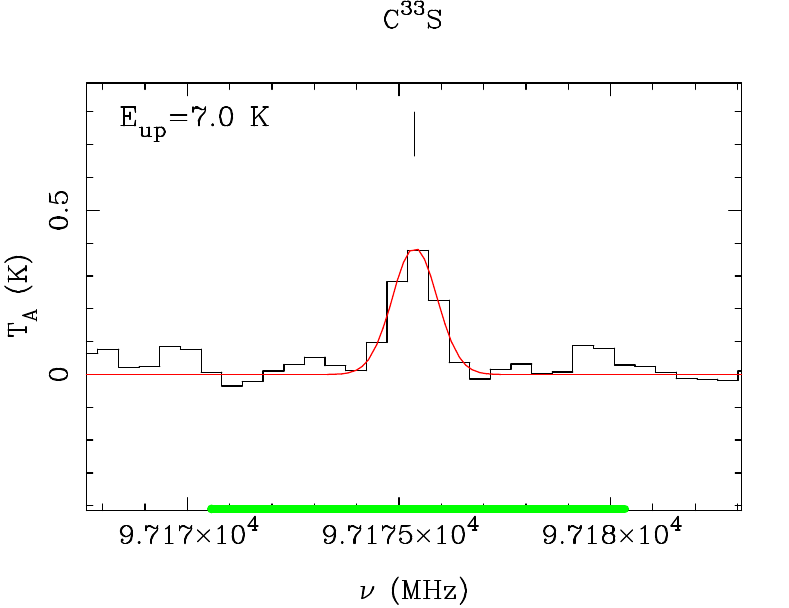}}\\
\subfloat[][]{\includegraphics[angle=-90,ext=.pdf,width= 0.25 \textwidth]{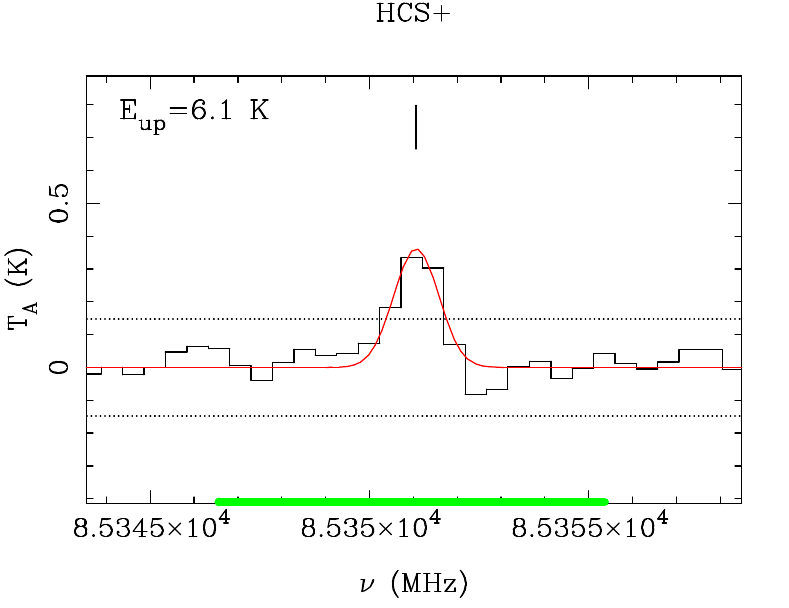}}
\subfloat[][]{\includegraphics[angle=-90,ext=.pdf,width= 0.25 \textwidth]{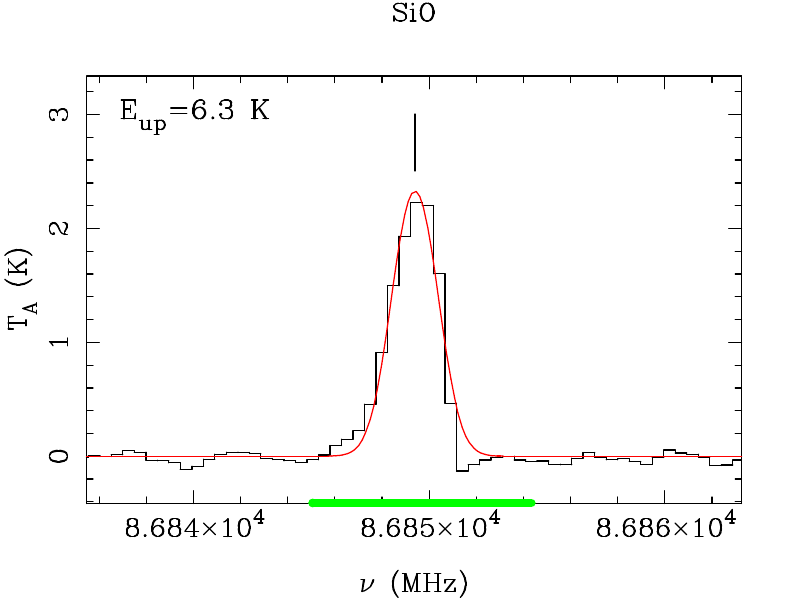}}
\caption{\textbf{\hspace{-0.2 em}(cont.)}\ Same as Figure \ref{fig-SpecDRb}. Panels (h) to (q) show lines of CCH, \hcqn, \htcn, HNCO, $^{34}$SO, SO, CS, C$^{33}$S, \hcsp, and SiO, respectively.}
\end{figure}

\begin{figure}
\subfloat[][]{\includegraphics[angle=-90,ext=.pdf,width= 0.24 \textwidth]{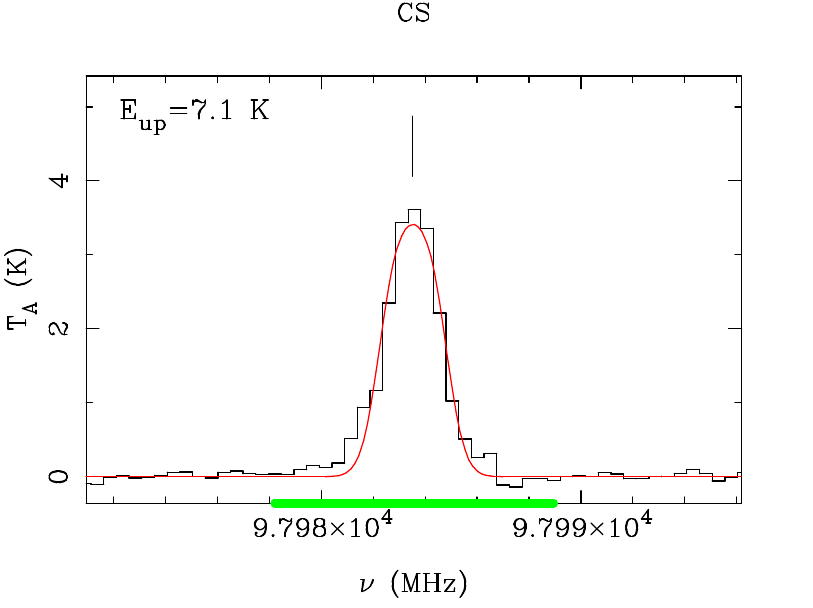}}
\subfloat[][]{\includegraphics[angle=-90,ext=.pdf,width= 0.24 \textwidth]{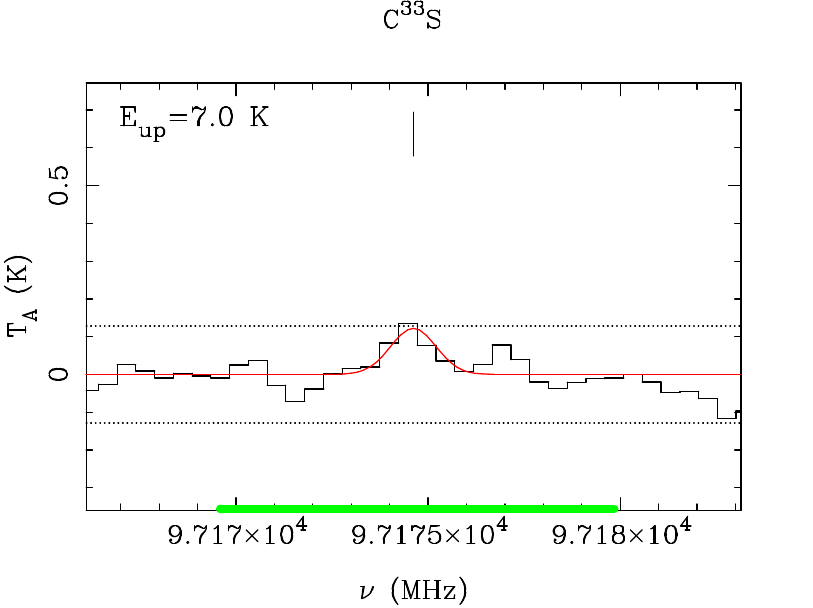}}
\subfloat[][]{\includegraphics[angle=-90,ext=.pdf,width= 0.24 \textwidth]{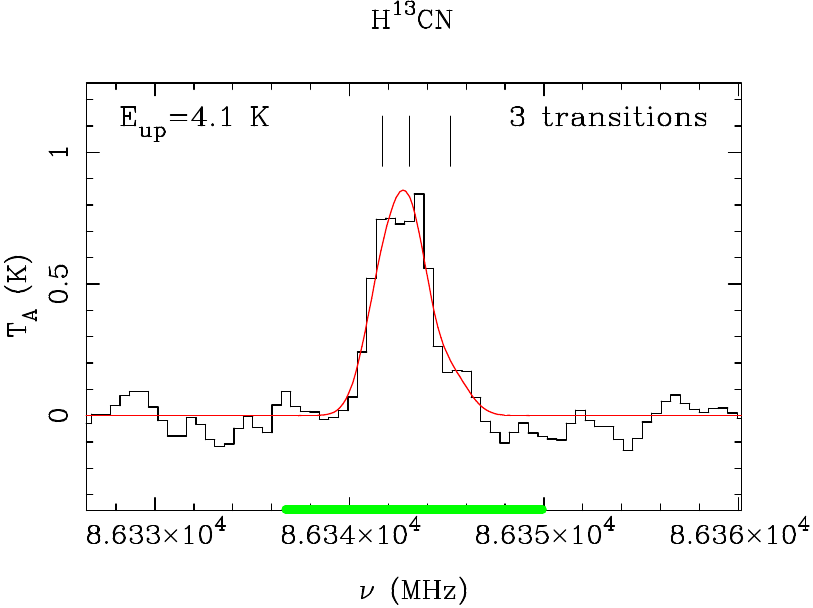}}
\subfloat[][]{\includegraphics[angle=-90,ext=.pdf,width= 0.24 \textwidth]{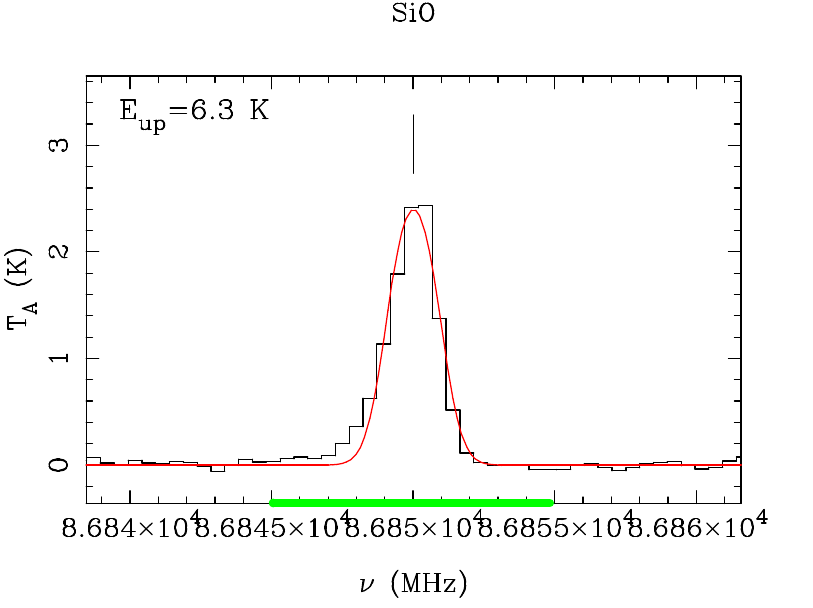}}\\
\subfloat[][]{\includegraphics[angle=-90,ext=.pdf,width= 0.24 \textwidth]{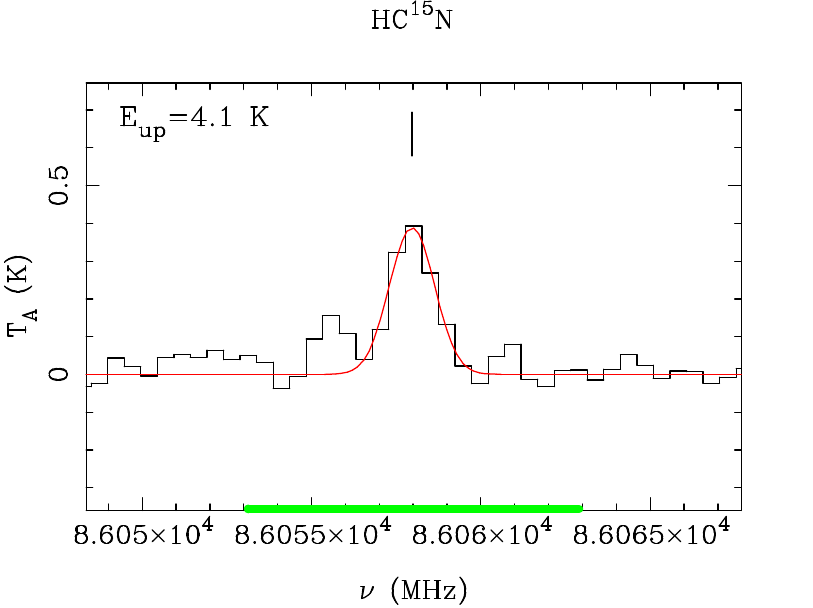}}
\subfloat[][]{\includegraphics[angle=-90,ext=.pdf,width= 0.47 \textwidth]{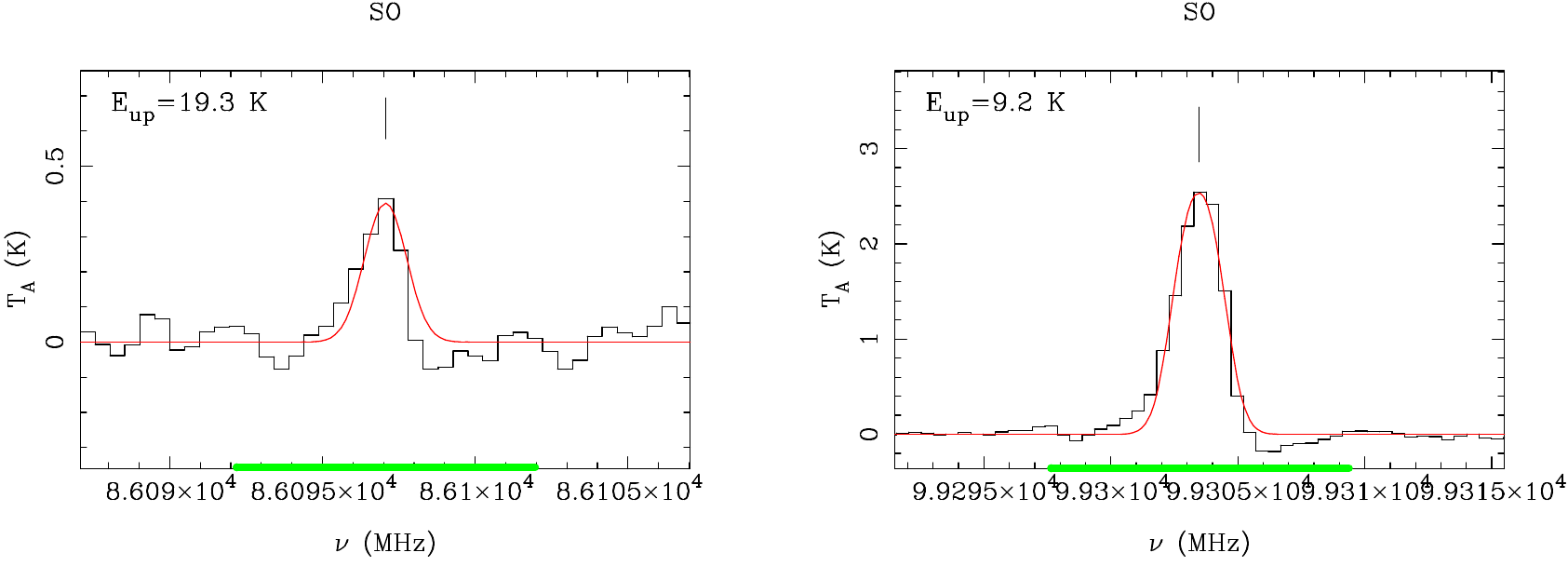}}
\subfloat[][]{\includegraphics[angle=-90,ext=.pdf,width= 0.24 \textwidth]{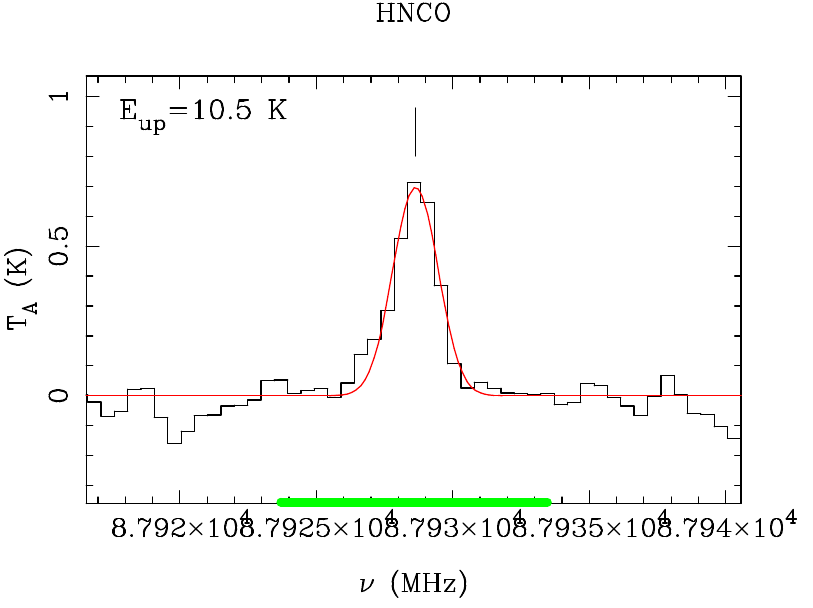}}\\
\subfloat[][]{\includegraphics[angle=-90,ext=.pdf,width= 0.47 \textwidth]{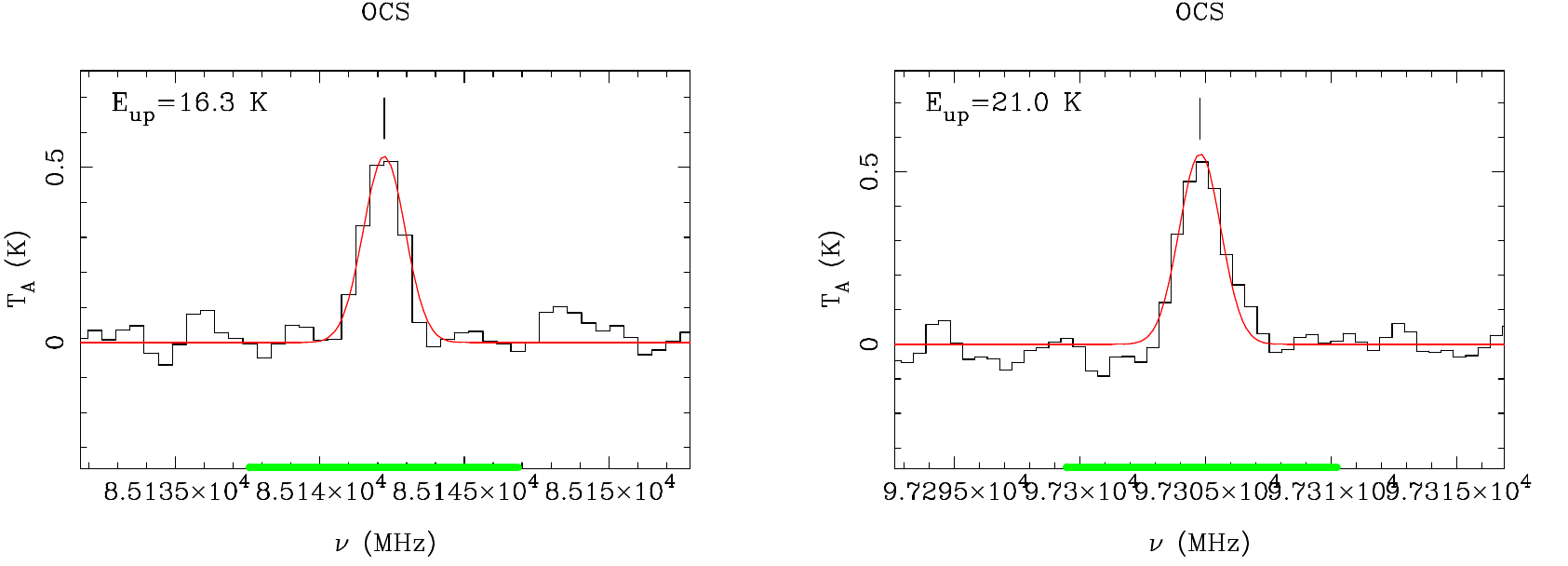}}
\subfloat[][]{\includegraphics[angle=-90,ext=.pdf,width= 0.24 \textwidth]{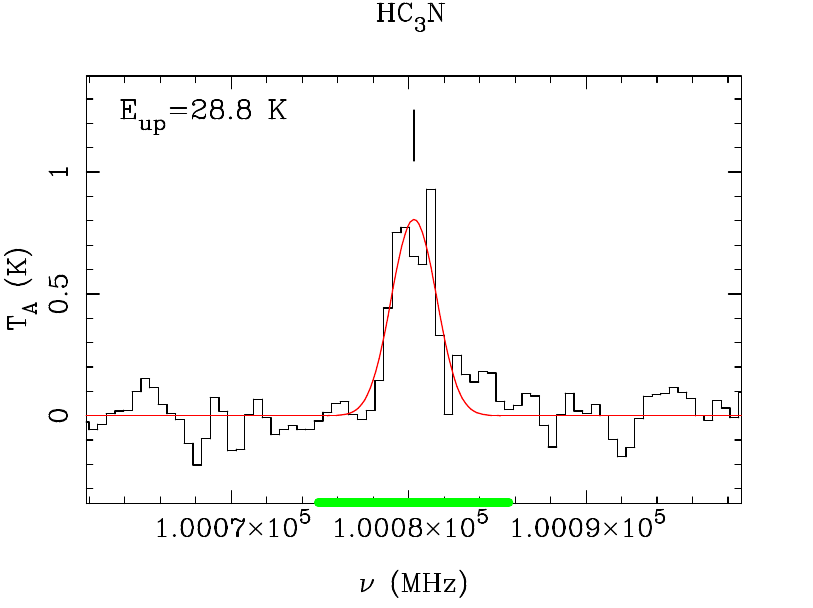}}
\subfloat[][]{\includegraphics[angle=-90,ext=.pdf,width= 0.24 \textwidth]{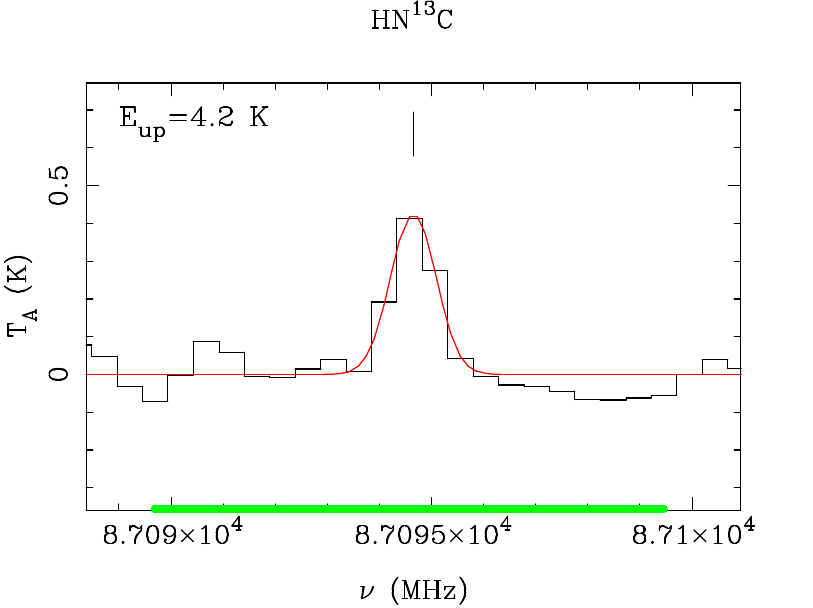}}\\
\subfloat[][]{\includegraphics[angle=-90,ext=.pdf,width= 0.95 \textwidth]{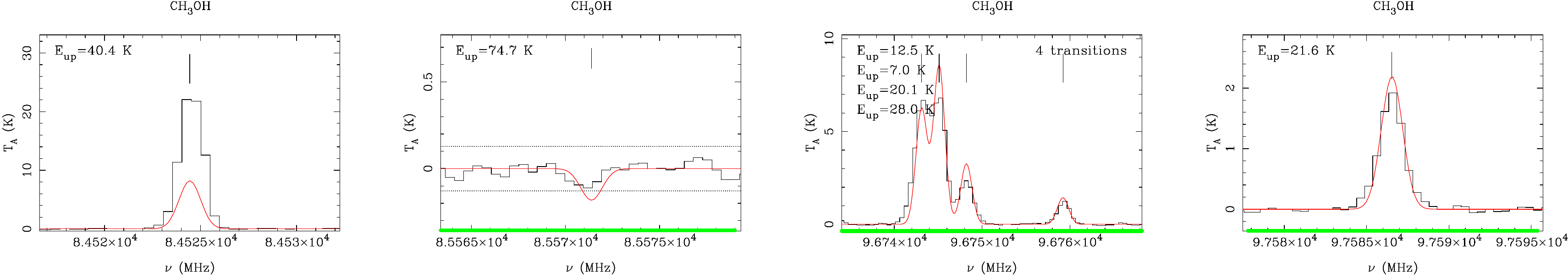}}\\
\subfloat[][]{\includegraphics[angle=-90,ext=.pdf,width= 0.47 \textwidth]{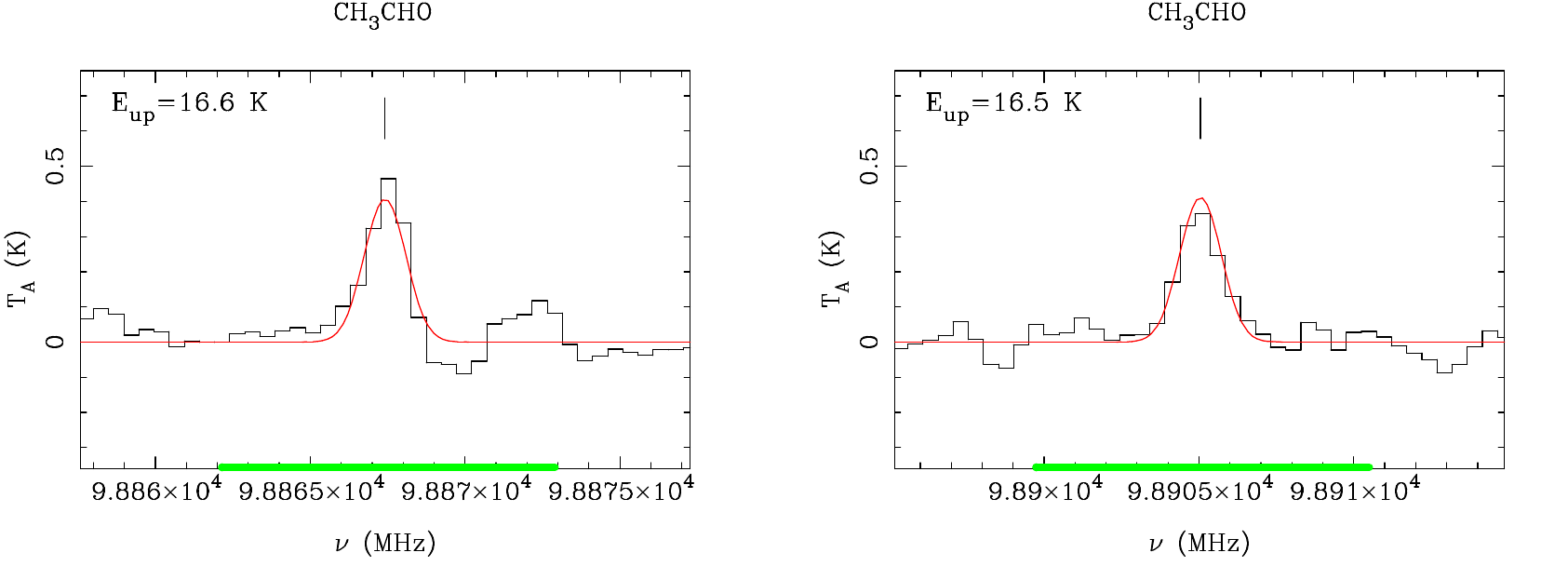}}
\caption{Emission lines from several molecules toward the point (c) in the Diffuse ridge (DR (c)). 
  Line types and colors as described in Figure
    \ref{fig-SpecCCcore1} with $\sigma=0.05$ K. Panels (a) to (l) show lines of CS, C$^{33}$S, \htcn, SiO, \hcqn, SO, HNCO, OCS, \hctn, 
\hntc, \met, and  \acet, respectively. \added{We note that the strong non-LTE \met, \maser\ emission shown in (k) is not well reproduced with the models used in this work.} \label{fig-SpecDRc}}
\end{figure}
 \clearpage
\begin{figure}
\subfloat[][]{\includegraphics[angle=-90,ext=.pdf,width= 0.25 \textwidth]{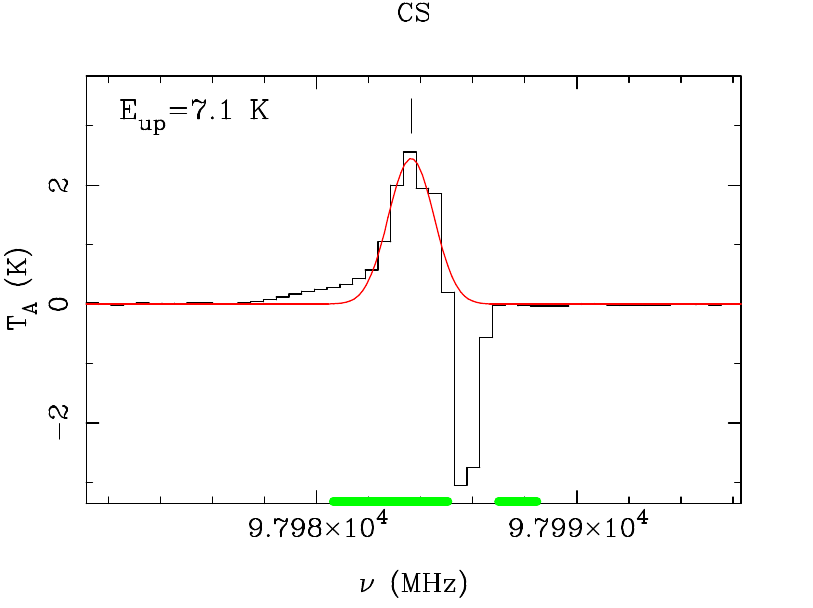}}
\subfloat[][]{\includegraphics[angle=-90,ext=.pdf,width= 0.25 \textwidth]{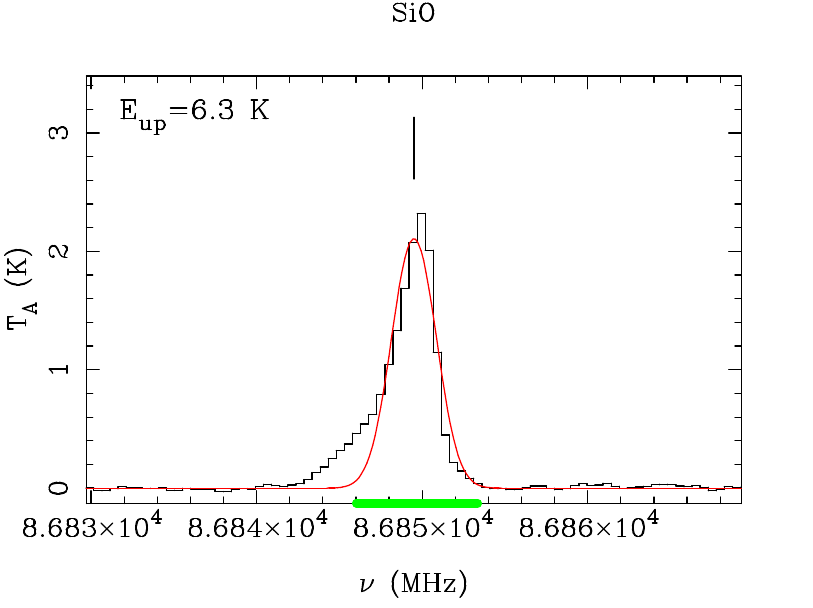}}
\subfloat[][]{\includegraphics[angle=-90,ext=.pdf,width= 0.25 \textwidth]{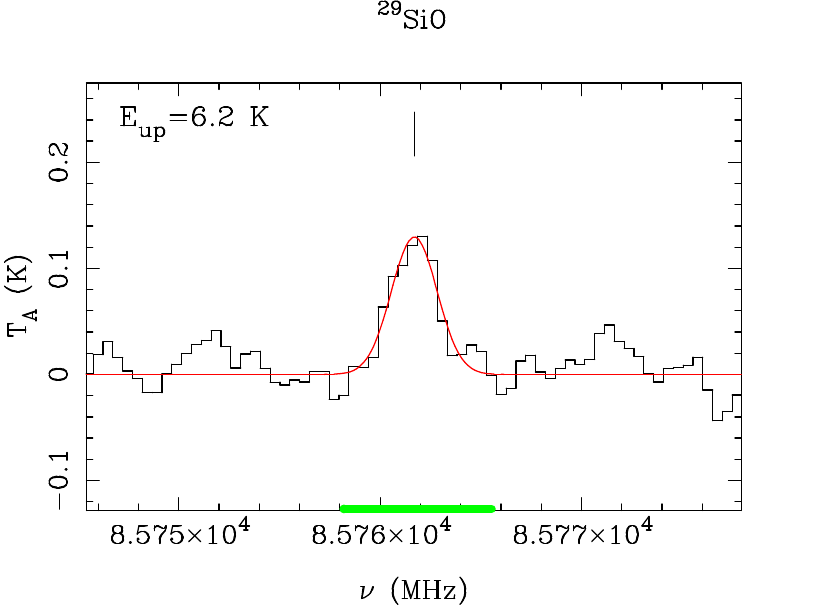}}
\subfloat[][]{\includegraphics[angle=-90,ext=.pdf,width= 0.25 \textwidth]{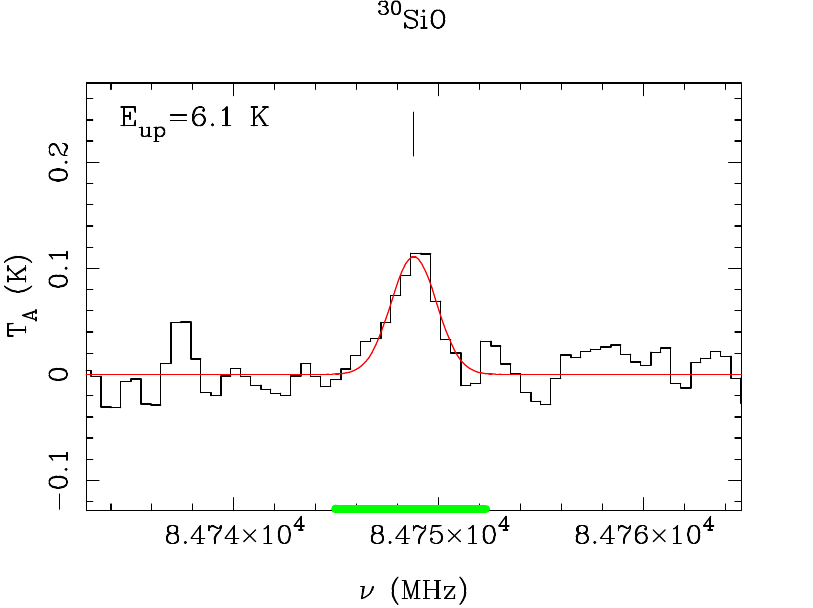}}\\
\subfloat[][]{\includegraphics[angle=-90,ext=.pdf,width= 0.25 \textwidth]{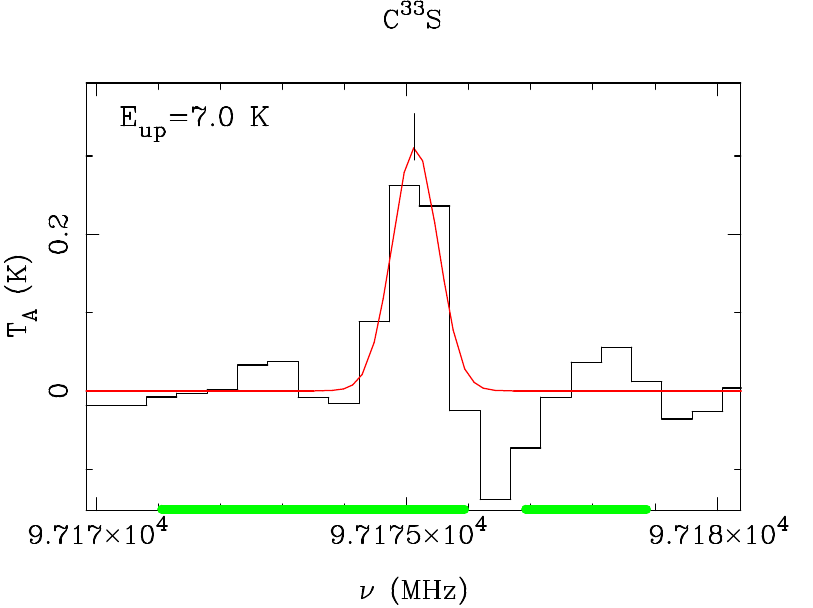}}
\subfloat[][]{\includegraphics[angle=-90,ext=.pdf,width= 0.75 \textwidth]{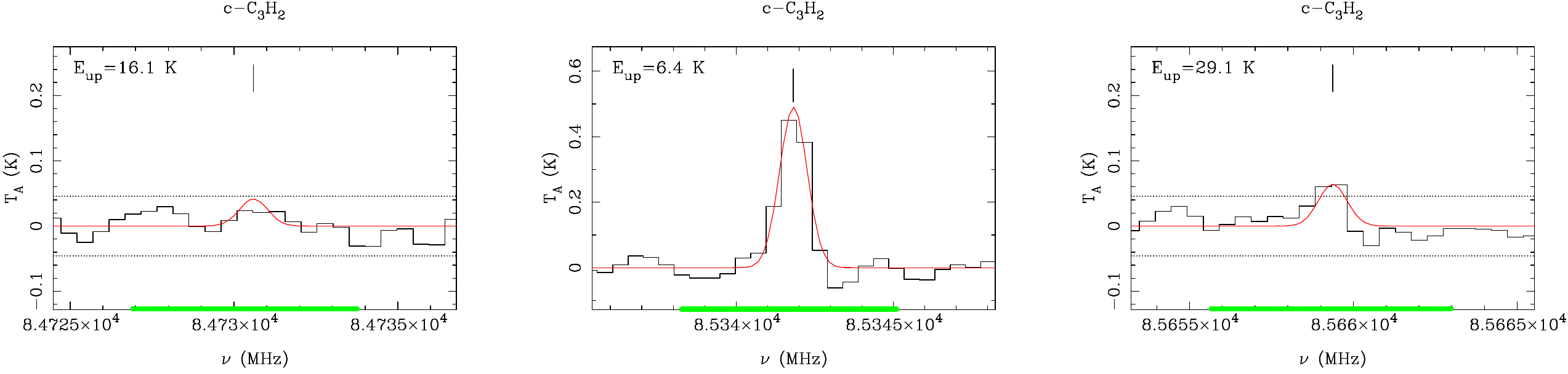}}\\
\subfloat[][]{\includegraphics[angle=-90,ext=.pdf,width= 1.00 \textwidth]{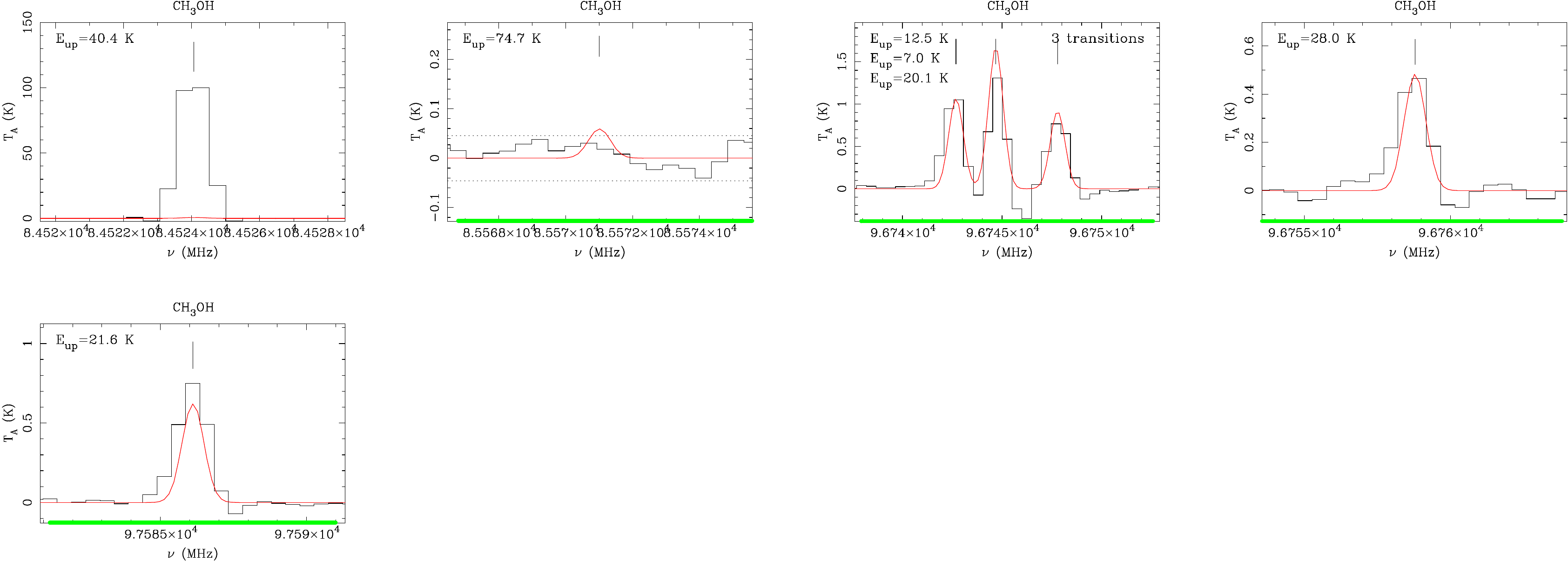}}\\
\subfloat[][]{\includegraphics[angle=-90,ext=.pdf,width= 0.25 \textwidth]{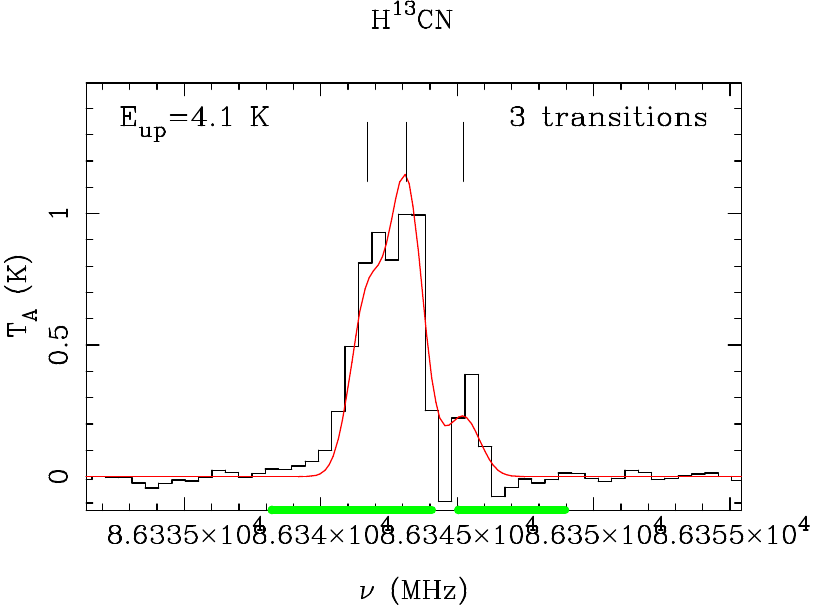}}
\subfloat[][]{\includegraphics[angle=-90,ext=.pdf,width= 0.25 \textwidth]{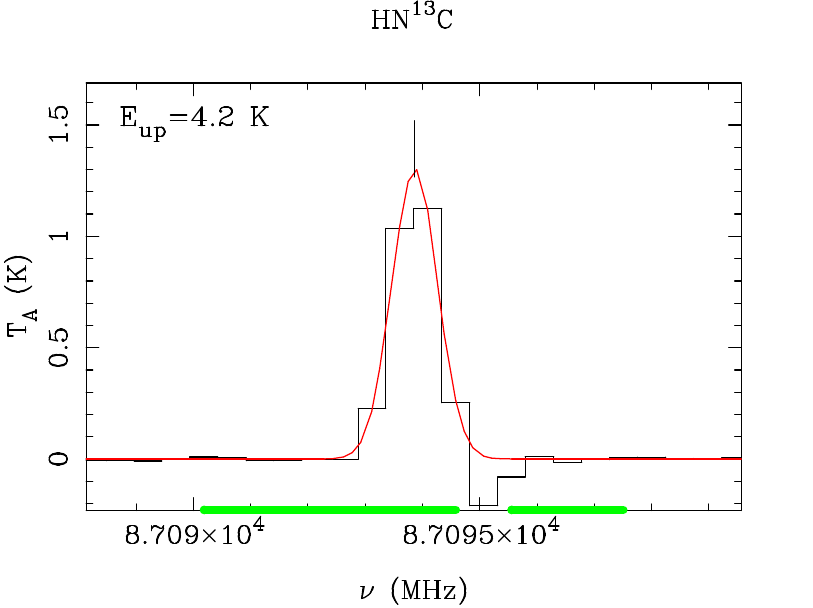}}
\subfloat[][]{\includegraphics[angle=-90,ext=.pdf,width= 0.25 \textwidth]{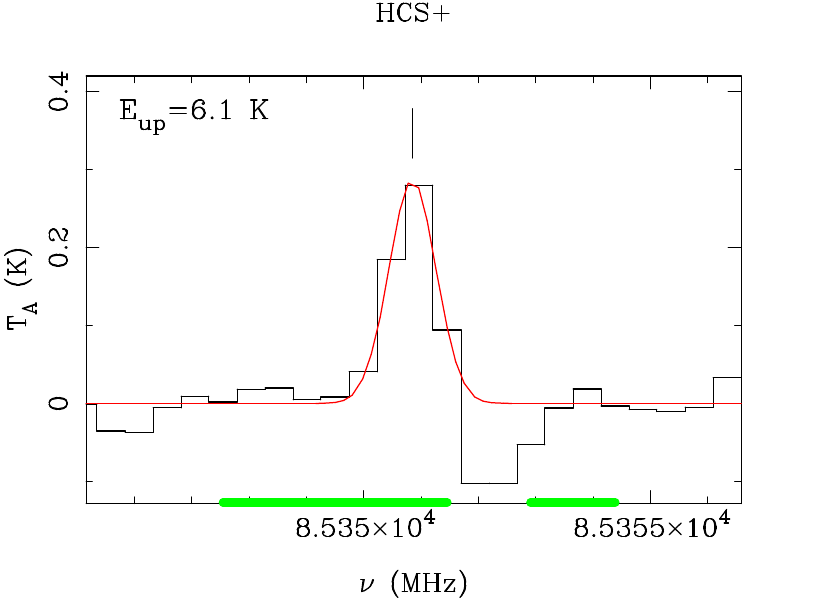}}
\subfloat[][]{\includegraphics[angle=-90,ext=.pdf,width= 0.25 \textwidth]{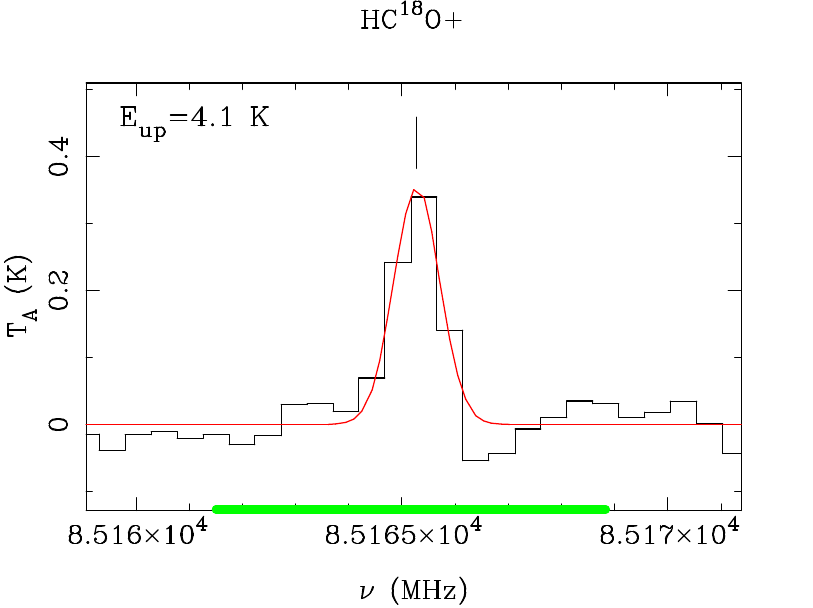}}
\caption{Emission lines from several molecules toward the point S8E maser. 
  Line types and colors as described in Figure
    \ref{fig-SpecCCcore1} with $\sigma=0.02$ K. Panels (a) to (k) show lines of CS, SiO, $^{29}$SiO, $^{30}$SiO, C$^{33}$S, \cyc, \met, \htcn, \hntc, \hcsp, and \hcdop, respectively. \added{The \met, \maser\ maser transition shown in (g) cannot be reproduced with the models used in this work.} \label{fig-SpecS8Em}}
\end{figure}
 \begin{figure}
\ContinuedFloat
\subfloat[][]{\includegraphics[angle=-90,ext=.pdf,width= 1.00 \textwidth]{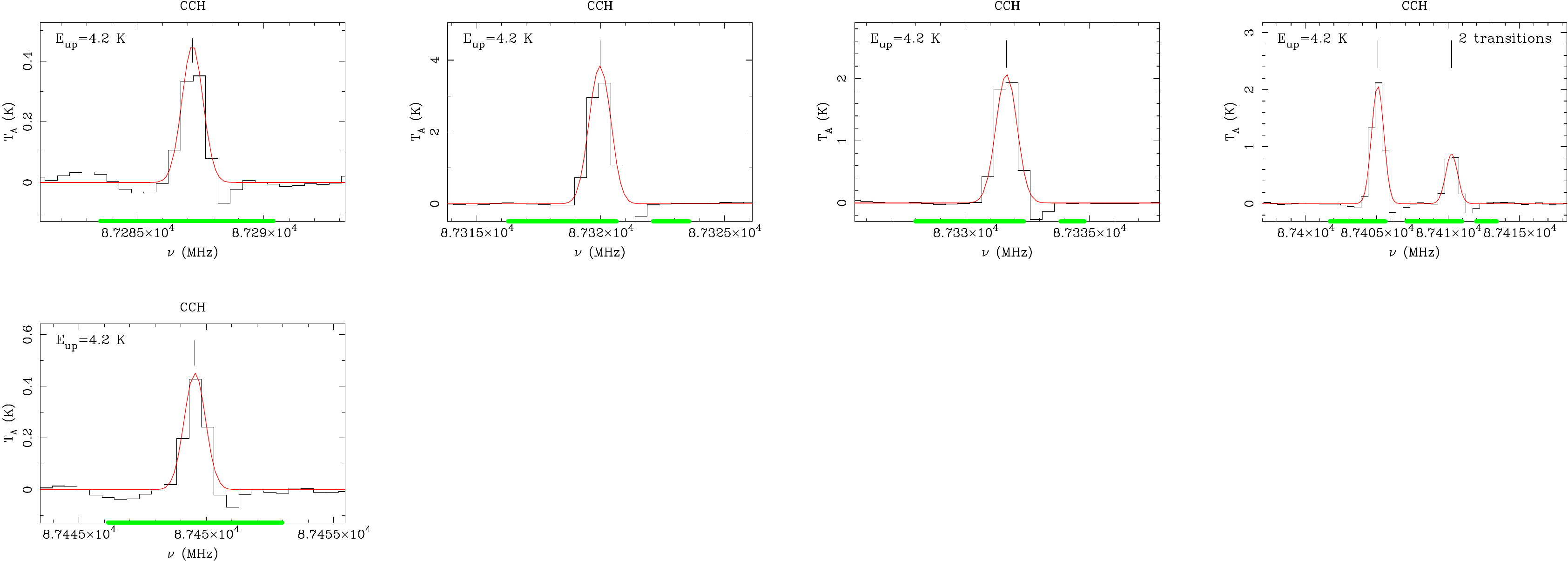}}\\
\subfloat[][]{\includegraphics[angle=-90,ext=.pdf,width= 0.75 \textwidth]{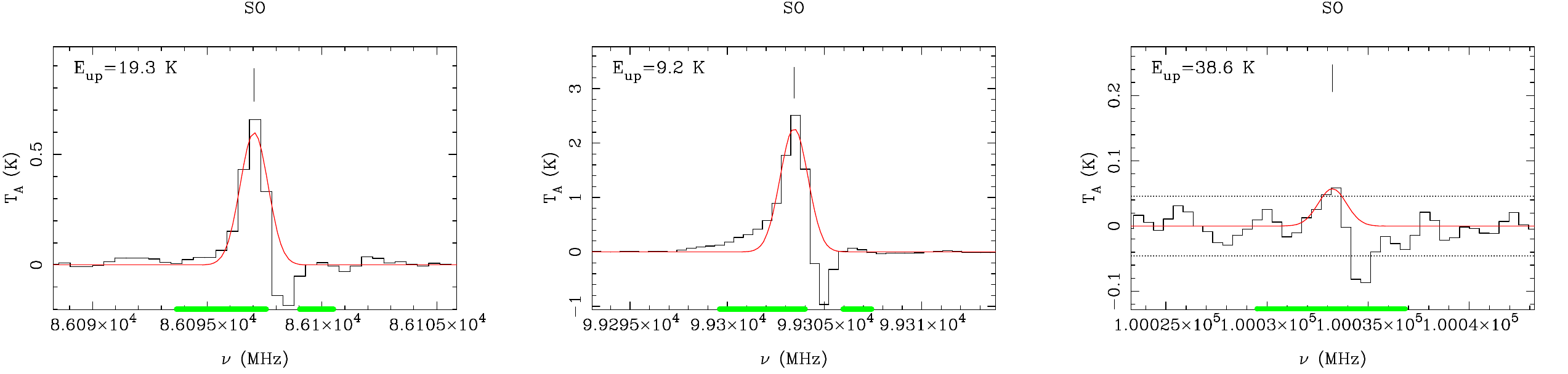}}
\subfloat[][]{\includegraphics[angle=-90,ext=.pdf,width= 0.25 \textwidth]{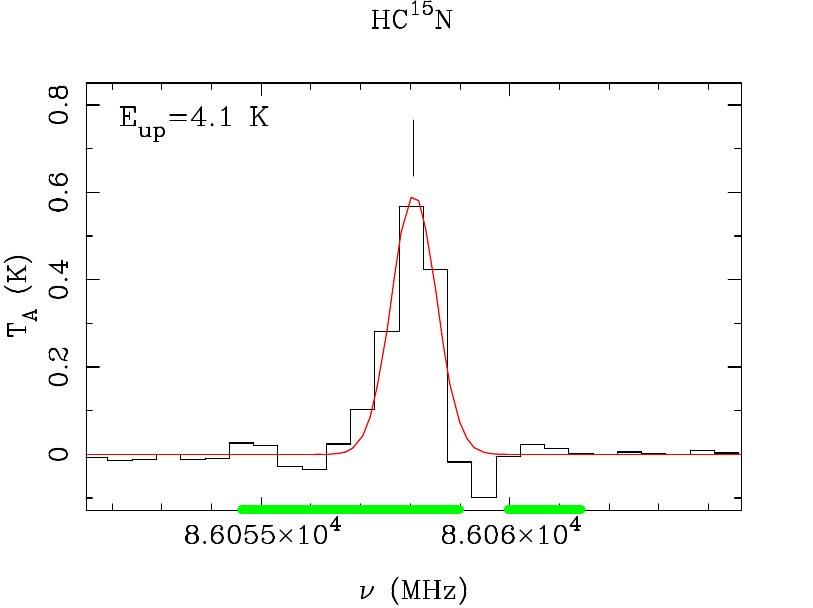}}\\
\subfloat[][]{\includegraphics[angle=-90,ext=.pdf,width= 0.50 \textwidth]{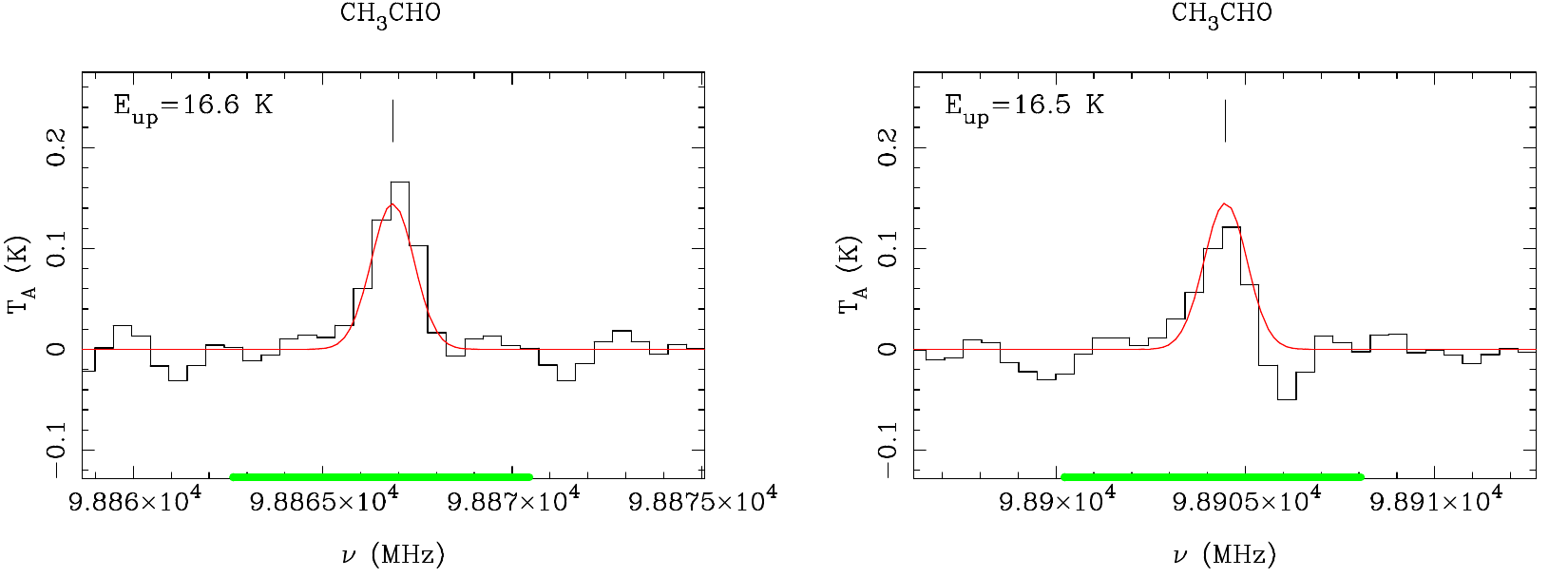}}
\subfloat[][]{\includegraphics[angle=-90,ext=.pdf,width= 0.50 \textwidth]{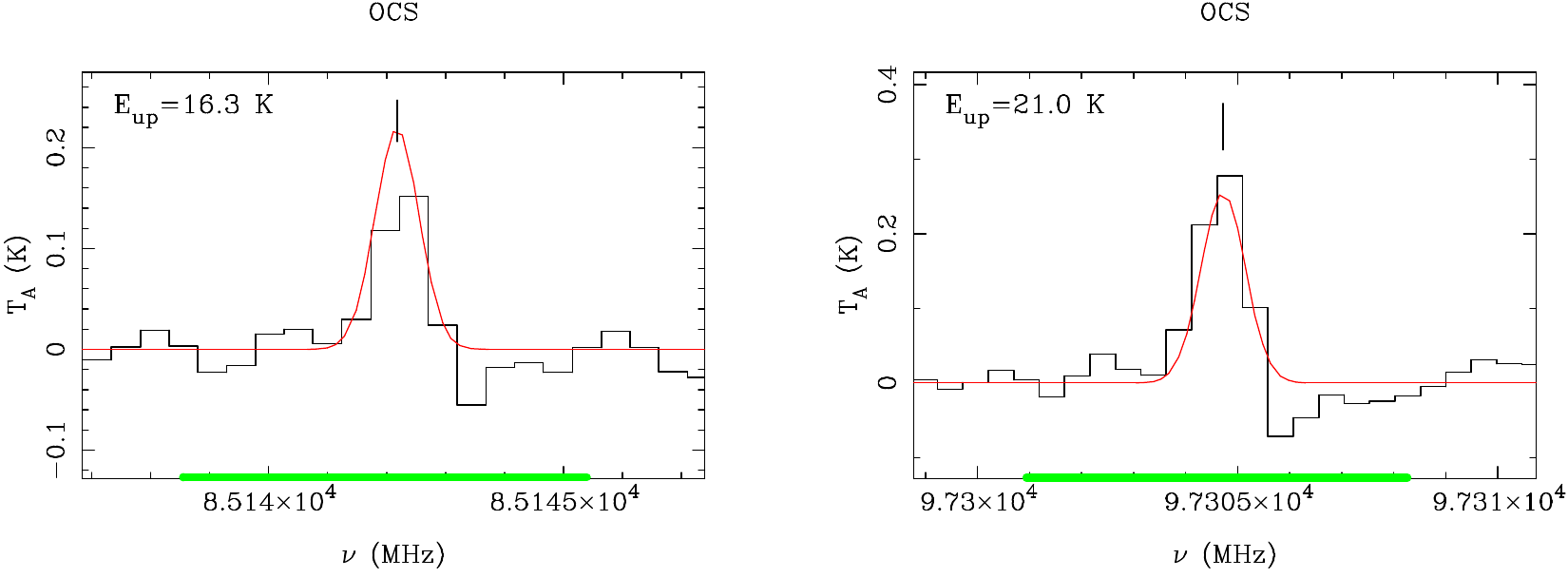}}\\
\subfloat[][]{\includegraphics[angle=-90,ext=.pdf,width= 0.75 \textwidth]{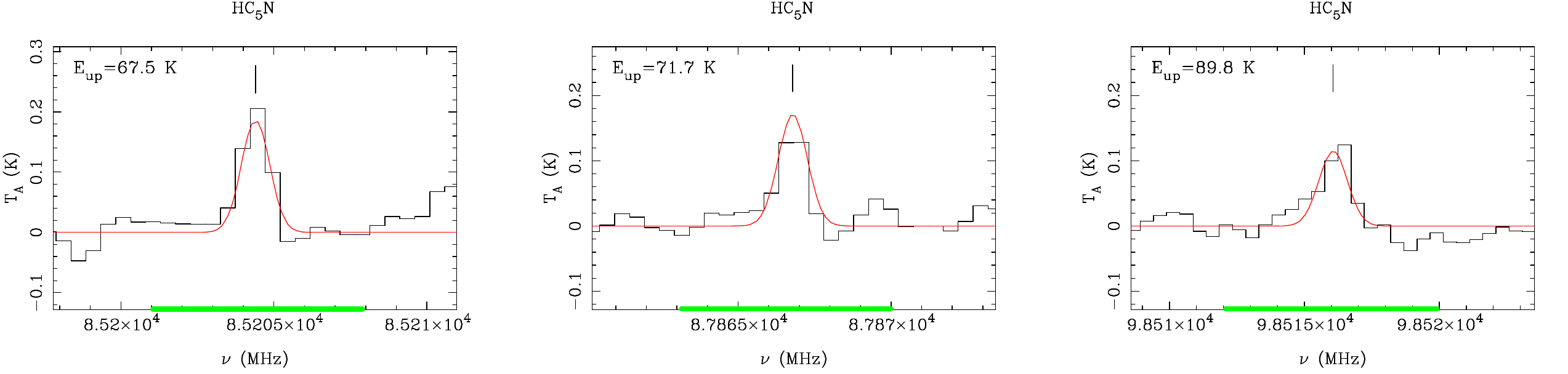}}
\subfloat[][]{\includegraphics[angle=-90,ext=.pdf,width= 0.25 \textwidth]{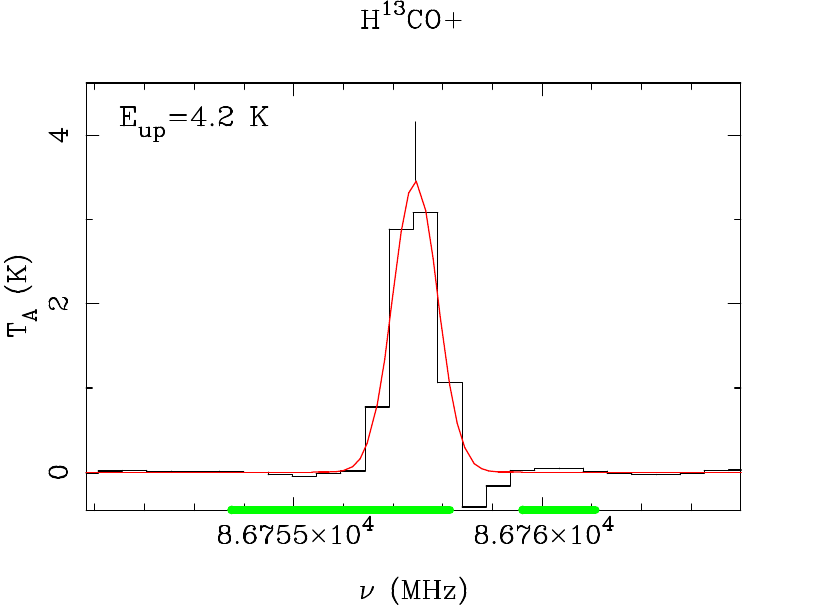}}
\caption{\textbf{\hspace{-0.2 em}(cont.)}\ Same as Figure \ref{fig-SpecS8Em}. Panels (l) to (r) show lines of CCH, SO, \hcqn,
\met, OCS, \hccn, and \htcop, respectively.}
\end{figure}
\begin{figure}
\ContinuedFloat
\subfloat[][]{\includegraphics[angle=-90,ext=.pdf,width= 0.25 \textwidth]{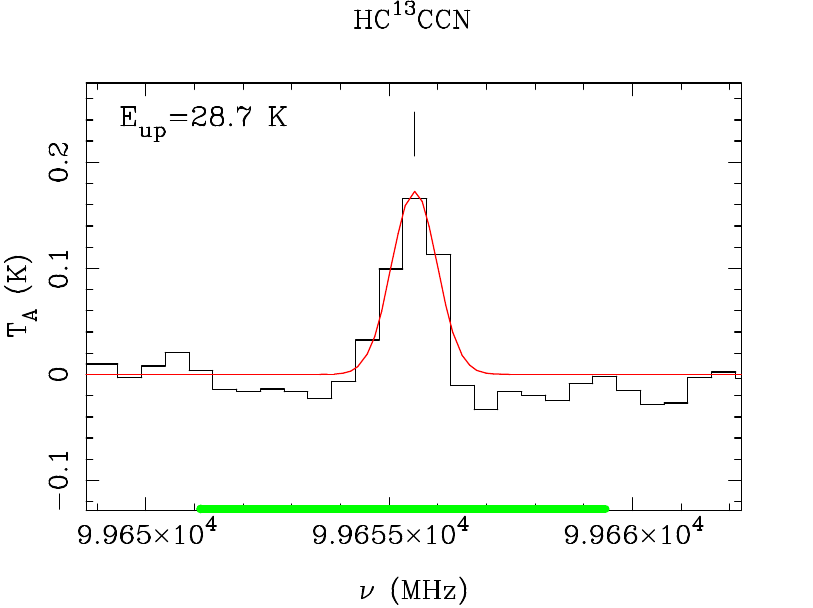}}
\subfloat[][]{\includegraphics[angle=-90,ext=.pdf,width= 0.75 \textwidth]{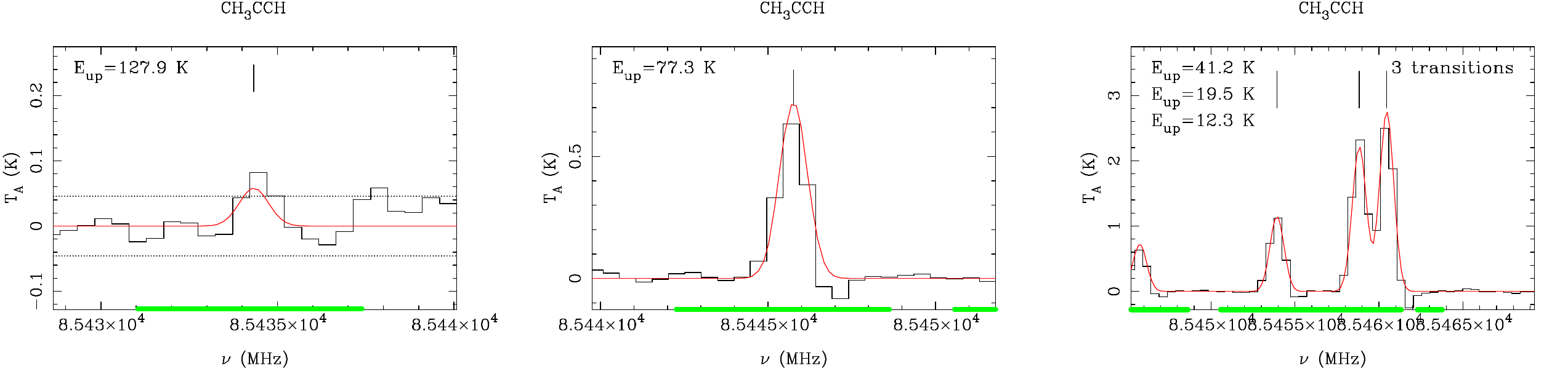}}\\
\subfloat[][]{\includegraphics[angle=-90,ext=.pdf,width= 0.25 \textwidth]{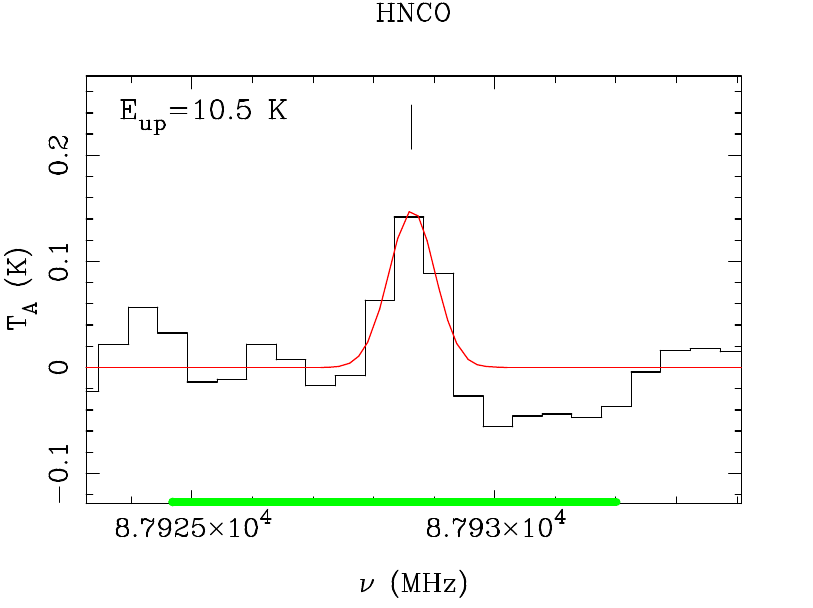}}
\subfloat[][]{\includegraphics[angle=-90,ext=.pdf,width= 0.25 \textwidth]{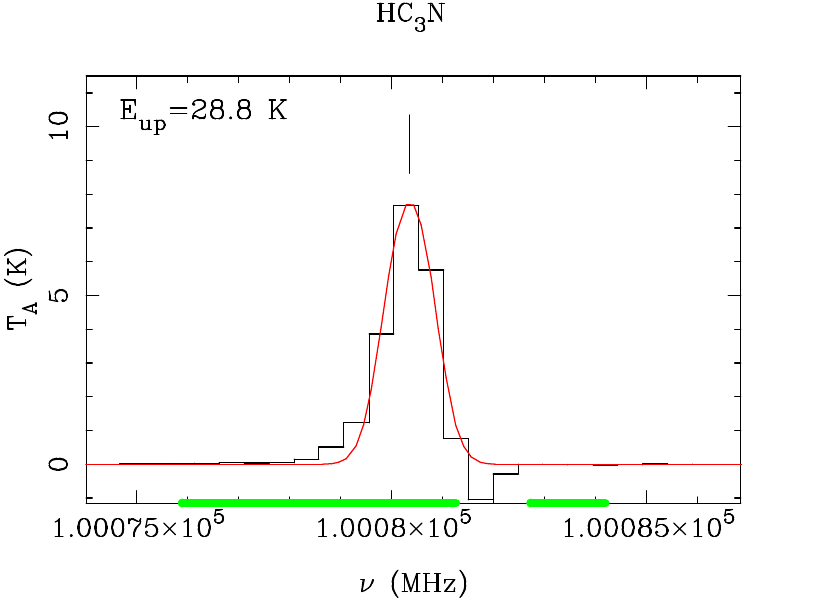}}
\caption{\textbf{\hspace{-0.2 em}(cont.)}\ Same as Figure \ref{fig-SpecS8Em}. Panels (s) to (v) show lines of HC$^{13}$CCN, \propyne, HNCO, and \hctn, respectively.}
\end{figure}

\begin{figure}
\subfloat[][]{\includegraphics[angle=-90,ext=.pdf,width= 0.25 \textwidth]{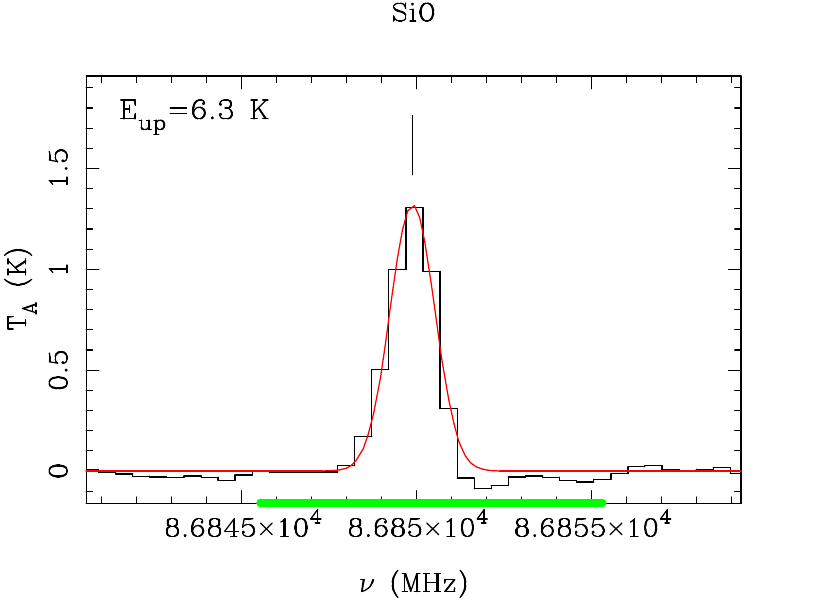}}
\subfloat[][]{\includegraphics[angle=-90,ext=.pdf,width= 0.50 \textwidth]{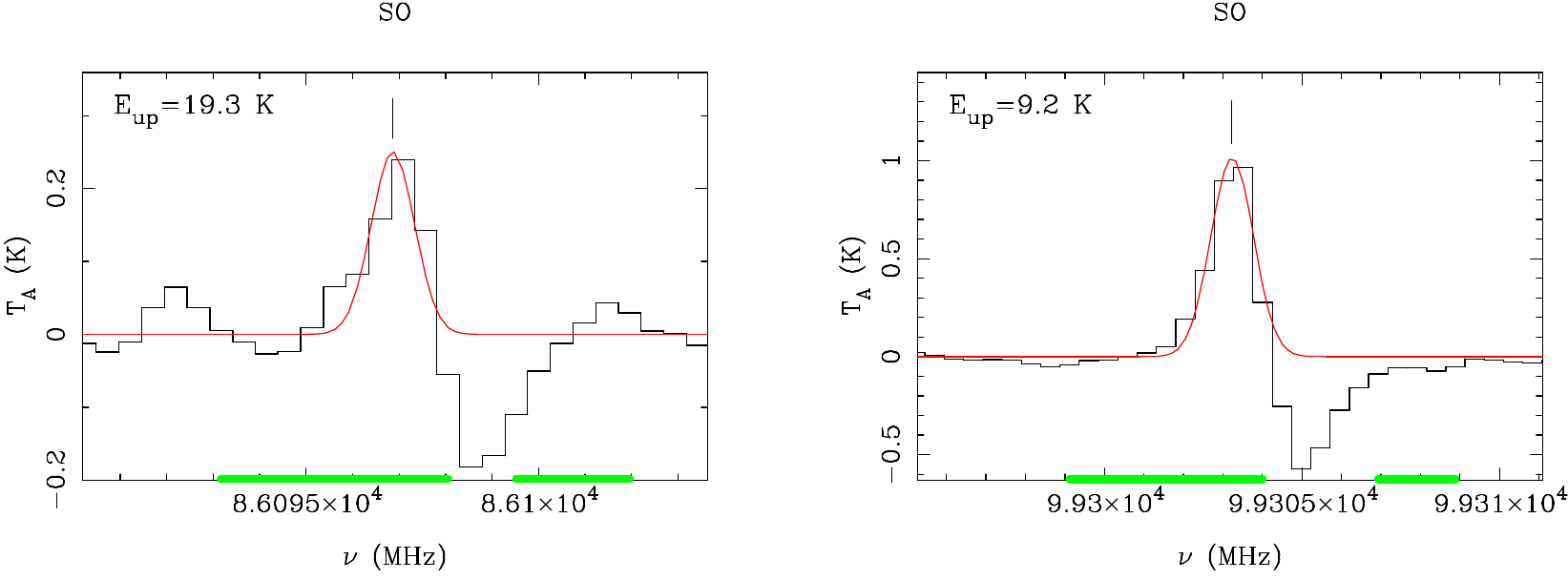}}
\subfloat[][]{\includegraphics[angle=-90,ext=.pdf,width= 0.25 \textwidth]{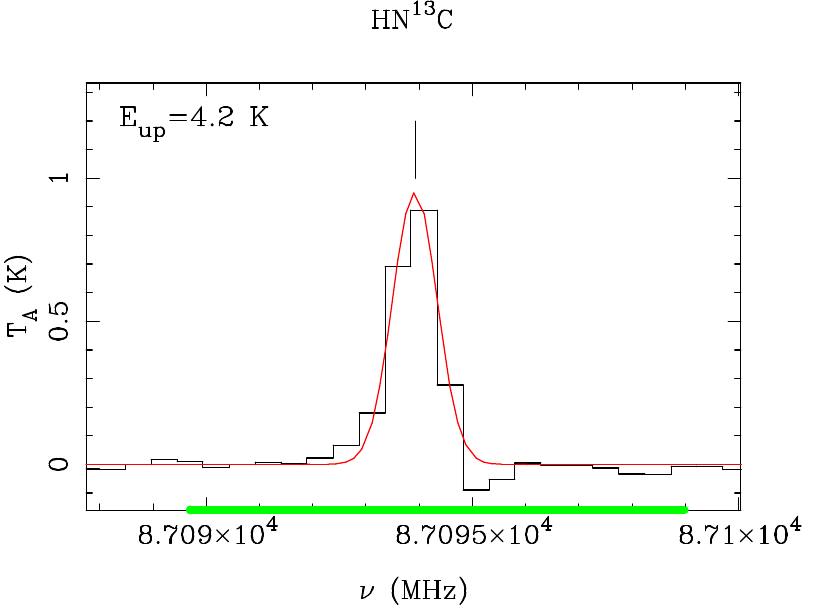}}
\caption{Emission lines from several molecules toward the point S8E 1\farcs5 radius. 
  Line types and colors as described in Figure
    \ref{fig-SpecCCcore1} with $\sigma=0.02$ K. Panels (a) to (c) show lines of  SiO, SO, and \hntc, respectively.\label{fig-SpecS8E15}}
\end{figure}
\begin{figure}
\ContinuedFloat
\subfloat[][]{\includegraphics[angle=-90,ext=.pdf,width= 1.00 \textwidth]{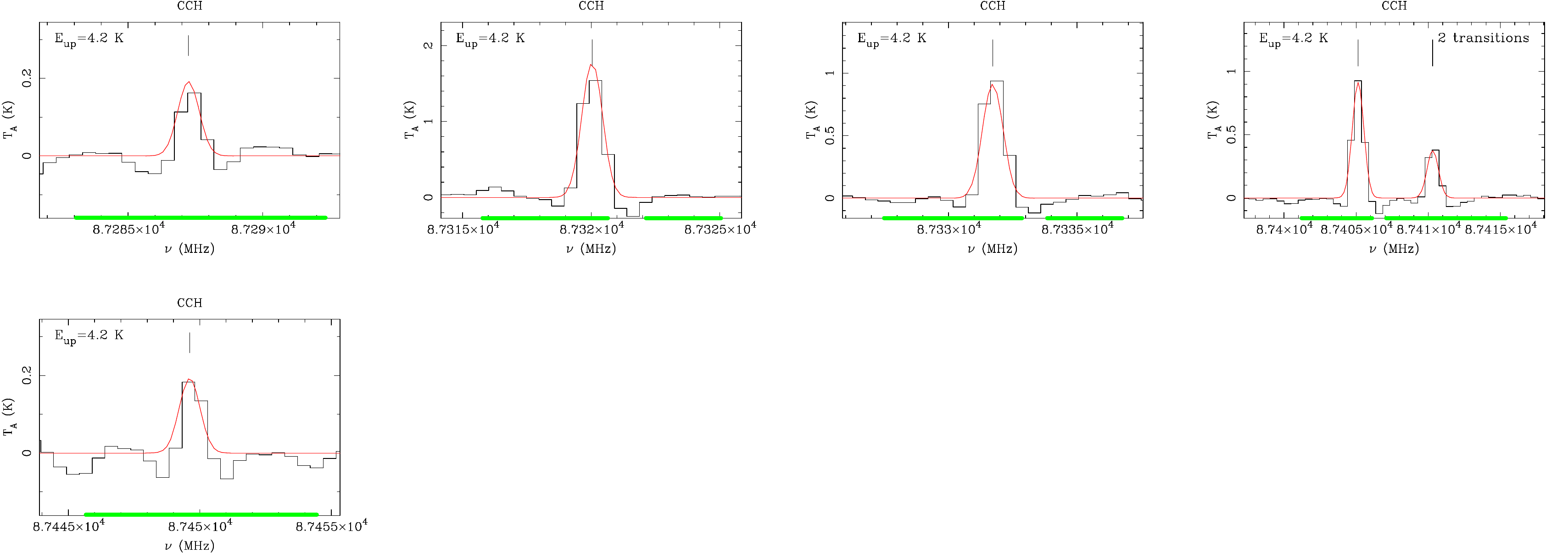}}\\
\subfloat[][]{\includegraphics[angle=-90,ext=.pdf,width= 1.00 \textwidth]{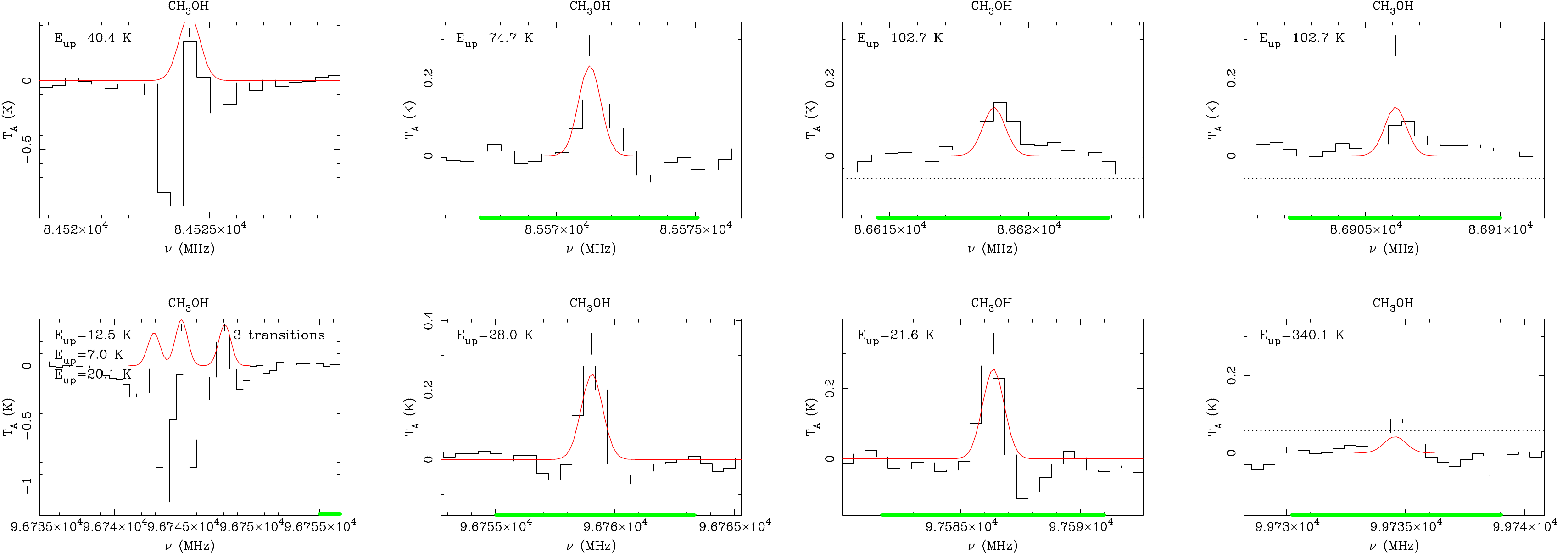}}
\caption{\textbf{\hspace{-0.2 em}(cont.)}\ Same as Figure \ref{fig-SpecS8E15}. Panels (d) and (e) show the CCH and 
\met\ lines, respectively. \added{Note that the model does not reproduce well the strong absorption features associated with the lowest energy  \met\ transitions and the  \met, \maser\ line. The latter is usually affected by strong non-LTE effects.}}
\end{figure}

\begin{figure}
\ContinuedFloat
\subfloat[][]{\includegraphics[angle=-90,ext=.pdf,width= 1.00 \textwidth]{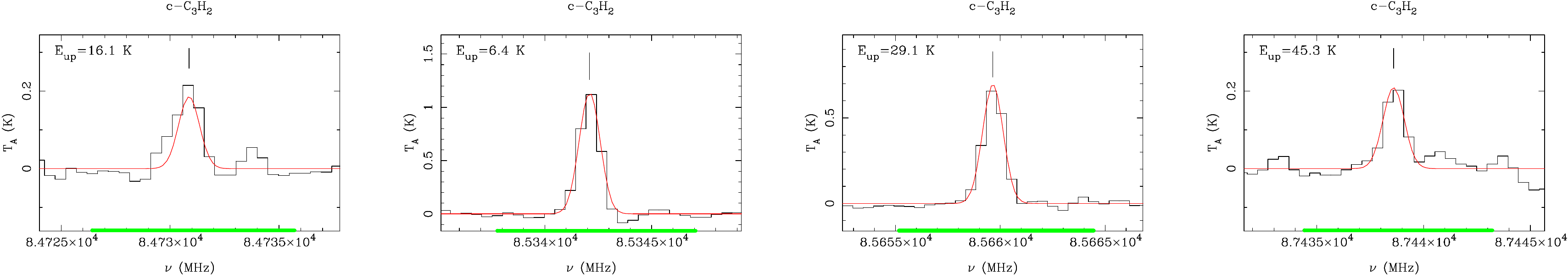}}\\
\subfloat[][]{\includegraphics[angle=-90,ext=.pdf,width= 0.25 \textwidth]{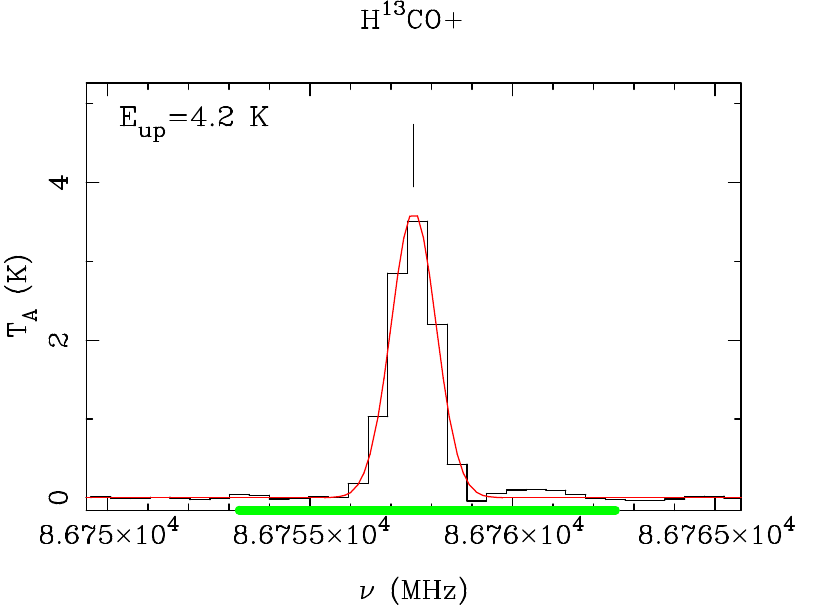}}
\subfloat[][]{\includegraphics[angle=-90,ext=.pdf,width= 0.50 \textwidth]{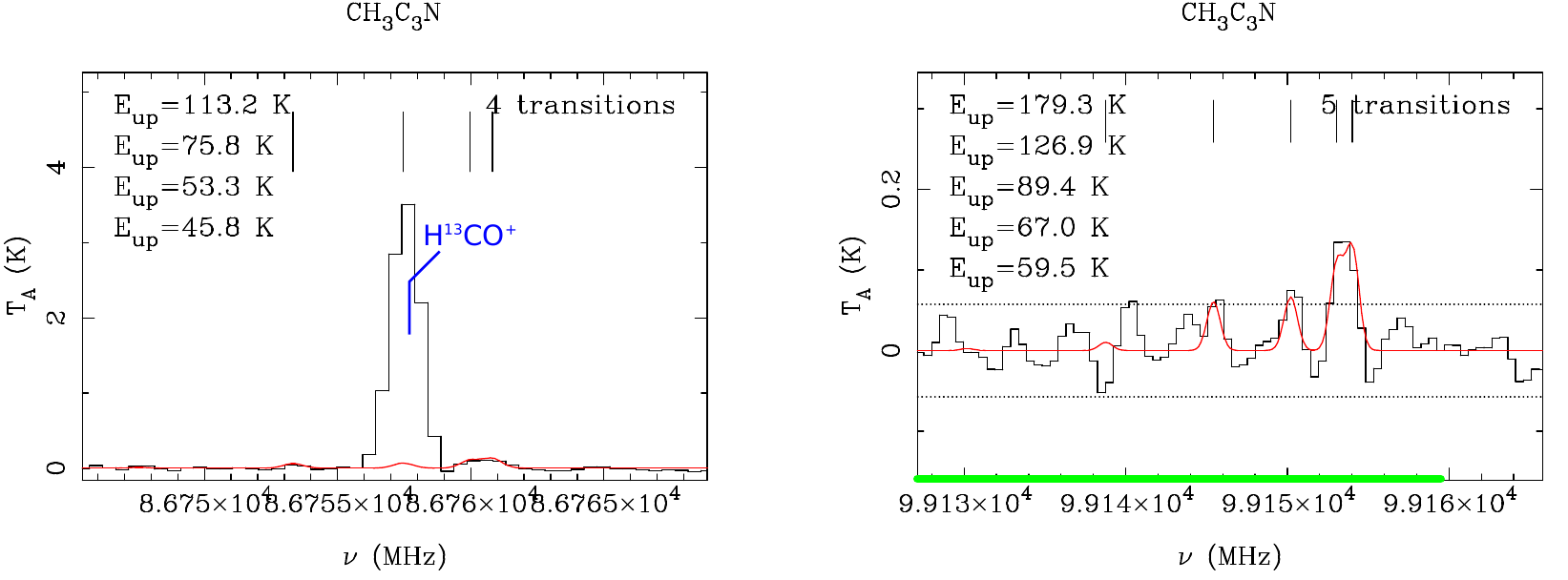}}
\subfloat[][]{\includegraphics[angle=-90,ext=.pdf,width= 0.25 \textwidth]{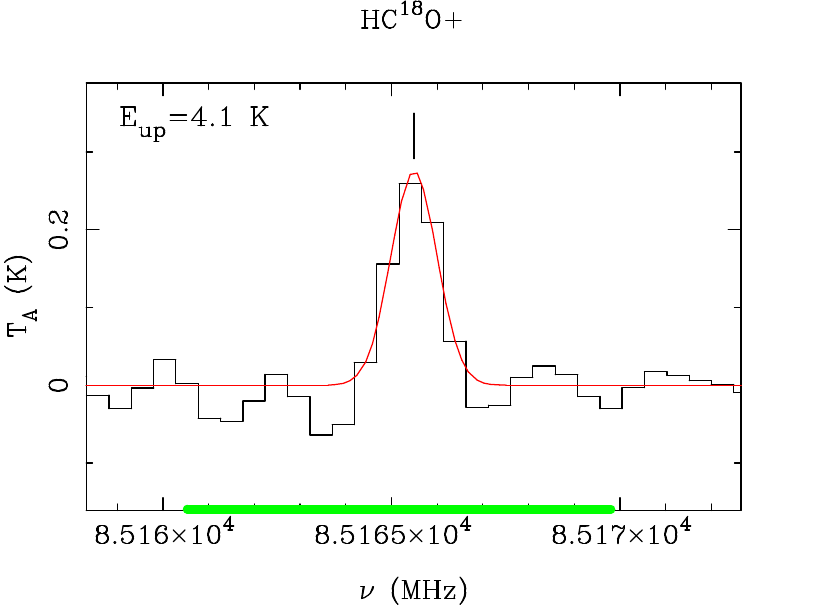}}
\caption{\textbf{\hspace{-0.2 em}(cont.)}\ Same as Figure \ref{fig-SpecS8E15}. Panels (f) and (i) show the lines of \cyc, \htcop, \cyan, and \hcdop, respectively. \added{A strong  feature due to  \htcop\ within one of the displayed frequency windows of \cyan\ is  marked in blue.}}
\end{figure}

\begin{figure}
\ContinuedFloat
\subfloat[][]{\includegraphics[angle=-90,ext=.pdf,width= 0.50 \textwidth]{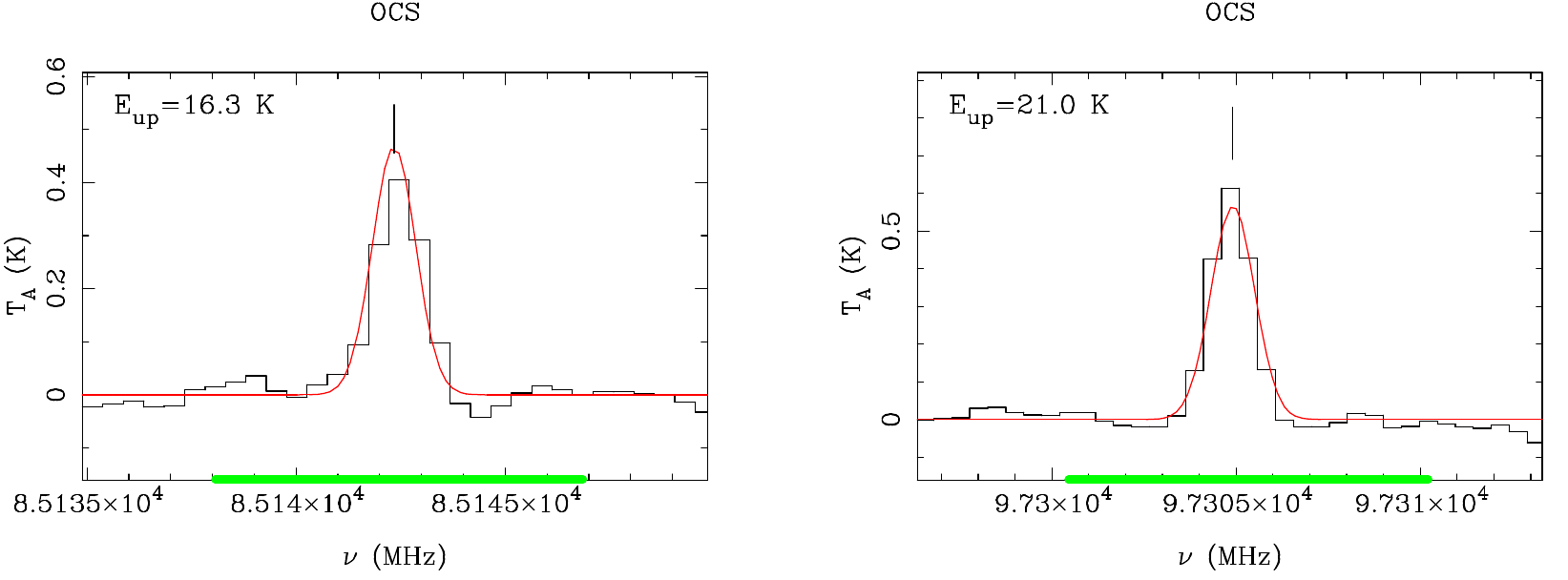}}
\subfloat[][]{\includegraphics[angle=-90,ext=.pdf,width= 0.25 \textwidth]{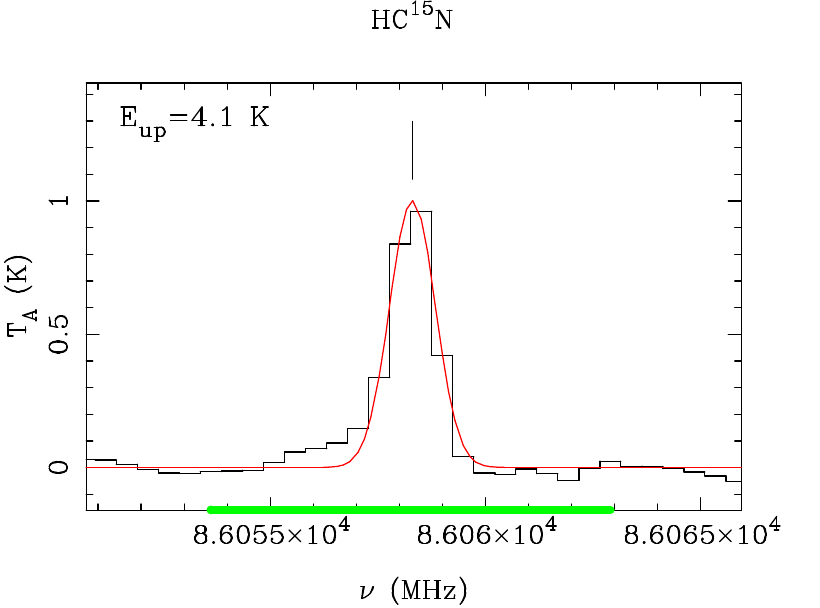}}
\subfloat[][]{\includegraphics[angle=-90,ext=.pdf,width= 0.25 \textwidth]{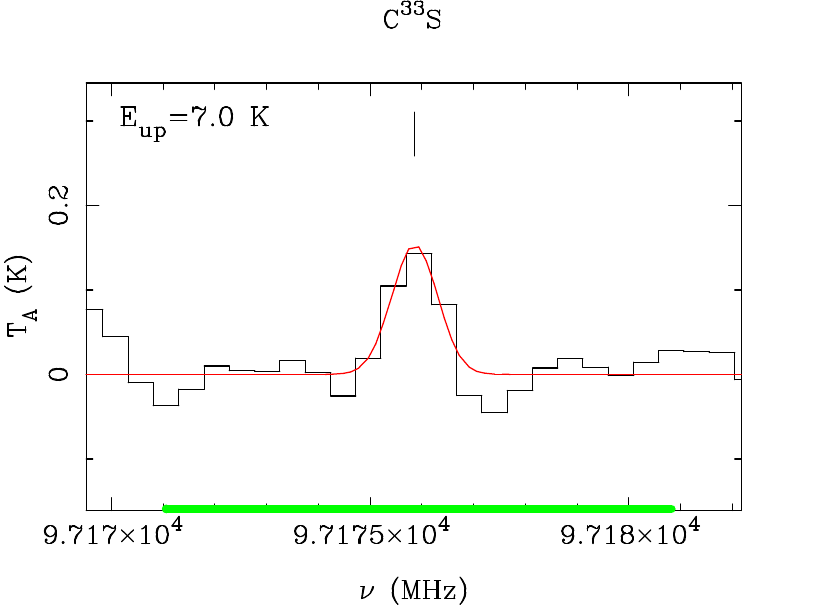}}\\
\subfloat[][]{\includegraphics[angle=-90,ext=.pdf,width= 0.75 \textwidth]{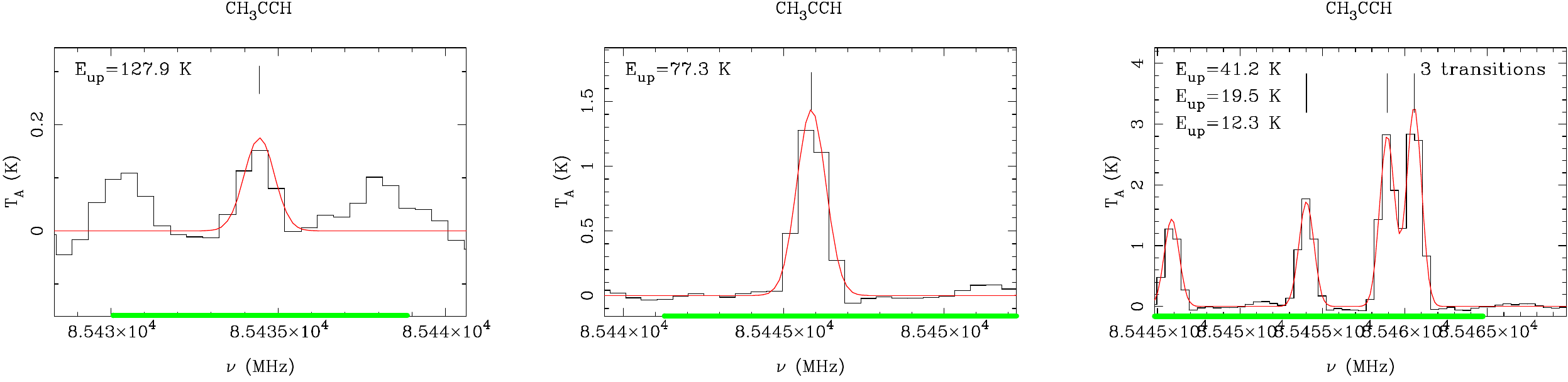}}
\subfloat[][]{\includegraphics[angle=-90,ext=.pdf,width= 0.25 \textwidth]{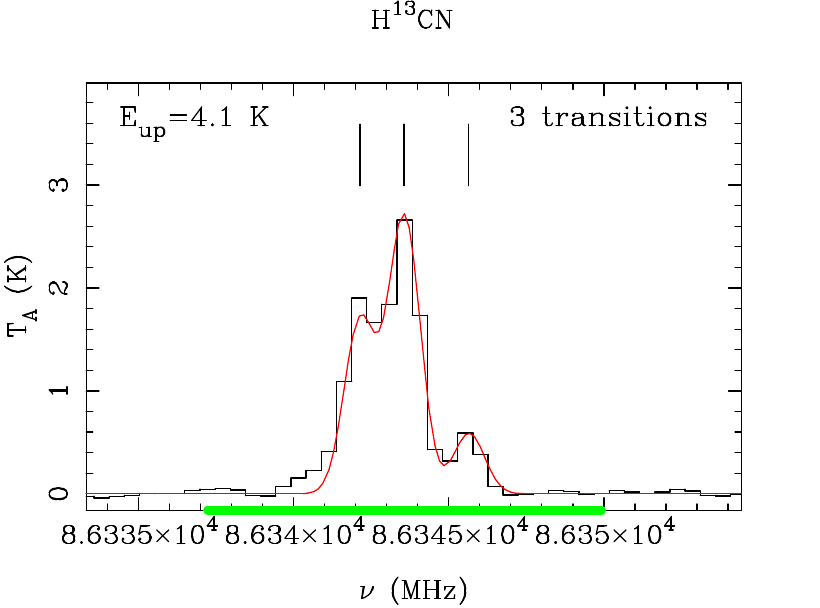}}
\caption{\textbf{\hspace{-0.2 em}(cont.)}\ Same as Figure
  \ref{fig-SpecS8E15}. Panels (j) to (n) show the OCS,  \hcqn, C$^{33}$S,  \propyne, and \htcn\  lines,
  respectively.}
\end{figure}
\begin{figure}
\ContinuedFloat
\subfloat[][]{\includegraphics[angle=-0,ext=.pdf,width= 1.00 \textwidth]{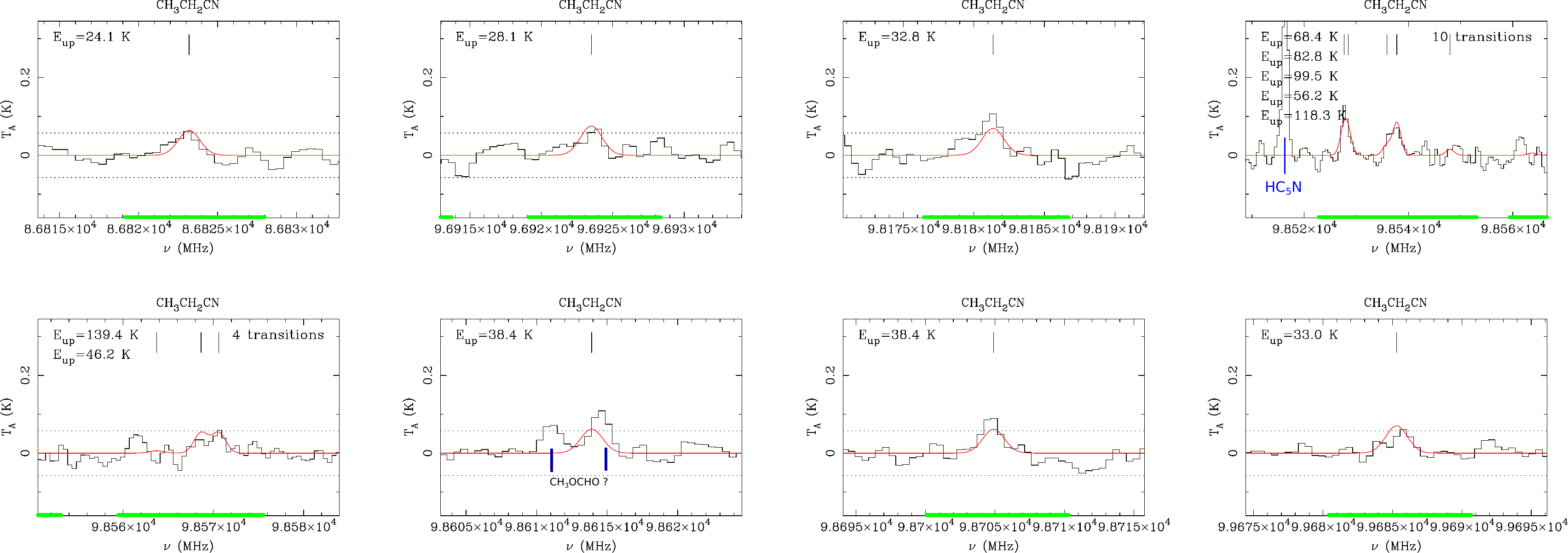}}\\
\subfloat[][]{\includegraphics[angle=-0,ext=.pdf,width= 1.00 \textwidth]{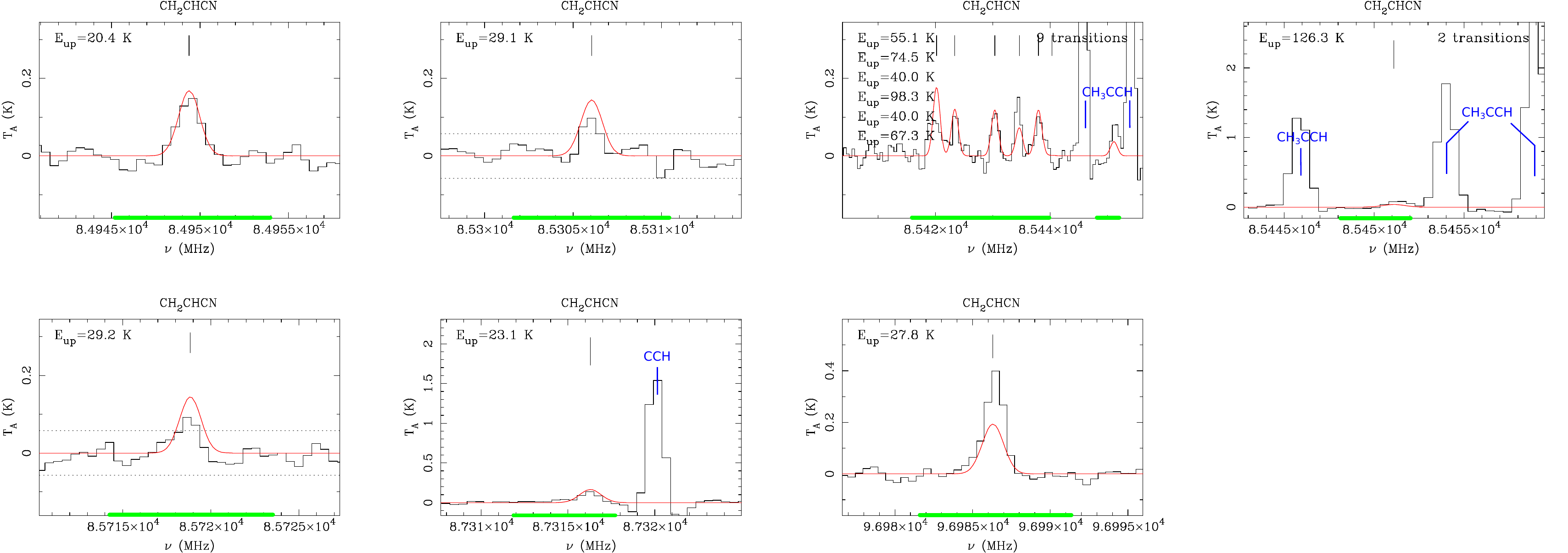}}\\
\caption{\textbf{\hspace{-0.2 em}(cont.)}\ Same as Figure
  \ref{fig-SpecS8E15}. Panels (o) and (p) show the  \prop\ and \vinyl\ lines, 
  respectively. Panel (o) shows also the location of two possible lines of CH$_3$OCHO. \added{Strong lines from other species within the displayed frequency windows are marked in blue.}}
\end{figure}

\begin{figure}
\ContinuedFloat
\subfloat[][]{\includegraphics[angle=-90,ext=.pdf,width= 0.25 \textwidth]{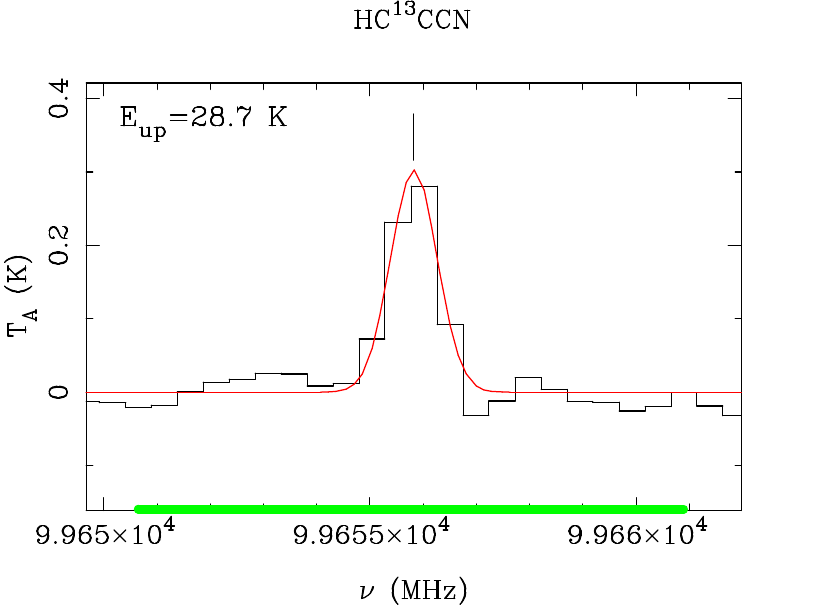}}
\subfloat[][]{\includegraphics[angle=-90,ext=.pdf,width= 0.25 \textwidth]{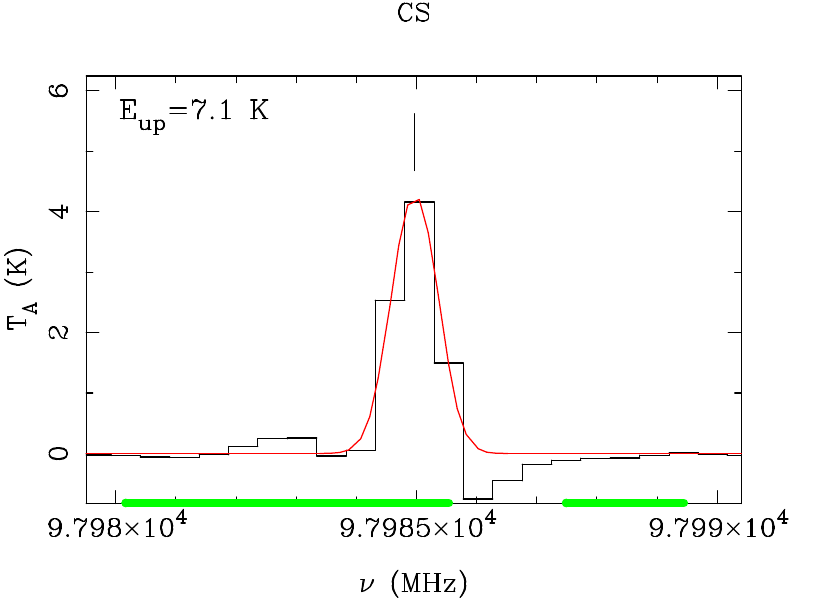}}
\subfloat[][]{\includegraphics[angle=-90,ext=.pdf,width= 0.25 \textwidth]{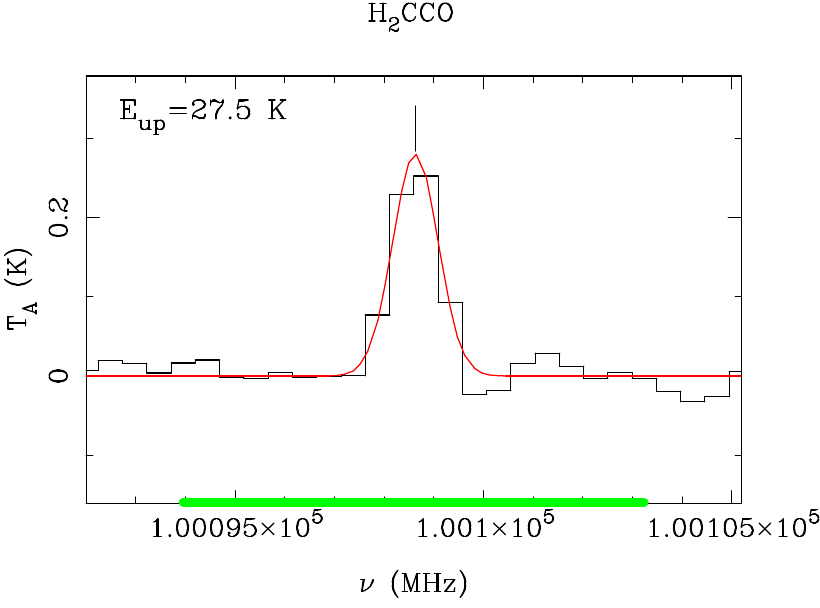}}
\subfloat[][]{\includegraphics[angle=-90,ext=.pdf,width= 0.25 \textwidth]{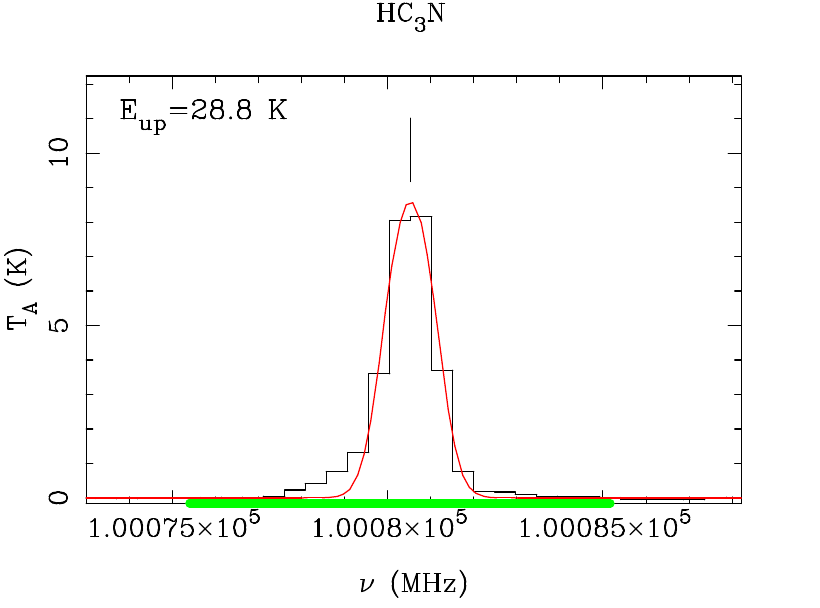}}\\
\subfloat[][]{\includegraphics[angle=-90,ext=.pdf,width= 1.00 \textwidth]{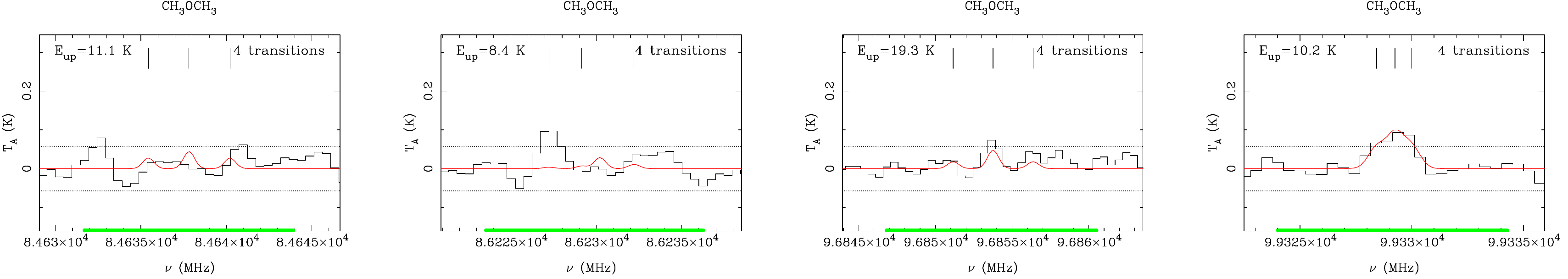}}
\caption{\textbf{\hspace{-0.2 em}(cont.)}\ Same as Figure
  \ref{fig-SpecS8E15}. Panels (q) to (u) show the lines of HC$^{13}$CCN, CS, \ethe, \hctn, and CH$_3$OCH$_3$, 
  respectively.}
\end{figure}

\begin{figure}
\subfloat[][]{\includegraphics[angle=-90,ext=.pdf,width= 0.25 \textwidth]{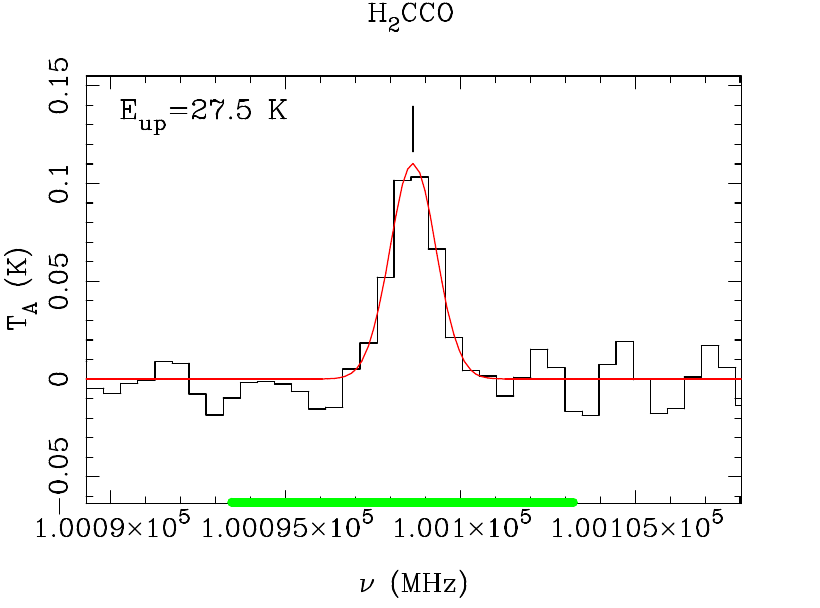}}
\subfloat[][]{\includegraphics[angle=-90,ext=.pdf,width= 0.25 \textwidth]{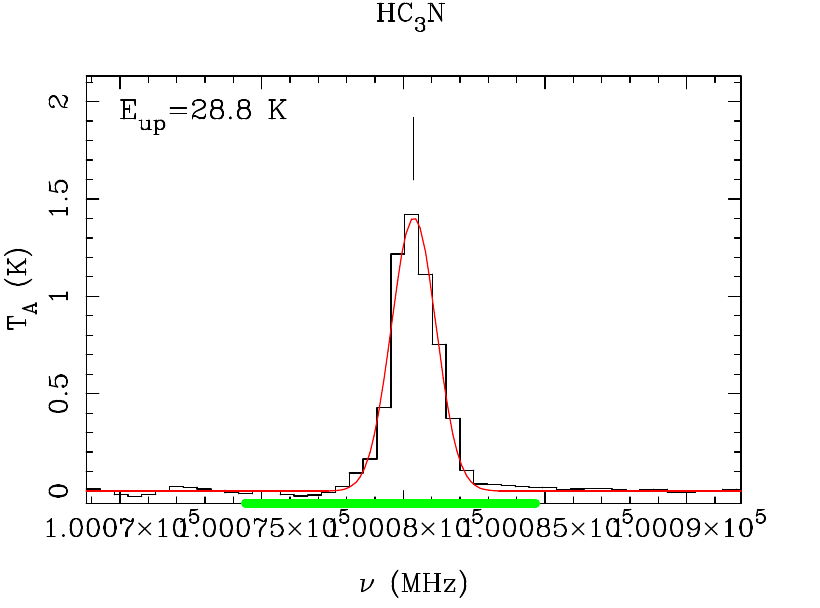}}
\subfloat[][]{\includegraphics[angle=-90,ext=.pdf,width= 0.25 \textwidth]{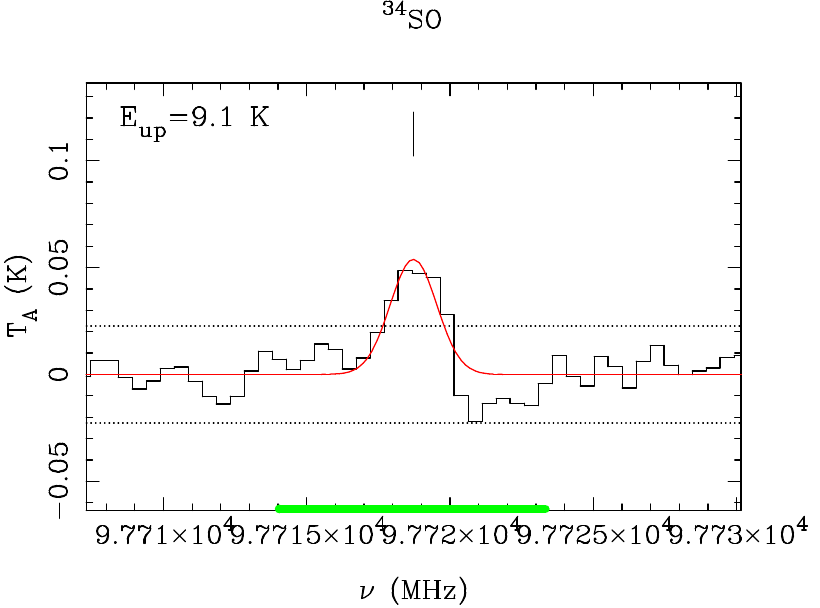}}
\subfloat[][]{\includegraphics[angle=-90,ext=.pdf,width= 0.25 \textwidth]{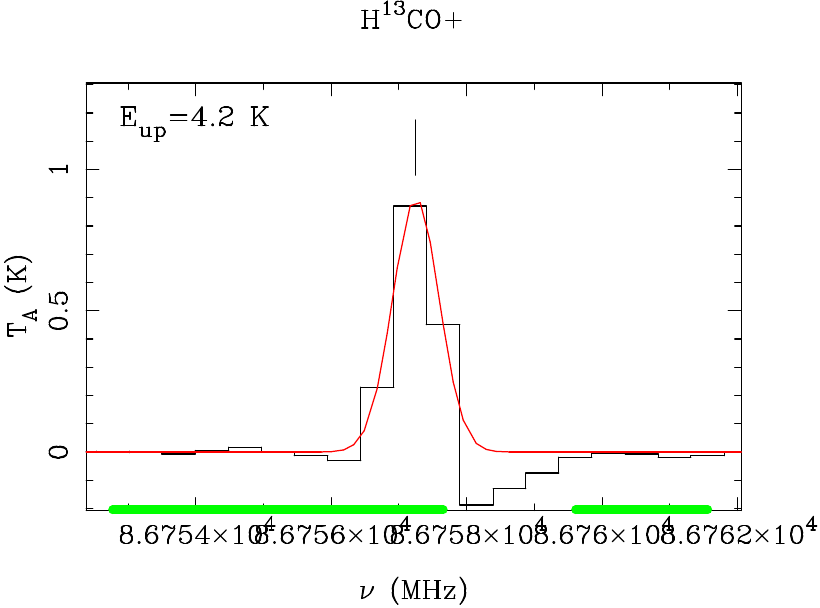}}\\
\subfloat[][]{\includegraphics[angle=-90,ext=.pdf,width= 1.00 \textwidth]{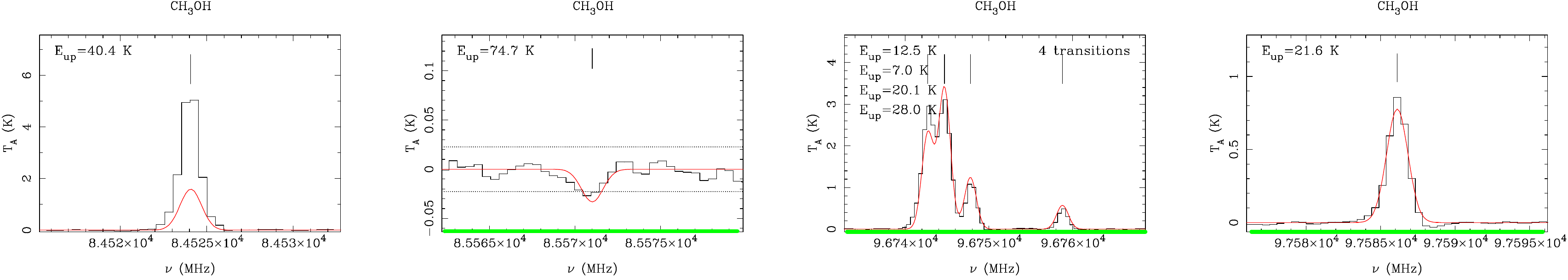}}\\
\caption{Emission lines from several molecules toward the integrated NW cloud. 
  Line types and colors as described in Figure
    \ref{fig-SpecCCcore1} with $\sigma=0.009$ K. Panels (a)  to (e) show lines of \ethe, \hctn, $^{34}$SO, \htcop, and \met, respectively. \added{The strong non-LTE \met, \maser\ emission shown in panel (e) cannot be reproduced with the models used in this work.}\label{fig-SpecNWC}}
\end{figure}

\begin{figure}
\ContinuedFloat
\subfloat[][]{\includegraphics[angle=-90,ext=.pdf,width= 0.75 \textwidth]{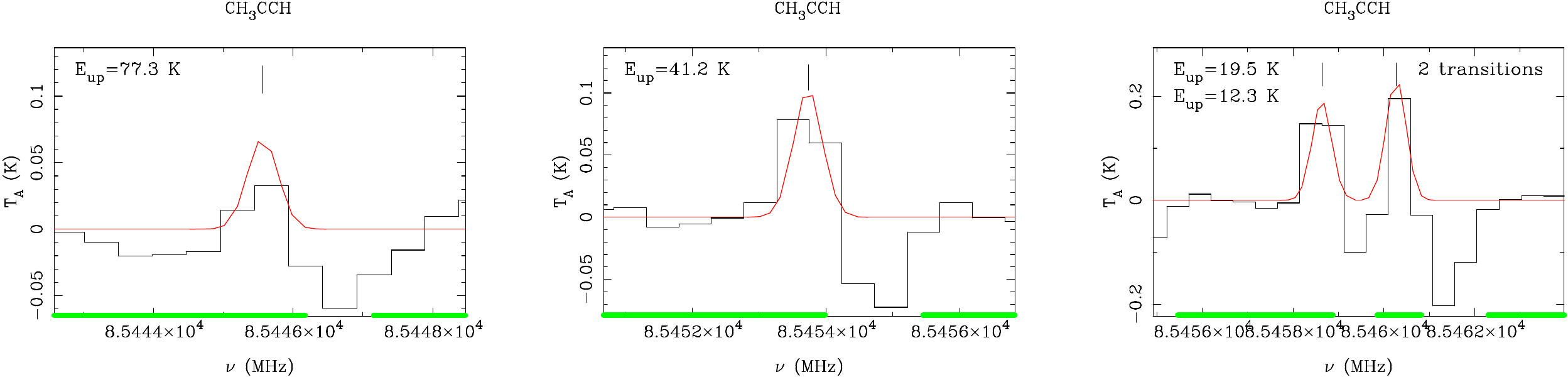}}
\subfloat[][]{\includegraphics[angle=-90,ext=.pdf,width= 0.25 \textwidth]{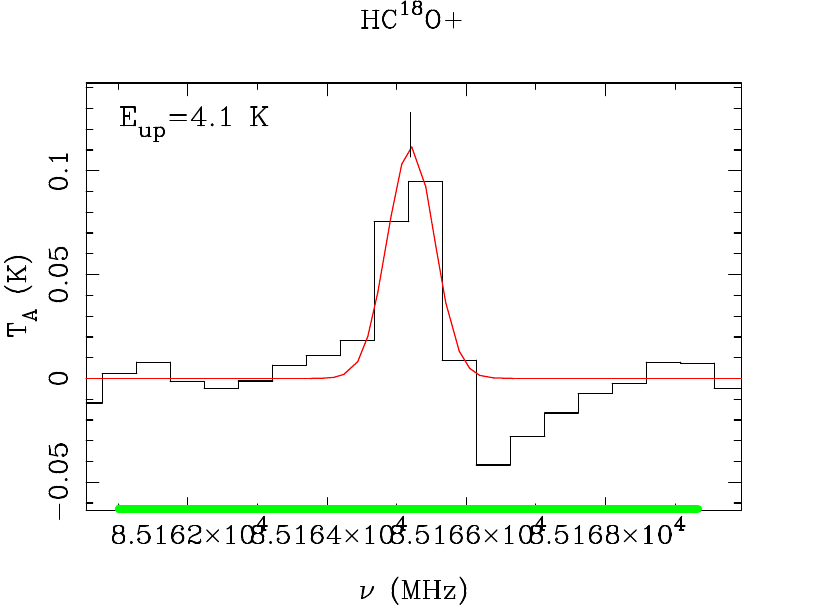}}\\
\subfloat[][]{\includegraphics[angle=-90,ext=.pdf,width= 0.75 \textwidth]{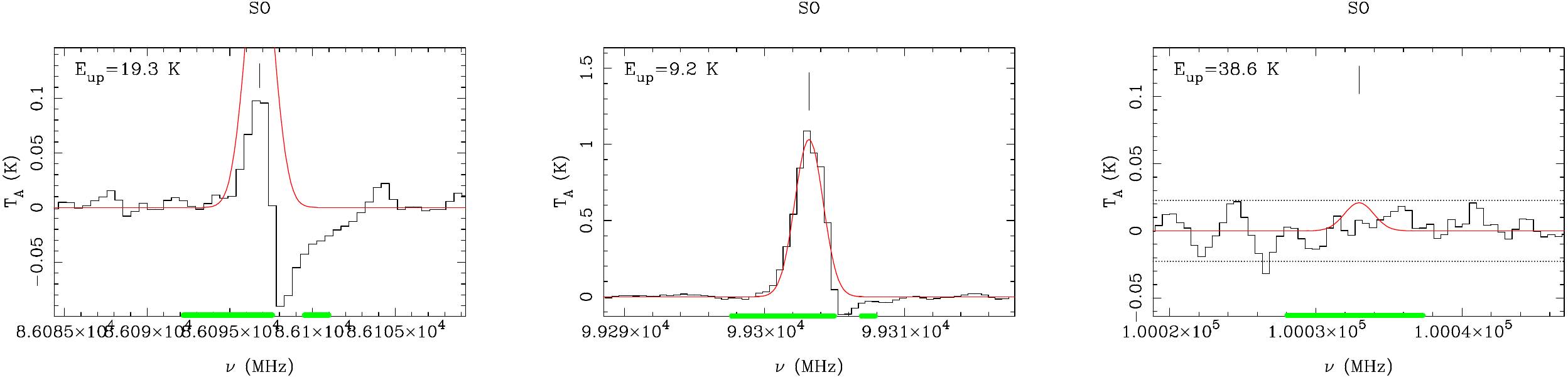}}
\subfloat[][]{\includegraphics[angle=-90,ext=.pdf,width= 0.25 \textwidth]{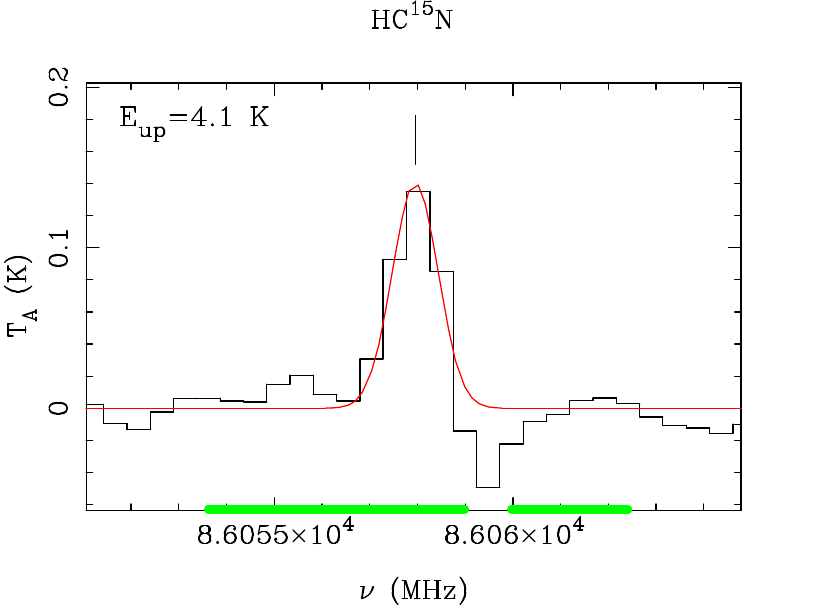}}\\
\subfloat[][]{\includegraphics[angle=-90,ext=.pdf,width= 0.50 \textwidth]{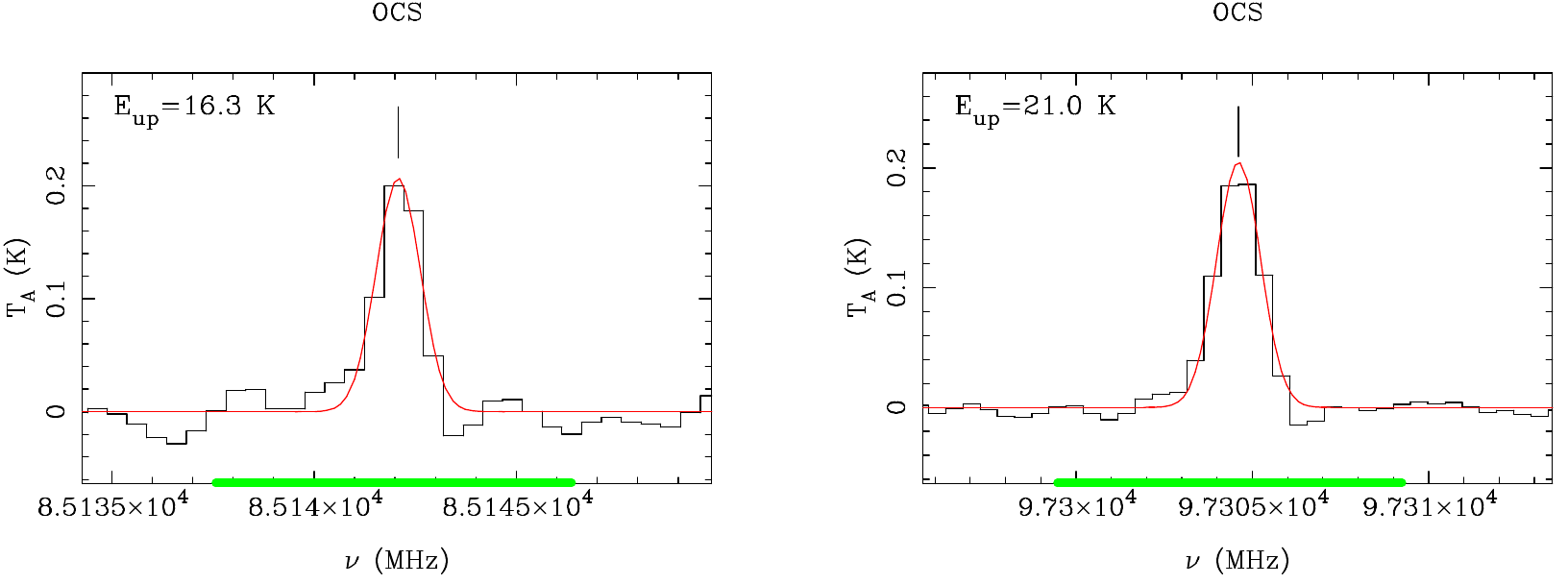}}
\subfloat[][]{\includegraphics[angle=-90,ext=.pdf,width= 0.25 \textwidth]{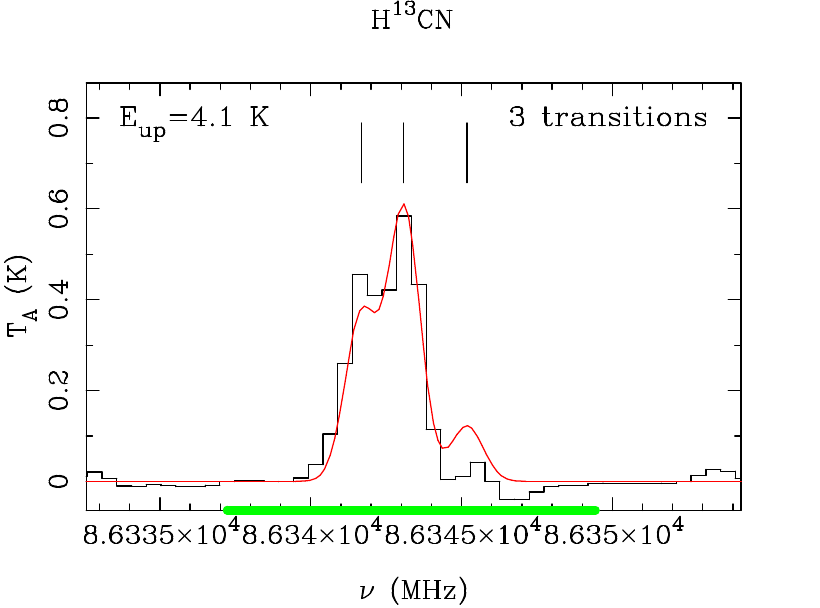}}
\subfloat[][]{\includegraphics[angle=-90,ext=.pdf,width= 0.25 \textwidth]{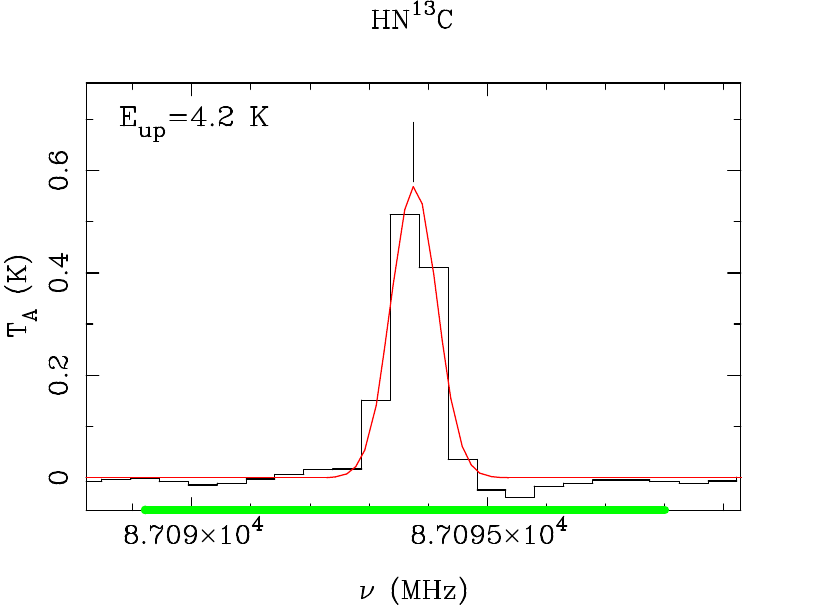}}\\
\subfloat[][]{\includegraphics[angle=-90,ext=.pdf,width= 1.00 \textwidth]{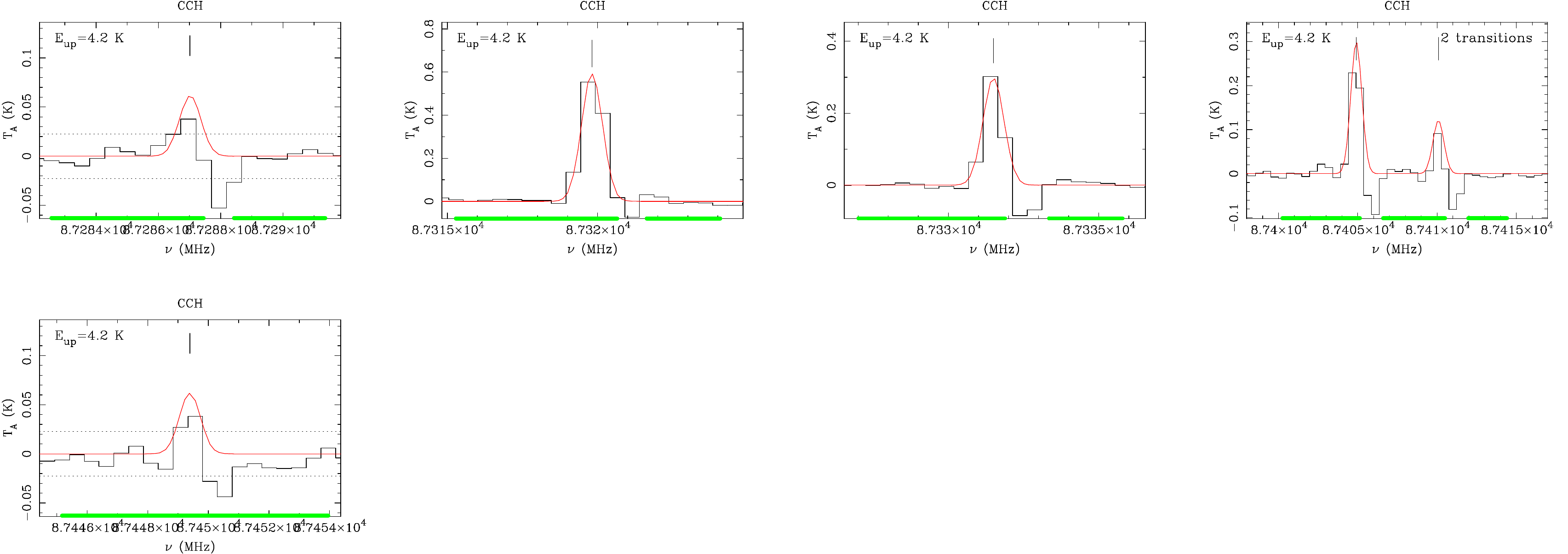}}
\caption{\textbf{\hspace{-0.2 em}(cont.)}\ Same as Figure \ref{fig-SpecNWC}. Panels (f) to (m) show lines of \propyne, \hcdop, SO, \hcqn, OCS, \htcn, \hntc, and CCH, respectively.}
\end{figure}
\clearpage
\begin{figure}
\ContinuedFloat
\subfloat[][]{\includegraphics[angle=-90,ext=.pdf,width= 1.00 \textwidth]{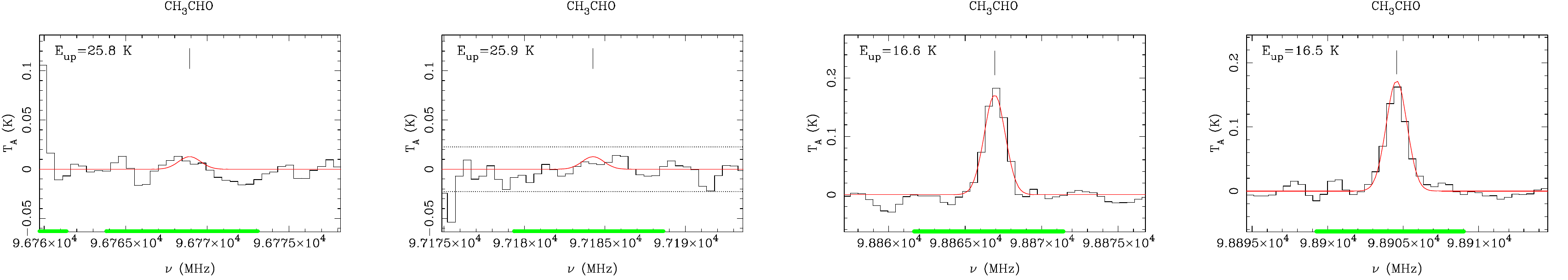}}\\
\subfloat[][]{\includegraphics[angle=-90,ext=.pdf,width= 0.50 \textwidth]{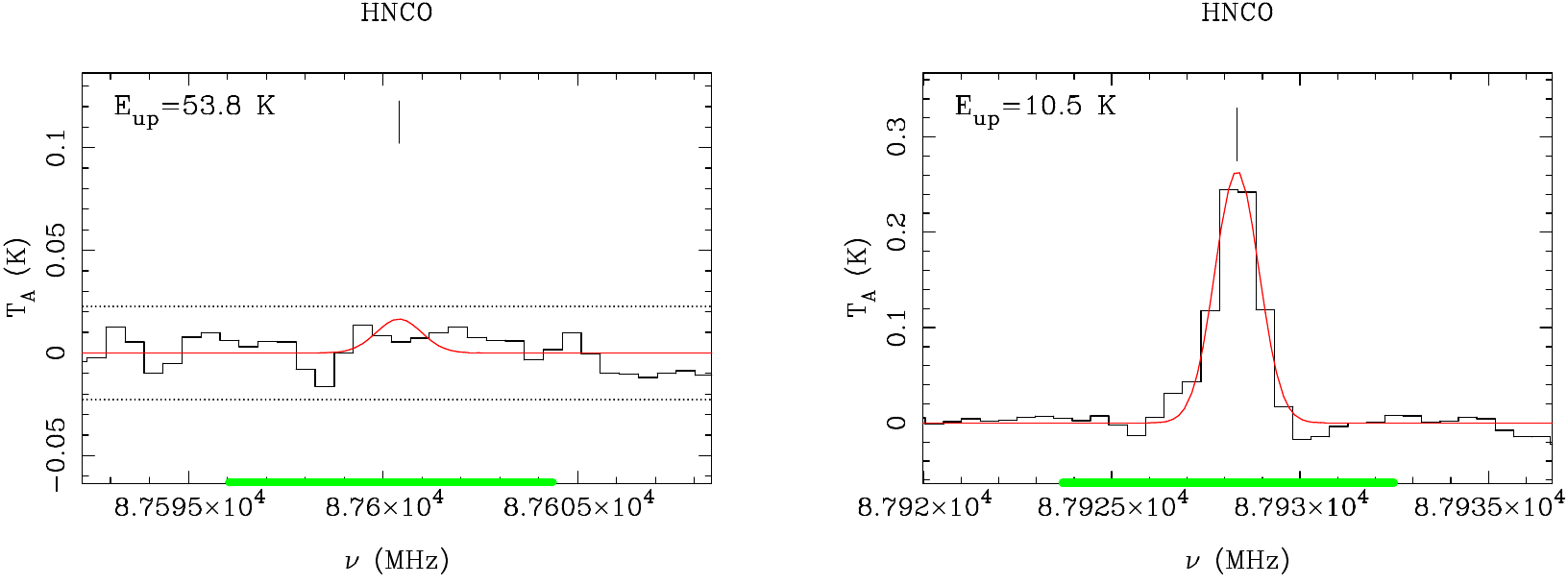}}
\subfloat[][]{\includegraphics[angle=-90,ext=.pdf,width= 0.25 \textwidth]{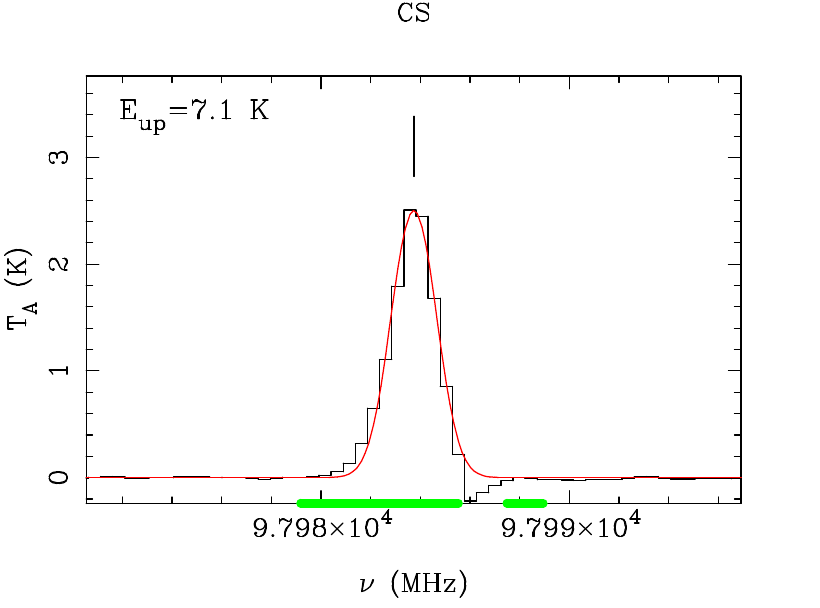}}
\subfloat[][]{\includegraphics[angle=-90,ext=.pdf,width= 0.25 \textwidth]{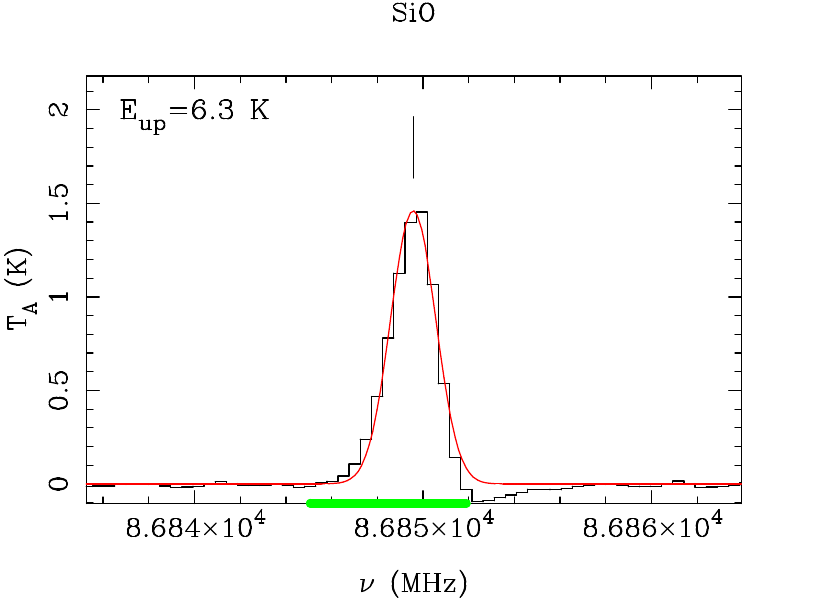}}\\
\subfloat[][]{\includegraphics[angle=-90,ext=.pdf,width= 0.25 \textwidth]{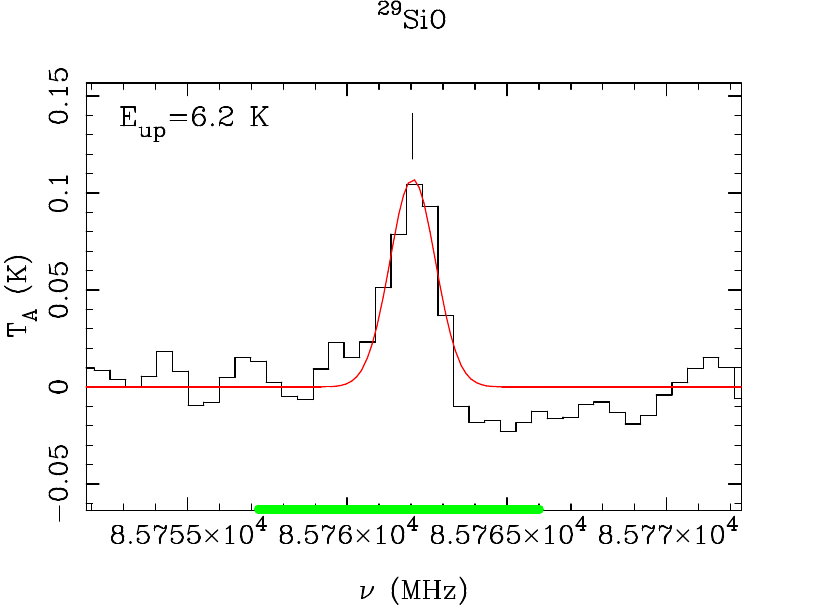}}
\subfloat[][]{\includegraphics[angle=-90,ext=.pdf,width= 0.25 \textwidth]{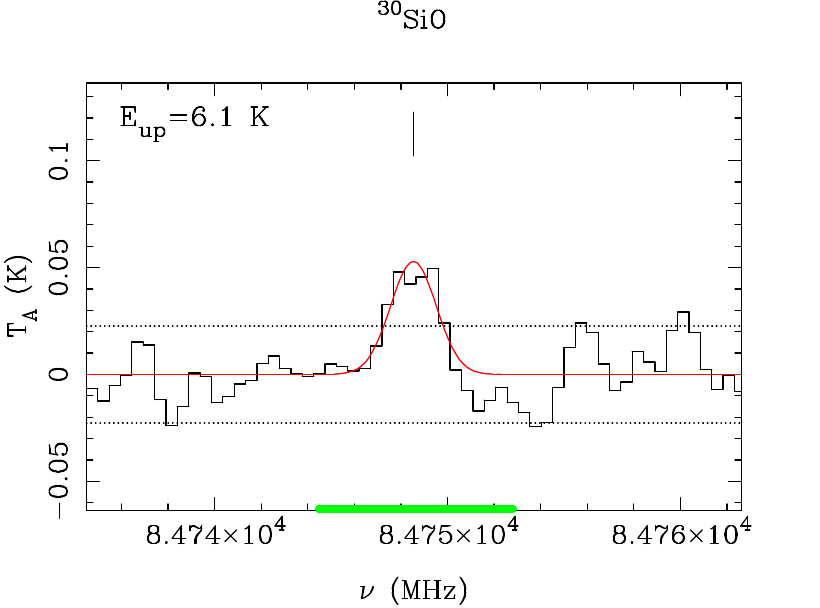}}
\caption{\textbf{\hspace{-0.2 em}(cont.)}\ Same as Figure \ref{fig-SpecNWC}. Panels (n) to (s) show lines of \acet, HNCO, CS, SiO, $^{29}$SiO, and $^{30}$SiO, respectively.}
\end{figure}

\begin{figure}
\subfloat[][]{\includegraphics[angle=-90,ext=.pdf,width= 1.00 \textwidth]{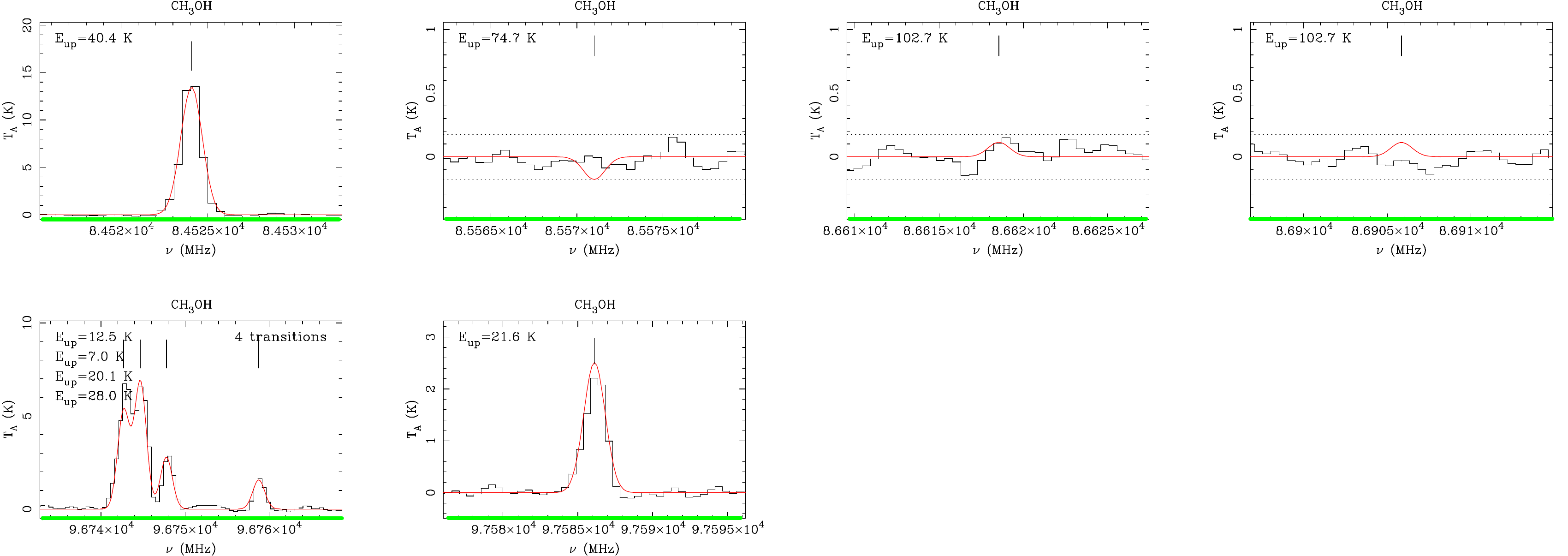}}\\
\caption{\met\ lines from several molecules toward the point NW cloud (a), the methanol peak. 
  Line types and colors as described in Figure
    \ref{fig-SpecCCcore1} with $\sigma=0.08$ K. \label{fig-SpecNWCa}}
\end{figure}
\begin{figure}
\ContinuedFloat
\subfloat[][]{\includegraphics[angle=-90,ext=.pdf,width= 0.50 \textwidth]{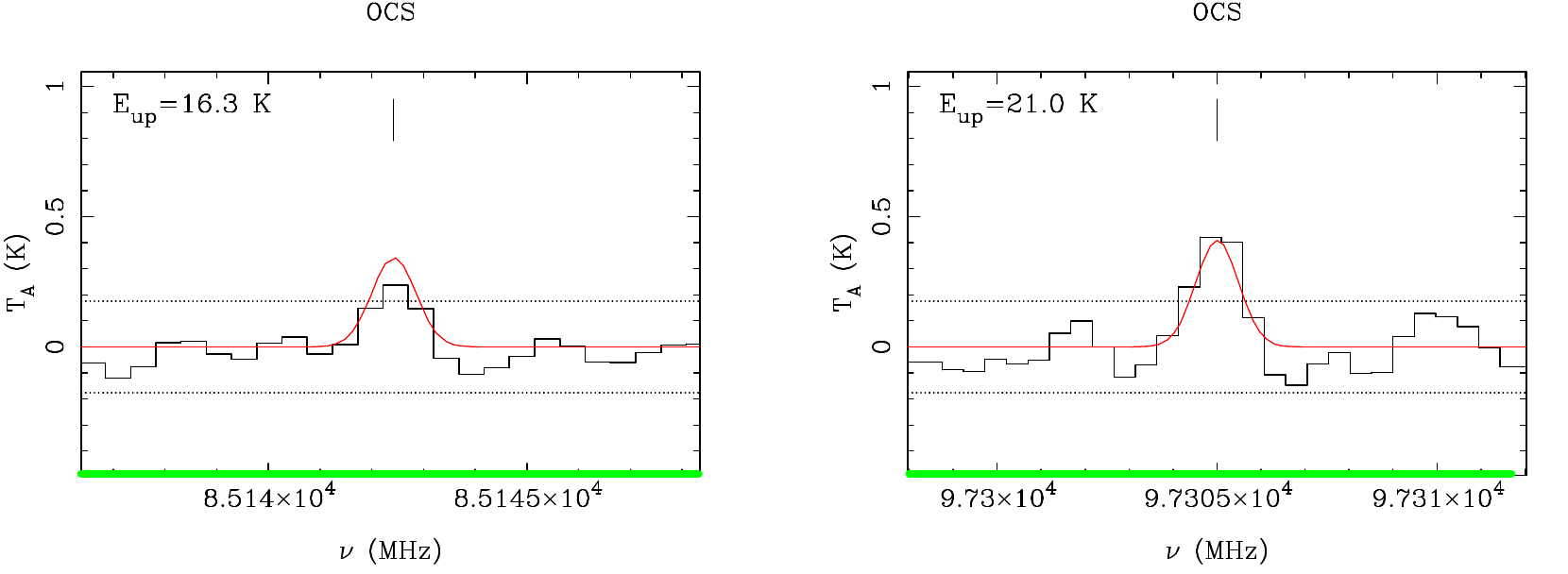}}
\subfloat[][]{\includegraphics[angle=-90,ext=.pdf,width= 0.50 \textwidth]{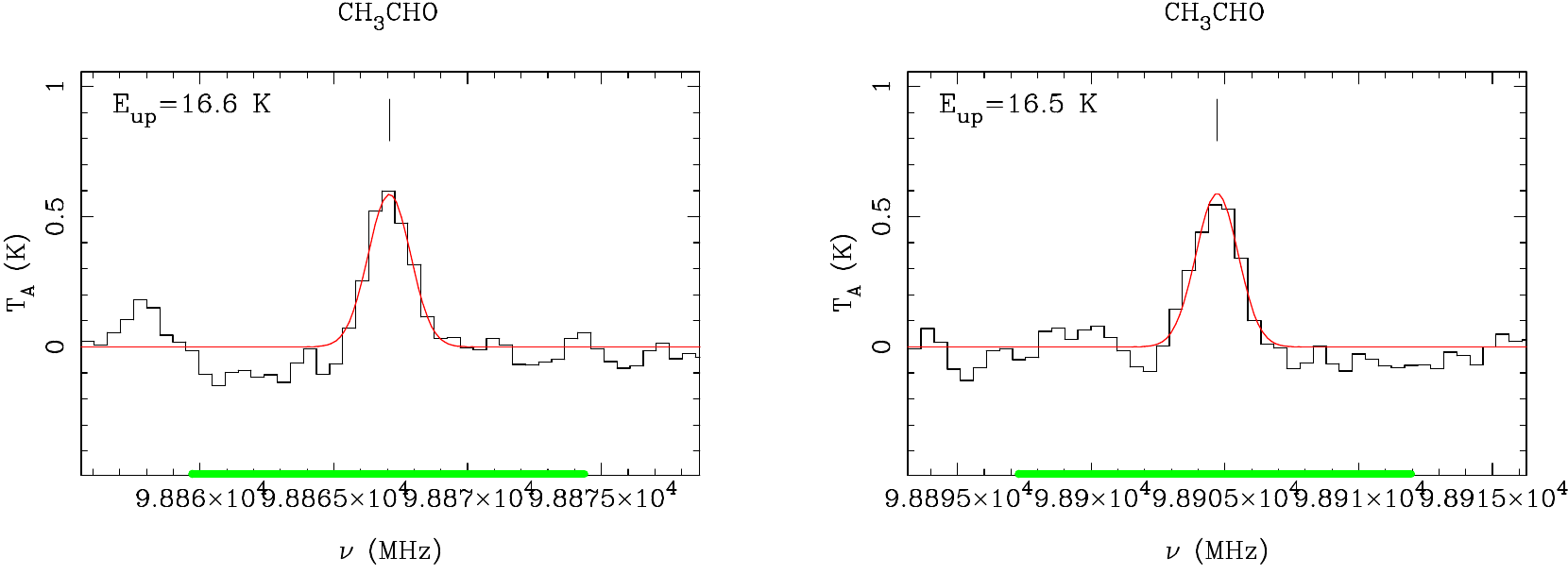}}\\
\subfloat[][]{\includegraphics[angle=-90,ext=.pdf,width= 0.25 \textwidth]{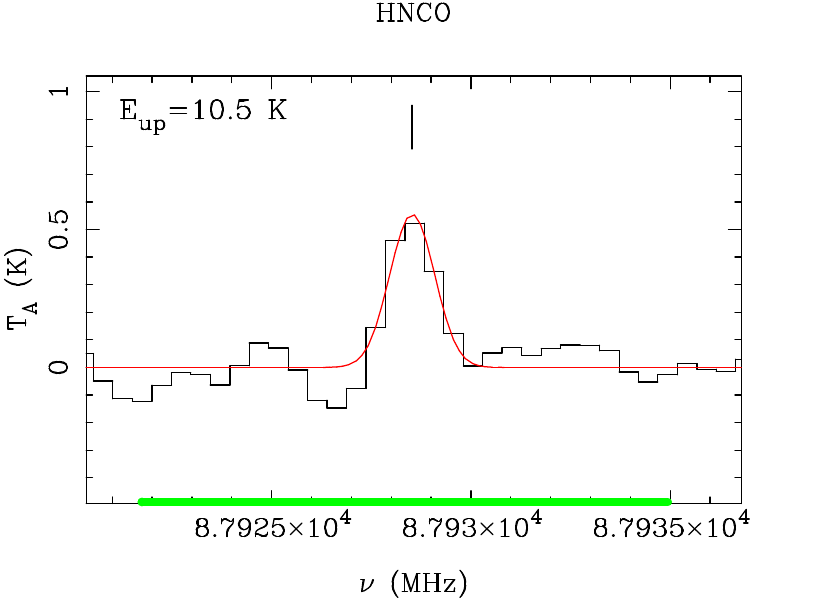}}
\subfloat[][]{\includegraphics[angle=-90,ext=.pdf,width= 0.25 \textwidth]{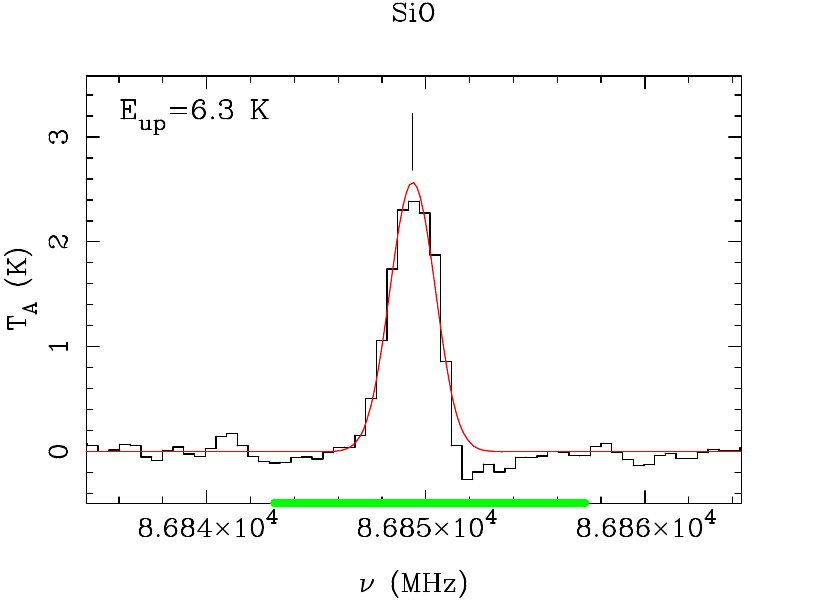}}
\subfloat[][]{\includegraphics[angle=-90,ext=.pdf,width= 0.25 \textwidth]{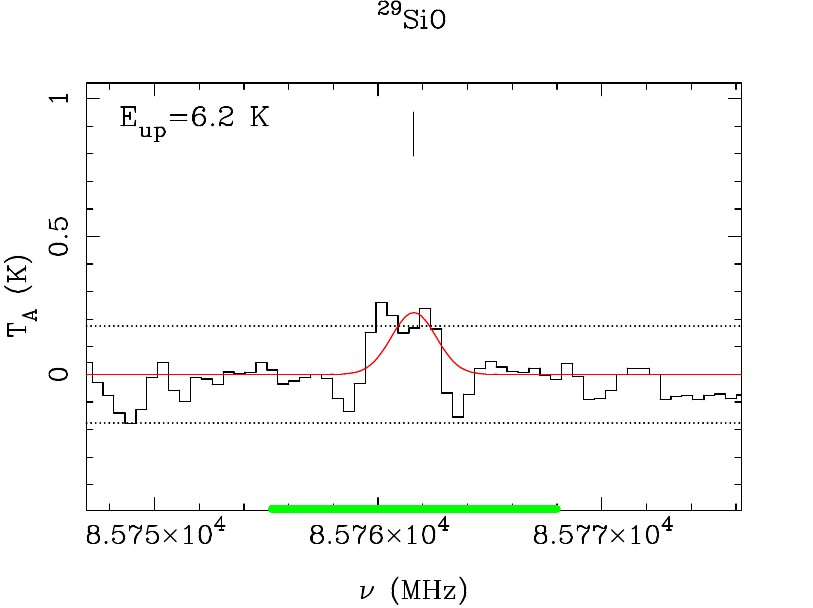}}
\caption{\textbf{\hspace{-0.2 em}(cont.)}\ Same as Figure
  \ref{fig-SpecNWCa}. Panels (b) to (f) show lines of OCS, \acet,
  HNCO, SiO, and $^{29}$SiO, respectively.}
\end{figure}
\begin{figure}
\includegraphics[angle=-90,ext=.pdf,width= 1.00 \textwidth]{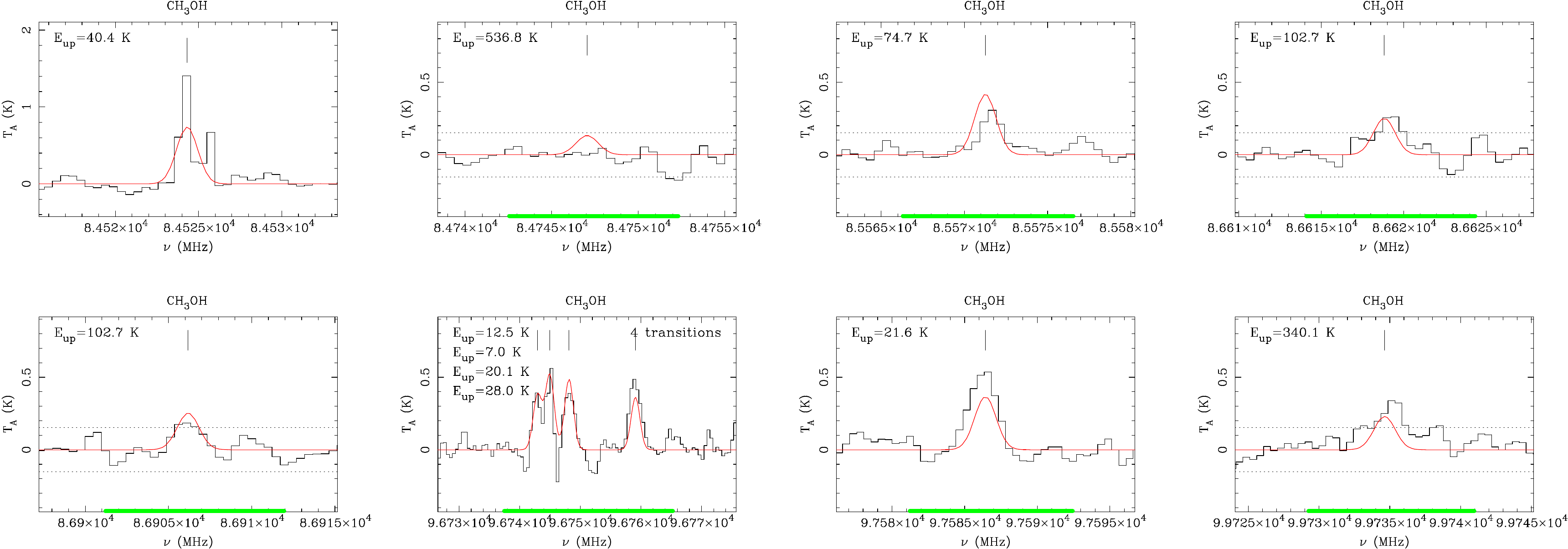}
\caption{Emission lines from \met\ toward Source 3.  Line types and colors
  as described in Figure \ref{fig-SpecCCcore1} with $\sigma=0.05$ K. \label{fig-SpecS3}}
\end{figure}
\begin{figure}
\includegraphics[angle=-90,ext=.pdf,width= 1.00 \textwidth]{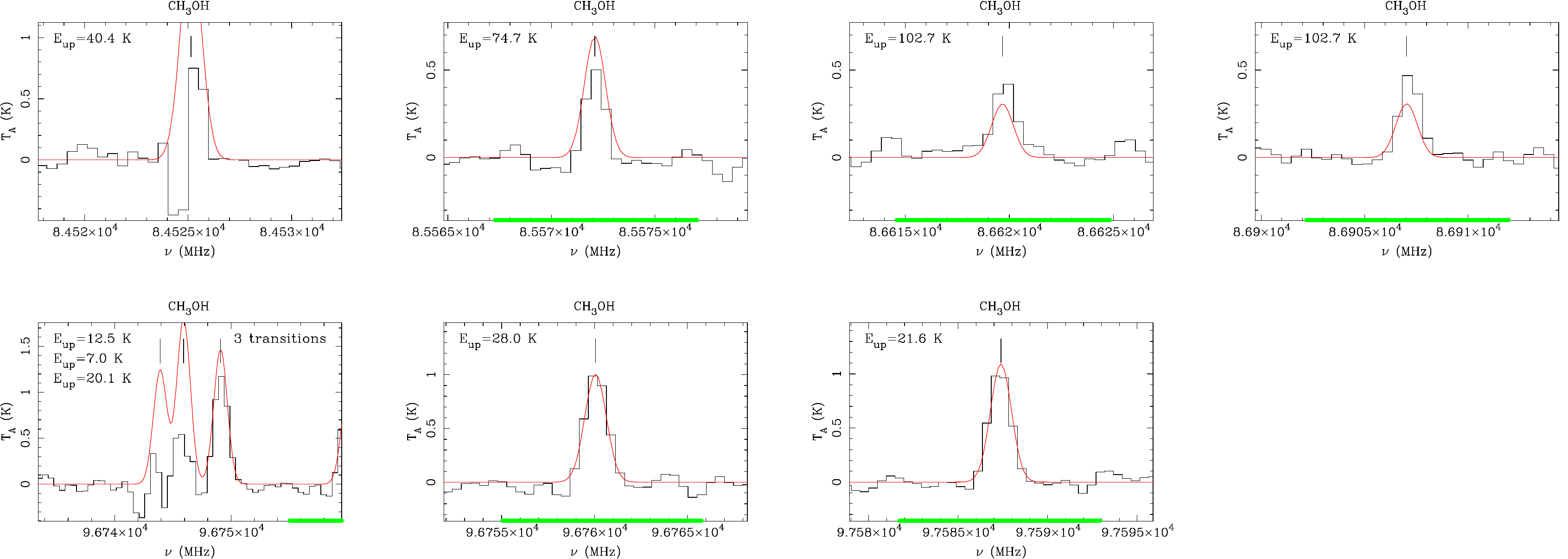}
\caption{Emission lines from \met\ toward Source 18.  Line types and colors
  as described in Figure \ref{fig-SpecCCcore1} with $\sigma=0.05$ K. \added{The model used in this work does not reproduce well the strong absorption features associated with the lowest energy  \met\ transitions and the  \met, \maser\ line.} \label{fig-SpecS18}}
\end{figure}

%%%%
\clearpage
\bibliography{bibliografia}
\listofchanges
\end{document}